\documentclass[%
 reprint,
superscriptaddress,
 amsmath,amssymb,
 aps,
 prx,
]{revtex4-2}

\usepackage{subfigure}
\usepackage{hyperref}
\usepackage{xcolor}
\usepackage{cleveref}
\crefname{appendix}{App.}{Apps.}
\crefname{equation}{Eq.}{Eqs.}
\crefname{figure}{Fig.}{Figs.}
\crefname{table}{Tab.}{Tabs.}
\crefname{section}{Sec.}{Secs.}

\usepackage{verbatim}

\usepackage{graphicx}% Include figure files
\usepackage{dcolumn}% Align table columns on decimal point
\usepackage{bm}% bold math
\begin{document}

% \preprint{APS/123-QED}

\title{Time Direction Entanglement Entropy of Free Fermion Ground States}

\author{Xinyu Sun}
\thanks{These two authors contributed equally.}
\affiliation{Chen-Ning Yang Institute for Advanced Study (YIAS), Tsinghua University, Beijing 100084, China}
\author{Bowei Liu}
\thanks{These two authors contributed equally.}
\affiliation{Department of Physics, Princeton University, Princeton, New Jersey 08544, USA}
\author{Joshua Yi}
\affiliation{Department of Physics, University of California, Santa Barbara, CA 93106}
\author{Biao Lian}
\email{biao@princeton.edu}
\affiliation{Department of Physics, Princeton University, Princeton, New Jersey 08544, USA}

%Institute for Advanced Study, Tsinghua University, Beijing 100084, China
%Department of Physics, Princeton University, Princeton, New Jersey 08544, USA

\date{\today}% It is always \today, today,

\newcommand{\bl}[1]{\textcolor{cyan}{[BL]: #1}}

\begin{abstract}
While the entanglement entropy of a spatial subregion has been extensively studied, a properly defined entanglement entropy of quantum states in a time direction interval remains underexplored. We investigate the scaling behaviors of the recently defined time direction entanglement entropy (TDEE) for free fermion quantum states, which is computed from the reduced density matrix of a sub-Hilbert space that minimally supports all the time-evolved operators in the time direction interval. Based on the Fisher--Hartwig conjecture for Toeplitz matrices, we develop an analytical framework for understanding the TDEE in a time interval with discrete time step $\tau$. For disorderless one-dimensional (1D) systems (gapless fermions or band insulators), in different regimes of gap, bandwidth and $\tau$, we find the TDEE can exhibit area law (for nonzero gap), logarithmic law or volume law in the length of time. For 1D Anderson insulators, we numerically find the TDEE shows a Page-curve like behavior, growing at short time, and decaying in power-law at long time exceeding the crossover time determined by localization. In the continuum time limit $\tau\rightarrow 0$, the TDEE of disorderless gapless fermions, band insulators and Anderson insulators exhibit logarithmic law, area law and sub-area law, respectively. We also show that the sub-Hilbert space of a time-direction interval stabilizes in the $\tau\rightarrow 0$ limit, which is understandable as a subspace of the Hilbert space of the Lieb-Robinson causal diamond.

%The entanglement entropy is a fundamental property of a quantum state. However, how to properly define a Hilbert space along the time direction---and the corresponding notion of entanglement entropy---remains an open question. 
%In this work, we adopt the definition of \textit{time-direction entanglement entropy} (TDEE) and calculate the TDEE for three representative systems: gapless free fermions, band insulators, and Anderson insulators.
%For the band insulator, with gapless fermions recovered as a special limiting case, we analytically compute the TDEE using the Fisher--Hartwig conjecture and verify the results through numerical simulations, obtaining a consistent phase diagram. Interestingly, we find that the TDEE of the band insulator can exhibit \textit{volume-law}, \textit{area-law}, or \textit{log-law} scaling, depending on the system parameters—a counterintuitive outcome compared to conventional spatial entanglement entropy. We also derive an effective Lieb-Robinson bound for TDEE of band insulators. For the Anderson insulator, we discover a \textit{dynamical Page transition} in the TDEE, arising from the localization of the wavefunction. Moreover, we show that the critical transition time is directly related to the localization length.

\end{abstract}

\maketitle

%\tableofcontents

\section{Introduction}

Entanglement entropy plays a central role in condensed matter physics~\cite{Vidal2003Entanglement,Calabrese2004Entanglement,casini2004finite,Amico2008Entanglement}, quantum information theory~\cite{vedral2002role,eisert2006entanglement,Horodecki2009Quantum,islam2015measuring}, and quantum gravity~\cite{ryu2006aspects,Veronika2007covariant,nishioka2009holographic,rangamani2017holographic,nakata2021new}, which characterizes the intrinsic properties of many-body quantum states such as correlation functions~\cite{hastings2007entropy,Wolf2008Area,hartman2013entanglement} and topological orders~\cite{Alioscia2005Ground,kitaev2006topological,levin2006detecting,li2008entanglement,Pollmann2010Entanglement,jiang2012identifying,Zhang2012Quasiparticle}, etc.
The most well-studied entanglement entropy is defined for a spatial subregion of a quantum system, which characterizes the quantum correlations between the subregion and its complement. It is known that the entanglement entropy of a one-dimensional (1D) gapped ground state of a local Hamiltonian obeys an area law~\cite{hastings2007area,arad2013area,brandao2013area}, which is linear in the area of the boundary of the subregion. 
In higher dimensions, the entanglement entropy of gapped ground states also typically follows an area law, or at least exhibits sub-volume-law scaling in certain known examples~\cite{Plenio2005Entropy,Cramer2006Entanglement,eisert2010colloquium}. 
For gapped ground states with topological orders, the subleading constant in the area-law scaling of entanglement entropy, known as the \emph{topological entanglement entropy}, serves as a robust indicator of topological order~\cite{Furukawa2007Topological,Dong2008Topological,isakov2011topological}. 
For gapless systems, the entanglement entropy typically scales as the area of subregion boundary multiplied by a factor logarithmic in the subregion size~\cite{Christoph1994Geometric,Li2006Scaling,Wolf2006Violation,gioev2006entanglement,swingle2010entanglement}. 
Furthermore, the so-called \emph{boundary entropy} provides insight into the boundary properties of a system~\cite{Affleck2009Entanglement,Cardy2016Entanglement}. 
The celebrated $g$-theorem relates this quantity to the renormalization group flow between different boundary conditions~\cite{affleck1991universal,affleck1993exact,friedan2004boundary,casini2016g}.

%For gapless systems, however, additional logarithmic corrections can arise. For instance, in a one-dimensional gapless system, the ground-state entanglement entropy scales as $S \sim \log L$, whereas for a $d>1$ dimensional system with a Fermi surface, it behaves as $S \sim L^{d-1} \log L$~\cite{gioev2006entanglement,swingle2010entanglement}. In topological systems, the subleading constant in the area-law scaling, known as the \emph{topological entanglement entropy}, serves as a robust indicator of topological order~\cite{kitaev2006topological,levin2006detecting,isakov2011topological}. Furthermore, the so-called \emph{boundary entropy} provides insight into the boundary properties of a system. The celebrated $g$-theorem relates this quantity to the renormalization group (RG) flow between different boundary conditions~\cite{affleck1991universal,affleck1993exact,friedan2004boundary}.

Recently, generalizations of quantum entanglement entropy in non-constant time subregions, such as a time direction interval at a fixed position (which will be the focus of this work), have been explored in multiple approaches. The first approach is to characterize the temporal entanglement through \emph{operator growth} as a measure of complexity of quantum simulation~\cite{hosur2016chaos,von2018Operator,Khemani2018Operator,nahum2018operator,zhou2020entanglement}, or by the \emph{operator entanglement entropy}~\cite{Zanardi2001Entanglement,Prosen2007Operator,Prosen2007Is,Pi2009Operator,Zhou2017Operator,Dubail2017Entanglement,Pal2018Entangling,Bertini2020Operator}.
The second approach is the temporal entanglement entropy defined in the \emph{influence matrix approach}~\cite{Lerose2021Influence,Lerose2021Scaling,Michael2021Influence,Giudice2022Temporal,Thoenniss2023Nonequilibrium,Lerose2023Overcoming,Foligno2023Temporal,Vilkoviskiy2026Temporal}, where the time-direction influence matrix is viewed as a ``wavefunction" with space replaced by time, for which the entanglement entropy is calculated.
The third approach is the analytical continuation of entanglement entropy from a spatial slice to a timelike slice, and this leads to a pseudo-entropy \cite{Narayan2023de,Doi2023Pseudoentropy,Doi2023Timelike,afrasiar2024timelike,Narayan2024Further,heller2025geometric,giataganas2025holographic,nunez2025holographic,Heller2025Temporal,Guo2025Relation}, which is a complex number understandable as the entropy of a non-Hermitian density matrix and manifests in the dS/CFT duality. 
The fourth approach, as proposed in \cite{Liu_2024} (and a similar idea in \cite{Nebabu_2023,nebabu2026two,akhtar2026blast} for generalized free fermions), identifies a natural sub-Hilbert space for the time-direction interval at fixed position, in which the conventional entanglement entropy is computed and denoted as a time-direction entanglement entropy (TDEE); however, this generalization is so far only fully resolved for free fermion systems. 

We note that the first three approaches of generalizations above do not yield an entanglement entropy defined in the conventional way for a sub-Hilbert space of the physical quantum states, but should be understood as an entanglement entropy of certain effective ``quantum states", the relation of which to the physical quantum states is obscure. In contrast, the fourth approach of generalization identifies the sub-Hilbert space supporting the Heisenberg picture operators in a time-direction interval, and then defines the entanglement entropy (the TDEE) in this sub-Hilbert space in the standard way \cite{Liu_2024}. This gives an entanglement entropy of the physical quantum states in a physical sub-Hilbert space. The sub-Hilbert space identified can be understood as the minimal subspace accommodating the spreading of operators in the time-direction interval, which is approximately upper bounded by the Lieb-Robinson causal diamond. While this is the most faithful generalization of entanglement entropy, the identification of this sub-Hilbert space of a time-direction interval is only solved for free fermion systems, and remains unsolved for interacting quantum systems to date. Moreover, even if the TDEE can be numerically calculated for free fermions, the analytical understanding of their behaviors is far from clear.

In this paper, we investigate the TDEE for 1D free fermion systems defined in the fourth approach mentioned above \cite{Liu_2024}, and we provide a semi-analytical understanding of the scaling behaviors of TDEE for various 1D free fermion ground states. The TDEE is defined in a time interval of length $T$ at one or more fixed sites, consisting of $K=T/\tau$ discrete time points equally spaced by time $\delta t=\tau$. Taking $\tau\rightarrow 0$ gives the continuum time limit. We investigate its scaling behaviors with respect to $T$ and $\tau$ for three classes of 1D free fermion many-body ground states. For disorderless gapless free fermions and band insulators, we develop a generic analytical understanding of the TDEE by applying the Fisher-Hartwig conjecture for Toeplitz matrices. The results show that the TDEE crucially depends on two dimensionless parameters $\tau/\tau_0$ and $\tau\Delta$, where $\Delta$ is the band gap, and $\tau_0=2\pi/E_0$ is a characteristic time given by the bandwidth $E_0$. Interestingly, in different regimes, the TDEE can show volume law, logarithmic law or area law in time $T=K\tau$. We also show that the 2-site TDEE shows distinct behaviors for topological and trivial phases in the example of the Su-Schrieffer-Heeger (SSH) chain. We further investigate the Anderson insulators, in which we find the TDEE shows a universal Page-curve like behavior as a function of $T=K\tau$, which increases at small $T$, and shows a power law decay in $T$ when $T$ is larger than the localization length.

Particularly, in the continuum time limit $\tau\rightarrow 0$, the TDEE distinguishes the three classes of free fermion ground states. The disorderless gapless free fermions show a logarithmic law TDEE, with $S=\frac{1}{6}\ln T$ for one flavor of nonchiral fermion. The band insulators show an area law TDEE with $S\simeq \text{Const}$. The Anderson insulator shows a sub-area law TDEE, with $S\propto T^{-\alpha}$ for certain $\alpha>0$ at large $T$. Interestingly, the TDEE distinguishes between band insulators and Anderson insulators, which indicates that the TDEE contains information of the Hamiltonian dynamics. This is in contrast to the conventional spatial entanglement entropy, which shows area law for both types of insulators.

Furthermore, we show evidences and arguments that, the sub-Hilbert space of a time direction interval, in which the TDEE is defined, converges to a stabilized Hilbert space in the continuum time limit $\tau\rightarrow 0$. This sub-Hilbert space can be roughly understood as a subspace of the Hilbert space of the Lieb-Robinson causal diamond, which accommodates the spreading of the operators in the time interval.

The rest of this paper is organized as follows. 
In Sec.~\ref{sec:TDEE in a general free fermion model}, we introduce TDEE, and explain the analytical framework we use to understand the TDEE of free fermion ground states. 
In Sec.~\ref{sec:TDEE of gapless free fermions}, we develop an analytical understanding of the TDEE for 1D disorderless gapless free fermion ground state numerically studied in Ref.~\cite{Liu_2024}, and extend the discussion to multi-site TDEE. We also show that the sub-Hilbert space of the time interval is stabilized in the continuum time limit. 
Sec.~\ref{TDEE of 1D Band Insulator (Disorderless)} is devoted to the numerical study and analytical understanding of TDEE of band insulators in various regimes. 
In Sec.~\ref{sec:TDEE of Anderson insulators}, we numerically study the TDEE of the Anderson insulator, and provide an understanding of its Page-curve like behavior. 
Lastly, Sec.~\ref{sec:Conclusions} gives the summary and potential directions to be explored in the future.

\section{General formulation}
\label{sec:TDEE in a general free fermion model}

We first review the definition of time-direction entanglement entropy (TDEE) for free fermions in a lattice. We consider a generic time-independent free fermion Hamiltonian in a real space lattice with $N$ fermion modes, in the bilinear form 
\begin{equation}\label{eq:free_fermion_H}
    H=\sum_{i,j=1}^N h_{ij} c_i^\dagger c_j\ ,
\end{equation}
where $c_i^\dagger$ ($c_i$) are the fermion creation (annihilation) operators on site $i$, $h_{ij}$ is the $N\times N$ matrix of hopping amplitudes. 
% \bl{When log means ln, I use ln (I think the journal also has such a convention).} {\color{red}XS: figures to be improved later}
%{\color{red} Here the site index $i$ may label sites in a lattice of arbitrary spatial dimension. In the following discussion, however, we mainly specialize to one-dimensional systems.}

\subsection{Conventional spatial entanglement entropy}

The conventional entanglement entropy is defined for a sub-Hilbert space 
\begin{equation}
\mathcal{H}_{A_\text{eq}}=\otimes_{i\in A_\text{eq}}\mathcal{H}_i
\end{equation}
of an equal time (in the Heisenberg picture, e.g. $t=0$) spatial subsystem $A_\text{eq}$ of the lattice, where $\mathcal{H}_i$ is the Hilbert space of site $i$ at the chosen time. For a Fock state $|\Psi\rangle$, for example the ground state of \cref{eq:free_fermion_H}, by \cite{peschel2003}, the reduced density matrix $\rho_{A_\text{eq}}=\text{Tr}_{\mathcal{H}^C}(|\Psi\rangle\langle\Psi|)$ in subsystem $A_\text{eq}$ (where $\mathcal{H}^C$ denotes the complement of sub-Hilbert space $\mathcal{H}_{A_\text{eq}}$) takes the Gaussian form
\begin{equation}
\rho_{A_\text{eq}}=Z_A^{-1}e^{-H^E_{A_\text{eq}}}\ ,\quad H^E_{A_\text{eq}}=\sum_{i,j\in A_\text{eq}} h^E_{ij} c_i^\dagger c_j\ ,
\end{equation}
where $Z_A=\text{Tr}(e^{-H^E_{A_\text{eq}}})$ is a constant, and $H_{A_\text{eq}}^E$ is known as the entanglement Hamiltonian or modular Hamiltonian of sub-Hilbert space $\mathcal{H}_{A_\text{eq}}$. It is proved \cite{peschel2003} that the matrix $h^E_{ij}$ is given by the equal-time two-point correlation matrix $C$ in subregion $A_\text{eq}$ as
\begin{equation}
h^E=[\ln(C^{-1}-I)]^T,\ \ C_{ij}=\langle \Psi| c_i^\dagger c_j |\Psi \rangle,\ \ (i,j\in A_\text{eq})
\end{equation}
where $I$ is the identity matrix. Generically, the eigenvalues of $C$ are within $[0,1]$. Accordingly, the von Neumann entanglement entropy $S(A_\text{eq})=-\text{Tr}(\rho_{A_\text{eq}}\ln \rho_{A_\text{eq}})$ is given by
\begin{equation}\label{eq:EE_C}
S(A_\text{eq})=-\text{Tr}[C\ln C+(I-C)\ln(I-C)]\ .
\end{equation}

\subsection{Time-direction entanglement entropy (TDEE)}
\label{sec:Definition and physical meanings}

\begin{figure}
    \centering
    \subfigure[]{\includegraphics[width=0.47\linewidth]{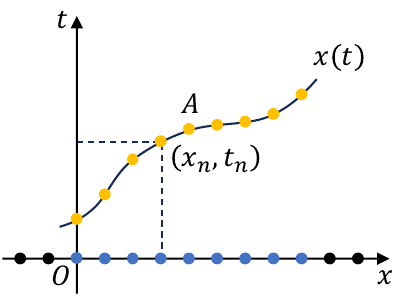}}\quad
    \subfigure[]{\includegraphics[width=0.475\linewidth]{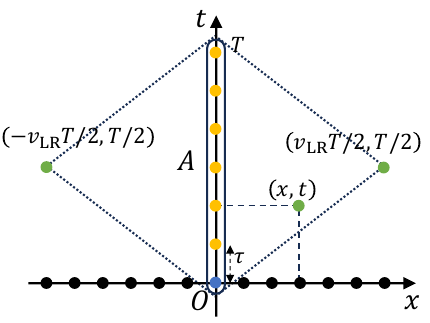}}
    \caption{(a) We plot a general spacetime slice $x(t)$ in a 1D lattice, and choose the region $A$ shown with yellow dots, which are under the Heisenberg time evolution from fermions shown with blue dots.
    (b) The subsystem $A$ is a fixed-site time interval at $x=0$, with discrete time points $t_n=n\tau$ and total duration $T$. 
    The dashed lines show the Lieb--Robinson causal diamond, with left and right vertices at approximately $(\pm v_{\rm LR}T/2,T/2)$. 
    The green point $(x,t)$ denotes a generic spacetime point inside the diamond.
    % (\bl{Yes roughly, but let's follow my definition of time interval above \cref{eq:time-length}, where the boundary of the interval is not at a time point. Also, let's label both the left/right edge, by $(\pm v_\text{LB}T/2,T/2)$, and explain the diamond and these notations in the caption. Also, pick a point in the time interval and give it a label $(x_n,t_n)$. I suggest coloring (grey) of the dashed diamond is not needed.}{\color{red}[XS]: Do you mean the figure now? And if we exactly use the definition, the middle point should be $(T-\tau/2)/2$.} \bl{Almost, the only remaining point is, I thought $(x_n,t_n)$ should label a point in the time interval. A general spacetime point would be $(x,t)$. Also, in panel (a), since in this paper we didn't introduce the complement $A^c$ (the grey points), perhaps we can just keep the spacetime region $A$ (the yellow points).})
    }
    \label{fig:config_x_t}
\end{figure}

For free fermions, the above entanglement entropy can be generalized and calculated \cite{Liu_2024} for a generic spacetime subsystem $A=\{(\mathbf{r}_i,t_i)|i\in A\}$, which consists of a set of distinct spacetime points $(\mathbf{r}_i,t_i)$ with time $t_i$ not equal (see \cref{fig:config_x_t} for examples). Here $\mathbf{r}_i$ are lattice sites. Without loss of generality, we assume there is one fermion mode per site (multiple modes per site can be relabeled by multiple sites in a unit cell).

In this generalization, time $t$ plays a role by evolving the local operators and the on-site Hilbert spaces in the Heisenberg picture. For a generic operator $O$ (e.g., $c_i$ and $c_i^\dagger$), we define the Heisenberg picture operator at time $t$ as
\begin{eqnarray}
\label{eq:haisenberg_evolution_O}
    O(t)=e^{{\rm i}Ht}Oe^{-{\rm i}Ht}\ .
\end{eqnarray}
Accordingly, if $i$ denotes a site with a fermion mode $c_i,c_i^\dagger$, the Hilbert space on site $i$ at time $t$ becomes $\mathcal{H}_i(t)=\text{span}\{|0\rangle\ ,c_i^\dagger(t)|0\rangle\}$, which is time dependent. Here $|0\rangle$ refers to the particle vacuum.

For a spacetime subsystem $A=\{(\mathbf{r}_n,t_n)|n\in A\}$, we define its sub-Hilbert space $\mathcal{H}_A$ as the Hilbert space spanned by all many-particle states of the form $\prod_{n\in A} [c^\dagger_{\mathbf{r}_n}(t_n)]^{n_n}|0\rangle$, where $n_n\in\{0,1\}$, and all possible orders of products of $c^\dagger_{\mathbf{r}_n}(t_n)$ are considered. Intuitively, $\mathcal{H}_A$ is the physical Hilbert space that the operators $O(t)$ in time interval $A$ can influence, which is approximately upper bounded by the sub-Hilbert space of the spatial range within the celebrated Lieb-Robinson bound \cite{lieb1972finite,Bravyi2006Lieb,nachtergaele2006lieb,Eisert2006General}. Importantly, with times $t_n$ being unequal, such a Hilbert space  $\mathcal{H}_A$ is not equal to the tensor product $\otimes_{n\in A}\mathcal{H}_{\mathbf{r}_n}(t_n)$ of Hilbert spaces at spacetime points $(\mathbf{r}_n,t_n)$, since the Hilbert spaces $\mathcal{H}_{\mathbf{r}_n}(t_n)$ are not orthonormal to each other due to time evolutions.

For free fermions, such a Hilbert space $\mathcal{H}_A$ can be identified in a simple way. With a bilinear $H$ in \cref{eq:free_fermion_H}, the annihilation operator $c_i(t)=e^{{\rm i}Ht}c_ie^{-{\rm i}Ht}=\sum_{j}\zeta_{ij}(t)c_j$ at time $t$ is always a linear superposition of fermion annihilation operators $c_j$ at time $0$, with the coefficients $\zeta_{ij}(t)$ depending on the Hamiltonian $H$. Therefore, the fermion operators $c_{\mathbf{r}_n}(t_n),c^\dagger_{\mathbf{r}_n}(t_n)$ have commutation relations
\begin{equation}
\{c_{\mathbf{r}_m}(t_m),c_{\mathbf{r}_n}^{\dagger}(t_n)\}=B_{mn}\ ,\qquad (m,n\in A)
\label{eq:anti_commutation_matrix}
\end{equation}
where $B_{mn}$ are numbers which form a $|A|\times |A|$ non-negative Hermitian matrix $B=B^\dagger$, where $|A|$ denotes the number of spacetime points in subsystem $A$. The matrix $B$ can be diagonalized as $B=Q_1\Lambda Q_1^\dagger$, where $Q_1$ is a $|A|\times |A|$ unitary matrix, and 
\begin{equation}\label{eq:rankB}
 \Lambda=\text{diag}(b_1,\cdots,b_{d_A},0,\cdots,0)\ ,\quad    d_A=\operatorname{rank}B
\end{equation}
is a diagonal matrix, with $b_n>0$ ($1\le n\le d_A$) denoting the nonzero eigenvalues. Here the rank $d_A\le |A|$ can be smaller than $|A|$, if the fermion operators $c_{\mathbf{r}_n}(t_n)$ in subsystem $A$ are not all linearly independent. By defining $\overline{\Lambda}=\text{diag}(b_1,\cdots,b_{d_A})$ as the nonzero part of $\Lambda$, and $\overline{Q}_1$ as the first $d_A$ columns of matrix $Q_1$, we have $B=\overline{Q}_1 \overline{\Lambda}\ \overline{Q}_1^\dagger$. We then define a $d_A\times |A|$ matrix
\begin{equation}\label{eq:M-matrix}
M=Q_2\overline{\Lambda}^{-1/2}\overline{Q}_1^\dagger\ ,
\end{equation}
where $Q_2$ is an arbitrary $d_A\times d_A$ unitary matrix. It is easy to see that $M B M^\dagger =I$ is an identity matrix. We can then define a new fermion basis
\begin{equation}
\label{eq:orthonormal_basis}
    d_m=\sum_{n=1}^{|A|} M_{mn}c_{\boldsymbol{r}_n}(t_n)\ .  \qquad (1\le m\le d_A)  
\end{equation}
It is easy to verify that they form an orthonormal fermion basis:
\begin{equation}
\{d_{m},d_n^\dagger\}=\delta_{mn}\ ,
\end{equation}
and all fermion operators 
\begin{equation}\label{eq:c-in-terms-of-d}
c_{\boldsymbol{r}_n}(t_n)=\sum_{m=1}^{d_A} (BM^\dagger)_{nm} d_m
\end{equation}
in subsystem $A$ are linear superpositions of $d_m$. As a result, the sub-Hilbert space $\mathcal{H}_A$ of subsystem $A$ we defined earlier is spanned by all Fock states generated by basis $d_m$:
\begin{equation}\label{eq:sub-H-A}
\mathcal{H}_A=\text{span}\{\prod_{m=1}^{d_A}(d_m^\dagger)^{n_m}|0\rangle\Big|n_m\in\{0,1\}\}\ ,
\end{equation}
which has a many-body Hilbert space dimension $2^{d_A}$. For later convenience, we call $d_A$ the \emph{single-particle} dimension of Hilbert space $\mathcal{H}_A$.

The entanglement entropy in subsystem $A$ can then be defined for sub-Hilbert space $\mathcal{H}_A$ in the conventional way. For a Fock state $|\Psi\rangle$ (which is time-independent in the Heisenberg picture), we can define the correlation matrix in the $d_m$ basis
\begin{equation}\label{eq:correlation_matrix_orthonormal}
D_{mn}=\langle \Psi|d_m^\dagger d_n|\Psi\rangle=(M^* C M^T)_{mn}\ ,
\end{equation}
where $M^*$ and $M^T$ stand for the complex conjugation and transpose of $M$, respectively, and $C$ is the correlation matrix in the original spacetime fermion basis:
\begin{equation}\label{eq:generalized_correlation_matrix}
C_{mn}=\langle\Psi|c_{\mathbf{r}_m}^\dagger(t_m)c_{\mathbf{r}_n}(t_n)|\Psi\rangle\ .\qquad (m,n\in A)
\end{equation}
Similar to \cref{eq:EE_C}, the generalized entanglement entropy $S(A)=-\text{Tr}(\rho_A\ln\rho_A)$ in the sub-Hilbert space $\mathcal{H}_A$ of spacetime subsystem $A$ is then given by the formula
\begin{equation}
\label{eq:n_1_TDEE_modeified_D}
S(A)=-{\rm Tr}[D\ln{D}+(I-D)\ln{(I-D)}]\ .
\end{equation}
One may also calculate other quantities associated with the reduced density matrix $\rho_A$ of subsystem $A$, for instance, the mutual information between sub-Hilbert space of $A$ and its complement Hilbert space \cite{Liu_2024}.

Specifically, when the subsystem $A$ is a set of time points in a time interval at a fixed site (or multiple fixed sites), we denote $S(A)$ in \cref{eq:n_1_TDEE_modeified_D} as the TDEE \cite{Liu_2024}. More concretely, as shown in \cref{fig:config_x_t}(b), we assume the time direction subsystem $A$ consists of $K$ points $(\mathbf{r}_0,t_n)$ at uniformly distributed discrete times
\begin{equation}\label{eq:uniform-time-points}
t_n=n\tau\ ,\qquad (0\le n\le K-1)
\end{equation}
where $\tau$ is the time step separating neighboring points, and $\mathbf{r}_0$ is a fixed site (which can be generalized to a fixed set of sites for a thick time interval). The time interval $A$ is defined to be $[-\frac{1}{2}\tau,(K-\frac{1}{2})\tau]$, which has a total length of time
\begin{equation}\label{eq:time-length}
T=K\tau\ .
\end{equation}
Here $\tau$ plays the role of a chosen ultraviolet (UV) cutoff of the continuous time $t$ in defining the TDEE.

Explicitly, assume the correlation matrix $D$ have the eigenvalues $\frac{\nu_l+1}{2}$, where $1\le l\le d_A$, and by definition we have $\nu_l\in[-1,1]$.  \cref{eq:n_1_TDEE_modeified_D} can then be written as
\begin{equation}
\label{eq:EE_eigenvalue}
\begin{split}
&S(A)=\sum_{l=1}^{d_A} e(\nu_l) \ ,\\
&e(x)=-\Big(\frac{1+x}{2}\Big)\ln\Big(\frac{1+x}{2}\Big)-\Big(\frac{1-x}{2}\Big)\ln\Big(\frac{1-x}{2}\Big).
\end{split}
\end{equation}
% \bl{(Keeping it below for now. Is it sufficient to use the above expression throughout the paper? If so, we can move Renyi to appendix?) For the R\'enyi-$q$ entanglement entropy $S_q(A)=\frac{1}{1-q}\ln{{\rm Tr}(\rho_{A}^q)}$, we have
% \begin{equation}
% \label{eq:Renyi_n_EE_eigenvalue}
% \begin{split}
%     S_q(A)=&\sum_{l=1}^{d_A} e_q(\nu_l), \\
%     e_q(x)=&\frac{1}{1-q}\ln{\left[\left(\frac{1+x}{2}\right)^q+\left(\frac{1-x}{2}\right)^q\right]}.
% \end{split}
% \end{equation}
% Taking the $q\rightarrow 1$ limit gives the von Neumann entropy $S(A_\text{eq})$ in \cref{eq:EE_eigenvalue}. 
% }

We add a few remarks. First, the entanglement entropy $S(A)$ is independent of the choice of $Q_2$. Secondly, by QL decomposition, $Q_2$ in \cref{eq:M-matrix} can always be chosen such that $M$ is a lower triangular matrix, which is useful in giving the basis $d_i$ a spacetime understanding \cite{Liu_2024,Nebabu_2023}. Lastly, for a full rank matrix $B$, one can choose $Q_2=Q_1$, such that \cref{eq:M-matrix} can be understood as $M=B^{-1/2}$. This gives a correlation matrix in the $d_i$ basis
\begin{equation}\label{eq:correlation_matrix_D_with_BC}
D=B^{*-1/2}C B^{*-1/2}\ .
\end{equation}

\subsection{Toeplitz matrix and analytical method}
\label{sec:Analytical}

The TDEE defined above can be easily calculated numerically by \cref{eq:n_1_TDEE_modeified_D}, as readily demonstrated in \cite{Liu_2024}. 
However, the analytical understanding of the TDEE numerical results is still far from clear. Fortunately, utilizing the fact that the correlation matrix $D$ in \cref{eq:correlation_matrix_orthonormal} can be formulated as a Toeplitz matrix, we can derive a partial analytical understanding of the TDEE of the ground states of one-dimensional (1D) free fermion Hamiltonians, which is one of the central results of this paper. Here we briefly outline this analytical method we will employ.

For the conventional ground state entanglement entropy of an equal time subregion of 1D lattice models, there are two common analytical methods for calculating their entanglement entropy. The first method is to analytically solve the eigenvalues of the correlation matrix under appropriate boundary conditions, by transforming the eigenvalue equation of the matrix into a functional form~\cite{Peschel_2009}. This method has been extensively applied to calculate the entanglement spectrum for both free fermion systems~\cite{Ingo_Peschel_2004} and free boson systems~\cite{callan1994geometric}, and more recently employed to study free fermion systems under weak measurements~\cite{tang2024criticalstateweakmeasurement}. It is however difficult to apply this method to TDEE of ground states, because of the complicated form of the correlation matrix $D$ in \cref{eq:correlation_matrix_orthonormal}. 
The second method is based on the Fisher--Hartwig conjecture~\cite{Fisher1969Toeplitz,Jin_2004,Fagotti_2011} for Toeplitz matrix, which we will employ to analyze the TDEE of free fermion ground states, as described below and in more details later.

We first note that the entanglement entropy expression in \cref{eq:EE_eigenvalue} can be expressed by the residue theorem as a complex contour integral: 
% (\bl{Please note I changed $g_L(\lambda)$ to $g(\lambda)$, since $L$ is total system size; it should be $d_A$.})
\begin{equation}
\label{eq:contour_integral_of_EE}
\begin{split}
    S(A)=\oint \frac{{\rm d}\lambda}{2\pi {\rm i}} \sum_{l=1}^{d_A} \frac{e(\lambda)}{\lambda-\nu_l}
    =\oint \frac{{\rm d}\lambda}{2\pi {\rm i}}\ e(\lambda)\frac{{\rm d}\ln{g(\lambda)}}{{\rm d}\lambda}\ ,
\end{split}
\end{equation}
where 
\begin{equation}
\label{eq:deternimant_g}
    g(\lambda)=\prod_{l=1}^{d_A}(\lambda-\nu_l)={\rm Det}(G)\ ,
\end{equation}
is a polynomial function of degree $d_A$, and we have defined a $d_A\times d_A$ character matrix
\begin{equation}\label{eq:G-def}
G=(\lambda+1)I-2D
\end{equation}
in terms of the correlation matrix $D$.

For TDEE, it is key to observe that the subsystem $A$ is a fixed-site time interval with uniformly distributed time points $t_j$ in \cref{eq:uniform-time-points}. Because of the time translation symmetry ($H$ is time-independent), \cref{eq:anti_commutation_matrix} gives $B_{mn}=B_{0,m-n}$ only depending on $m-n$. 
% (\bl{I changed $B(m-n)$ to $B_{0,m-n}$, to save the notation $B(\theta)$ as the Fourier trans. Please check other places.}) 
A matrix with such a property is known as a \emph{Toeplitz matrix}. Furthermore, for a Fock state $|\Psi\rangle$ being an eigenstate of $H$ (thus including the ground state), \cref{eq:generalized_correlation_matrix} also gives $C_{mn}=C_{0,m-n}$ as a Toeplitz matrix. 
Assuming the full rank matrix $B$ and translation-invariant approximation discussed in \cref{sec:Background of correlation matrix and entanglement entropy}, it is easy to see that the $d_m$ basis correlation matrix $D=B^{*-1/2}C B^{*-1/2}$ in \cref{eq:correlation_matrix_D_with_BC} is also a Toeplitz matrix satisfying
\begin{equation}
D_{mn}=D_{0,m-n}\ .
\end{equation}
If $B$ is not full rank, it is still possible to choose $Q_2$ to make $D$ a Toeplitz matrix, due to the time translation symmetry. Accordingly, $G_{mn}=G_{0,m-n}$ defined in \cref{eq:G-def} is also a Toeplitz matrix.

For the Toeplitz matrix elements $G_{0,n}$ which are defined on integers $n\in\mathbb{Z}$, we can define its Fourier transform $G(\theta)$ for $\theta\in[0,2\pi)$, which satisfies
\begin{equation}\label{eq:Fourier_transf_G}
G(\theta)=\sum_n e^{-in\theta}G_{0,n}\ ,\quad G_{0,n}=\int_0^{2\pi} \frac{{\rm d}\theta}{2\pi} e^{{\rm i}n\theta}G(\theta)\ .
\end{equation}
The Fourier transform $G(\theta)$ is referred to as the \emph{symbol} of the Toeplitz matrix $G$. The \emph{Fisher--Hartwig conjecture}~\cite{Fisher1969Toeplitz,Jin_2004,Fagotti_2011,deift2011asymptotics} for Toeplitz matrix states that, near its singularities, the symbol $G(\theta)$ can be represented in the form
\begin{equation}
\label{eq:FT_special_form_G}
    G(\theta)=f(\theta)\prod_{r=1}^R e^{{\rm i}b_r(\theta-\theta_r-\pi{\rm sgn}(\theta-\theta_r))}(2-2\cos{(\theta-\theta_r)})^{a_r},
\end{equation}
where ${\rm sgn}(x)$ is the sign function, $\theta_r$ are the locations of discontinuities of $G(\theta)$, and the parameters $a_r$ and $b_r$ are determined such that $f(\theta)$ is a smooth function with zero winding number. This determines the asymptotic behaviors of the function $g(\lambda)$, by which the entanglement entropy in \cref{eq:contour_integral_of_EE} can be estimated to be given by the following three parts:
% (see Appendix \bl{I omitted some steps here, which can go to appendix, please check by comparing with main3.1})
\begin{equation}\label{eq:SA_with_FH_conj}
S(A)=\alpha_V d_A+\alpha_\text{log}\ln{d_A}+\alpha_A\ ,
\end{equation}
where the coefficient $\alpha_A$ is non-universal, while
\begin{equation}\label{eq:general_EE_with_FH_conj}
\begin{split}
    \alpha_V=&\frac{1}{2\pi {\rm i}}\oint {\rm d}\lambda \ e(\lambda)\frac{{\rm d}}{{\rm d}\lambda}\ln{F[f(\theta)]},\\
    \alpha_{\rm log}=&\frac{1}{2\pi {\rm i}}\oint {\rm d}\lambda \ e(\lambda)\frac{{\rm d}}{{\rm d}\lambda}\sum_{r=1}^R(a_r^2-b_r^2)\ ,
\end{split}
\end{equation}
with the function $F[f(\theta)]=\exp{\left(\int_0^{2\pi} \frac{{\rm d}\theta}{2\pi}\ \ln{f(\theta)} \right)}$.

\cref{eq:SA_with_FH_conj,eq:general_EE_with_FH_conj} are the main formulas for analytically understanding the TDEE hereafter. If matrix $B$ is full rank (which is typically true when $H$ has no symmetry), the rank $d_A=K=T/\tau$. When $B$ is not full rank, $d_A$ is also often proportional to $K=T/\tau$. Therefore, $\alpha_V$, $\alpha_\text{log}$ and $\alpha_A$ in \cref{eq:SA_with_FH_conj} correspond to the volume law, log law and area law coefficients in the time interval length $T$, respectively.

By \cref{eq:correlation_matrix_D_with_BC,eq:G-def}, one can show (for full rank $B$) that (see \cref{sec:Background of correlation matrix and entanglement entropy})
\begin{equation}
\label{eq:Fourier_transf_D}
\begin{split}
    G(\theta)=\lambda+1-2D(\theta)\ ,\quad D(\theta)=&\frac{C(\theta)}{B^*(\theta)}\ ,
\end{split}
\end{equation}
where $D(\theta)$, $C(\theta)$ and $B^*(\theta)$ are the symbol of matrices $D$, $C$ and $B^*$, respectively. 

Heuristically, the symbol $D(\theta)$ approaches the eigenvalues (labeled by $\theta$) of a Toeplitz matrix $D$ in the infinite matrix size limit $d_A\rightarrow\infty$, given that $D_{mn}$ decays sufficiently fast as $|m-n|$ increases (see \cref{sec:reduced_hilbert_gapless} for a more detailed discussion).

%\bl{cmts kept temporarily:} where we slightly abuse the notation $C$ and $\hat{C}$. \bl{Why two notations are necessary?} {\color{red}[XS]:Because one is the total correlation matrix and another is the submatrix.} \bl{OK. Using the same notation, for $i,j$ in $A$ only, seems sufficient.}

\subsection{Regimes of Interest}

Hereafter, we will focus on investigating the TDEE of the ground states of 1D free fermion models, and show they exhibit distinct scaling behaviors with the time interval length $T$. All our discussion in the main text is assumed to be in the regime 
\begin{equation}
K\ll L\ ,
\end{equation}
where $K$ is the number of time points in the time-interval $A$ (see \cref{eq:uniform-time-points}), and $L$ is the total number of sites of the 1D lattice. Particularly, for a fixed time interval length $T=K\tau$, taking the continuum time limit means taking $\tau\rightarrow 0$ while keeping $L\gg K=T/\tau$.

When $K>L$, the sub-Hilbert space $\mathcal{H}_A$ will approach the Hilbert space of the entire system, and thus $S(A)$ will generically approach zero \cite{Liu_2024}. In \cref{sec:Proof of vanish of time direction entanglement entropy for large time sites}, we show that for models with spatial translation symmetry, when $K>L$, the TDEE is identically zero if there is no energy degeneracy, and at most $\mathcal{O}(1)$ if there is accidental degeneracy. This is not the regime we are interested in.

% For $K>L$, the TDEE vanishes when all sampled phases $e^{-iE_j\tau}$ are distinct, as shown in Appendix B. When different energy eigenstates share the same sampled phase, a nonzero residual entropy may remain if the corresponding sector contains both occupied and unoccupied modes.

% \bl{Note a few notation differences above: the number of time points is called $K$, and defined $T=K\tau$. ALso, $\hat{T}_x c_i = c_{i+1}$ should be written as $\hat{T}_x c_i \hat{T}_x^{-1} = c_{i+1}$ ? Please correct in the rest of paper/appendix.}

\section{TDEE of 1D Gapless Fermion (Disorderless)}
\label{sec:TDEE of gapless free fermions}

We first consider the 1D translationally invariant tight-binding model with one fermion mode per site and a nearest neighbor hopping amplitude $-u<0$:
\begin{equation}
\label{eq:tight_binding_model}
    H=-u\sum_{i=0}^{L-1} (c_i^\dagger c_{i+1}+c_{i+1}^\dagger c_i)\ ,
\end{equation}
where $L$ is the length of the 1D lattice. We impose periodic boundary conditions (PBC) for the models considered below. The model has a cosine single-particle energy dispersion 
\begin{equation}\label{eq:cosine-Ek}
    E(k)=-2u\cos{k}
\end{equation}
as a function of quasi-momentum $k$, the eigenmodes of which are given by $\widetilde{c}_k = \frac{1}{\sqrt{L}}\sum_j e^{-{\rm i}kj}c_j$.  
% (\bl{I removed lattice constant $a_0$, which seems only appeared here. Also, I saw $a_0$ refers to a different quantity at a later place.})
This spans a total bandwidth
\begin{equation}\label{eq:gapless-bandwidth}
E_0=4u\ .
\end{equation}
This gives a characteristic time scale of the system:
\begin{equation}
\label{eq:characteristic_time}
    \tau_0=\frac{2\pi}{E_0}\ .
\end{equation}

We assume the system has a fermion filling factor $\nu\in (0,1)$, namely the total number of fermions is $N_e=\nu L$. The ground state $|\Psi_0\rangle$ of the system is thus a gapless Fermi liquid with a Fermi momentum $k_F=\nu\pi$ and Fermi energy $E_F=-2u\cos k_F$.
Unless otherwise specified, all numerical results for gapless fermions and band insulators in \cref{sec:TDEE of gapless free fermions,TDEE of 1D Band Insulator (Disorderless)} are obtained at half filling, $\nu=0.5$.

\begin{figure}
    \centering
    \subfigure[]{\includegraphics[width=0.45\linewidth]{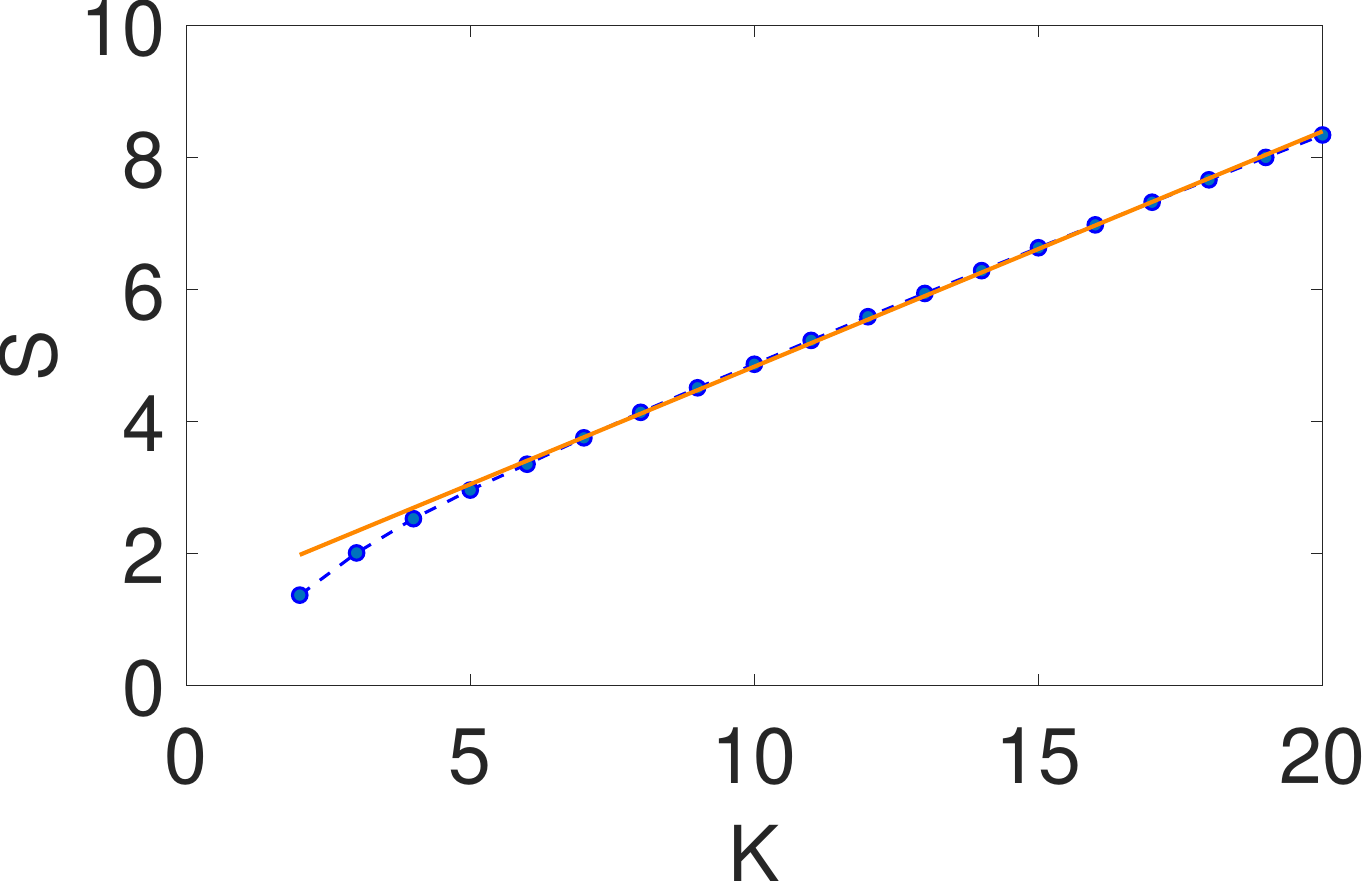}}
    \subfigure[]{\includegraphics[width=0.45\linewidth]{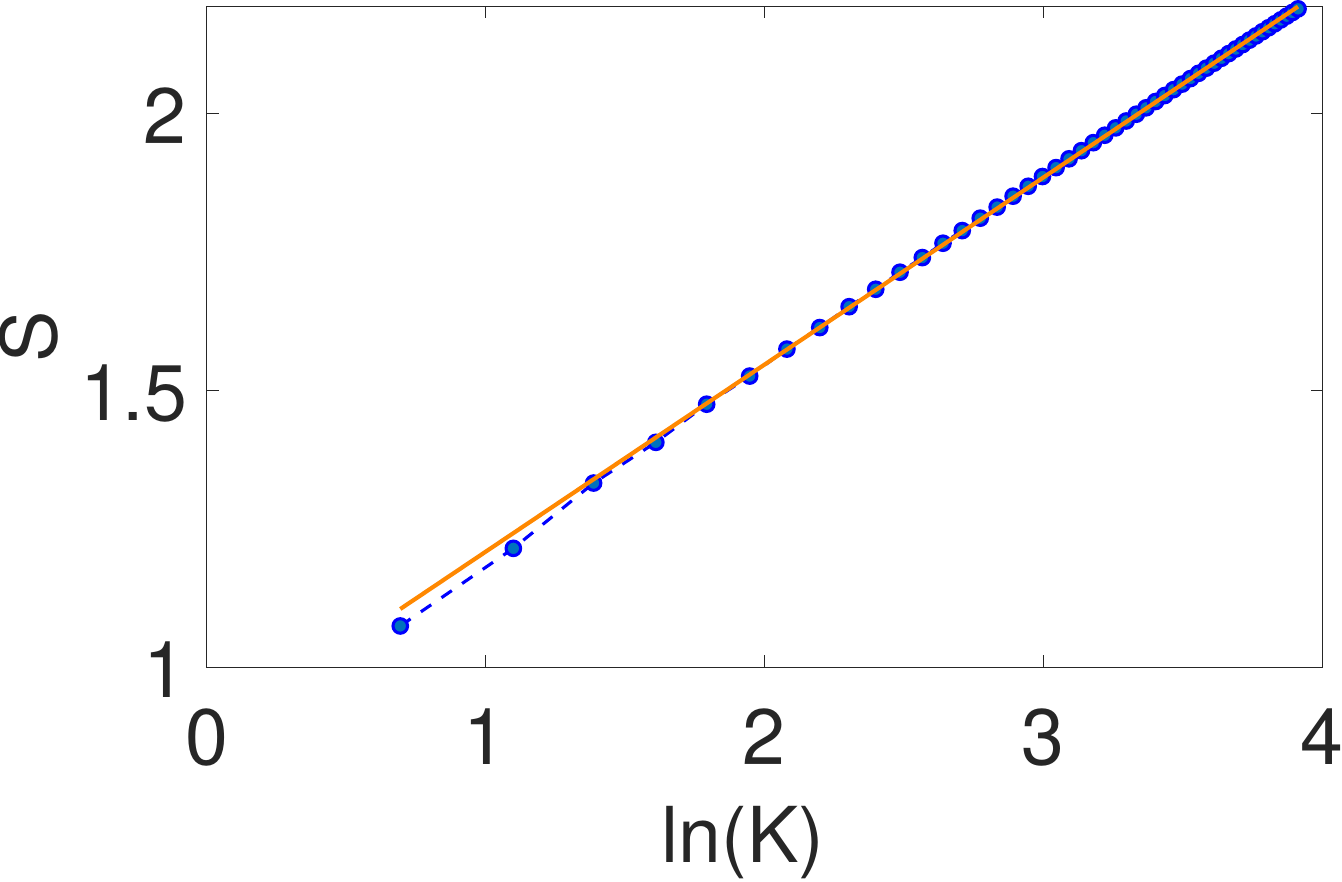}}
    \subfigure[]{\includegraphics[width=0.45\linewidth]{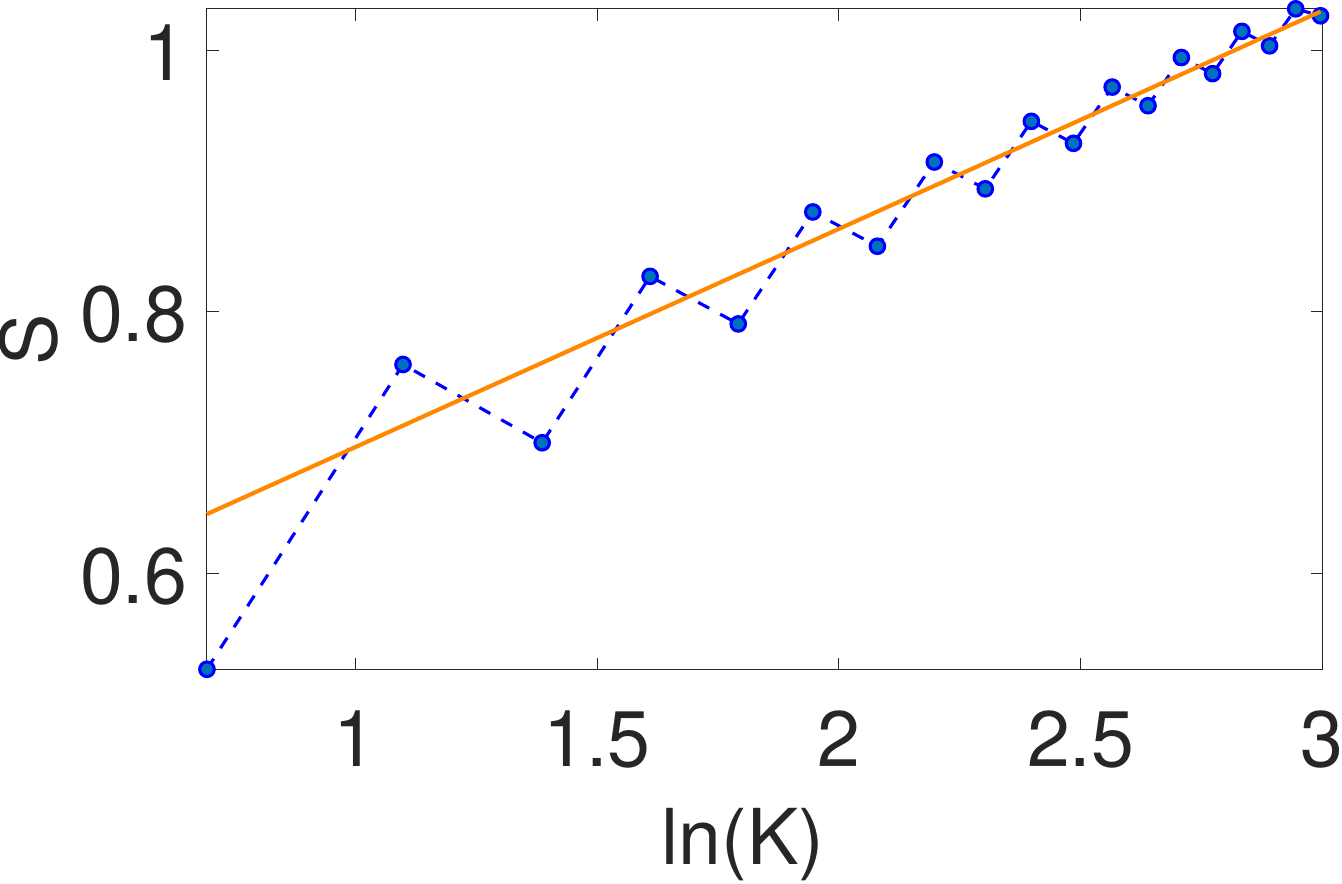}}
    \subfigure[]{\includegraphics[width=0.45\linewidth]{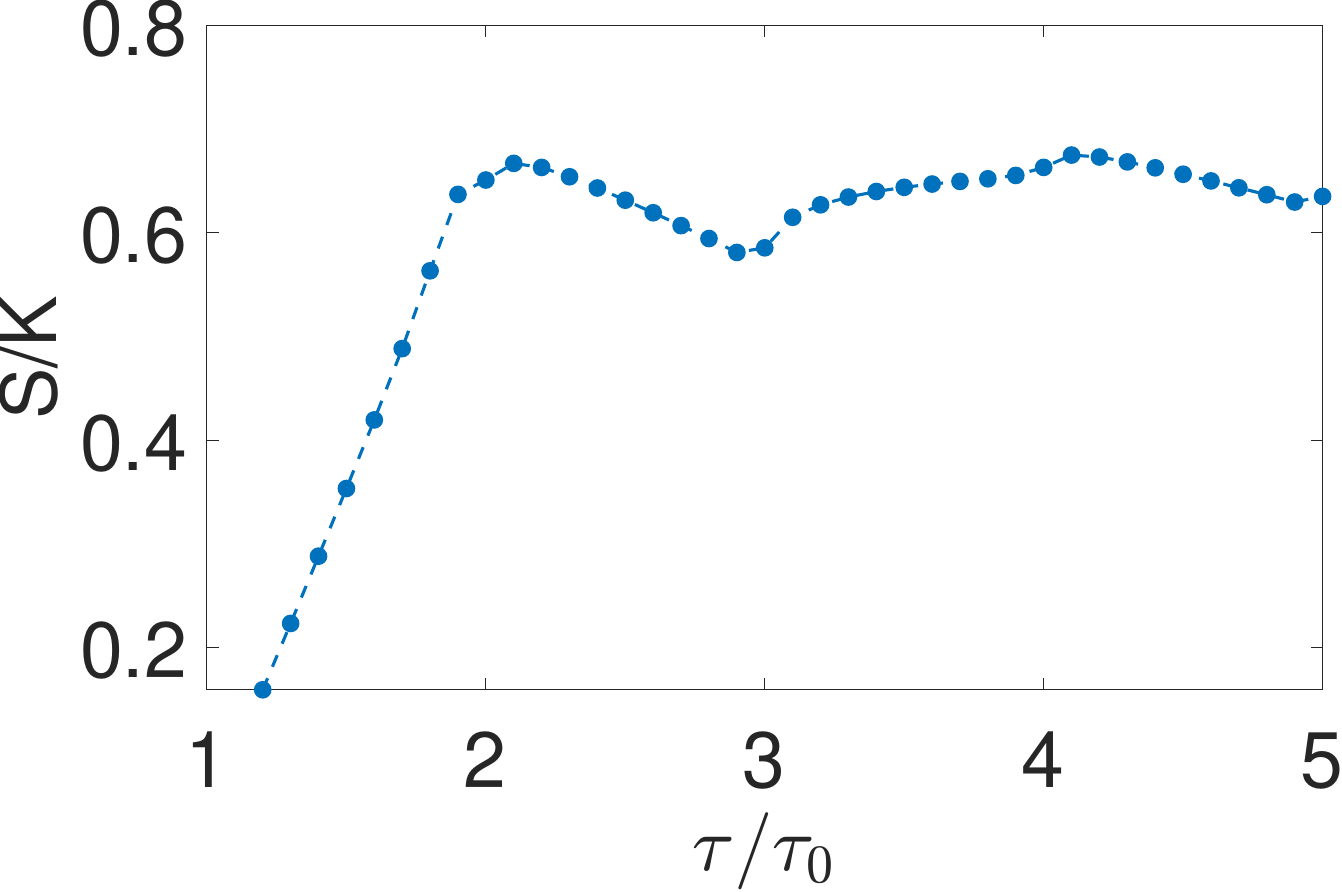}}
    \caption{The TDEE for gapless free fermions with $L=1000$, filling $\nu=0.5$, and (a) $\tau=1.5\tau_0$, (b) $\tau=\tau_0$ and (c) $\tau=0.5\tau_0$. The volume-law coefficient at $\tau=1.5\tau_0$ is $0.356$. (d) The prefactor of volume-law term $\alpha_V\approx S/K$ for $\tau>\tau_0$. 
    % \bl{Calculated for which filling $\nu$, and which $L$? Please note in the caption.}
    % {\color{red} All figures are calculated for $\nu=0.5$. May be added to the main text later.}
    % % \textcolor{red}{add fit for (a)-(c). And show the concrete value of volume-law coefficient at $\tau=1.5\tau_0$.}
    }
    \label{fig:gapless_EE}
\end{figure}

For the fixed-site time interval $A$ with $K$ time points as imposed in \cref{eq:uniform-time-points}, Ref. \cite{Liu_2024} discovered numerically that the TDEE of ground state of the model in \cref{eq:tight_binding_model} (with filling $0<\nu<1$) has the following leading scaling behaviors with $K=T/\tau$ for fixed time step $\tau$ in different regimes:
\begin{equation}
\label{eq:gapless_log_law}
    S(A)=\begin{cases}
    \alpha_V K + \text{Const.}\ ,\qquad\  (\tau>\tau_0)\\
    &\\
    \dfrac{1}{3}\ln{K} + \text{Const.}\ ,\qquad (\tau=\tau_0)\\
    &\\
    \dfrac{1}{6}\ln{K} + \text{Const.}\ ,\qquad (\tau<\tau_0)
    \end{cases}
\end{equation}
with the volume law coefficient $\alpha_V>0$. 
% For $\tau>2\tau_0$, one has $\alpha_V\approx e(1-2\nu)$ in terms of the function $e(x)$ in \cref{eq:EE_eigenvalue}. 
For $\tau>2\tau_0$, the coefficient $\alpha_V$ is approximately given by $e(1-2\nu)$, defined in \cref{eq:EE_eigenvalue}, and approaches this value in the large-$\tau$ limit.
\cref{fig:gapless_EE} shows our numerical results verifying \cref{eq:gapless_log_law}.

It is useful to recall the celebrated Cardy-Calabrese formula \cite{Calabrese2004Entanglement} for the 1D spatial entanglement entropy of one flavor of gapless nonchiral fermion (or equivalently, a conformal field theory (CFT) with central charge $1$), which states that the ground state entanglement entropy $S(\ell)=\frac{1}{3}\ln \ell +\text{Const.}$ for an equal time spatial interval of length $\ell$. In comparison, the TDEE in \cref{eq:gapless_log_law} in the continuum time limit $\tau\rightarrow 0$ scales as $S(A)=\frac{1}{6}\ln T +\text{Const}.$, which has half the log coefficient compared with the Cardy-Calabrese formula. It is argued in \cite{Liu_2024} that, this is because the Fermi sea ground state has two Fermi points $\pm k_F$ in the momentum space contributing to the spatial entanglement entropy, while it has only one Fermi point in the energy space, the Fermi energy $E_F$, contributing to the TDEE. However, a more rigorous analytical derivation of \cref{eq:gapless_log_law} was not established, which is the main goal of \cref{sec:derive-gapless} below.

\subsection{Analytical derivation of TDEE}
\label{sec:derive-gapless}

We apply the method of Fisher--Hartwig conjecture as outlined in \cref{sec:Analytical} to establish an analytical derivation of \cref{eq:gapless_log_law} for the model in Eq.~\eqref{eq:tight_binding_model}. For simplicity, we set the filling to $\nu=1/2$, and the hopping $u=1$. 

We choose a time interval $A$ at site $\mathbf{r}_0=0$ and time points $t_n=n\tau$ as defined in \cref{eq:uniform-time-points}. In terms of the momentum eigenbasis $\widetilde{c}_k$, the fermion operators in the time interval $A$ read
\begin{equation}
\label{eq:operator Heisenberg time evolution}
c_0(t_n)=\frac{1}{\sqrt{L}}\sum_k e^{-{\rm i}E_k n\tau}\ \widetilde{c}_k\ .
\end{equation}
% \bl{I kept the original expression below, which has notation problem: sum over $n$ and $n$ in $n\tau$ mean different things. Please check if there are similar problems elsewhere. It seems the second equality is not needed so I removed it. Also, I used $\widetilde{c}_k$ for Fourier modes, to distinguish from $c_j$ in real space. If OK, please check elsewhere.
% \begin{equation*}
% c_0(n\tau)=\frac{1}{\sqrt{L}}\sum_k e^{-{\rm i}E_k n\tau}c_k=\frac{1}{L}\sum_{j,k} e^{-{\rm i}E_k n\tau}e^{-{\rm i}kj}c_j,
% \end{equation*}}
This gives the following ground state correlation matrix $C$ and anti-commutation matrix $B$:
\begin{eqnarray}
\label{eq:Toeplitz_matrix_translation_symm_C}
    C_{mn}=&&\frac{1}{L}\sum_k e^{{\rm i}E_k (m-n)\tau}n_F(E_k),\\
\label{eq:Toeplitz_matrix_translation_symm_B}
    B_{mn}=&&\frac{1}{L}\sum_k e^{-{\rm i}E_k (m-n)\tau},
\end{eqnarray}
where $n_F(E_k)=\Theta(-E_k)$ is the zero-temperature Fermi-Dirac distribution at filling $\nu=1/2$, where $\Theta(x)$ is the Heaviside step function. Applying a Fourier transformation yields their symbols (see Appendix~\ref{sec:Background of correlation matrix and entanglement entropy}):
\begin{eqnarray}
\label{eq:Toeplitz_matrix_translation_symm_C_FT}
    &&C(\theta)=\sum_k\frac{\sin{[L(E_k\tau-\theta)/2]}}{L\sin{[(E_k\tau-\theta)/2]}}n_F(E_k)\ ,\\
\label{eq:Toeplitz_matrix_translation_symm_B_FT}
    &&B^*(\theta)=\sum_k\frac{\sin{[L(E_k\tau-\theta)/2]}}{L\sin{[(E_k\tau-\theta)/2]}}\ ,
\end{eqnarray}
from which $D(\theta)=C(\theta)/B^*(\theta)$ in \cref{eq:Fourier_transf_D} can be calculated. In general, the summations above do not have closed-form analytical solutions. 
However, as shown in Ref.~\cite{Liu_2024}, the scaling behavior of the TDEE is insensitive to details of the dispersion $E_k$. Therefore, as an approximation, we replace the dispersion with a nonchiral linear function 
\begin{equation}\label{eq:linear-Ek}
E_k=\frac{4|k|}{\pi}-2  \ ,\qquad (k\in(-\pi,\pi])
\end{equation}
which has the same bandwidth $E_0=4$ (for $u=1$) as the model in \cref{eq:tight_binding_model}, and has a uniform density of states. This allows us to analytically calculate $D(\theta)$ and derive the TDEE as follows.

(i) The $\tau>\tau_0$ case. Ref.~\cite{Liu_2024} already established an understanding of the saturated value $\alpha_V=e(1-2\nu)$ of the volume-law TDEE coefficient when $\tau > 2\tau_0$. Here, by analytically computing $D(\theta)$, we can further show that the volume-law coefficient $\alpha_V$ grows linearly in the interval $\tau\in[\tau_0,2\tau_0]$ before saturating at $\tau\approx 2\tau_0$. 
The details are presented in Sec.~\ref{sec:SSH model with the particle-hole symmetry}, where the TDEE of gapped free fermions is studied, and the gapless case here is included as the gap \(\Delta\rightarrow 0\) limit.

(ii) The $\tau=\tau_0$ case. In this case, in the $L\rightarrow\infty$ limit, the summation in $k$ in Eqs.~(\ref{eq:Toeplitz_matrix_translation_symm_C}) and (\ref{eq:Toeplitz_matrix_translation_symm_B}) can be rewritten as an integral of $p=E_k\tau=2|k|-\pi$ in the interval $(-\pi,\pi]$. This gives $B_{mn}=\delta_{mn}$, and $C_{mn}$ as the Fourier transform $n_F(\frac{p}{\tau_0})$ with respect to $p$. Therefore, one finds $D(\theta)=C(\theta)$ simply given by
%by defining an effective `momentum' $p=2|k|-\pi\in(-\pi,\pi]$, one can rewrite $e^{{\rm i}E_k (m-n)\tau}=e^{{\rm i}p(m-n)}$ in Eqs.~(\ref{eq:Toeplitz_matrix_translation_symm_C}) and (\ref{eq:Toeplitz_matrix_translation_symm_B}), which gives $B_{mn}=\delta_{mn}$, and $C_{mn}=D_{mn}$ equivalent to a spatial correlation matrix in a length $K$ interval. Thus, we can utilize the existing analytical result for the conventional spatial entanglement entropy \cite{Jin_2004}, which gives 
% (\bl{The formula below doesn't seem right. First, $|k_F-\theta|$ is always non-negative. Secondly, shouldn't $D(\theta)=n_F(\theta)=\Theta(-\theta)$? For $\theta\in(-\pi,\pi]$. Please correct.})
%(\bl{Pls check again: are you sure it is not $D(\theta)=\Theta(-\theta)$? For $\tau<\tau_0$ below, the discontinuity is at $\theta=0$, not $\theta=\pm k_F$.} {\color{red}Sorry. I consider it to be the same as the real-space $D(\theta)$ before. And you are right, it should be $D(\theta)=\Theta(-\theta)$.})
\begin{equation}
\label{eq:FT_D_gapless_tau0}
    D(\theta)=n_F\left(\frac{\theta}{\tau_0}\right)=\Theta(-\theta)\ ,\qquad \theta\in(-\pi,\pi]
\end{equation}
This gives two discontinuous jumps of $G(\theta)$ (between $\lambda+1$ and $\lambda-1$) at $\theta=0,\pi$. 
Via a detailed calculation using Eq.~\eqref{eq:general_EE_with_FH_conj}, we find $\alpha_V=0$ and $\alpha_\text{log}=1/3$, which agrees with Eq.~\eqref{eq:gapless_log_law}. In particular, each discontinuity contributes $1/6$ to $\alpha_\text{log}$. We note that in this case, the matrix $D$ is the same as the correlation matrix of an equal-time spatial interval of length $K$ for a Fermi sea with momenta $p\in(-\pi,0]$ occupied, for which $\alpha_\text{log}=1/3$ (the Cardy-Calabresse coefficient) is known from the existing analytical result \cite{Jin_2004}.

\begin{figure}
    \centering
    \subfigure[]{\includegraphics[width=0.48\linewidth]{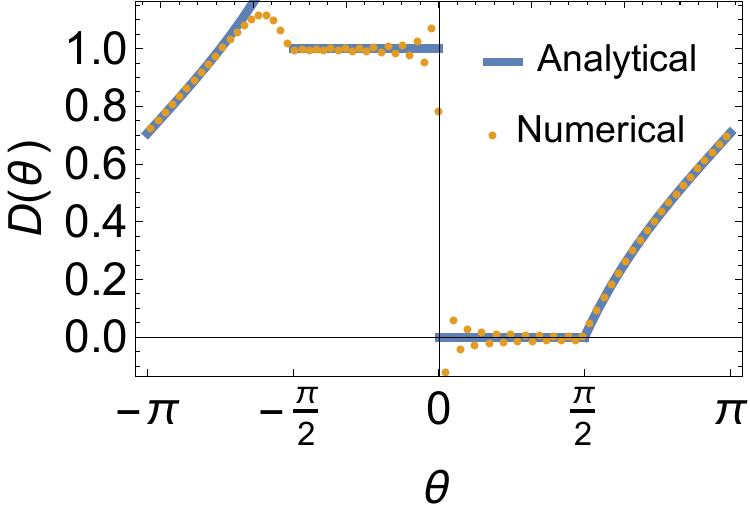}}
    \subfigure[]{\includegraphics[width=0.48\linewidth]{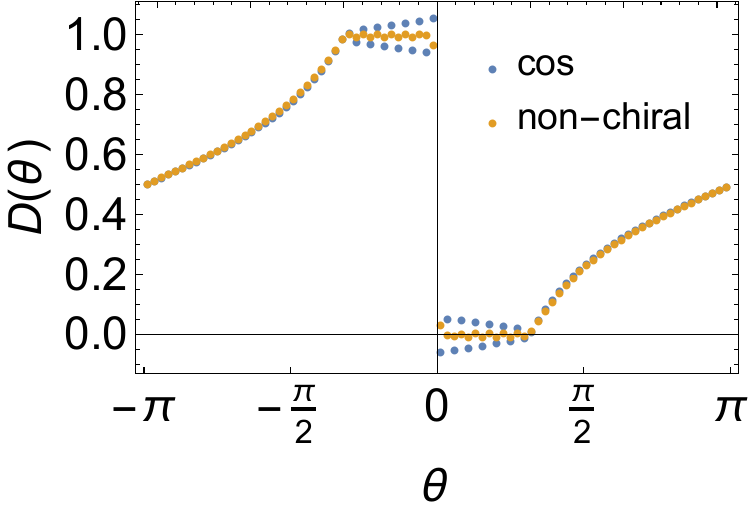}}
    \caption{(a) The numerical result and analytical result~\eqref{eq:D_result_gapless_tau02} of the Fourier transformation $D(\theta)$ of matrix $D$ for the non-chiral linear dispersion, with $\tau=\frac{\tau_0}{2}=\frac{\pi}{4}$ and $L=81$.
    (b) The numerical results of $D(\theta)$ for the cosine and nonchiral linear dispersion with $\tau_0=\frac{\pi}{2}$, $\tau=1/2$, and $L=84$.
    }
    \label{fig:gapless_FT_D}
\end{figure}

(iii) The $\tau<\tau_0$ case. A generic analytical result in this case is unavailable. However, in a representative case with $\tau=\tau_0/2=\pi/4$ and $L=8\mathbb{Z}+1$, we find $D(\theta)$ can be calculated exactly in the $L\rightarrow\infty$ limit. After a lengthy calculation (see \cref{sec:Free fermion with uniform hopping}), we find
\begin{equation}
\label{eq:D_result_gapless_tau02}
D(\theta)=\begin{cases}
    &-\dfrac{\cos{(\frac{\theta}{2}+\frac{\pi}{4})}}{\sin{\frac{\theta}{2}}}, \quad  \theta\in(-\pi,-\frac{\pi}{2})\cup(\frac{\pi}{2},\pi]\\
    &\\
    &\Theta(-\theta)\ ,\qquad\qquad \theta\in[-\frac{\pi}{2},\frac{\pi}{2}]\ .
    \end{cases}
\end{equation}
% and $D(\theta)=\Theta(-\theta)$ for $\theta\in[-\frac{\pi}{2},\frac{\pi}{2}]$, where $\Theta(x)$ is the Heaviside function.
It has a robust intrinsic discontinuity at $\theta=0$, which corresponds to the Fermi energy $E_F=0$. It also has a spurious discontinuity at $\theta=-\frac{\pi}{2}$, which depends on the way $L$ approaches infinity. (\cref{sec:Free fermion with uniform hopping} gives another way $L$ approaches infinity, which removes the discontinuity at $\theta=-\frac{\pi}{2}$). To further see this, in Fig.~\ref{fig:gapless_FT_D}(a) we compare the analytical formula in \cref{eq:D_result_gapless_tau02} with the numerical $D(\theta)$ with $L=81$ (for $\theta\in\frac{2\pi}{L}\left(\frac{-L+1}{2}+\mathbb{Z}\right)$), which match well, except that the numerical $D(\theta)$ is continuous at $\theta=-\frac{\pi}{2}$~\footnote{Here $D(\theta)$ is obtained from the ratio of finite-cutoff Fourier sums rather than the exact finite-size correlation matrix, so its values are not necessarily restricted to $[0,1]$.}. 
% {\bl{Check if the above sentences are good.}}

The above features are generic for other ratios $\tau/\tau_0<1$ and for the cosine band dispersion $E_k=-2\cos k$. Fig.~\ref{fig:gapless_FT_D}(b) plots the numerical $D(\theta)$ with $\tau=\tau_0/\pi=1/2$, for the non-chiral linear dispersion in \cref{eq:linear-Ek} and for the cosine dispersion, respectively. The two $D(\theta)$ are almost the same (except for UV oscillations), and only have a discontinuity at $\theta=0$.

Via a calculation using Eq.~\eqref{eq:general_EE_with_FH_conj}, the single discontinuity at $\theta=0$ contributes only a coefficient $\alpha_\text{log}=1/6$, while $\alpha_V=0$. This leads to the TDEE in Eq.~\eqref{eq:gapless_log_law}. Heuristically, the discontinuity at $\theta=0$ indeed corresponds to the Fermi energy $E_F$, which verifies the argument proposed in \cite{Liu_2024}.

\subsection{The time interval sub-Hilbert space in the continuum time limit}
\label{sec:reduced_hilbert_gapless}

The dependence of the TDEE $S(A)$ on the time discretization step $\tau$ implies that the sub-Hilbert space $\mathcal{H}_A$ of the time interval $A$ is $\tau$-dependent. This is different from the sub-Hilbert space of an equal-time spatial interval of a lattice, which is unambiguously defined. This is intrinsically because time $t$ is continuous. This is similar to the case of quantum field theory in continuous spacetime, where it is known that the entanglement entropy of an equal-time spatial subregion is well-defined only if a spatial UV cutoff $\epsilon$ is introduced, and the associated Hilbert space is $\epsilon$ dependent. For TDEE, $\tau$ plays the role of the time-direction UV cutoff.

However, as we shall argue below, with the space being a discrete lattice, the sub-Hilbert space $\mathcal{H}_A$ of a time-direction interval $A$ will stabilize in the continuum time limit $\tau\rightarrow 0$. This is also consistent with the fact that the TDEE $S(A)$ in \cref{eq:gapless_log_law} stabilizes to $\frac{1}{6}\ln K+\text{Const.}$ for any $\tau<\tau_0$. 

We consider 1D gapless free fermions with a generic dispersion $E_k\in [-\frac{E_0}{2},\frac{E_0}{2}]$ of bandwidth $E_0$. Taking the $L\rightarrow\infty$ limit in Eqs. (\ref{eq:Toeplitz_matrix_translation_symm_C}) and (\ref{eq:Toeplitz_matrix_translation_symm_B}) gives the matrix elements $C_{mn}=C_{0,m-n}$ and $B_{mn}=B_{0,m-n}$ as
\begin{equation}
\label{eq:continuous_B_C}
C_{0,n}=\int_{-\pi}^{\pi}\frac{{\rm d}k}{2\pi}e^{{\rm i}n E_k\tau}n_F(E_k),\quad B_{0,n}=\int_{-\pi}^{\pi}\frac{{\rm d}k}{2\pi}e^{-{\rm i}n E_k\tau} .
\end{equation}
Therefore, their symbols (by \cref{eq:Fourier_transf_G}) in the $L\rightarrow\infty$ limit are formally given by
\begin{equation}\label{eq:Linfty-symbolC}
C(\theta)=\int_{-\pi}^{\pi} {\rm d}k \sum_{l\in\mathbb{Z}}\delta(\theta-E_k\tau-2\pi l)n_F(E_k)\ ,
\end{equation}
and 
% \bl{Note that I replaced $a(\theta)$ by $B^*(\theta)$, which are essentially the same quantity}
\begin{equation}\label{eq:Linfty-symbolB}
B^*(\theta)=\int_{-\pi}^{\pi} {\rm d}k \sum_{l\in\mathbb{Z}}\delta(\theta-E_k\tau-2\pi l)\ ,
\end{equation}
where we restrict $\theta\in(-\pi,\pi]$. Note that there may be $\theta$ values where $C(\theta)=B^*(\theta)=0$, which would give an ill-defined $D(\theta)=C(\theta)/B^*(\theta)$. This is why the $L\rightarrow\infty$ limit need be taken carefully (for $C(\theta)$ and $B^*(\theta)$ simultaneously) in calculating $D(\theta)$ in \cref{eq:D_result_gapless_tau02}.

However, $B^*(\theta)$ in \cref{eq:Linfty-symbolB} alone readily reveals important information on the sub-Hilbert space $\mathcal{H}_A$. We first note that, the $K\times K$ Toeplitz matrix $B_{mn}$ resembles a translationally-symmetric Hamiltonian for a single particle in a 1D lattice of $K$ sites with open boundary condition (OBC), where the $K$ sites correspond to the $K$ time points $t_n$ in time interval $A$. If we replace the OBC by PBC, and take the $K\rightarrow\infty$ limit, the eigenvalues of the Hermitian matrix $B$ would be exactly given by $B^*(\theta)$, with $\theta\in[-\pi,\pi]$ playing the role of ``quasi-momentum" of the particle. Therefore, we can regard the symbol $B^*(\theta)$ in \cref{eq:Linfty-symbolB} as a good approximation of the eigenvalues of the actual anti-commutation matrix $B$ (which has OBC), provided that $B_{mn}$ decays sufficiently fast with $|m-n|$. By this argument, we can estimate the spectral density $\rho(\lambda)$ of matrix $B$ as:
\begin{equation}\label{eq:density_distribution_rho_general}
\rho(\lambda)=\frac{1}{K}\sum_{i}\delta(\lambda-\lambda_i)\approx\int_{-\pi}^{\pi} \frac{{\rm d}\theta}{2\pi}\delta(\lambda-B^*(\theta))\ ,
\end{equation}
where $\lambda_i$ ($1\le i\le K$) are eigenvalues of matrix $B$.

\begin{figure}
    \centering
    \subfigure[]{\includegraphics[width=0.48\linewidth]{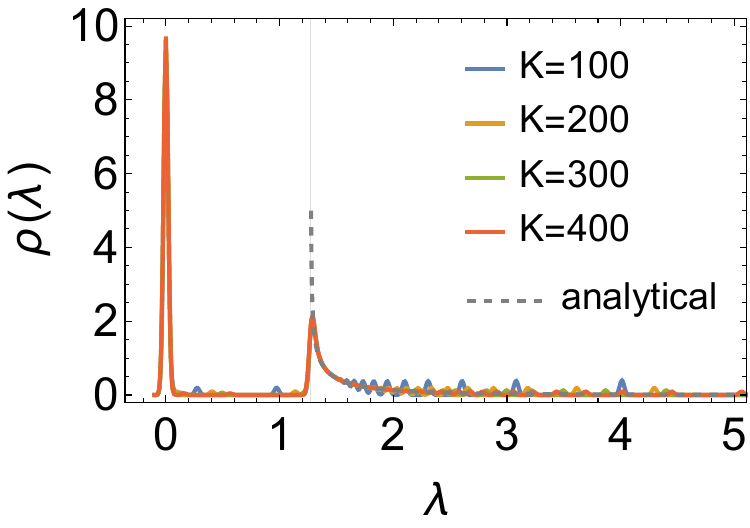}}
    \subfigure[]{\includegraphics[width=0.48\linewidth]{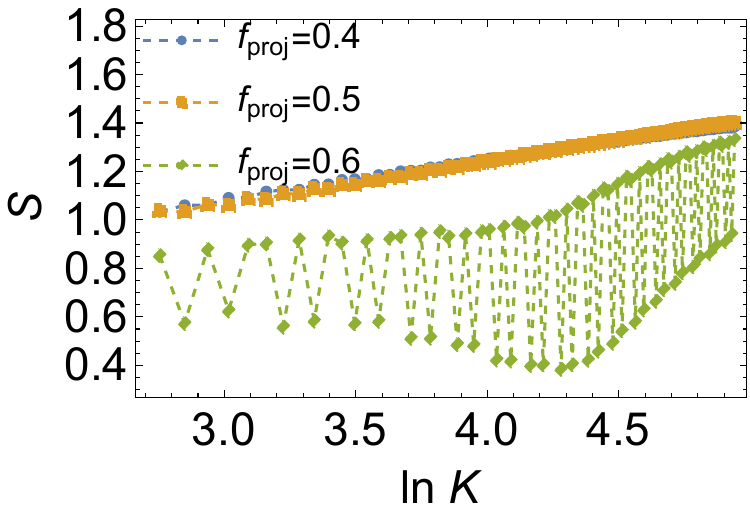}}
    \caption{For the cosine dispersion $E_k=-2\cos k$ (thus $\tau_0=\pi/2$), we plot: (a) Spectral density of the anti-commutation matrix $B_{mn}$ for $K=100,200,300,400$ time points, with $\tau=\pi/4$.
    The spectral density is plotted with a Gaussian broadening, $\rho_h(\lambda)=\frac{1}{Kh}\sum_{i=1}^{K}\mathcal{K}((\lambda-\lambda_i)/h)$, where $\mathcal{K}(u)=\frac{e^{-u^2/2}}{\sqrt{2\pi}}$ and the broadening $h=0.02$. 
    The small apparent weight at $\lambda<0$ arises from this broadening, although $B$ is positive semidefinite. 
    (b) TDEE for gapless free fermions with $\tau=\pi/4$ and cutoffs $f_{\rm proj}=0.4,0.5,0.6$. 
    The cutoff $f_{\rm proj}$ denotes the fraction of eigenmodes of $B$ with the smallest eigenvalues that are projected out.
    For $f_{\rm proj}=0.4,0.5$, the TDEE follows $S(A)=c\ln K+{\rm const.}$ with coefficients $c=0.157077,0.176206$, respectively.}
    \label{fig:reduced_Hilbert_space}
\end{figure}

When $\tau<\tau_0=\frac{2\pi}{E_0}$, importantly, \cref{eq:Linfty-symbolB} implies that $B^*(\theta)=0$ when $\frac{\tau}{\tau_0}\pi<|\theta|\le \pi$. By \cref{eq:density_distribution_rho_general}, this implies that the matrix $B$ has roughly $(1-\frac{\tau}{\tau_0})K$ eigenvalues $\lambda_i$ being nearly zero. Generically, the other eigenvalues are nonzero and larger than a positive spectral gap. In the example of cosine dispersion $E_k=-2\cos k$, which yields $\tau_0=\pi/2$, we find from \cref{eq:density_distribution_rho_general} 
% (\bl{show details (which was in this section) in appendix and cite here})
that for $\tau<\tau_0$ (see \cref{sec:Spectral density of anti-commutation matrix}):
\begin{equation}
\label{eq:density_distribution_rho}
\begin{split}
    \rho(\lambda)\approx &\left(1-\frac{\tau}{\tau_0}\right)\delta(\lambda)+\frac{\Theta(\lambda-\tau^{-1})}{\tau_0 \lambda^2\sqrt{\tau^2\lambda^2-1}}\ ,
\end{split}
\end{equation}
which has a spectral gap $1/\tau$ between the $(1-\frac{\tau}{\tau_0})K$ nearly-zero eigenvalues and other eigenvalues. \cref{fig:reduced_Hilbert_space}(a) shows that the estimation in \cref{eq:density_distribution_rho} matches well with the numerical results at large finite $K$.

When $\tau\ge \tau_0$, an argument similar to the above implies that all the eigenvalues of matrix $B$ are nonzero (positive) in general.

By the construction in \cref{sec:Definition and physical meanings}, the sub-Hilbert space $\mathcal{H}_A$ has a single-particle dimension $d_A=\operatorname{rank}B$. If we regard the nearly zero eigenvalues of matrix $B$ given by \cref{eq:density_distribution_rho_general} as exactly zero, and note that the length of time interval is $T=K\tau$, we conclude that the single-particle dimension of sub-Hilbert space $\mathcal{H}_A$ of a single-site time interval $A$ is given by
\begin{equation}\label{eq:HA-dimension}
d_A\approx
\begin{cases}
& K=\dfrac{T}{\tau}\ ,\qquad (\tau\ge \tau_0)\\
&\\
& \dfrac{\tau}{\tau_0}K=\dfrac{T}{\tau_0}\ .\qquad (\tau <\tau_0)
\end{cases}
\end{equation}
Interestingly, for fixed $T$ in the continuum time limit $\tau\rightarrow0$, the above single-particle Hilbert space dimension $d_A$ is stably $T/\tau_0$, which is independent of $\tau$.

In the actual $B$ matrix, when $\tau<\tau_0$, the lowest $(1-\frac{\tau}{\tau_0})K$ eigenvalues are never exactly zero. Therefore, matrix $B$ is still nearly full rank, which would give a large single-particle Hilbert space dimension $d_A\approx K$. However, the spectral gap $1/\tau$ in the above picture suggests that, the nearly zero eigenvalues of matrix $B$ are negligible. Indeed, by \cref{eq:c-in-terms-of-d}, the fermion operators $d_i$ corresponding to nearly zero eigenvalues of matrix $B$ barely contribute to the fermion operators $c_\mathbf{r_0}(t_n)$ in time interval $A$. Therefore, we conjecture that these nearly zero eigenvalues can be rounded to exactly zero, after which the physical single-particle dimension of sub-Hilbert space $\mathcal{H}_A$ is given by \cref{eq:HA-dimension}.

To test our conjecture above, in \cref{fig:reduced_Hilbert_space}(b), we plotted the ground state TDEE $S(A)$ by removing the space of the lowest $f_{\rm proj} K$ eigenvalues of matrix $B$, or equivalently, setting $d_A=\operatorname{rank}B=(1-f_{\rm proj})K$ given by the highest $(1-f_{\rm proj})K$ eigenvalues. 
% (\bl{It's better to change $\alpha$ to a different notation? Say $\gamma$? I saw $\alpha$ used for a different quantity around Fig. 9 (e.g. panel b)})
In \cref{fig:reduced_Hilbert_space}(b), the dispersion is $E_k=-2\cos k$, and we set $\tau=\tau_0/2=\pi/4$, which gives $\frac{\tau}{\tau_0}K=0.5K$ nearly zero eigenvalues in $B$. For $f_{\rm proj}=0.4$ and $0.5$, we find $S(A)=c\ln K + {\rm const.}$ with $c=0.157077$ and $0.176206$, respectively, both close to the expected value $1/6$ in \cref{eq:gapless_log_law}. In contrast, for $f_{\rm proj}=0.6$, the TDEE deviates significantly from \cref{eq:gapless_log_law}. This verifies our conjecture in \cref{eq:HA-dimension}.

It is worthwhile to compare $\mathcal{H}_A$ with the Hilbert space of the spatial range within the Lieb-Robinson bound of the time-interval $A$. The Lieb-Robinson bound \cite{lieb1972finite} states that information in a local lattice model can at most propagate at the Lieb-Robinson velocity $v_\text{LR}$ (up to exponentially small errors), which plays the role of a ``speed of light". This implies that the length $T$ time-interval $A$ is approximately influenced by a spatial interval of length $v_\text{LR}T$ at the bisector of time-interval $T$, forming a causal diamond as shown in \cref{fig:config_x_t}(b). Therefore, due to the Lieb-Robinson bound, the Hilbert space $\mathcal{H}_A$ is upper bounded by the Hilbert space of the length $v_\text{LR}T$ spatial interval, namely, $d_A\le v_\text{LR}T$. In free fermion models, $v_\text{LR}$ is given by the largest Fermi velocity, which is no smaller than $E_0/\pi$. Therefore, we find
\begin{equation}
v_\text{LR}T\ge \frac{E_0}{\pi}T=\frac{2T}{\tau_0}\ .
\end{equation}
Therefore, by \cref{eq:HA-dimension}, we find the single-particle dimension $d_A$ of sub-Hilbert space $\mathcal{H}_A$ of the single-site time interval $A$ satisfies
\begin{equation}\label{eq:dA-upperlimit}
\frac{d_A}{v_\text{LR} T}\le \frac{1}{2}\ .
\end{equation}
Namely, the single-particle dimension of $\mathcal{H}_A$ is at most $1/2$ of that of the Lieb-Robinson bound spatial region of time interval $A$. Intuitively, this is because fermion operators in the time interval $A$ can only propagate into the mirror symmetric single-particle sub-Hilbert space about the site of $A$, the single-particle dimension of which is at most $1/2$ of that of the full Hilbert space within the Lieb-Robinson bound. Therefore, $\mathcal{H}_A$ for defining the TDEE can be understood as the actual sub-Hilbert space the time interval influences within the Lieb-Robinson bound.
% ~\cite{kac2005solitons,Yasuyuki2006Classification}
% This is because we have c_0(n\tau)=1/\sqrt{L} \sum_j \alpha_j(n\tau) c_j, \alpha_j(n\tau)1/\sqrt{L}\sum_k e^{-i E_k n\tau}e^{-ik j}. So \alpha_j(n\tau)=\alpha_{-j}(n\tau). Therefore, we only use the symmatric combination of c_j and c_{-j}, which decrease the effective Hilbert space with factor 1/2.

\subsection{Multi-site TDEE}

In this subsection, we briefly study the ground state TDEE in multiple time intervals at different sites for the 1D gapless free fermion model in \cref{eq:tight_binding_model}. Part of the motivation is to explore the time-direction analog to the entanglement entropy of disjoint spatial intervals in CFT \cite{Pasquale2009Entanglement}.

\begin{figure}
    \centering
    \subfigure[]
    {\includegraphics[width=0.4\linewidth]{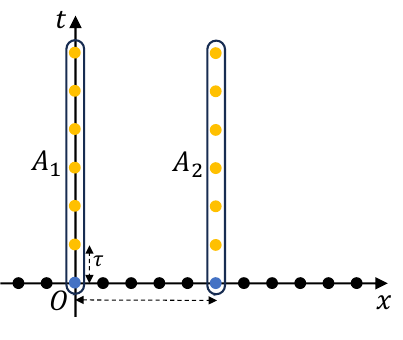}}\qquad
    \subfigure[]
    {\includegraphics[width=0.4\linewidth]{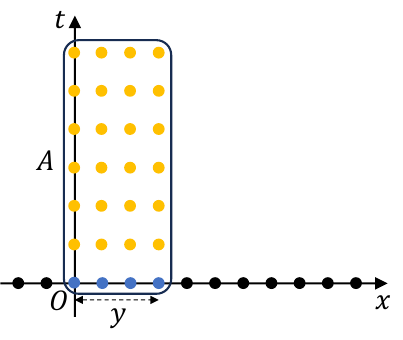}}
    \caption{(a) The subsystem $A=A_1\cup A_2$ with fixed sites $0,x$, and $t_{2n}=t_{2n+1}=n\tau$.
    (b) The subsystem $A$ generated by $y$ sites, with $\boldsymbol{r}_{yn+i}=i$ and $t_{yn+i}=n\tau$ for $i=0,1,\cdots,y-1$.}
    \label{fig:config_x_t_2}
\end{figure}

We first consider a spacetime region $A=A_1\cup A_2$ being the combination of two time intervals $A_1$ and $A_2$ located at sites $0$ and $x$, respectively, as shown in Fig.~\ref{fig:config_x_t_2} (a). More concretely, $A=\{(x_n,t_n)|0\le n\le 2K-1\}$ contains spacetime points $(x_n,t_n)$ with $x_{2n}=0$, $x_{2n+1}=x$ and $t_{2n}=t_{2n+1}=n\tau$. We then numerically calculate the (half-filled) ground state TDEE in $A$ according to \cref{sec:Definition and physical meanings}.

\begin{figure}
    \centering
    \subfigure[]{\includegraphics[width=0.48\linewidth]{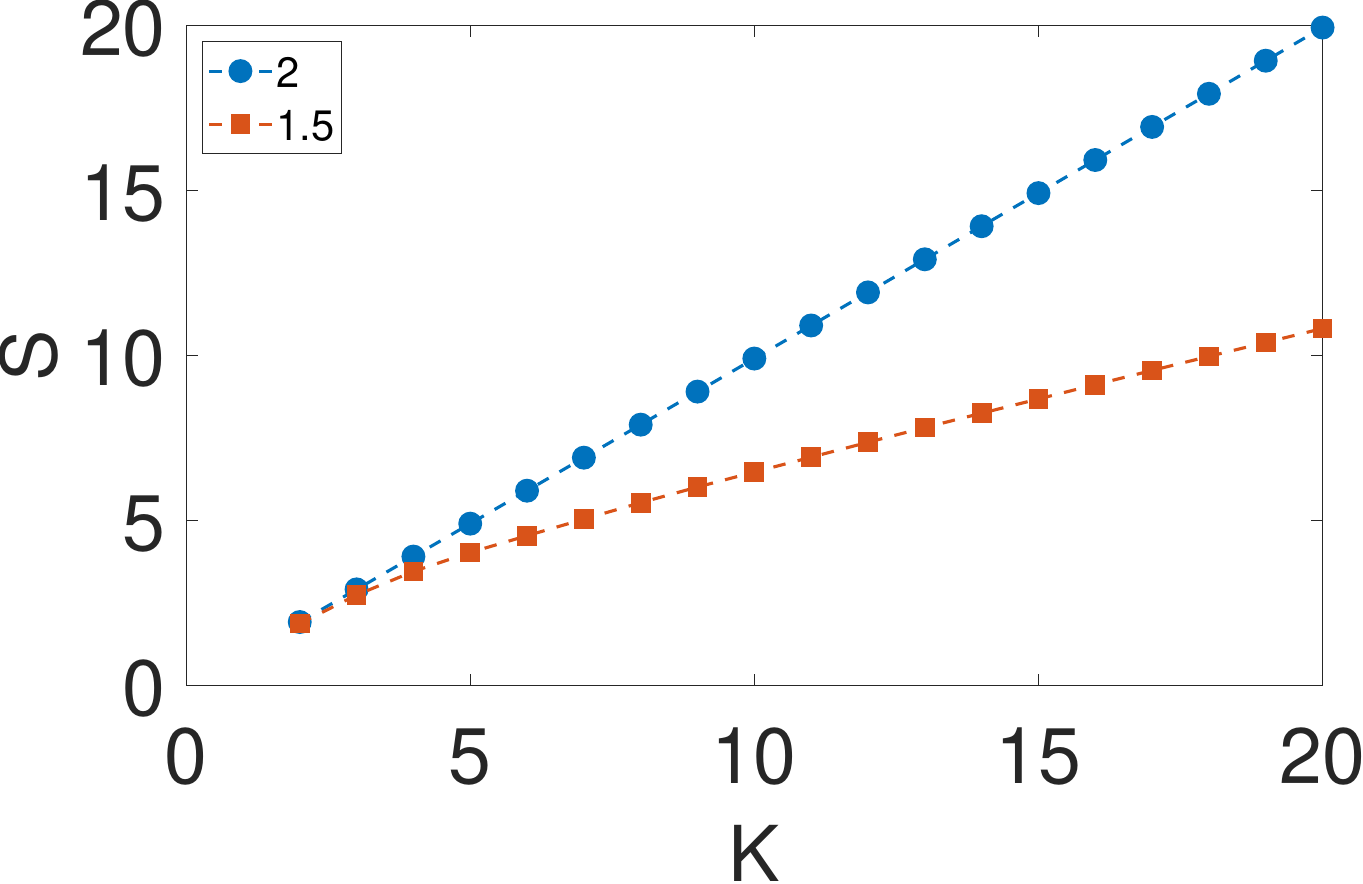}}
    \subfigure[]{\includegraphics[width=0.48\linewidth]{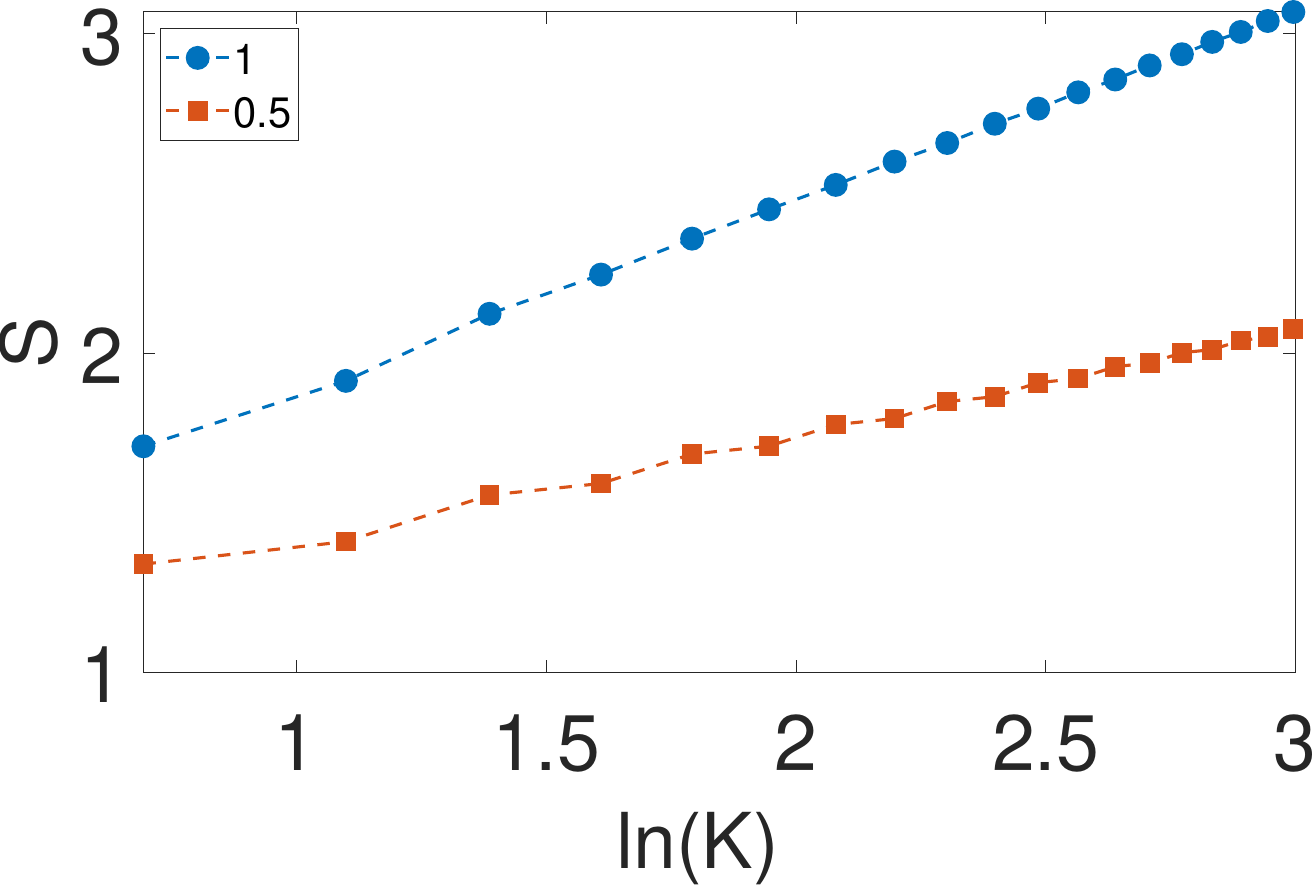}}
    \subfigure[]{\includegraphics[width=0.48\linewidth]{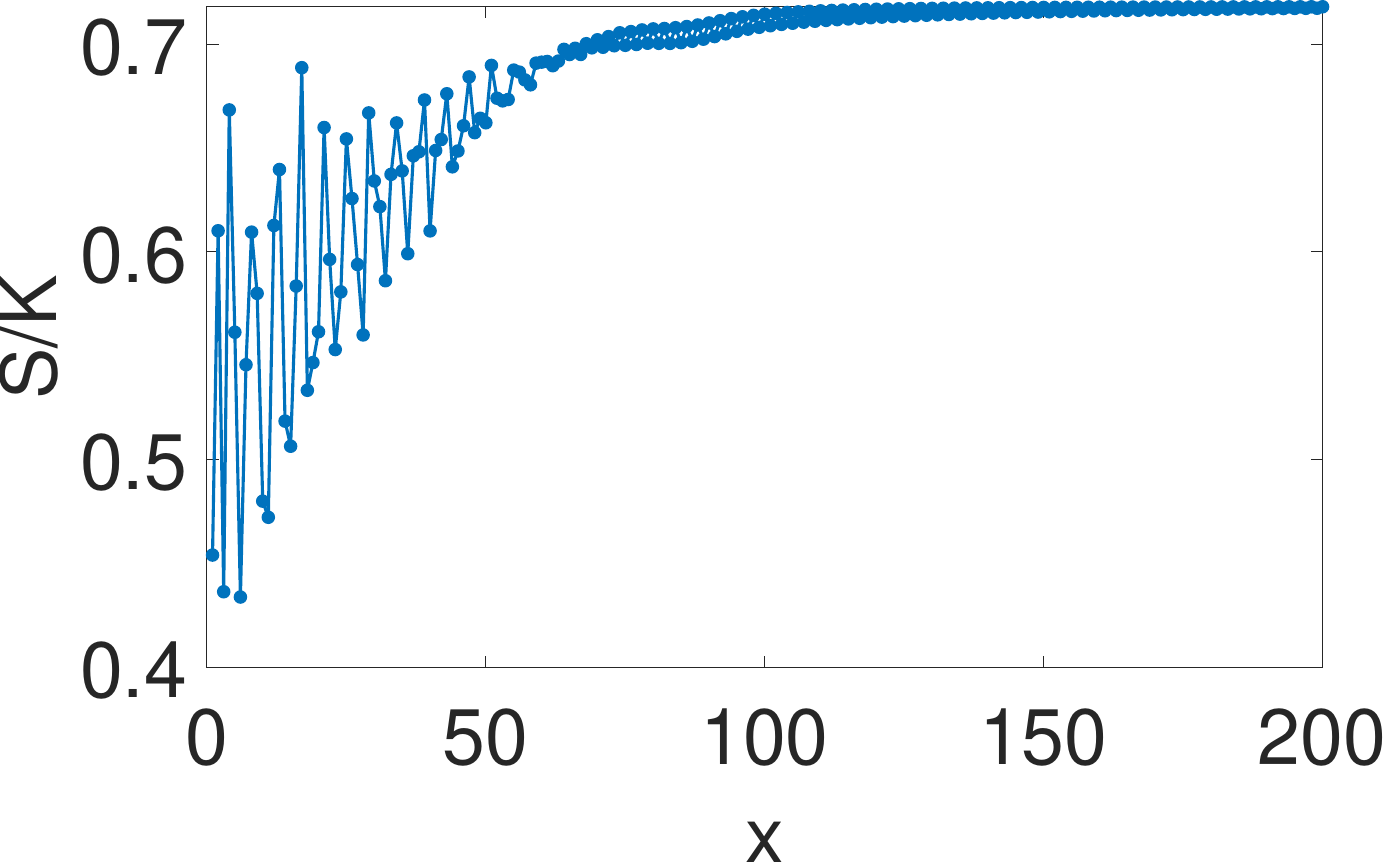}} % 1 20
    \subfigure[]{\includegraphics[width=0.48\linewidth]{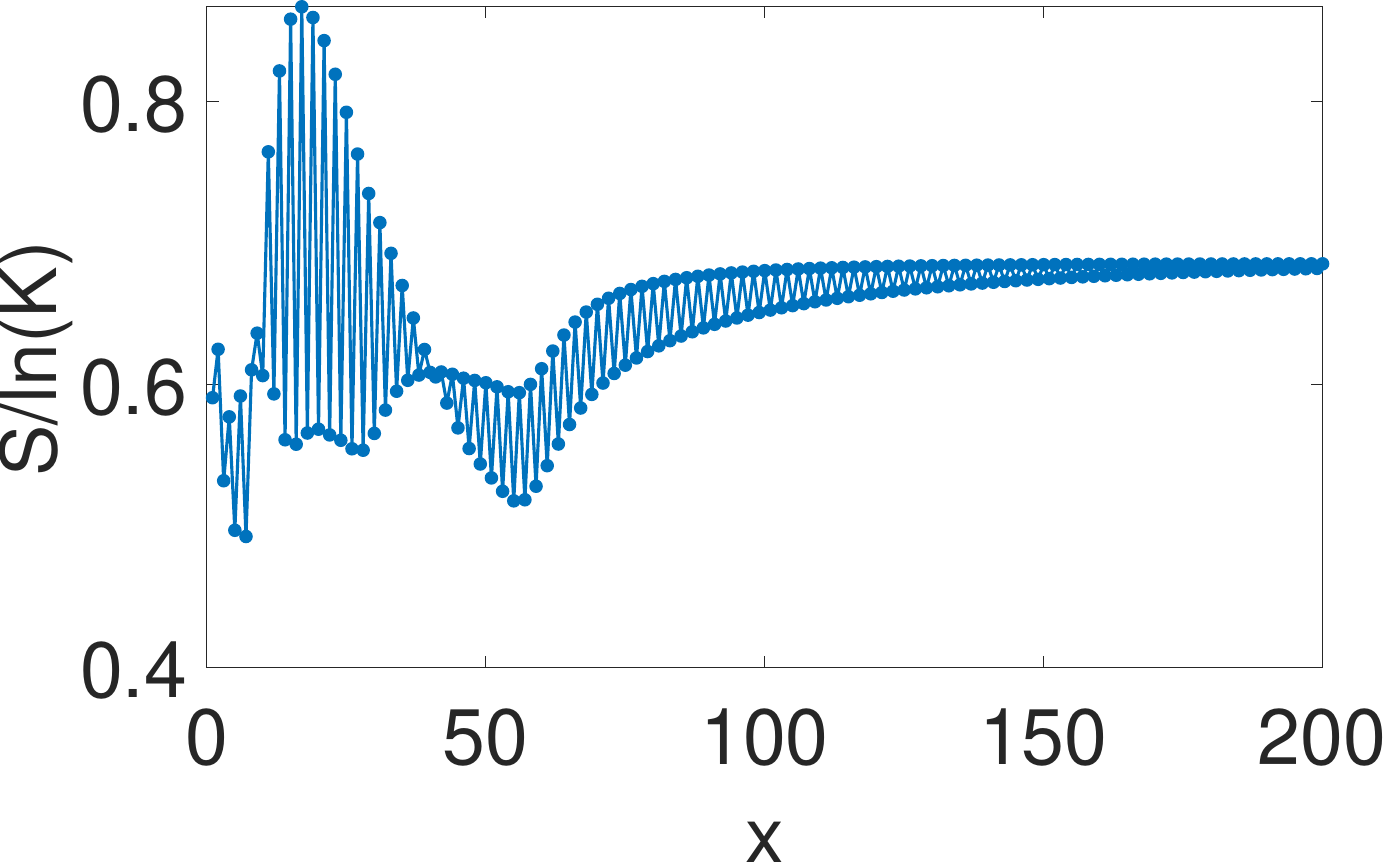}} % 1 20
    \subfigure[]{\includegraphics[width=0.48\linewidth]{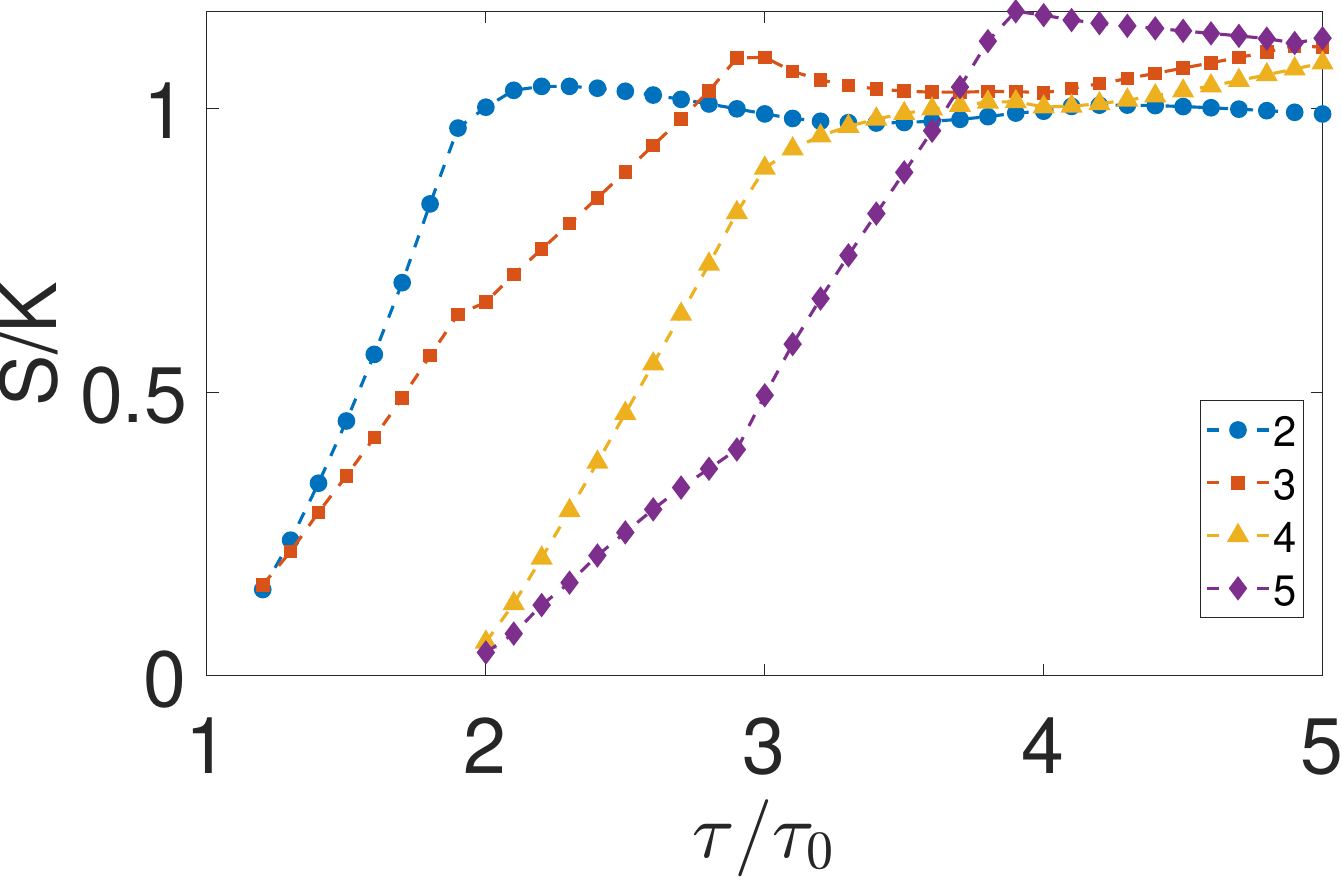}} % 2 20
    \subfigure[]{\includegraphics[width=0.48\linewidth]{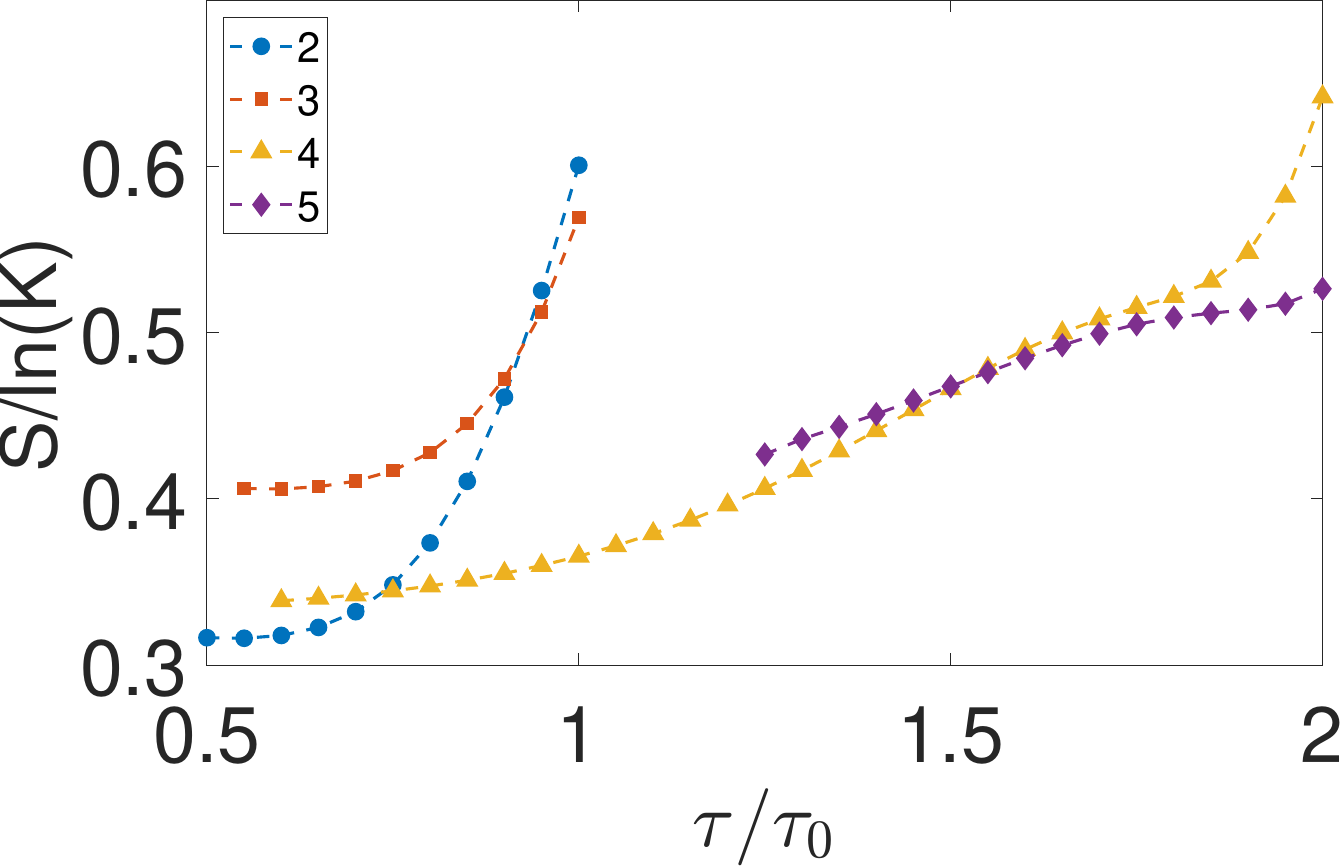}} % 2 6
    \caption{(a) The 2-site TDEE for gapless free fermions with $\tau=2\tau_0,1.5\tau_0$ and $x=1$. The legend indicates $\tau/\tau_0$.
    (b) The 2-site TDEE for gapless free fermions with $\tau=\tau_0,0.5\tau_0$ and $x=1$. The legend indicates $\tau/\tau_0$. The logarithmic coefficients are $0.592$ and $0.337$, respectively.
    (c) The volume-law coefficient for $\tau=1.5\tau_0$ as a function of the distance $x$ between two sites, which saturates to $0.72$ at large $x$.
    (d) The logarithmic coefficient for $\tau=\tau_0$ as a function of $x$, which saturates to $0.68$ (approximately $2/3$) at large $x$.
    (e) The volume-law coefficient for $\tau>\tau_c$ with widths $y=2,3,4,5$.
    (f) The logarithmic coefficient for $\tau<\tau_c$ with widths $y=2,3,4,5$.
    The coefficients in (c)--(f) are obtained by linear fits over $1\le K\le 20$ for (c) and (d), $2\le K\le20$ for (e), and $2\le K\le 6$ for (f).
    All calculations use $L=1000$.
    % \bl{What are the $K$ value for (c)-(f)?} {\color{red}[XS]: To be confirmed later.}
    }
    \label{fig:gapless_2site}
\end{figure}

For $x=1$, the region $A$ can be viewed as a thick time-direction interval of length $T=K\tau$. Accordingly, this $2$-site TDEE $S(A)$ behaves qualitatively similar to that of the ``thin" time-direction interval in \cref{eq:gapless_log_law}: 

(i) for $\tau>\tau_0$, $S(A)= \alpha_V K+\text{Const.}$ shows a volume law with $K$ (see \cref{fig:gapless_2site}(a)), where the coefficient $\alpha_V$ increases for $\tau\in(\tau_0,2\tau_0)$, and saturates for $\tau>2\tau_0$. 

(ii) For $\tau=\tau_0$, $S(A)$ shows a logarithmic law with $K$ (see \cref{fig:gapless_2site}(b)), with the coefficient $\alpha_\text{log}\approx 0.6$. Note that this value is smaller than $2/3$, namely, smaller than $2$ times of the single-site TDEE. This implies non-negligible correlations between the two neighboring time intervals in $A$.

(iii) For $\tau<\tau_0$, $S(A)$ shows a logarithmic law with $K$ with the coefficient $\alpha_\text{log}$ further decreasing towards approximately $1/3$, as shown in \cref{fig:gapless_2site}(b) and (f). This is indeed $2$ times of the single-site TDEE, despite the correlations between the two neighboring time intervals in $A$.

For $x>1$, one expects that the correlation between the two time intervals decreases as $x$ increases. Indeed, \cref{fig:gapless_2site}(c) and (d) show how the $2$-site TDEE $S(A)$ varies and saturates as the distance $x$ between two time intervals increases, for the volume law coefficient $\alpha_V$ at $\tau=1.5\tau_0$ and the logarithmic coefficient $\alpha_\text{log}$ at $\tau=\tau_0$, respectively. In both cases, the coefficient at large $x$ saturates to $2$ times of that of the single-site TDEE, as expected. Similarly, it was shown that the entanglement entropy of two disjoint spatial intervals approaches $2$ times of that of a single spatial interval \cite{Calabrese2009Entanglement,Pasquale2009Entanglement}.

More generally, we calculate the ground state TDEE of a spacetime region $A$ being the combination of $y$ time intervals of length $T=K\tau$ on $y$ consecutive sites from $0$ to $y-1$, as illustrated in \cref{fig:config_x_t_2}(b). Concretely, we choose $A=\{(x_n,t_n)|0\le n\le yK-1\}$, where the spacetime coordinates are given by $x_{yn+i}=i$ and $t_{yn+i}=n\tau$ ($0\le i\le y-1$). Generically, such a TDEE $S(A)$ always shows a volume law (logarithmic law) when $\tau$ is above (below) a threshold $\tau_c$. As shown in \cref{fig:gapless_2site}(e) which calculates the TDEE up to $y\le 5$, we find this threshold is $\tau_c=\tau_0$ for $y\le 3$, and is $\tau_c=2\tau_0$ for $y=4$ and $y=5$. For $\tau>\tau_c$, \cref{fig:gapless_2site}(e) shows that the volume law coefficient $\alpha_V\approx S(A)/K$ saturates at larger $\tau$ for larger $y$. For $\tau<\tau_c$, \cref{fig:gapless_2site}(f) shows how the logarithmic coefficient $\alpha_\text{log}\approx S(A)/\ln K$ decreases (for finite $K$) as $\tau$ decreases. In the $\tau\rightarrow 0$ limit, \cref{fig:gapless_2site}(f) suggests that $\alpha_\text{log}$ of different number of sites $y$ saturate to different values.

It is instructive to compare the above $y$-site TDEE in the continuum time limit $\tau\rightarrow 0$ with the timelike tube theorem~\cite{borchers1961vollstandigkeit,Witten2018APS,strohmaier2023timelike} in relativistic quantum field theories. The timelike tube theorem states that, a timelike tube, or equivalently a time-direction interval of finite width, contains the same amount of information as the spatial region within its relativistic causal diamond. When applied to 1D relativistic gapless fermions, this would imply that the entanglement entropy in a timelike tube of length $T$ should scale as $S(A)=\frac{1}{3}\ln T+\text{Const}.$ with a universal coefficient $\alpha_\text{log}=1/3$, the same as the entanglement entropy in its causal diamond spatial interval given by the Cardy-Calabrese formula. In our calculations, the $y$-site time interval $A$ with $y>1$ in the continuum time $\tau\rightarrow 0$ can be viewed as a time tube. However, \cref{fig:gapless_2site}(f) suggests that $\alpha_\text{log}$ is $y$-dependent, although the $y=2$ case gives $\alpha_\text{log}=1/3$ in agreement with the prediction from the timelike tube theorem. We conjecture this is because of the loss of Lorentz symmetry in the non-relativistic lattice model we consider. 

We note that a single-site time interval would have zero width, which is not a timelike tube. Indeed, as we argued below \cref{eq:dA-upperlimit}, a single-site time interval can only access the mirror symmetric single-particle sub-Hilbert space of its causal diamond (i.e. Lieb-Robinson bound) spatial interval, and thus the TDEE in the $\tau\rightarrow 0$ limit has a coefficient $\alpha_\text{log}=1/6$, only one half of that of the Cardy-Calabrese formula for spatial entanglement entropy.
% ~\cite{kac2005solitons,Yasuyuki2006Classification}

\section{TDEE of 1D Band Insulator (Disorderless)}
\label{TDEE of 1D Band Insulator (Disorderless)}

\begin{figure}
    \centering
    \subfigure[]{\includegraphics[width=0.49\linewidth]{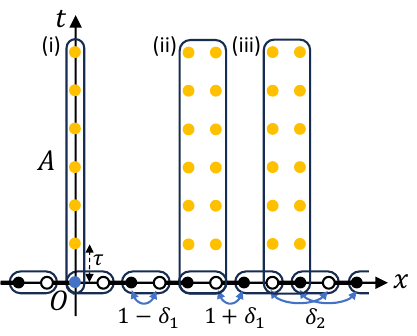}}
    \subfigure[]{\includegraphics[width=0.49\linewidth]{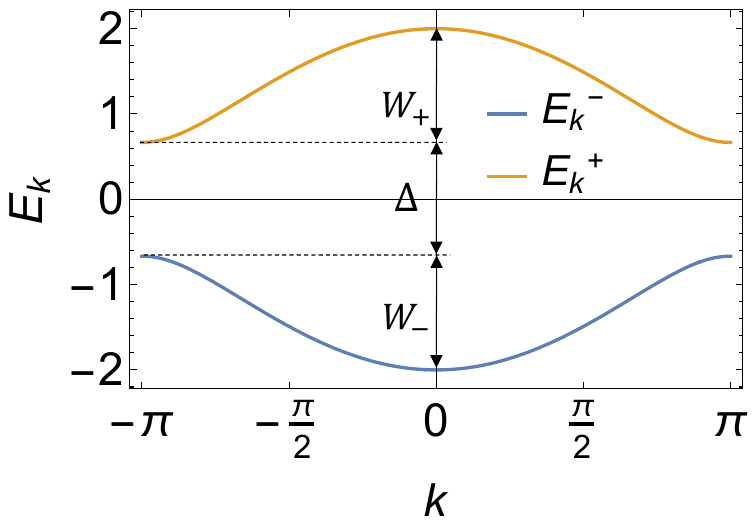}}
    \caption{(a) Schematic illustration of the subsystems used to compute the band-insulator TDEE in three cases: (i) single-site TDEE, and two-site TDEE in the (ii) topological and (iii) trivial phases (for $\delta=\delta_1>0$ and $\delta_2=0$). 
    (b) Energy dispersion of the particle-hole-symmetric SSH model with $\delta_1=\delta=\frac{1}{3}$ and $\delta_2=0$.}
    \label{fig:PH_symmetric_band}
\end{figure}

In this section, we investigate the TDEE of 1D gapped band insulator ground states, which has not been studied before. For simplicity, we restrict ourselves to two-band free fermion models with translation symmetry, where the lower (valence) band is fully occupied, and the upper (conduction) band is fully empty. We assume the conduction band and valence band have bandwidths $W_+$ and $W_-$, respectively, and there is a gap $\Delta$ between them (see \cref{fig:PH_symmetric_band}(b)). We define a total bandwidth $E_0=W_++W_-$. We then define three characteristic times
\begin{equation}\label{eq:insulator-characteristic-times}
\tau_0^+=\frac{\pi}{W_+}\ , \quad \tau_0^-=\frac{\pi}{W_-}\ ,\quad \tau_0=\frac{2\pi}{E_0}=\frac{2\tau_0^+\tau_0^-}{\tau_0^++\tau_0^-}\ .
\end{equation}
In this way, $E_0$ and $\tau_0$ reduce to those we defined for gapless free fermions in \cref{sec:TDEE of gapless free fermions} when $\Delta=0$.

\subsection{The particle-hole symmetric SSH model}
\label{sec:SSH model with the particle-hole symmetry}

As a concrete example, we consider the Su-Schrieffer-Heeger (SSH) model with PBC, which has the following Hamiltonian:
\begin{equation}
\label{eq:SSH_hamiltonian_unit_cell}
    H=-\sum_i\big[(1-\delta)c_{i,1}^\dagger c_{i,2}+(1+\delta)c_{i,2}^\dagger c_{i+1,1}+h.c.\big],
\end{equation}
where $\delta$ controls the dimerization strength, while $c_{i,1}$ and $c_{i,2}$ are the two sites in the $i$-th unit cell. Particularly, the SSH model has a particle-hole symmetry, which enforces the bandwidths $W_+=W_-=E_0/2$, and thus $\tau_0=\tau_0^+=\tau_0^-$ in \cref{eq:insulator-characteristic-times}.

For a time-direction interval with time step $\tau$ and length $T=K\tau$, a particle-hole symmetric band insulator has two dimensionless parameters, $\tau/\tau_0$ and $\tau\Delta$. Both parameters affect the scaling behaviors of the ground state TDEE. In this subsection, we focus on the single-site TDEE (the time interval (i) in \cref{fig:PH_symmetric_band}(a)), where the site is chosen as site $1$ of the $0$-th unit cell without loss of generality.

\begin{figure}
    \centering
    % \subfigure[]{\includegraphics[width=0.45\linewidth]{ssh_tauR3.pdf}}
    \subfigure[]{\includegraphics[width=0.45\linewidth]{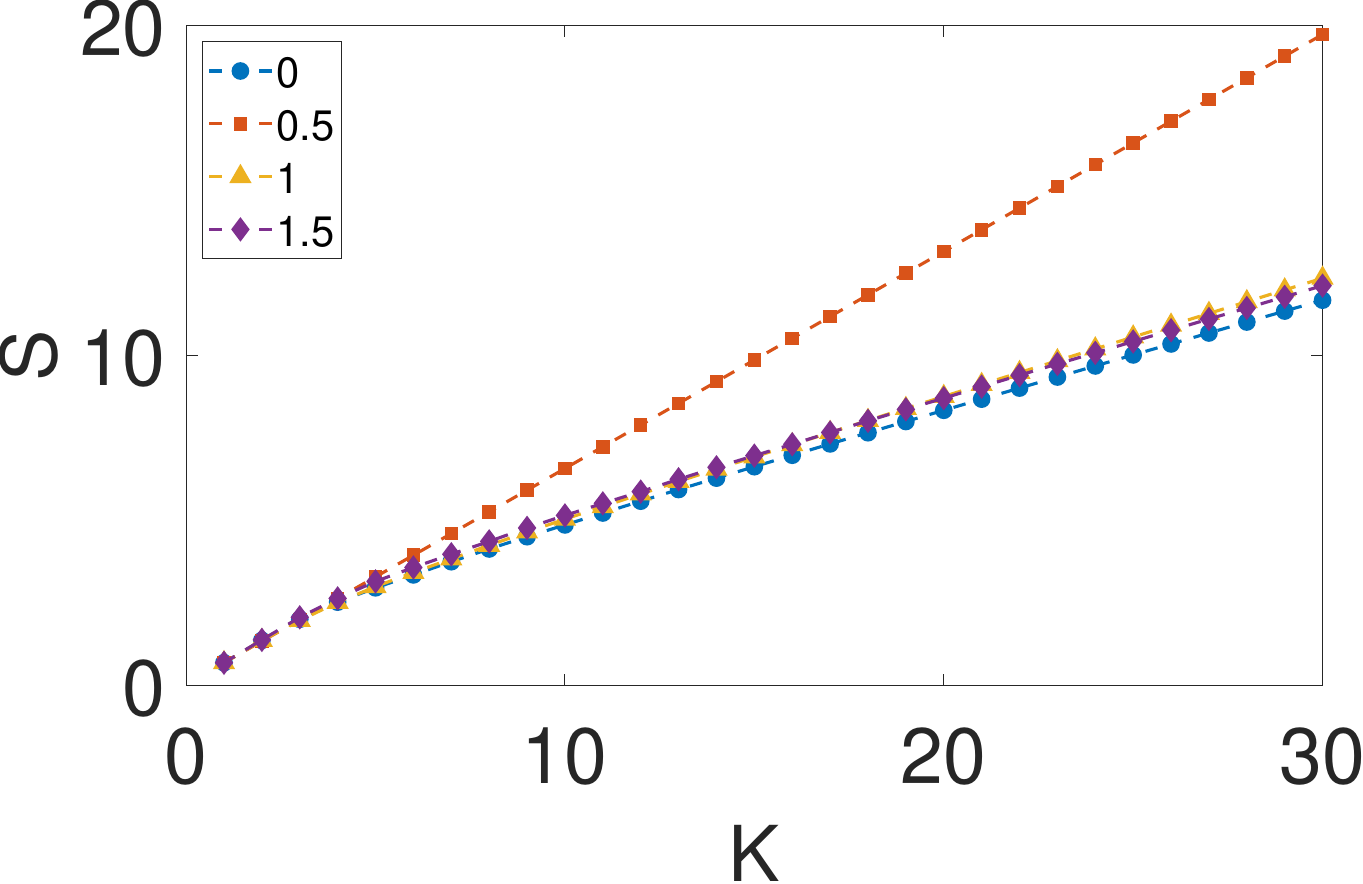}}
    \subfigure[]{\includegraphics[width=0.45\linewidth]{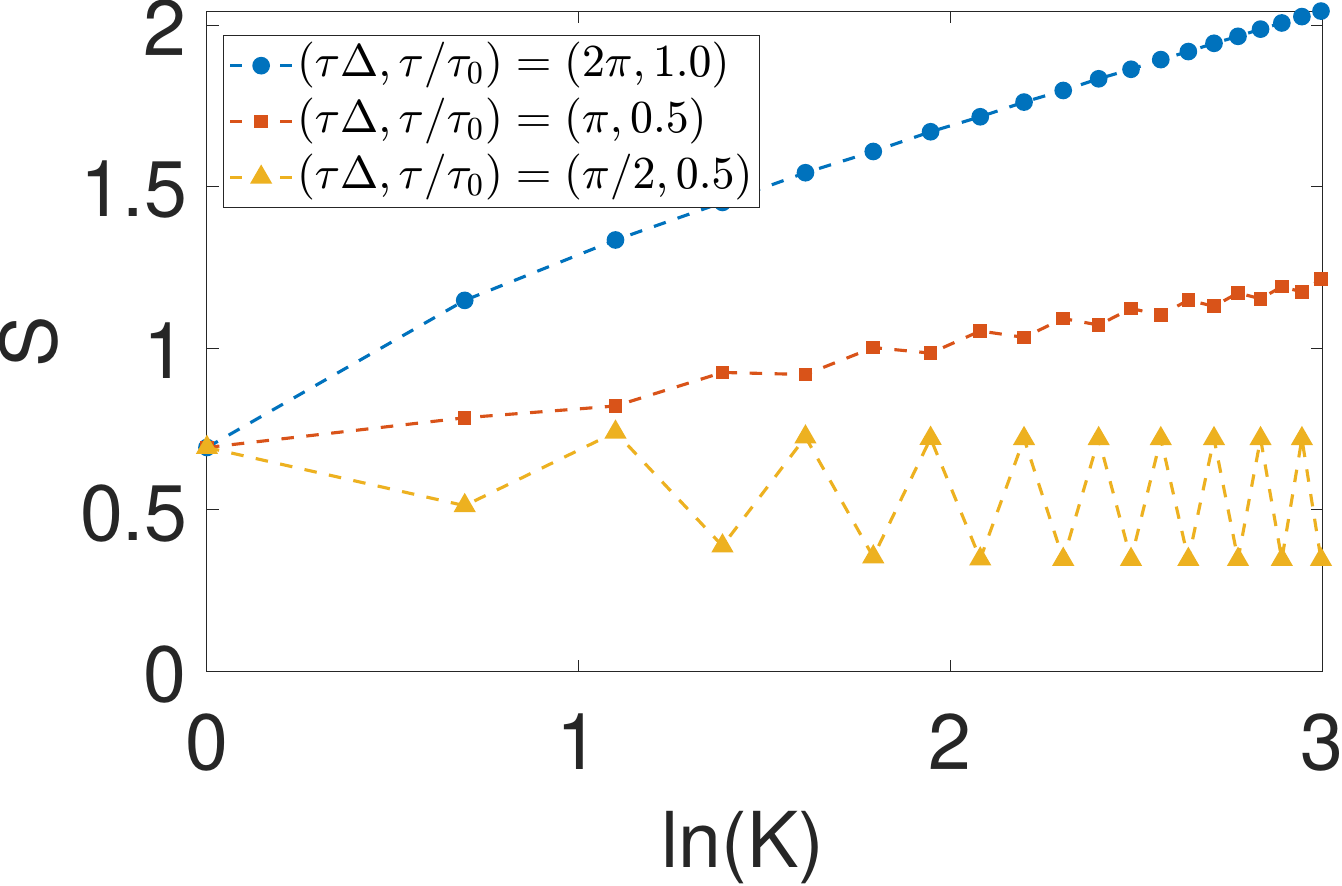}}
    % \subfigure[]{\includegraphics[width=0.45\linewidth]{Figure/ssh_tauR1.pdf}}
    \subfigure[]{\includegraphics[width=0.45\linewidth]{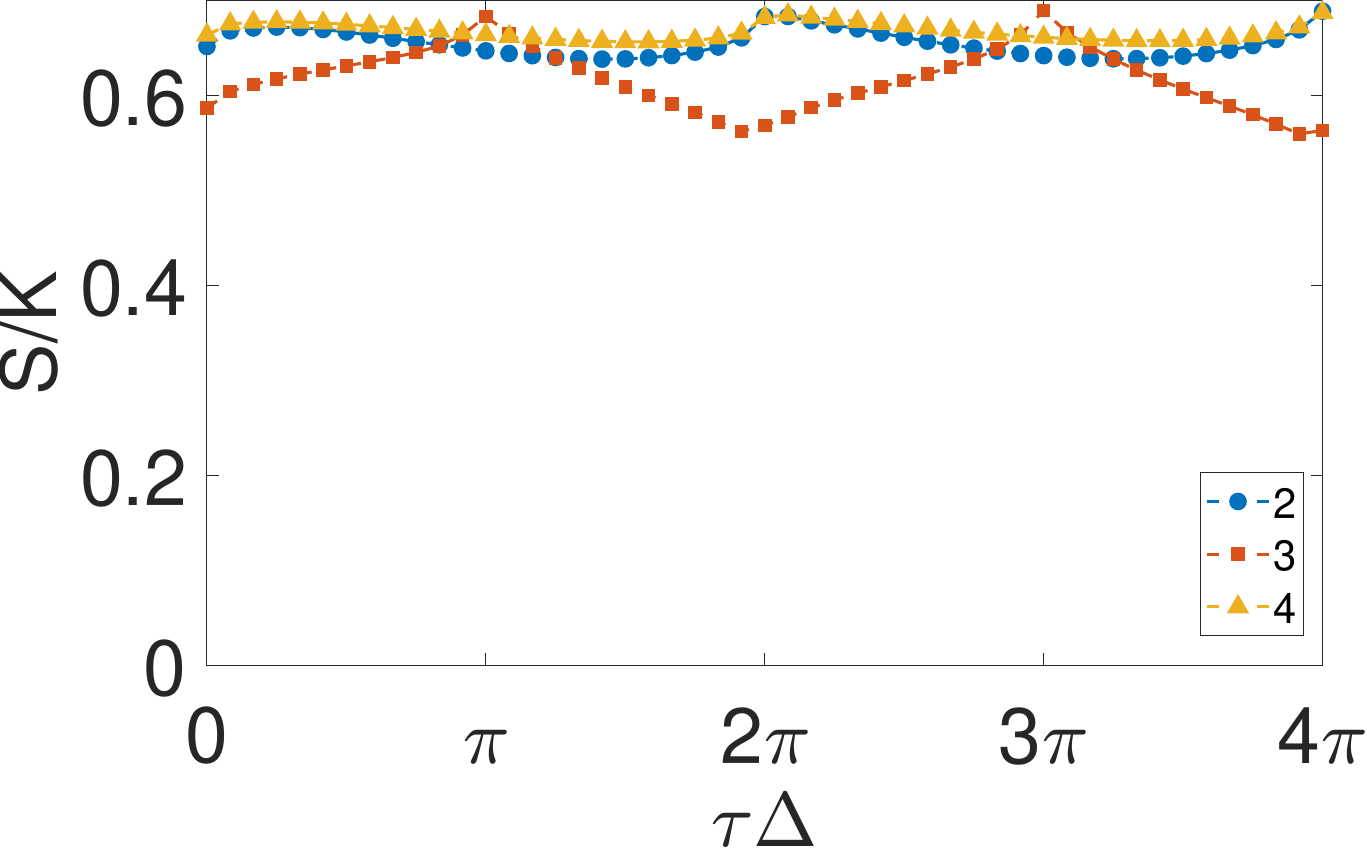}}
    \subfigure[]{\includegraphics[width=0.45\linewidth]{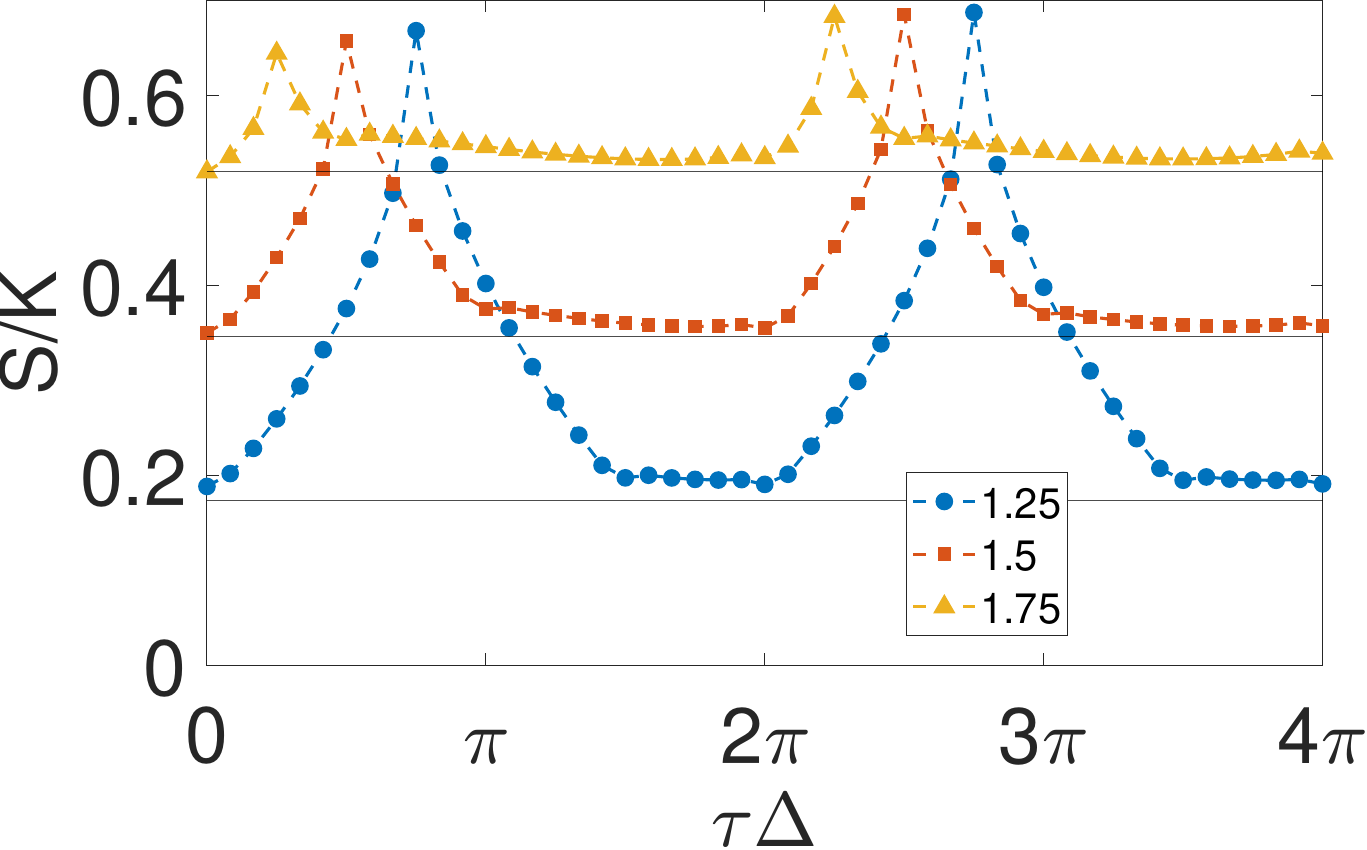}}
    \subfigure[]{\includegraphics[width=0.45\linewidth]{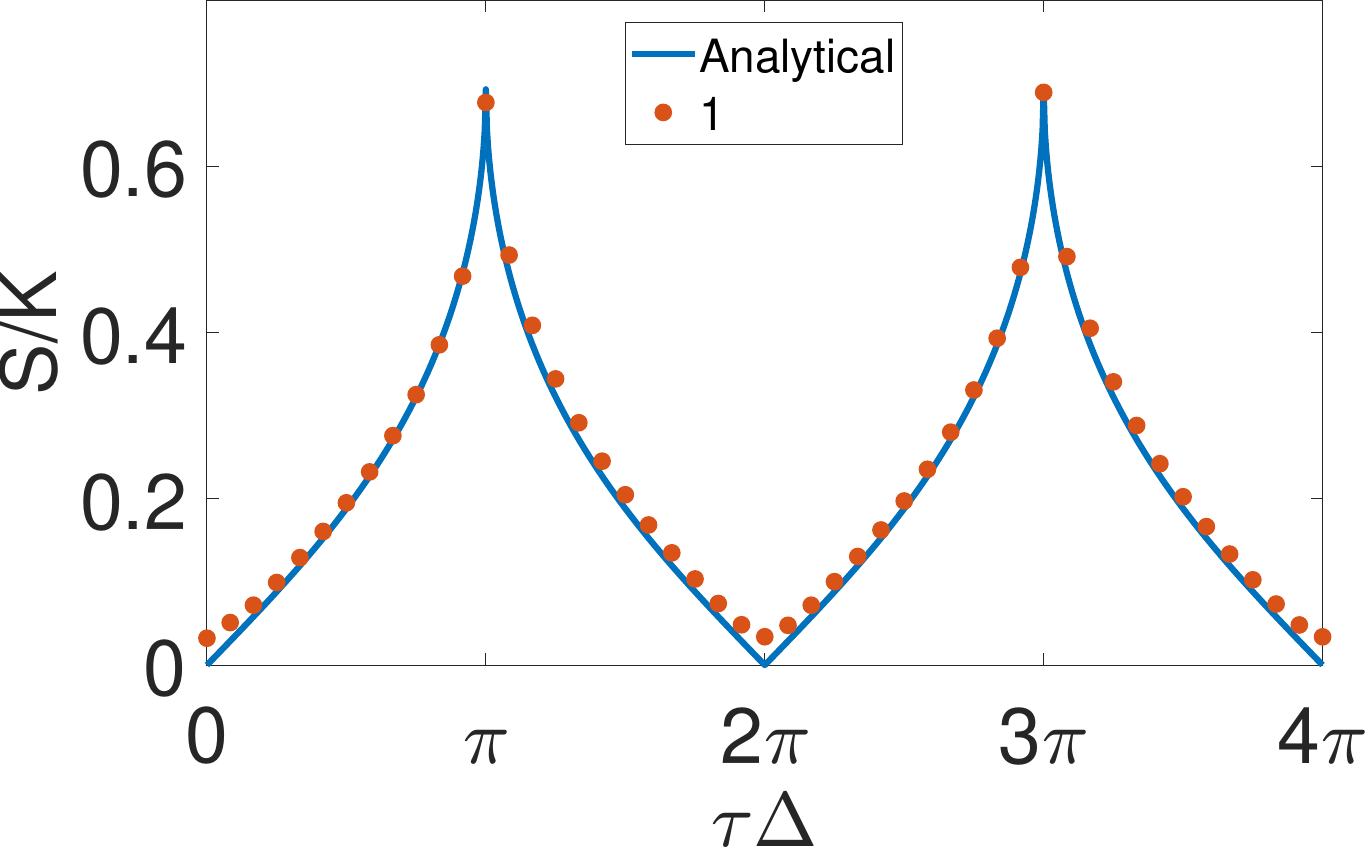}}
    % \subfigure[]{\includegraphics[width=0.45\linewidth]{Figure/ssh_tauR1-2.pdf}}
    \subfigure[]{\includegraphics[width=0.45\linewidth]{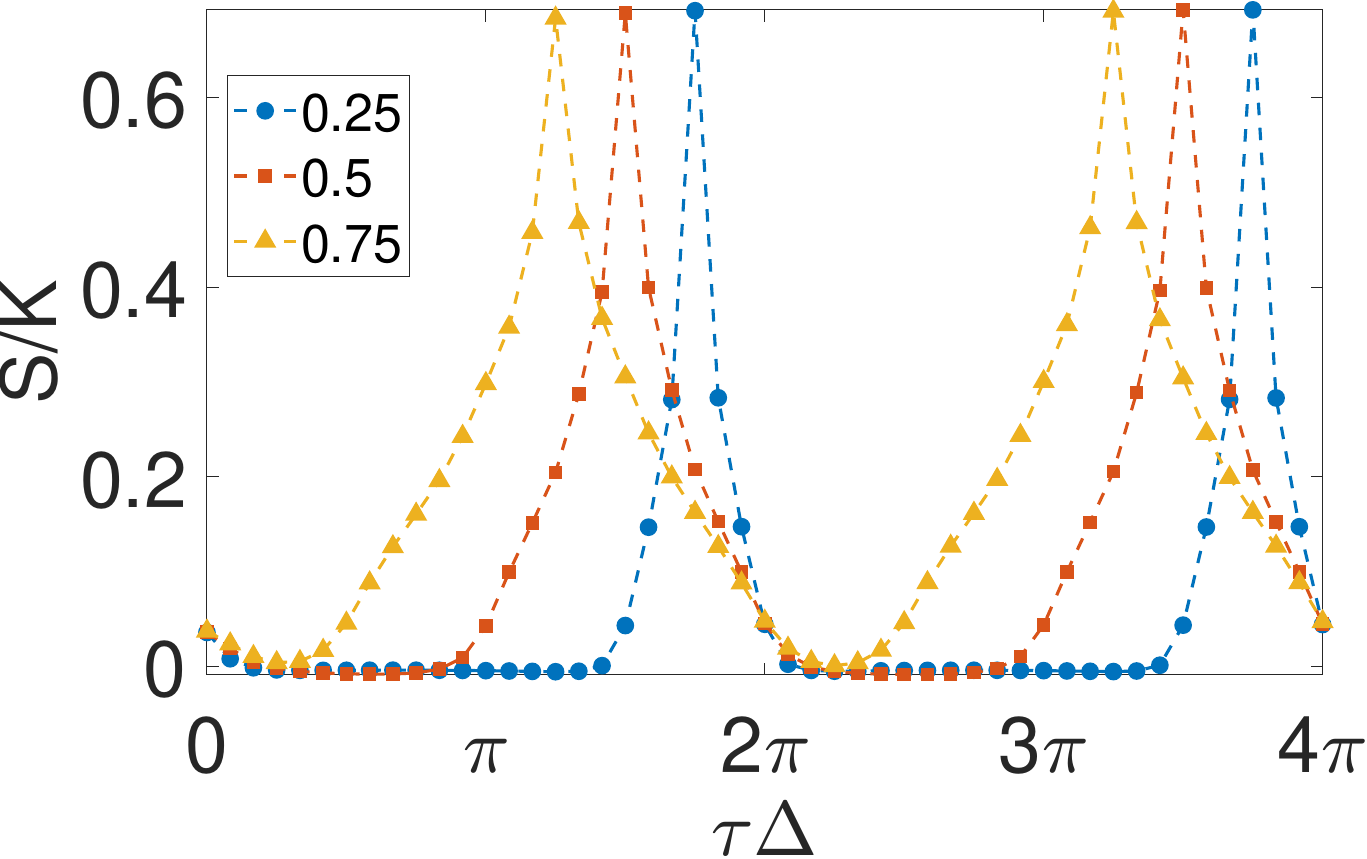}}
    \caption{
    % (a) The TDEE for $\frac{\tau}{\tau_0}=3$ and $L=1000$, which show the volume-law entanglement. $\tau\Delta$ is shown in legend.
    (a) The TDEE in the volume-law regime with $\frac{\tau}{\tau_0}=\frac{3}{2}$ and $\tau\Delta/\pi=0,\frac{1}{2},1,\frac{3}{2}$ (legend).
    (b) The TDEE in the log-law and area-law cases. The slopes are $0.365, 0.181, 0$, respectively.
    (c) The volume-law term prefactor $S/K$ for the TDEE, with $\tau\Delta\in[0,4\pi]$, $\frac{\tau}{\tau_0}=2, 3, 4$ (legend) and $L=1000$.
    (d) The volume-law term prefactor $S/K$ for the TDEE, with $\tau\Delta\in[0,4\pi]$, $\frac{\tau}{\tau_0}=\frac{5}{4}, \frac{3}{2},\frac{7}{4}$. The horizontal lines are $(1/4,1/2,3/4)\cdot\ln{2}$.
    (e) The volume-law term prefactor $S/K$ for the TDEE, with $\tau\Delta\in[0,4\pi]$, $\frac{\tau}{\tau_0}=1$.
    The blue line shows the analytical result in Eq.~\eqref{eq:detailed_volume}, with details in Appendix~\ref{app:restricted_support}.
    % (g) The TDEE for $\frac{\tau}{\tau_0}=\frac{1}{2}$ and $\tau\Delta=0,\frac{\pi}{2},\pi,\frac{3\pi}{2}$ (legend).
    (f) The volume-law term prefactor $S/K$ for the TDEE, with $\tau\Delta\in[0,4\pi]$, $\frac{\tau}{\tau_0}=\frac{1}{4}, \frac{1}{2}, \frac{3}{4}$ (legend).}
    \label{fig:gapped_PH}
\end{figure}

For the insulator ground state of \cref{eq:SSH_hamiltonian_unit_cell}, we numerically calculate the TDEE $S(A)$ as a function of $\tau\Delta$ for a fixed $\tau/\tau_0$ (which requires tuning $\tau$ and $\delta$ simultaneously). As shown in \cref{fig:gapped_PH}(a) and (b), we find the TDEE can generically exhibit volume law, logarithmic law or area law in $K$, depending on the two dimensionless parameters. Our numerical observation is summarized as follows.

First of all, for any fixed $\tau/\tau_0$, the TDEE $S(A)$ is approximately periodic in $\tau\Delta$ with a period $2\pi$. This can be seen from the volume law coefficient $\alpha_V\approx S(A)/K$ as a function of $\tau\Delta$ at different fixed $\tau/\tau_0$ plotted in \cref{fig:gapped_PH}(c)-(f).

Then, the behavior of the TDEE can be divided into the following four regimes:

(i) $\tau\ge 2\tau_0$. In this case, the TDEE always shows a volume-law
\begin{equation}\label{eq:volume-band-ins}
S(A)\approx \alpha_V K \ ,
\end{equation}
where the volume law coefficient $\alpha_V$ approaches $\ln 2$ and is nearly independent of $\tau\Delta$, as shown in Fig.~\ref{fig:gapped_PH} (c).

(ii) $\tau_0<\tau<2\tau_0$. In this case, the TDEE still always shows a volume-law of the form \cref{eq:volume-band-ins} with $\alpha_V>0$. However, the coefficient $\alpha_V$ now depends on $\tau\Delta$. As shown in Fig.~\ref{fig:gapped_PH} (d), for a fixed $\tau/\tau_0$, we find $\alpha_V$ is a triangular-shape function for $\tau\Delta$ within the interval $[0,a]$ (mod $2\pi$), and saturates to a plateau in interval $[a,2\pi]$ (mod $2\pi$), where $a\approx 2\pi(2-\frac{\tau}{\tau_0})$. The maximum of $\alpha_V$ in the triangular regime $[0,a]$ is approximately $\ln 2$, while the plateau value of $\alpha_V$ in $[a,2\pi]$ is approximately $(\frac{\tau}{\tau_0}-1)\ln 2$.

(iii) $\tau=\tau_0$. At this critical point, as shown in Fig.~\ref{fig:gapped_PH} (e), the volume law coefficient $\alpha_V$ as a function of $\tau\Delta$ exhibits a full triangle shape with period $2\pi$, where $\alpha_V=0$ at $\tau\Delta=0$ and $2\pi$, and $\alpha_V$ reaches its maximum, approximately $\ln 2$, at $\tau\Delta=\pi$. At $\tau\Delta=0\ (\text{mod }2\pi)$ where $\alpha_V$ vanishes, we find the TDEE follows a logarithmic law
\begin{equation}
S(A)=\frac{1}{3}\ln K +\text{Const.}\ ,\quad \tau\Delta=0\ (\text{mod }2\pi)
\end{equation}
This agrees with the result of gapless systems with $\Delta=0$ in \cref{eq:gapless_log_law}. Non-trivially, for gapped insulators with gap $\Delta=2n\pi/\tau_0$ with integer $n>0$, the TDEE also shows a logarithmic law.

(iv) $\tau<\tau_0$. The volume law coefficient $\alpha_V$ in this case is shown in Fig.~\ref{fig:gapped_PH} (f). The behavior of TDEE for $\tau<\tau_0$ can be summarized as
\begin{equation}
\label{eq:gapped_volume_log_law}
    S(A)=\begin{cases}
    \text{Const.}\ ,\qquad\  \tau\Delta\in (0,a')\\
    &\\
    \dfrac{1}{6}\ln{K} + \text{Const.}\ ,\qquad \tau\Delta=0\text{ or }a'\\
    &\\
    \alpha_V K + \text{Const.}\ ,\qquad \tau\Delta\in (a',2\pi)
    \end{cases}
\end{equation}
where $a'\approx 2\pi(1-\frac{\tau}{\tau_0})$, and the pattern has a $2\pi$ period in $\tau\Delta$. Namely, the TDEE shows area law and volume law for $\tau\Delta$ in the interval $(0,a')$ and $(a',2\pi)$, respectively. For $\tau\Delta=0$ or $a'$, the TDEE shows a logarithmic law similar to the gapless case in \cref{eq:gapless_log_law}.

Therefore, unlike the spatial entanglement entropy which always shows an area law, the TDEE can show different behaviors in different regimes. In the following, we give an intuitive picture for understanding the TDEE shown above, and predict the TDEE for any $\frac{\tau}{\tau_0}$ and $\tau\Delta$.

\subsection{Analytical understanding in particle-hole symmetric cases}
\label{sec:Analytical understanding for gapped free fermions with the particle-hole symmetry}

For the above SSH model, we first Fourier transform into a momentum space basis 
$c_{i,a}=\frac{1}{\sqrt{L}}\sum_{j=0}^{L-1} e^{{\rm i}k_j i} \widetilde{c}_{k_j,a}$ 
with $k_j=\frac{2\pi}{L}j$. 
The Hamiltonian can then be written as $H=-\sum_{j=0}^{L-1}\psi_{k_j}^\dagger h_{k_j} \psi_{k_j}$, where $h_{k_j}=((1-\delta)+(1+\delta)\cos{k_j})\sigma_x+(1+\delta)\sin{k_j}\sigma_y$, and $\psi_{k_j}=(\widetilde{c}_{k_j,1}, \widetilde{c}_{k_j,2})^T$. To diagonalize $h_k$, we introduce a further unitary transformation
\begin{equation}
\label{eq:unitary_transf_U}
    U_{\theta_k}=\frac{1}{\sqrt{2}}\begin{pmatrix}
        e^{-{\rm i}\theta_k} & e^{-{\rm i}\theta_k}\\
        1 & -1
    \end{pmatrix},
\end{equation}
where $\tan{\theta_k}=\frac{(1+\delta)\sin{k}}{(1-\delta)+(1+\delta)\cos{k}}$.
This yields $U_{\theta_k}^\dagger h U_{\theta_k}={\rm Diag}(E_k, -E_k)$,
% And $E_k=\sqrt{((1-\delta)+(1+\delta)\cos{k})^2+(1+\delta)^2\sin^2{k}}$.
where $E_k>0$ is the eigenenergy. 
We define fermion operators $\widetilde{c}_{k,a}=U_{\theta_k} \widetilde{c}_{k,\pm}$, where $\widetilde{c}_{k,+}$ and $\widetilde{c}_{k,-}$ label the conduction and valence bands, in which basis the Hamiltonian is diagonal.
% \[
% H=\sum_{j=0}^{L-1} E_{k_j}(c_{k_j,\alpha_1}^\dagger c_{k_j,\alpha_1}-c_{k_j,\alpha_2}^\dagger c_{k_j,\alpha_2}).
% \]
%Thus, the total unitary transformation is $U\otimes U_{\theta_k}^{\dagger}$, where $(U^\dagger)_{ij}=\frac{1}{\sqrt{L}}e^{{\rm i}k_i j}$.

Focusing on the first unit cell, under time evolution we have 
\begin{equation}
\label{eq:time_evolution_c0_12}
\begin{split}
    c_{0,1}(t)=&\frac{1}{\sqrt{L}}\sum_{k_j}\frac{1}{\sqrt{2}}e^{-{\rm i}\theta_{k_j}}\big(e^{-{\rm i}E_{k_j}t}\widetilde{c}_{k_j,+}+e^{{\rm i}E_{k_j}t}\widetilde{c}_{k_j,-}\big),\\
    c_{0,2}(t)=&\frac{1}{\sqrt{L}}\sum_{k_j}\frac{1}{\sqrt{2}}\big(e^{-{\rm i}E_{k_j}t}\widetilde{c}_{k_j,+}-e^{{\rm i}E_{k_j}t}\widetilde{c}_{k_j,-}\big).
\end{split}
\end{equation}
The anti-commutation matrix $B$ and the correlation matrix $C$ for the first site are then given by
\begin{eqnarray}
    \label{eq:B_gapped_01}
    B_{mn}=&&\{c_{0,1}(m\tau),c_{0,1}^\dagger(n\tau)\} \nonumber\\
    =&&\frac{1}{L}\sum_{k_j}\frac{1}{2}\big(e^{{\rm i}E_{k_j}(m-n)\tau}+e^{-{\rm i}E_{k_j}(m-n)\tau}\big),\\
    \label{eq:C_gapped_01}
    C_{mn}=&&\langle c_{0,1}^\dagger(m\tau)c_{0,1}(n\tau)\rangle
    =\frac{1}{L}\sum_{k_j}\frac{1}{2}e^{-{\rm i}E_{k_j}(m-n)\tau}.
\end{eqnarray}
In fact, for a general system with particle-hole symmetry (including here), one can prove a simple relation $B = 2\,{\rm Re}\,C$.

For the SSH model in Eq.~\eqref{eq:SSH_hamiltonian_unit_cell}, the energy ranges of the two bands are 
$E\in[-2,-2\delta]\cup[2\delta,2]$. 
To simplify our discussion, we shift and rescale the energy spectrum so that the valence band lies in $E\in[-1,0]$ and the conduction band lies in $E\in[\Delta,\Delta+1]$. 
% \bl{I set $W=1$ in this section, to reduce the number of notations here. Generic $W$ only occurs in the next section. Also, please change 'negative band' to 'valence band', 'positive band' to 'conduction band'.}
This yields a characteristic time (see \cref{eq:insulator-characteristic-times})
\begin{equation}
\tau_0=\tau_0^+=\tau_0^-=\pi\ .
\end{equation}

From Eqs. \ref{eq:B_gapped_01} and \ref{eq:C_gapped_01}, one can show (see Appendix~\ref{Gapped free fermion with SSH Hamiltonian} as well as Eqs.~\eqref{eq:Toeplitz_matrix_translation_symm_C_FT} and \eqref{eq:Toeplitz_matrix_translation_symm_B_FT}) that the symbols we need are given by $B^*(\theta)=C_+(\theta)+C_-(\theta)$ and $D(\theta)=C_-(\theta)/B^*(\theta)$, where
\begin{equation}
\label{eq:Toeplitz_matrix_translation_symm_B_FT_pm}
\begin{split}
    C_{\pm}(\theta)=\sum_k\frac{\sin{[L(E_k^\pm\tau-\theta)/2]}}{L\sin{[(E_k^\pm\tau-\theta)/2]}}\ .
\end{split}
\end{equation}
Here $E_k^-\in[-1,0]$ and $E_k^+\in[\Delta,\Delta+1]$. It is straightforward to see that $C_{\pm}(\theta)$ has a $2\pi$-periodicity in $\tau\Delta$, and so does all the quantities derived from them.

\begin{figure*}[t]
    \centering
    \subfigure[]{\includegraphics[width=1\linewidth]{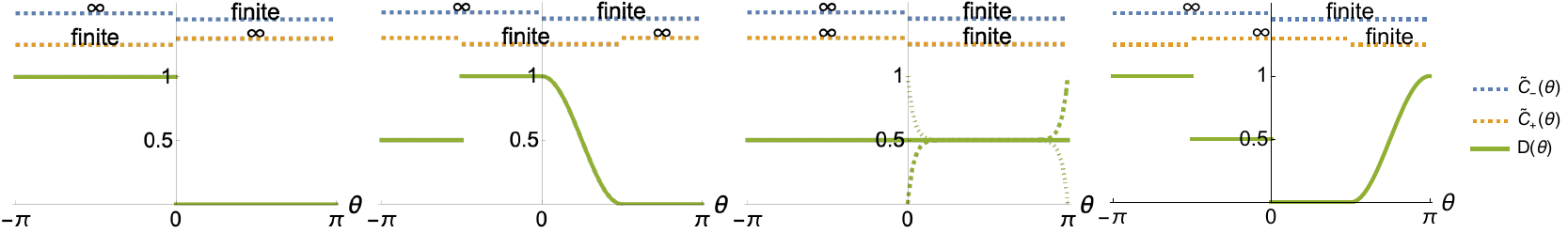}}
    \subfigure[]{\includegraphics[width=0.28\linewidth]{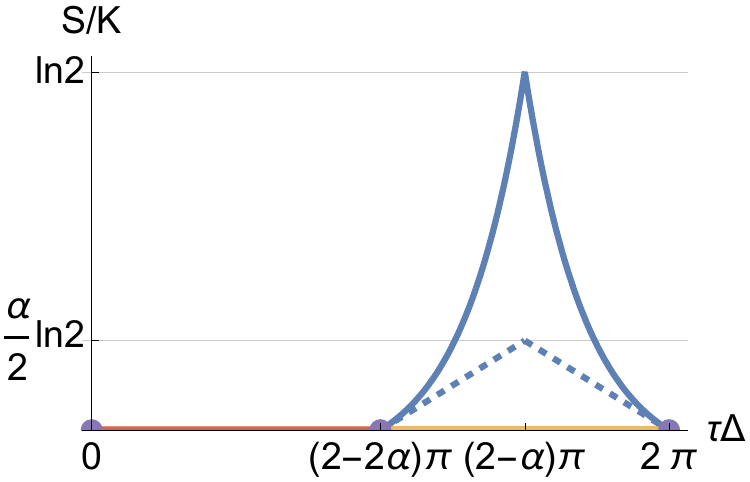}}
    \subfigure[]{\includegraphics[width=0.28\linewidth]{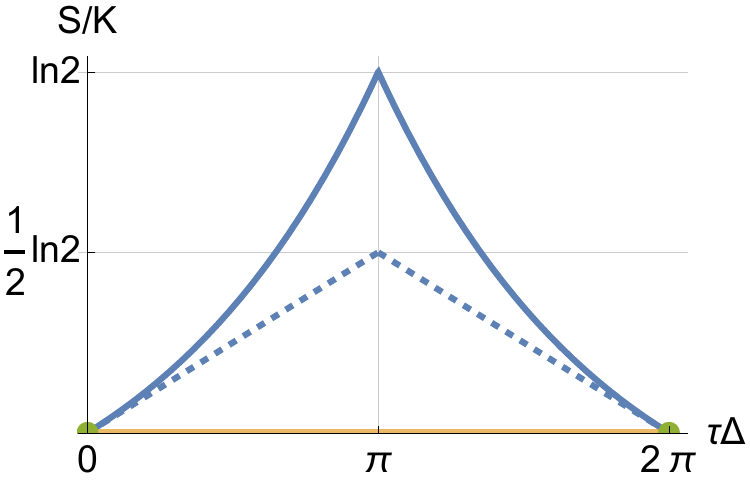}}
    \subfigure[]{\includegraphics[width=0.42\linewidth]{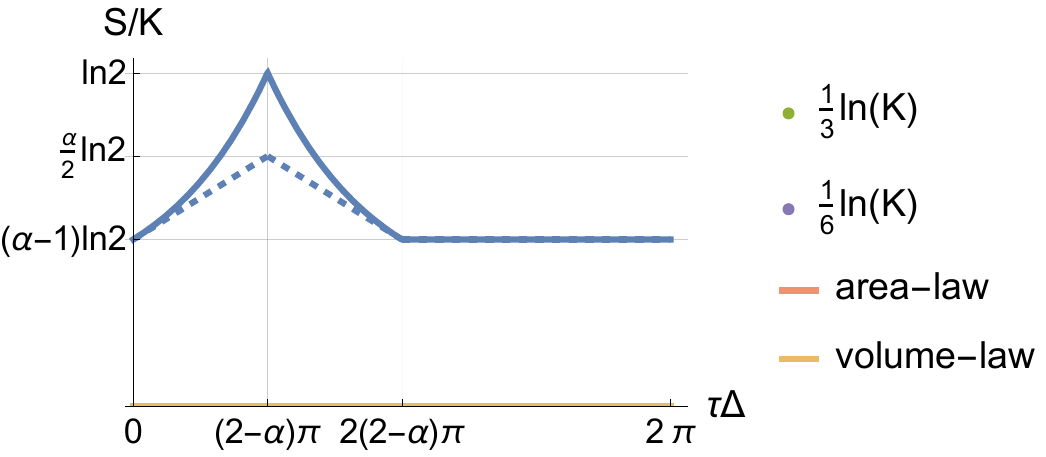}}
    \caption{(a) The Fourier transformed correlation matrix $\tilde{C}_\pm(\theta)$ (yellow and blue dashed lines) and corresponding $D(\theta)$ (green line) for $\tau=\tau_0$ and different $\tau\Delta$ with particle-hole symmetry. 
    $\tau\Delta=0$, $0<\tau\Delta<\pi$, $\tau\Delta=\pi$ and $\pi<\tau\Delta<2\pi$ for four panels.
    The TDEE volume-law coefficient $\alpha_V$ and scaling regime diagrams for (b) $\tau<\tau_0$, (c) $\tau=\tau_0$ and (d) $\tau_0<\tau<2\tau_0$, where the dashed lines are the analytical lower bounds from \cref{eq:constant_D_to_TDEE}, and the solid lines approximately illustrate the actual values.
    The spectral functions are shown schematically, while the coefficient curves illustrate the analytical plateau contributions and heuristic estimates.}
    \label{fig:PH_symmetric}
\end{figure*}

% \bl{See if the following paragraph is correct; you wrote $\sin f(\theta)$ earlier, which I think should be $f(\theta)$ by your appendix.} 
As shown in Appendix~\ref{sec:Background of correlation matrix and entanglement entropy}, we find that $C_{\pm}(\theta)=\frac{f(\theta)}{L}\Tilde{C}_\pm(\theta)$, where $f(\theta)$ is a rapidly oscillating function which cancels when we calculate $D(\theta)$. The remaining part $\Tilde{C}_\pm(\theta)$ is either finite or divergent depending on $\theta$. The divergence of $\Tilde{C}_\pm(\theta)$ happens when $\theta=E_k^\pm\tau$ in Eq.~\eqref{eq:Toeplitz_matrix_translation_symm_B_FT_pm}, which in the $L\rightarrow\infty$ limit corresponds to the interval $\theta\in[-\tau,0]$ and $\theta\in[\tau\Delta,\tau\Delta+\tau]$, respectively. Outside the interval, $\Tilde{C}_\pm(\theta)$ remains finite. Fig.~\ref{fig:PH_symmetric} (a) shows the behavior of $\Tilde{C}_\pm(\theta)$ and $D(\theta)$ for different values of $\Delta$ with $\tau=\tau_0$. 

(i) For $\Delta=0$, both $\Tilde{C}_-(\theta)$ and $\Tilde{C}_+(\theta)$ diverge in the ranges $\theta\in[-\pi,0]$ and $\theta\in[0,\pi]$, respectively, as shown in the first panel of Fig.~\ref{fig:PH_symmetric} (a). 
In this case, $D(\theta)=\Theta(-\theta)$ is a step function with two discontinuous jumps at $\theta=0$ and $\theta=\pi$, which gives a log-law TDEE with a coefficient $\frac{1}{6}\times 2=\frac{1}{3}$, according to our discussion in \cref{sec:derive-gapless}. This is the gapless case we studied earlier. 
In Appendix~\ref{Numerical calculation of D}, we also present numerical calculations of $D(\theta)$ using both cosine and linear dispersions, which agree well with this analytical picture.

For gap $\Delta>0$ systems, $\Tilde{C}_+(\theta)$ in the gapless case shifts to $\Tilde{C}_+(\theta-\tau\Delta)$. This gives the following cases.

(ii) For $0<\tau\Delta\le \pi$, corresponding to the second panel of Fig.~\ref{fig:PH_symmetric} (a), 
$D(\theta)=1$ for $\theta\in[-\pi+\tau\Delta,0]$ and $D(\theta)=0$ for $\theta\in[\tau\Delta,\pi]$, showing two intervals of plateaus.  
For the intermediate interval $\theta\in[0,\tau\Delta]$, we expect $D(\theta)=(1+\Tilde{C}_+(\theta)/\Tilde{C}_-(\theta))^{-1}$ to be a smooth function interpolating the plateaus, since both $\Tilde{C}_\pm(\theta))^{-1}$ are finite \footnote{This may not hold for a completely general model, but $D(\theta)$ should at least be piecewise continuous.}. For the remaining interval $\theta\in[-\pi,-\pi+\tau\Delta]$, we argue that the two divergent functions $\Tilde{C}_\pm(\theta)$ are equal up to subleading corrections. 
In general, when $\Tilde{C}_\pm(\theta)$ diverge at $E_k^\pm\tau-\theta=0$, their ratio is given by
\begin{eqnarray}
\label{eq:divergent_C_ratio}
    \frac{\Tilde{C}_+(\theta)}{\Tilde{C}_-(\theta)}=\frac{\varrho(E_k^+)}{\varrho(E_k^-)},
\end{eqnarray}
where $\varrho(E_k)=|\frac{{\rm d}E_k}{{\rm d}k}|^{-1}$ is the density of states at energy $E_k$. For particle-hole symmetric bands, and assuming linear dispersion approximation which gives constant $\varrho(E_k)$ for both bands, we find $D(\theta)=(1+\Tilde{C}_+(\theta)/\Tilde{C}_-(\theta))^{-1}=\frac{1}{2}$ for $\theta\in[-\pi,-\pi+\tau\Delta]$.

By detailed calculations in Appendix~\ref{sec:Calculation of time direction entanglement entropy with different D(theta)}, we show that the $D(\theta)\equiv\frac{1}{2}$ interval yields a volume-law entanglement entropy. 
For more generic systems, a length $\delta\theta$ interval with plateau value $D(\theta)=(1+\varrho(E_k^+)/\varrho(E_k^-))^{-1}\equiv a$ contributes
\begin{equation}
\label{eq:constant_D_to_TDEE}
    S(A)=\frac{\delta\theta}{2\pi}e(2a-1)\cdot K\ ,
\end{equation}
where the function $e(x)$ is defined in \cref{eq:EE_eigenvalue}. 
Particularly, when $\tau\Delta=\pi$, one reaches the maximal $\delta\theta=\pi$, as shown in the third panel of Fig.~\ref{fig:PH_symmetric} (a).

(iii) For $\pi<\tau\Delta<2\pi$, as shown in the fourth panel in Fig.~\ref{fig:PH_symmetric} (a), $D(\theta)=\frac{1}{2}$ for $\theta\in[\tau\Delta-2\pi,0]$, while $D(\theta)=0$ or $1$ for $\theta\in[0,\tau\Delta-\pi]$ and $\theta\in[-\pi,\tau\Delta-2\pi]$, respectively, which are continuously interpolated via the interval $\theta\in[\tau\Delta-\pi,\pi]$. Accordingly, one obtains a volume law similar to the former case.

From the above results at $\tau=\tau_0$, if we only include the contribution of the $D(\theta)=\frac{1}{2}$ interval, we obtain a volume-law coefficient $\alpha_V$ as a function of $\tau\Delta$ (by \cref{eq:constant_D_to_TDEE}) as shown by the blue dashed line in Fig.~\ref{fig:PH_symmetric} (c). This is however smaller than the numerical results in Fig.~\ref{fig:gapped_PH} (e). 
The more detailed calculation for the volume-law coefficient is shown in Appendix~\ref{app:restricted_support}. 
Therefore, the dashed line in Fig.~\ref{fig:PH_symmetric} (c) (namely \cref{eq:constant_D_to_TDEE}) only gives a lower bound of the TDEE volume-law coefficient $\alpha_V$, and the actual $\alpha_V$ from numerical results is given by Fig.~\ref{fig:gapped_PH} (e) (or heuristically, the solid line in Fig.~\ref{fig:PH_symmetric} (c)).

% {\color{orange} A particularly important case occurs when $\Tilde{C}_+(\theta)$ is simply a translation of $\Tilde{C}_-(\theta)$.  
% Near $\tau\Delta=\pi$, this implies that $D(\theta)$ for $\theta\in[0,\pi]$ may also take a constant value of $\frac{1}{2}$ over some finite range of $\theta$, generating additional volume-law entanglement entropy.  
% This is consistent with numerical results, where the observed volume-law prefactor is significantly larger than the lower bound near $\tau\Delta=\pi$.}

A particularly important case occurs when the conduction-band dispersion is a rigid translation of the valence-band dispersion. 
For $\tau=\tau_0=\pi$ and $\tau\Delta=\pi$, the energy shift satisfies $(\Delta+1)\tau=2\pi$, so the periodicity of the Fourier symbols gives $C_+(\theta)=C_-(\theta)$ and hence $D(\theta)=1/2$ throughout the Brillouin zone, as illustrated schematically in the third panel of Fig.~\ref{fig:PH_symmetric}(a).
Consequently, $\delta\theta=2\pi$, yielding the maximal volume-law entropy $S(A)\sim e(0)K=K\ln 2$. 
This explains why the numerical volume-law coefficient significantly exceeds the analytical lower bound and approaches its maximal value $\alpha_V=\ln 2$ near $\tau\Delta=\pi$.
% \bl{Is this orange paragraph in gray important? Or is the former paragraph already sufficient.}

For $\tau=\alpha\tau_0$ with $\alpha\in(0,1)$, a similar analysis can be carried out. 
When $\Delta=0$, $\Tilde{C}_\pm(\theta)$ diverge for $\theta\in[-\alpha\pi,0]$ and $\theta\in[0,\alpha\pi]$, while remaining finite elsewhere. 
Consequently, $D(\theta)=1$ for $\theta\in[-\alpha\pi,0]$ and $D(\theta)=0$ for $\theta\in[0,\alpha\pi]$, while $D(\theta)$ is continuous in the other intervals. 
This gives a single discontinuity, which yields a $\frac{1}{6}\ln{K}$ contribution to the TDEE.
For gapped systems with $0\leq\tau\Delta\leq2\pi$, there are three cases.
For $\tau\Delta\in(0,(2-2\alpha)\pi)$, $D(\theta)$ has no discontinuity and no intervals with $D(\theta)\equiv\frac{1}{2}$ plateau.  
At $\tau\Delta=(2-2\alpha)\pi$, there is exactly one discontinuity.  
For $\tau\Delta\in((2-2\alpha)\pi,2\pi)$, a region with $D(\theta)\equiv\frac{1}{2}$ emerges.
The detailed behaviors of $\Tilde{C}_\pm(\theta)$ and $D(\theta)$ for different $\tau\Delta$ with $\tau<\tau_0$ are shown in Appendix~\ref{sec:SSH equal W}. As a result, at $\tau\Delta=0$ and $(2-2\alpha)\pi$, the TDEE $S(A)=\frac{1}{6}\ln{K}+\text{Const.}$ follows a logarithmic law. For $\tau\Delta\in((2-2\alpha)\pi,2\pi)$, the TDEE exhibits a volume law, with maximal $\alpha_V$ reached at $\tau\Delta=(2-\alpha)\pi$. For $\tau\Delta\in(0,(2-2\alpha)\pi)$, the TDEE obeys an area law. This is summarized in Fig.~\ref{fig:PH_symmetric} (b), where the dashed line is the analytical lower bound, and the solid line approximates the actual coefficient $\alpha_V$. This agrees well with the numerical results in Fig.~\ref{fig:gapped_PH} (f).

For $\tau>\tau_0$, the TDEE generically shows a volume law $S(A)=\alpha_V K+\text{Const.}$ (see detailed analysis in Appendix~\ref{sec:SSH equal W}). When $\tau=\alpha\tau_0$ with $\alpha\in(1,2)$, by a similar analysis, we can derive the TDEE volume-law coefficient $\alpha_V$ as shown in Fig.~\ref{fig:PH_symmetric} (d), which is consistent with the numerical results in Fig.~\ref{fig:gapped_PH} (d). When $\tau>2\tau_0$, both $\Tilde{C}_\pm(\theta)$ diverge for all values of $\theta$, resulting in $D(\theta)\equiv\frac{1}{2}$ for all $\theta$. 
From Eq.~\eqref{eq:constant_D_to_TDEE}, the TDEE has 
$\alpha_V=\ln{2}$ for any $\tau\Delta$, which is consistent with the numerical data shown in Fig.~\ref{fig:gapped_PH} (c). 
The deviation of $\alpha_V$ from $\ln{2}$ arises from nonlinear dispersions. At large $\tau$, the existence of multiple $E_k$ satisfying $E_k\tau-\theta=0$ (mod $2\pi$) is effectively similar to the constant density of states of a linear dispersion, and thus $\alpha_V$ is closer to $\ln 2$.

In Appendix~\ref{Numerical calculation of D}, we present additional numerical plots of $D(\theta)$ for both the cosine and linear dispersions, which quantitatively agree with the analytical discussion provided above.
Furthermore, in Appendix~\ref{app:restricted_support}, we develop a more detailed approach to calculate the full volume-law coefficient by accounting for the different behavior of $D(\theta)$.
For the particle-hole-symmetric linear dispersion with $W=1$ and $\tau=\tau_0=\pi$, we obtain 
\begin{equation}
\label{eq:detailed_volume}
    \alpha_V=\frac{2\ln2}{\pi}\arcsin\left[\tan\left(\frac{\tau\Delta}{4}\right)\right]
\end{equation}
for $0\leq\tau\Delta\leq\pi$, with the remaining range determined by the symmetry $\tau\Delta\rightarrow2\pi-\tau\Delta$.
This result is larger than the coefficient lower bound $\alpha_V=\frac{\delta\theta}{2\pi}\ln2$ from Eq.~\eqref{eq:constant_D_to_TDEE}.
This analytical prediction is also plotted in Fig.~\ref{fig:gapped_PH}(e), showing agreement with the numerical results.

\subsection{Models without particle-hole symmetry}
\label{sec:SSH model without the particle-hole symmetry}

We now turn to the single-site TDEE (for a time interval with $K$ points) of band insulators without particle-hole symmetry. 
More concretely, we consider the SSH model with a next-nearest-neighbor (NNN) hopping $\delta_2$, which has a Hamiltonian
\begin{equation}
\label{eq:SSH_hamiltonian_unit_cell_NNN}
\begin{split}
    H'=-\sum_i &\big[(1-\delta_1)c_{i,1}^\dagger c_{i,2}+(1+\delta_1)c_{i,2}^\dagger c_{i+1,1} \\
    &- \delta_2 c_{i,1}^\dagger c_{i+1,1}-\delta_2 c_{i,2}^\dagger c_{i+1,2}+h.c.\big]\ .
\end{split}
\end{equation}
By rewriting the Hamiltonian into the momentum space as $H'=\sum_{j=0}^{L-1}\psi_{k_j}^\dagger v_{k_j}^\alpha\sigma_\alpha \psi_{k_j}$, where $\sigma_\alpha$ are Pauli matrices, $v_{k}^0=2\delta_2 \cos{k},\ v_{k}^x=-((1-\delta_1)+(1+\delta_1)\cos{k}),\ v_{k}^y=-(1+\delta_1)\sin{k}$, and $\psi_{k}=(\widetilde{c}_{k,1}, \widetilde{c}_{k,2})^T$. This gives band energies $E_k^\pm = v_k^0 \pm \sqrt{(v_k^x)^2+(v_k^y)^2}$, which have different bandwidths $W_\pm$ when $\delta_2\neq 0$.

\begin{figure}
    \centering
    \subfigure[]{\includegraphics[width=0.49\linewidth]{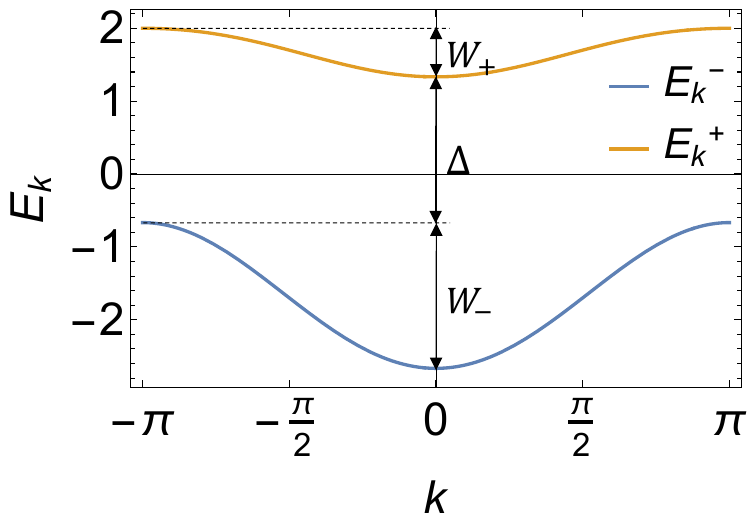}}
    \subfigure[]{\includegraphics[width=0.49\linewidth]{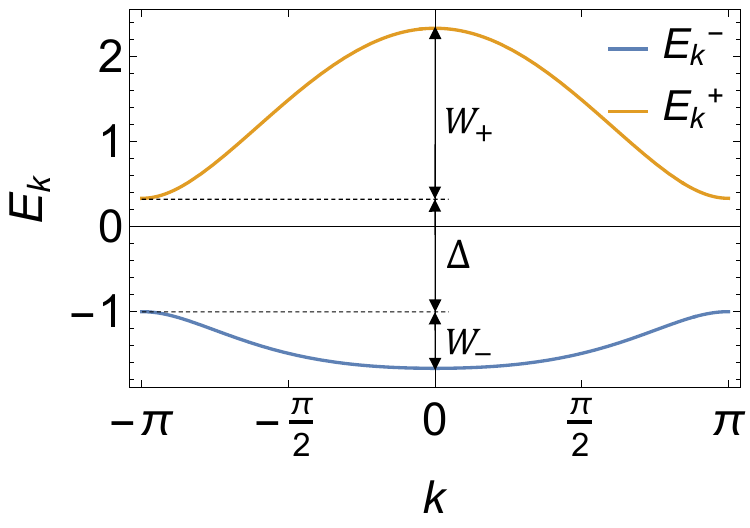}}
    \caption{Dispersion of SSH model with next-nearest-neighbor hopping breaking the particle-hole symmetry. (a) $\delta_1=-\frac{2}{3}$ and $\delta_2=-\frac{1}{3}$. (b) $\delta_1=\frac{1}{3}$ and $\delta_2=\frac{1}{6}$.}
    \label{fig:nonPH_symmetric_band}
\end{figure}

Fig.~\ref{fig:nonPH_symmetric_band} shows the band structures for two sets of parameters: $(\delta_1,\delta_2)=(-\frac{2}{3}, -\frac{1}{3})$ and $(\delta_1,\delta_2)=(\frac{1}{3}, \frac{1}{6})$. Their band energy ranges are 
$E\in[-\frac{8}{3},-\frac{2}{3}]\cup[\frac{4}{3},2]$ and 
$E\in[-\frac{5}{3},-1]\cup[\frac{1}{3},\frac{7}{3}]$, respectively. 

For convenience, we rescale and shift the band energy ranges, such that the valence and conduction bands are in the ranges $E_k^-\in[-1,0]$ and $E_k^+\in[\Delta,\Delta+W]$, respectively. Namely, the bandwidths of the two bands are $W_-=1$ and $W_+=W$, and the gap is $\Delta$. The two examples above have $W=\frac{1}{3}$, $\Delta=1$ and $W=3$, $\Delta=2$, respectively. Accordingly, their characteristic times are given in \cref{eq:insulator-characteristic-times} as 
\begin{equation}
\tau_0^-=\pi\ ,\qquad \tau_0^+=\frac{\pi}{W}\ ,\qquad \tau_0=\frac{2\pi}{1+W}\ .
\end{equation}
% (\bl{Pls note that your original $\tau_c$ should be changed to $\tau_0$, which have the same definition. Please change all $\tau_c$ in appendix, too.})

Without loss of generality, we focus on $W<1$, for which $\tau_0^+>\tau_0>\tau_0^-$. (The results for $W>1$ are qualitatively similar and are provided in Appendix~\ref{SSH model without particle-hole symmetry W}.) Following a similar analysis with respect to the functions $\Tilde{C}_\pm(\theta)$ developed in \cref{sec:Analytical understanding for gapped free fermions with the particle-hole symmetry} (see details in Appendix~\ref{sec:SSH small W}), we can derive the results summarized below, which are consistent with the numerical results. As a concrete example, we tune $\delta_1$ and $\delta_2$ such that $W=\frac{1}{3}$ remains fixed, and $\tau\Delta$ varies continuously from $0$ to $4\pi$
\footnote{The band extrema need not occur at $k=0,\pm\pi$.}. 
With $W=\frac{1}{3}$, we have $\tau_0^+=3\pi$, $\tau_0^-=\pi$, and $\tau_0=\frac{3}{2}\pi$.

(i) $\tau>2\tau_0^+$:
In this regime, $\Tilde{C}_\pm(\theta)$ diverge for all values of $\theta$ and $\Delta$, and lead to a volume-law TDEE with the same prefactor for different $\Delta$.  

(ii) $2\tau_0^-<\tau<2\tau_0^+$: 
In this case, $\Tilde{C}_-(\theta)$ diverges for all $\theta$, while $\Tilde{C}_+(\theta)$ diverges only for certain ranges of $\theta$. 
As a result, $D(\theta)$ has the same overall shape, leading to the same volume-law TDEE for any $\Delta$, but with a prefactor smaller than in the $\tau>2\tau_0^+$.  
This is consistent with the numerical results presented in Fig.~\ref{fig:gapped_nonPH} (b).

\begin{figure}
    \centering
    \subfigure[]{\includegraphics[width=0.99\linewidth]{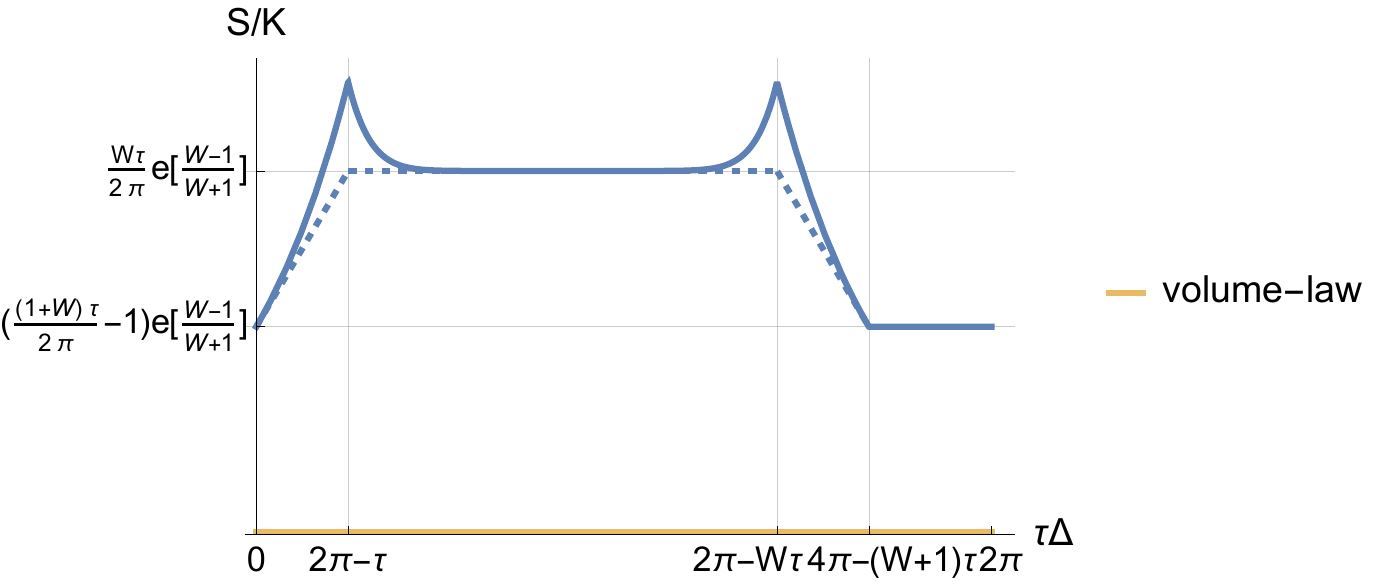}}
    \subfigure[]{\includegraphics[width=0.96\linewidth]{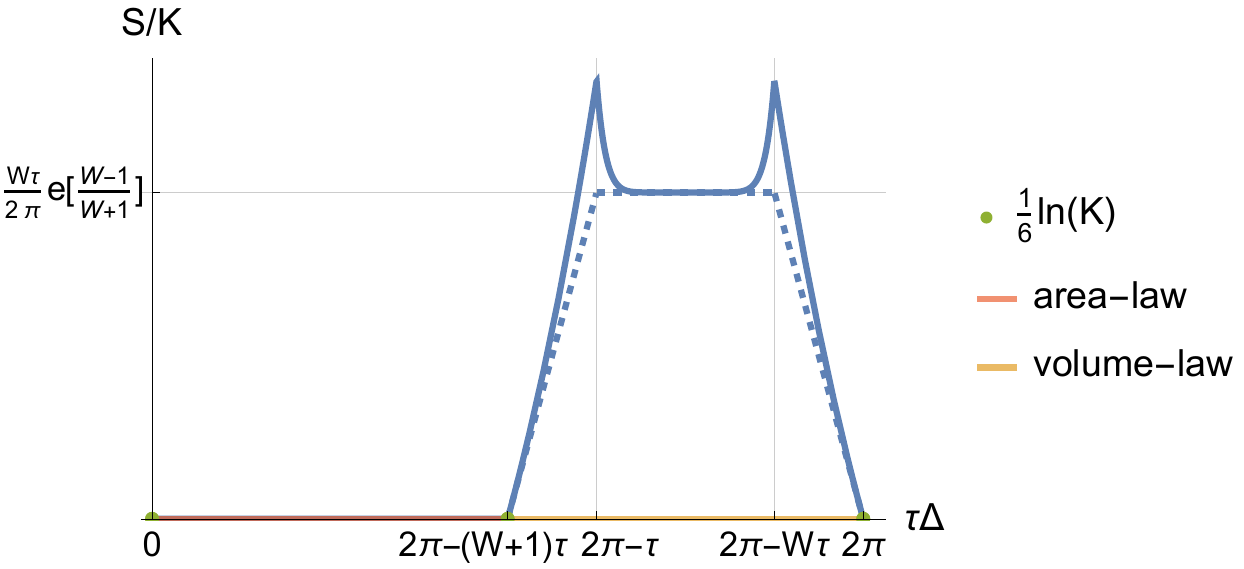}}
    \caption{Without particle-hole symmetry, the prefactors of volume-law term and scaling regime diagrams for (a) $\tau_0<\tau<2{\rm min}(\tau_0^-,\tau_0^+)$ and (b) $\tau<\tau_0$, where the analytical results shown with dashed lines are lower bounds.}
    %(a) (b) $\tau=\frac{7\pi}{4}$ (c) $\tau=\frac{3\pi}{4}$
    \label{fig:nonPH_symmetric_smallW}
\end{figure}

(iii) $\tau_0<\tau<2\tau_0^-$: By the detailed analysis of $\Tilde{C}_\pm(\theta)$ and $D(\theta)$ in Appendix~\ref{sec:SSH small W}, we find that the TDEE is in the volume-law regime for any $\tau\Delta$, but with varying coefficients. The volume-law coefficient $\alpha_V$ is shown in Fig.~\ref{fig:nonPH_symmetric_smallW} (a). These analytical predictions are in excellent agreement with the numerical results presented in Fig.~\ref{fig:gapped_nonPH} (d).

(iv) $\tau<\tau_0$. This gives three different regimes
as discussed in details in Appendix~\ref{sec:SSH small W}. When $\tau\Delta=0$ or $2\pi-(W+1)\tau$, there is a single discontinuity for $D(\theta)$, and the TDEE follows a logarithmic law $S(A)=\frac{1}{6}\ln{K}+\text{Const}$.
For $0<\tau\Delta<2\pi-(W+1)\tau$, there is no discontinuity, leading to an area-law TDEE.
For $2\pi-(W+1)\tau<\tau\Delta<2\pi$, there is always a nonzero volume-law coefficient $\alpha_V$ as a function of $\tau\Delta$. 
The scaling regime diagram and the volume-law coefficient $\alpha_V$ are shown in Fig.~\ref{fig:nonPH_symmetric_smallW} (b).  
These analytical predictions are in agreement with the numerical results presented in Fig.~\ref{fig:gapped_nonPH} (f).

\begin{figure}
    \centering
    \subfigure[]{\includegraphics[width=0.95\linewidth]{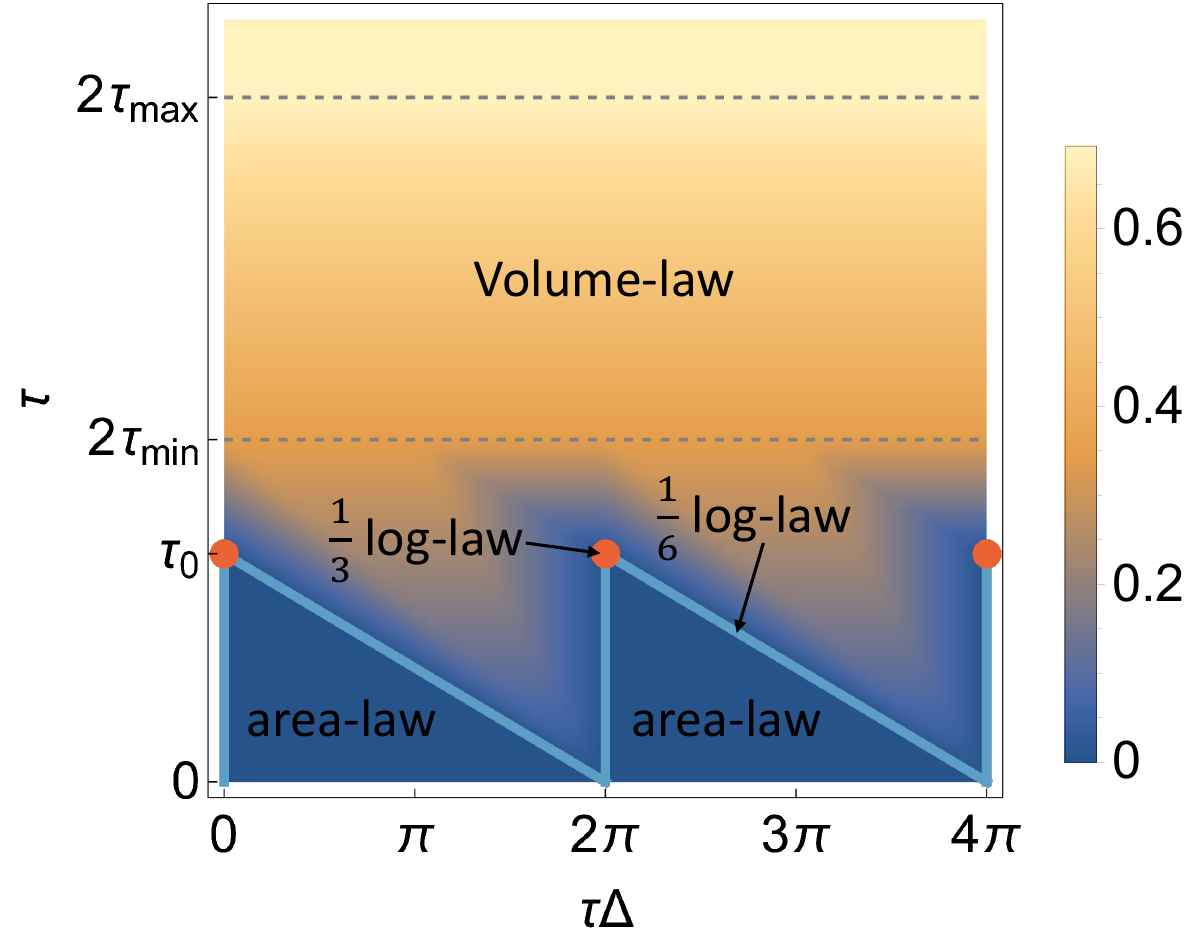}}
    \subfigure[]{\includegraphics[width=0.95\linewidth]{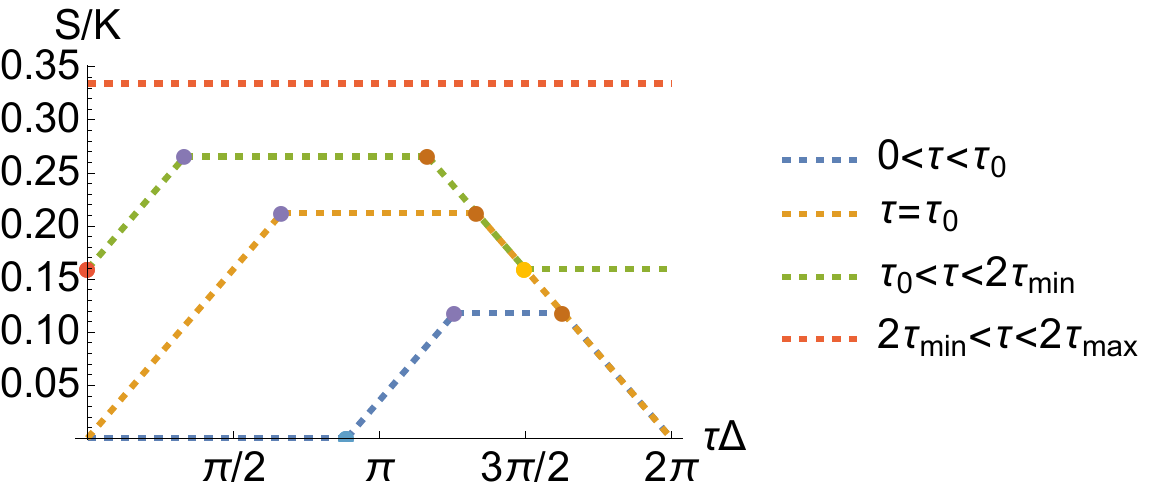}}
    \caption{
    (a) Scaling regime diagram of a general two-band model at half filling. 
    The volume-law coefficient is shown as a density plot, with a $2\pi$-periodicity. 
    Volume-law and area-law regimes appear as yellow and dark-blue regions, respectively, with the boundary between two scaling regimes shown as a blue line corresponding to a log-law TDEE with coefficient $\frac{1}{6}$. 
    Red dots indicate points with a log-law TDEE of coefficient $\frac{1}{3}$.
    (b) Volume-law coefficient as a function of $\tau\Delta$. 
    The red line is $a_1=\frac{\tau}{2\tau_{\rm max}}e(\frac{\tau_{\rm max}-\tau_{\rm min}}{\tau_{\rm max}+\tau_{\rm min}})$. 
    % \bl{There might be multiple $\alpha$ meaning different things (e.g., \cref{eq:anderson_insulator}). Let's change some of them to a different notation} 
    The purple and brown dots are $(\beta_1,a_1)$ and $(\beta_2,a_1)$, with $\beta_1=2\pi(1-\frac{\tau}{2\tau_{\rm min}})$ and $\beta_2=2\pi(1-\frac{\tau}{2\tau_{\rm max}})$. 
    The red and yellow dots are $(0,a_2)$ and $(\beta_1+\beta_2,a_2)$, with $a_2=(\frac{\tau}{\tau_0}-1)\cdot e(\frac{\tau_{\rm max}-\tau_{\rm min}}{\tau_{\rm max}+\tau_{\rm min}})$. 
    The blue dot is $(\beta_1+\beta_2-2\pi,0)$.}
    %(a) (b) $\tau=\frac{7\pi}{4}$ (c) $\tau=\frac{3\pi}{4}$
    \label{fig:general_solution_TDEE}
\end{figure}

In general, the behavior of single-site TDEE for a 1D disorderless two-band model can be summarized by the scaling regime diagram in Fig.~\ref{fig:general_solution_TDEE}(a), where the color indicates the (lower bound of the) volume-law coefficient $\alpha_V$, and we have defined $\tau_{\rm max}=\max{(\tau_0^+,\tau_0^-)}$ and $\tau_{\rm min}=\min{(\tau_0^+,\tau_0^-)}$. A more detailed description of the scaling regime diagram is given in Appendix~\ref{SSH model without particle-hole symmetry W} (see also Fig.~\ref{fig:general_solution_TDEE}(b) and the caption therein). 
% (\bl{Please put the description of the four regimes in original draft to the appendix as a small section.})

In the continuum time limit $\tau\rightarrow 0$, it is easy to see from the scaling regime diagram \cref{fig:general_solution_TDEE}(a) that, a band insulator with gap $\Delta>0$ always shows an area law TDEE, namely $S(A)\approx \text{Const}.$. In contrast, the gapless free fermions show a logarithmic TDEE with coefficient $\frac{1}{6}$ in the limit $\tau\rightarrow 0$ (see \cref{eq:gapless_log_law}).

Additionally, similar to Sec.~\ref{sec:reduced_hilbert_gapless}, for a gapped band insulator, the time-interval sub-Hilbert space in the continuum time limit $\tau\rightarrow 0$ has a stabilized single-particle Hilbert space dimension $d_A=\operatorname{rank}B$. The details can be found in Appendix~\ref{sec:reduced_hilbert_band_insulator}.
% (\bl{We can move the original section "reduced Hilbert space for band insulators" into an appendix. Check notations, e.g. I call $d_A$ the single-particle dimension of Hilbert space $\mathcal{H}_A$.})

\subsection{2-site TDEE for band insulator}
\label{sec:2-site TDEE}

In this section, we study the TDEE for a time interval on two adjacent sites (forming a unit cell) in the SSH band insulator, as shown in cases (ii) and (iii) in Fig.~\ref{fig:PH_symmetric_band}(a). Without loss of generality, we assume $\delta>0$ in the Hamiltonian~\eqref{eq:SSH_hamiltonian_unit_cell}.

Importantly, the SSH model has a topological phase and a trivial phase, depending on the sign of $\delta$. The two phases differ by a one-site translation. Here for $\delta>0$, we denote the TDEE setup in cases (ii) and (iii) in Fig.~\ref{fig:PH_symmetric_band}(a) as detecting the topological and trivial phases, respectively, in the sense that the 2-sites of the time interval as a unit cell contains a weak bond in case (ii) and a strong bond in case (iii).

\begin{figure}
    \centering
    \subfigure[]{\includegraphics[width=0.45\linewidth]{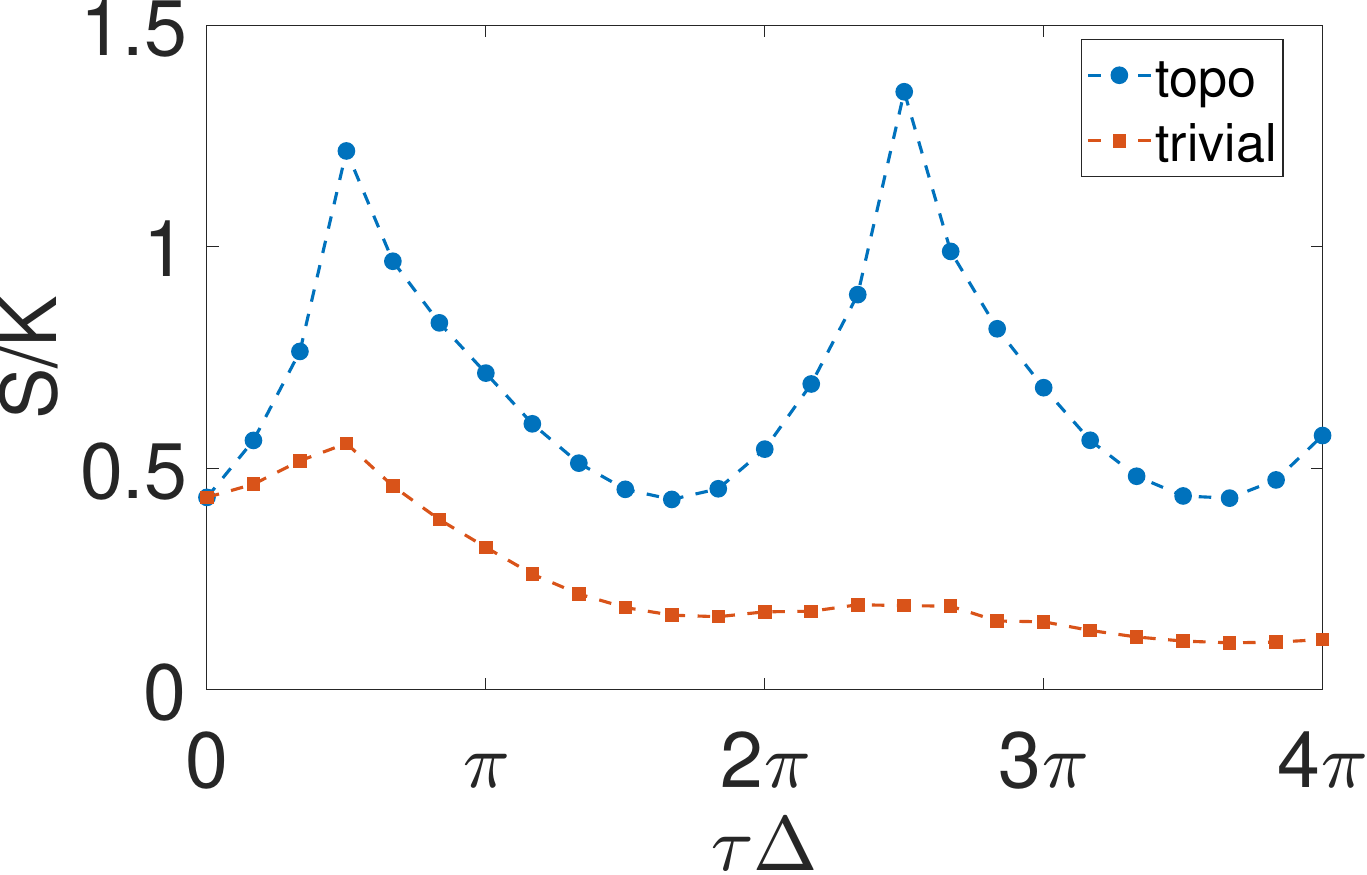}}
    \subfigure[]{\includegraphics[width=0.45\linewidth]{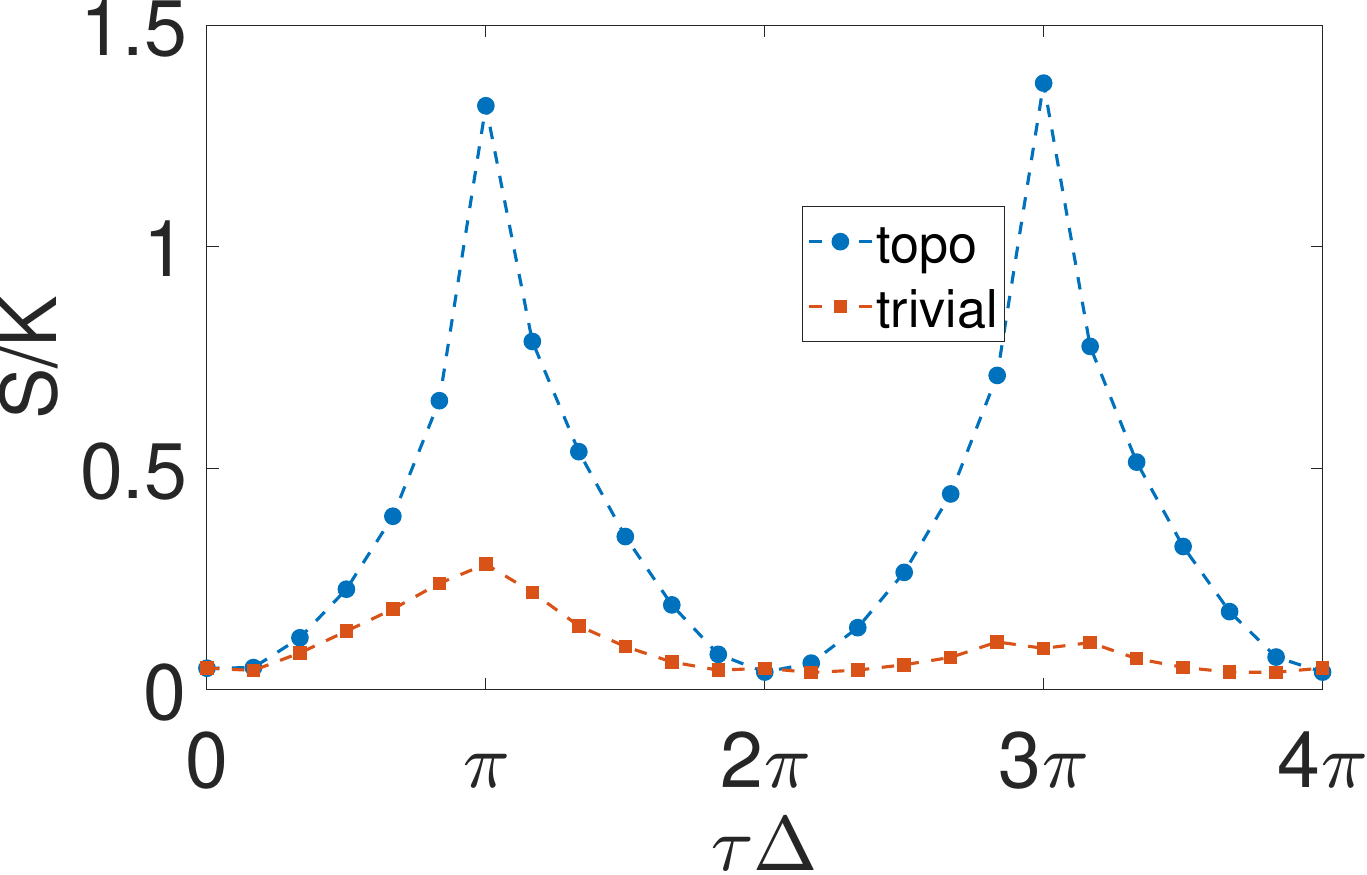}}
    \subfigure[]{\includegraphics[width=0.45\linewidth]{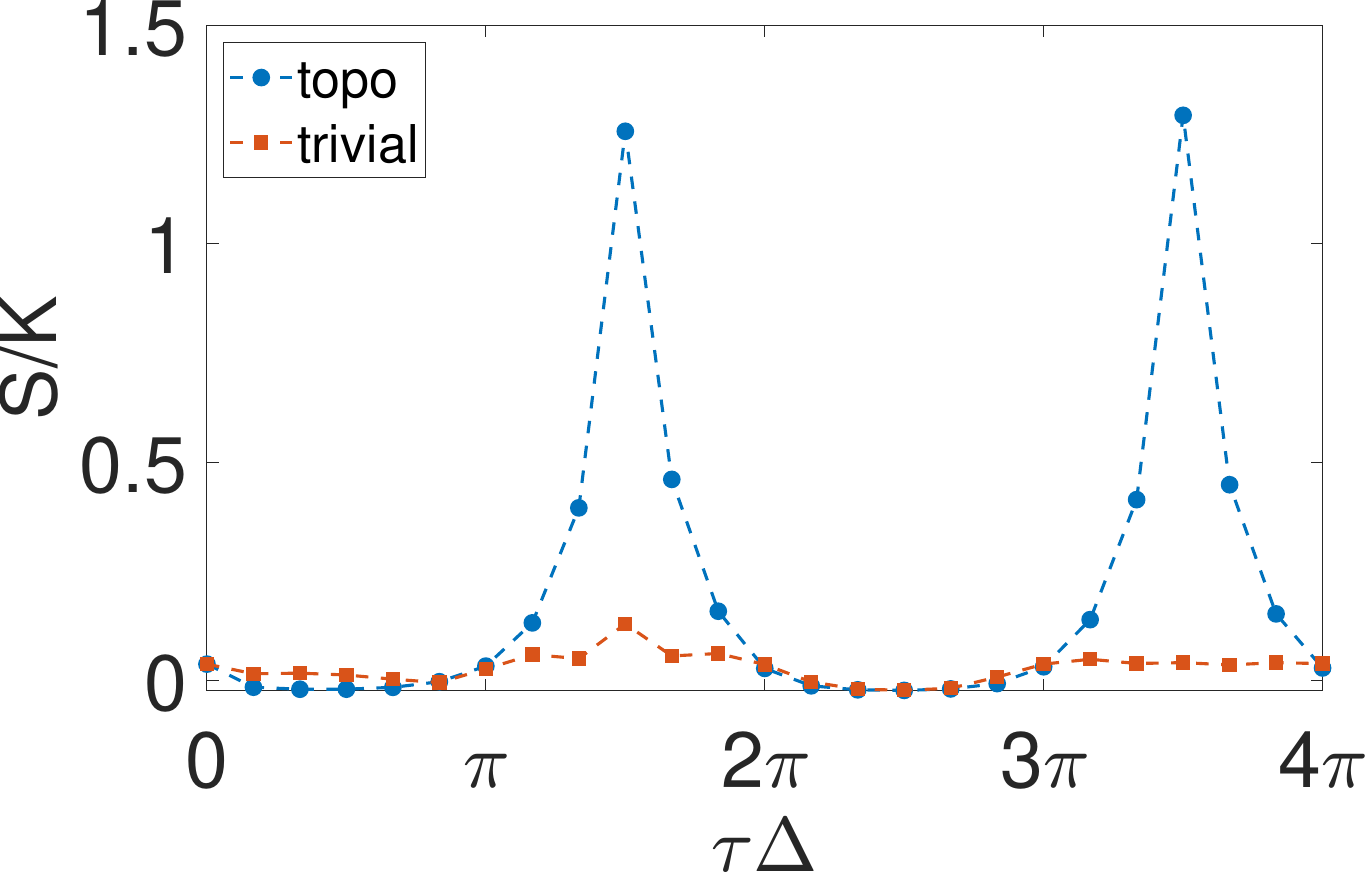}}
    \subfigure[]{\includegraphics[width=0.45\linewidth]{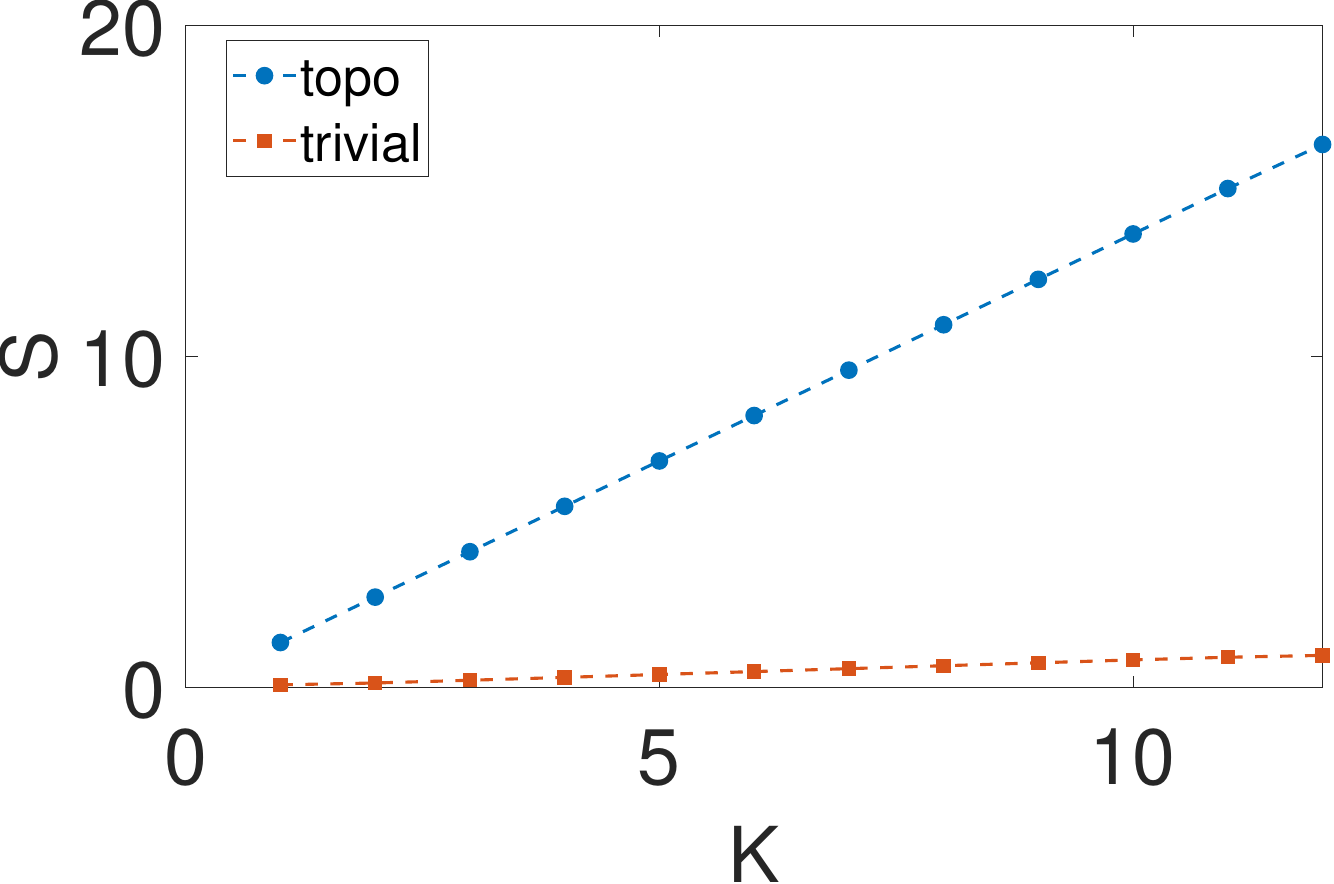}}
    \subfigure[]{\includegraphics[width=0.45\linewidth]{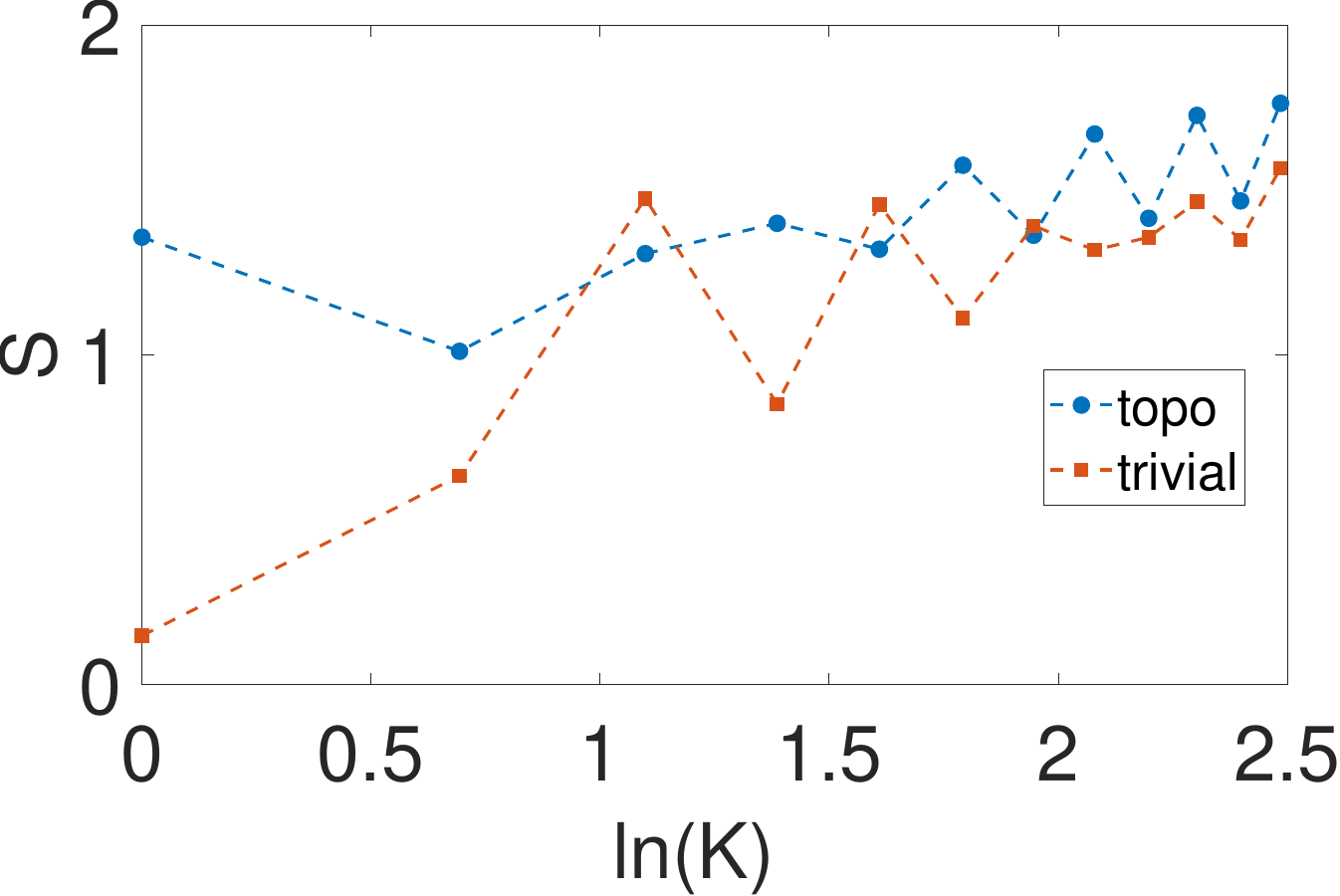}}
    \subfigure[]{\includegraphics[width=0.45\linewidth]{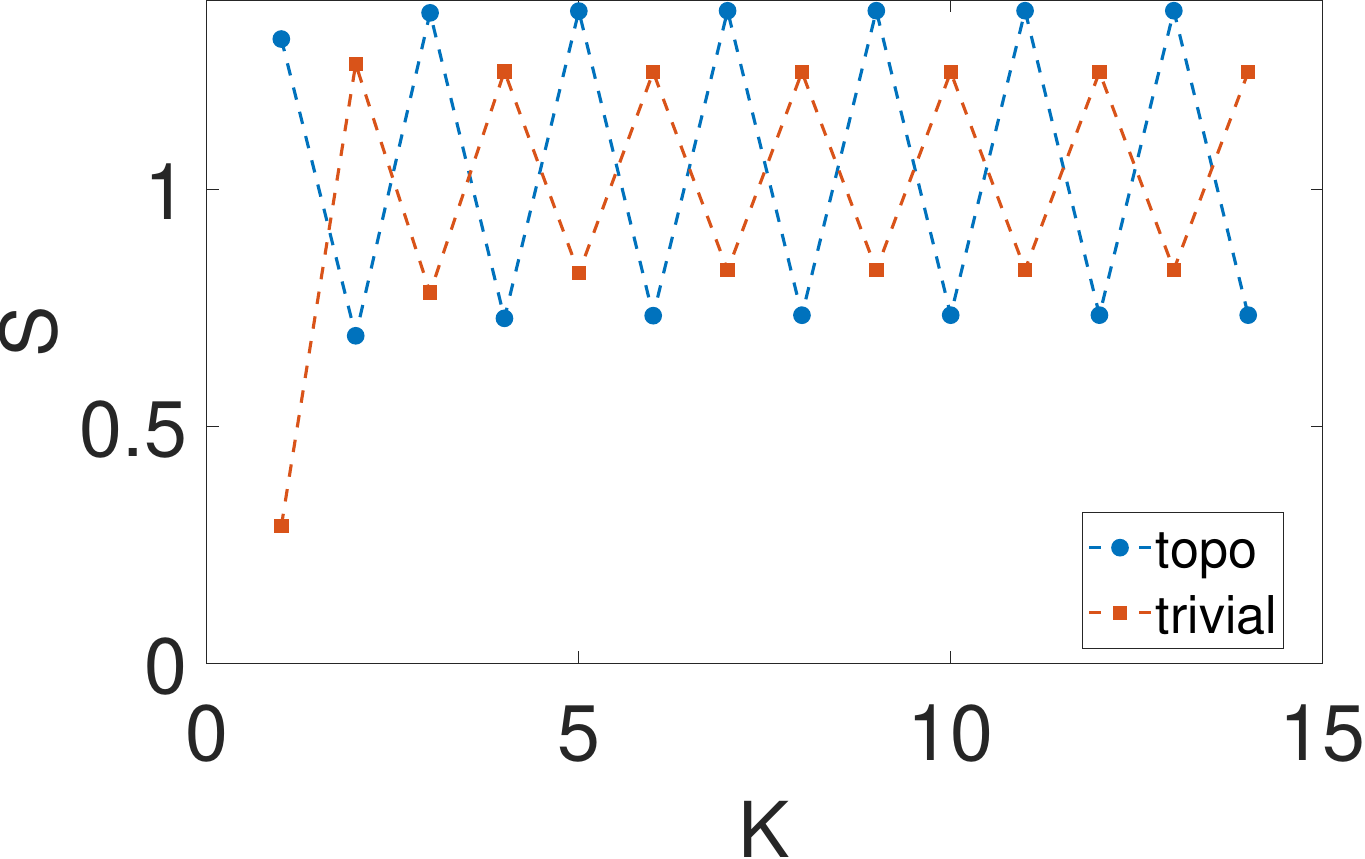}}
    \caption{The volume-law term prefactor $S/K$ for the TDEE of two phases, with $\tau\Delta\in[0,4\pi]$, and (a) $\frac{\tau}{\tau_0}=\frac{3}{2}$, (b) $\frac{\tau}{\tau_0}=1$, and (c) $\frac{\tau}{\tau_0}=\frac{1}{2}$.
    (d) The volume-law TDEE for two phases with $\tau\Delta=\frac{3\pi}{2}$, $\frac{\tau}{\tau_0}=\frac{1}{2}$.
    (e) The log-law TDEE for two phases with $\tau\Delta=\pi$, $\frac{\tau}{\tau_0}=\frac{1}{2}$.
    (f) The area-law TDEE for two phases with $\tau\Delta=\frac{\pi}{2}$, $\frac{\tau}{\tau_0}=\frac{1}{2}$.
    $L = 1000$ for all figures.
    % {\color{red}The data in figure a-c for $\tau\Delta>2\pi$ in trivial phase looks strange. Might be numerical inaccuracy that destroys the $2\pi$ period.}
    }
    \label{fig:gapped_2sites}
\end{figure}

The numerical TDEE for the topological and trivial phases as functions of $\tau$ and $\Delta$ (controlled by the positive $\delta$, with the two bandwidths still rescaled to $W_\pm=1$) are plotted in Fig.~\ref{fig:gapped_2sites}. In (a)–(c), we show the volume-law coefficient $\alpha_V\approx S/K$ at different $\tau/\tau_0$, which shows that the topological phase always has a higher $\alpha_V$, which we attribute to its stronger inter-unit-cell entanglement (or more nonlocal Wannier orbitals relative to the unit cell center). 
Additionally, the $2\pi$ periodicity in $\tau\Delta$ is clearer in the topological phase, while is significantly modulated in the trivial phase. Moreover, in Fig.~\ref{fig:gapped_2sites}(a)-(c), the maximal value of $\alpha_V$ in the topological phase is always close to $2\ln 2$, which is twice the maximum of single-site TDEE.

Similar to the single-site case, the 2-site TDEE also shows volume, log and area law regimes when $\tau<\tau_0$. As shown in Fig.~\ref{fig:gapped_2sites}(c) where $\frac{\tau}{\tau_0}=\frac{1}{2}$, the 2-site TDEE has an area law when $\tau\Delta\in(0,\pi)$ (where $\alpha_V$ vanishes), a volume law when $\tau\Delta\in(\pi,2\pi)$, and shows a log law when $\tau\Delta=0$ or $\pi$. Examples of these three regimes are shown in Fig.~\ref{fig:gapped_2sites}(d)–(f).

\section{TDEE of 1D Anderson Insulator}
\label{sec:TDEE of Anderson insulators}

Another interesting type of free fermion model is the Anderson insulator. Here we consider the following Anderson Hamiltonian \cite{Anderson1958,Mott1961The,AALR1979,Anderson1980New,Kramer1993Localization,Evers2008Anderson} in a 1D lattice of $L$ sites:
\begin{equation}
\label{eq:Anderson_insulator_Hamiltonian}
    H=-u\sum_{i=1}^{L}(c_i^\dagger c_{i+1}+ h.c.)+\sum_{i=1}^{L} w_i n_i,
\end{equation}
where $u$ is a uniform hopping, $w_i$ is a random potential on site $i$ uniformly distributed in the range $w_i\in[-W_0,W_0]$, and $n_i=c^\dagger_i c_i$ is the fermion number of site $i$.  
In our analysis, we fix $W_0=1$ and vary $u$ to tune the Anderson localization.
% (\bl{I changed $w_0$ to $W_0$, to distinguish from potential on site $0$. Pls check if OK.}) 

In 1D, it is known that arbitrarily weak disorders $W_0/u>0$ will lead to the localization of all single-particle eigenstates \cite{Anderson1958}, giving an Anderson insulator at any fermion filling factor $\nu$. The localization of a single-particle wavefunction $\psi$ can be characterized by the inverse participation ratio (IPR), defined as
\begin{equation}
\label{eq:IPR_definition}
    {\rm IPR}=\frac{\sum_i |\psi_i|^4}{(\sum_i |\psi_i|^2)^2},
\end{equation}
where $\psi_i$ is the amplitude of the wavefunction at site $i$, as shown in Fig.~\ref{fig:anderson_insulator} (a). 
For a normalized wavefunction, the denominator equals $1$. 
For a wavefunction localized around a site, e.g. $i=0$, one has $|\psi_i|^2\propto e^{-|x_i-x_0|/\xi}$, where $\xi$ is the localization length. This would yield an IPR proportional to $\xi^{-1}$. Therefore, for numerical purposes, for large enough system size $L$, we define the localization length of the Anderson insulator simply by
\begin{equation}
\label{eq:localization_wavefunction}
    \xi=\overline{\rm IPR}^{-1}\ ,
\end{equation}
where $\overline{\rm IPR}$ is the average of IPR of all single-particle eigenstates. We also note that the energy spectrum of an Anderson insulator has no gap, but it is insulating because of the localization.

\subsection{The single-site TDEE}

\begin{figure}
    \centering
    \subfigure[]{\includegraphics[width=0.45\linewidth]{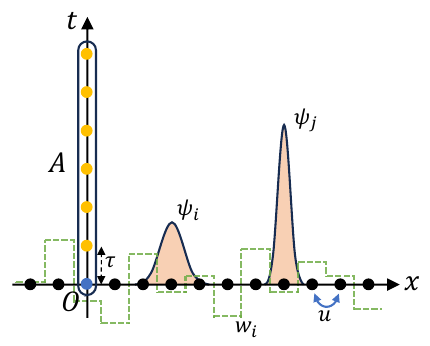}}
    \subfigure[]{\includegraphics[width=0.45\linewidth]{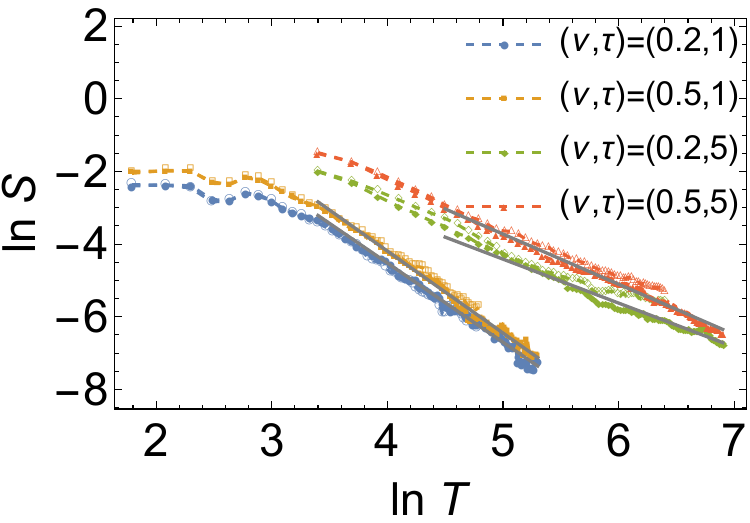}}
    \subfigure[]{\includegraphics[width=0.45\linewidth]{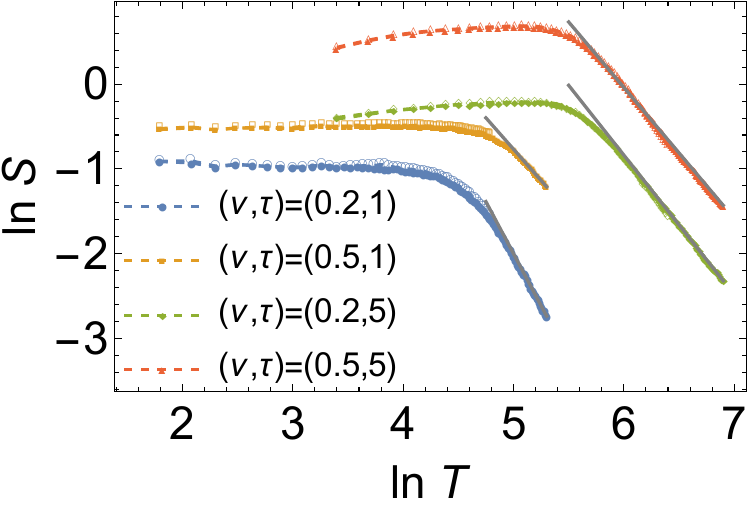}}
    \subfigure[]{\includegraphics[width=0.45\linewidth]{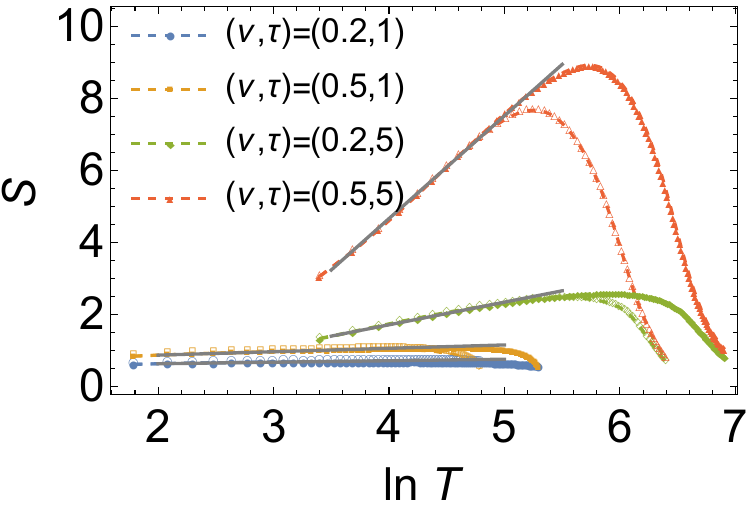}}
    \caption{(a) The Anderson insulator with random on-site potential shown with green dash line, and the localized function $\psi_{i,j}$. 
    (b)-(d) show the TDEE for (b) $u=0.03$, (c) $u=0.4$, and (d) $u=1.0$, with filling factors $\nu=0.2,0.5$ and time discretizations $\tau=1.0,5.0$. 
    Data for different system sizes, $L=120$ and $L=200$, are indicated by open and filled symbols, respectively. All the plots are averaged over 500 independent on-site potential disorder samplings. 
    % \bl{Change $S_t$ to $S$ in the figure y labels?} 
    }
    \label{fig:anderson_insulator}
\end{figure}

Below, we investigate the single-site TDEE of the Anderson insulator. 
There are three tuning parameters in this system: the hopping strength $u$ (with disorder strength $W_0=1$ fixed), the time step $\tau$ of the time-direction interval (of length $T=K\tau$), and the fermion filling factor $\nu$. 
Different from \cref{eq:characteristic_time}, we define the characteristic time here as determined by the disorder bandwidth $2W_0$ only, 
\begin{equation}
\tau_0=\frac{2\pi}{2W_0}=\pi\ .
\end{equation}

As examples, we consider two representative values $\tau=1$ and $\tau=5$. In Fig.~\ref{fig:anderson_insulator}(b)-(d), we plot the TDEE $S(A)$ as a function of time $T=\tau K$, for $\nu=0.2,0.5$ and $u=0.03,0.4,1.0$. Each TDEE is calculated from the average of 500 independent disorder samplings of on-site potentials.  

Fig.~\ref{fig:anderson_insulator}(b),(c) show $\ln S(A)$ versus $\ln T$ at small $u=0.03$ and $0.4$, which reveals a universal power-law decay of $S(A)$ as a function of $T$ at large $T$. 
The results for two different system sizes, $L=120$ and $L=200$, are indicated by open and filled symbols, respectively, and they overlap almost perfectly. This convergence with respect to $L$ confirms that $L\gg \xi$, and the power-law decay of TDEE is intrinsically due to the Anderson localization of the system.

Fig.~\ref{fig:anderson_insulator} (d) shows the semi-log plot of $S(A)$ versus $\ln{T}$ for a relatively large hopping $u=1$, in which one can see clearly that $S(A)$ grows linearly in $\ln T$ for $T$ until a crossover regime, after which $S(A)$ decays. The crossover regime, however, still depends on whether $L=120$ or $L=200$, and this is because the localization length $\xi$ is larger for this larger $u$ case, and $L$ is not yet much larger than $\xi$ (within our limit of computation). Nevertheless, the dependence of TDEE on $u$ shows the direct impact of the localized dynamics of the system.

The TDEE calculated in Fig.~\ref{fig:anderson_insulator}(b)-(d) suggests a universal understanding as follows. For an Anderson insulator with localization length $\xi$, its dynamics and information generated at a single site is restricted to propagate up to the range of $\xi$. Therefore, a single site effectively only sees the Hilbert space of a nearby spatial range up to distance $\xi$, which we denote as $\mathcal{H}_\xi$. Namely, the system size is effectively $\mathcal{O}(\xi)$ for operators on a single site. For a time-interval of $K$ time points at a single site, when $K<\xi$ (up to an order $1$ factor), one expects the sub-Hilbert space $\mathcal{H}_A$ to be smaller than its complement in $\mathcal{H}_\xi$. While when $K>\xi$, one expects $\mathcal{H}_A$ to approach $\mathcal{H}_\xi$, and thus is larger than its complement in $\mathcal{H}_\xi$. This leads to a Page curve like behavior for the TDEE $S(A)$, namely, $S(A)$ undergoes a Page crossover from growing into decaying around $K\sim \xi$. More quantitatively, Fig.~\ref{fig:anderson_insulator} suggests the following scaling behaviors of TDEE:
\begin{equation}
\label{eq:anderson_insulator}
    S(A)=\left\{\begin{aligned}
    &c\ln{K}+{\rm const.},&K<K_c,\\
    &\frac{c'}{K^\alpha},&K>K_c,\\
    \end{aligned}\right.
\end{equation}
where $K_c$ is the crossover point. When the total system size $L\gg\xi$, we expect
\begin{equation}
\label{eq:Kc_IPR}
    K_c\propto\xi\ ,
\end{equation}
where $\xi$ is defined by \cref{eq:localization_wavefunction}. When $L<\xi$, we expect $K_c\propto L$ to show the finite size effect.

To verify the scaling behavior in \cref{eq:Kc_IPR}, we calculate the TDEE $S(A)$ and the localization length $\xi$ defined by \cref{eq:localization_wavefunction} for various $u$, where we set $\tau=1$ or $\tau=5$, filling $\nu=0.2$ or $0.5$, and the system size $L=120$ or $200$. 

\begin{figure}
    \centering
    \subfigure[]{\includegraphics[width=0.47\linewidth]{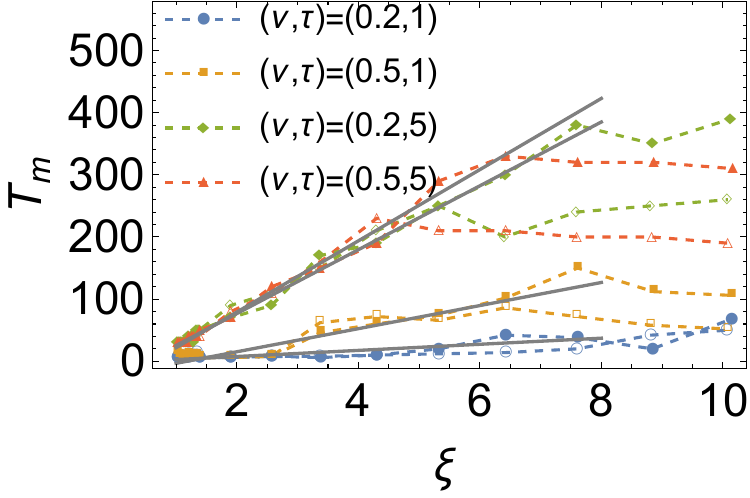}}
    \subfigure[]{\includegraphics[width=0.475\linewidth]{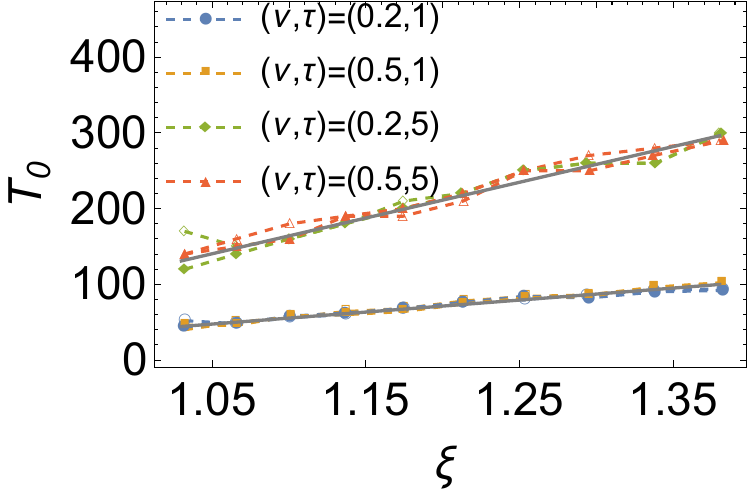}}
    \subfigure[]{\includegraphics[width=0.48\linewidth]{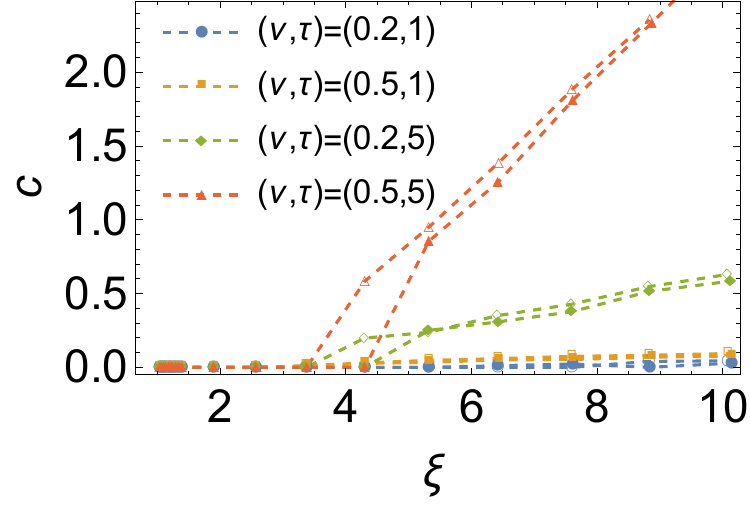}}
    \subfigure[]{\includegraphics[width=0.46\linewidth]{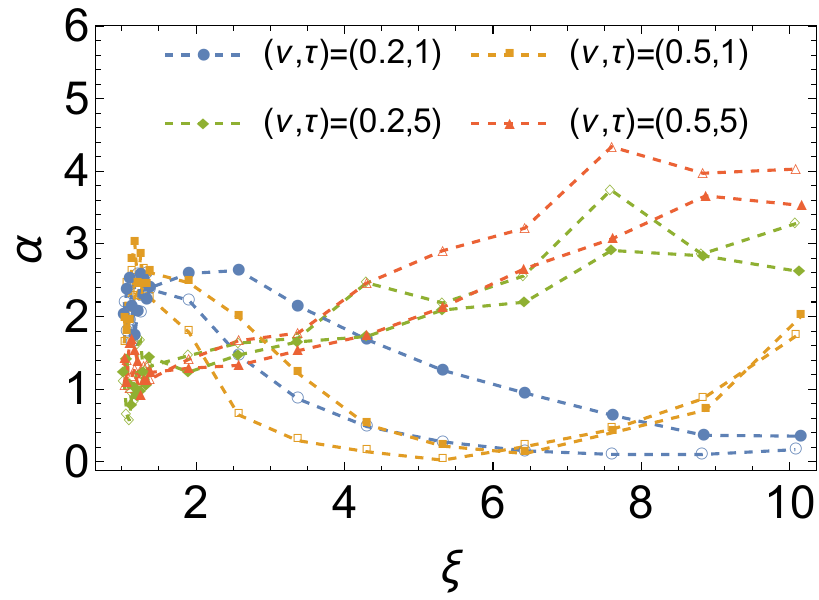}}
    \caption{The Page-curve like crossover time of the TDEE of Anderson insulator as a function of the localization length $\xi$, defined as 
    (a) $T_m=\tau K_m$ and (b) $T_0=\tau K_0$, for filling factors $\nu=0.2,0.5$ and time discretizations $\tau=1.0,5.0$. 
    Data for different system sizes, $L=120$ and $L=200$, are indicated by open and filled symbols, respectively.
    (c) The log-law coefficient $c$ and (d) the decay exponent $\alpha$ of the power-law behavior as functions of the localization length $\xi$, for filling factors $\nu=0.2,0.5$ and time discretizations $\tau=1.0,5.0$.}
    \label{fig:anderson_insulator_fit}
\end{figure}

In the relatively weak localization regime $\xi>1.5$, for which $u$ is not too small, we calculate the length of time interval $T_m=\tau K_m$ at which the TDEE reaches its maximum (for fixed $\tau$), and we approximately identify $K_c$ as $K_m$. As shown in Fig.~\ref{fig:anderson_insulator_fit}(a), $T_m$ is almost linear in $\xi$ up to $\xi\approx 7$, which supports the linear relation in \cref{eq:Kc_IPR}. For $\xi>7$, the effect of finite system size $L$ leads to a deviation from the linear behavior.

% \bl{Is this $\xi$ actually $\text{IPR}^{-1}$?} {\color{red}[XS]: Yes, in Fig.~\ref{fig:anderson_insulator_fit} the x axis is actually $\text{IPR}^{-1}$.}\bl{OK, then I defined $\xi$ using IPR directly, if OK, pls see \cref{eq:localization_wavefunction}.}

%We note that when the hopping $u$ is very small, one may have extremely small localization length $\xi\approx 1$, which yields $K_c\approx 1$. Accordingly, the TDEE will decay from the beginning as $K$ grows from $1$, according to \cref{eq:anderson_insulator}. This is the case shown in Fig.~\ref{fig:anderson_insulator}(b).

In the strong localization regime $\xi<1.5$, for which $u$ is small, \cref{eq:Kc_IPR} suggests a crossover $K_c$ around the order of $1$. According to \cref{eq:anderson_insulator}, the TDEE is almost always on the decaying side $K>K_c$, which is the case shown in Fig.~\ref{fig:anderson_insulator}(b). This makes it difficult to identify an accurate maximal TDEE position $T_m=\tau K_m$. Instead, we define $T_0=\tau K_0$ as the length of time interval at which the TDEE decreases to $10\%$ of its maximum value, and regard $K_c$ as proportional to $K_0$.
% \bl{Somehow starting some point, I used $t$ instead of $T$ for the length of time interval. I think we should still use $T$, what do you think? I've corrected some places, please help me double check this notation. Accordingly, $t_m$ and $t_0$ here better be changed to $T_m$ and $T_0$ (in the figures also), since $t_m$ earlier means the $m-th$ time point in the interval.} 
As shown in Fig.~\ref{fig:anderson_insulator_fit}(b), for $\xi<1.35$, we find a clear linear relation between $T_0$ and $\xi$. 
Moreover, the curves for different filling factors $\nu$ collapse to the same approximately linear curve. This provides further support to the linear relation of Eq.~\eqref{eq:Kc_IPR}.

Fig.~\ref{fig:anderson_insulator_fit}(c) and (d) show the fitted log-law coefficient $c$ at $K<K_c$ and power-law decay exponent $\alpha$ at $K>K_c$ in \cref{eq:anderson_insulator}, respectively.  
The log-law coefficient $c$ shows a linear growth in $\xi$, and grows as $\tau$ increases.  
The power-law exponent $\alpha$ shows different behaviors for different $\tau$. For $\tau=5$, $\alpha$ decreases as $\xi$ decreases, with $\alpha\rightarrow1$ as $\xi\rightarrow1$ regardless of fillings. For $\tau=1$, $\alpha$ increases as $\xi$ decreases, and $\alpha\rightarrow2.5$ as $\xi\rightarrow1$ regardless of fillings. This suggests potential universality as $\xi\rightarrow 1$, although $c$ and $\alpha$ generically depend on $\tau$ and $\xi$.

\subsection{Remark on the Page Curve Analogy}

The page curve analogy for the TDEE of Anderson insulator implies that, irrespective of the time step $\tau$ chosen, the TDEE $S(A)$ will decay towards zero as $T=K\tau$ increases when $K>K_c\propto \xi$ (\cref{eq:anderson_insulator}). We refer to such a decaying TDEE when $T$ grows as a \emph{sub-area law}. Interestingly, the TDEE thus distinguishes between the band insulator and Anderson insulator, which shows area law (for sufficiently small $\tau$) and sub-area law, respectively. This is in sharp contrast with the spatial entanglement entropy in a 1D spatial interval of length $\ell$, which shows an area law (constant independent of $\ell$) for insulators, regardless of whether it is an Anderson insulator or band insulator.

Similar to the discussion in \cref{sec:reduced_hilbert_gapless}, the page-curve like TDEE of Anderson insulator also implies that the Lieb-Robinson bound only extends to a finite spatial range of length $\xi$ as the time interval length $T$ grows. This agrees with the localized quantum dynamics in an Anderson insulator~\cite{Burrell2007Bounds,hamza2012dynamical}.

As a further remark, as shown in Fig.~\ref{fig:anderson_insulator} and Eq.~\eqref{eq:anderson_insulator}, at large $T=\tau K$ the TDEE of Anderson insulator exhibits a power-law decay, which yields a positive second derivative $d^2S(A)/dT^2>0$.

This behavior is different from the conventional Page-curve behavior of the spatial entanglement entropy. For the conventional spatial entanglement entropy $S(\ell)$ of a translationally invariant state in a 1D spatial interval of length $\ell$, one always has $d^2S(\ell)/d\ell^2\le 0$. The proof is as follows. Consider three adjacent subregions $a=[0,d\ell]$, $b=[d\ell,\ell]$, and $c=[\ell,\ell+d\ell]$. By strong subadditivity, $S_{ab}+S_{bc}-S_{abc}-S_b\ge 0$. By translational symmetry, $S_{ab}=S_{bc}=S(\ell)$, and thus the above inequality becomes $2S(\ell)-S(\ell+d\ell)-S(\ell-d\ell)\ge 0$, which gives $d^2S(\ell)/d\ell^2\le 0$.

Although a system with time-independent Hamiltonian has time translational symmetry, the above argument does not apply to the TDEE. The key difference is as follows. For different spatial subregions $a$, $b$, the Hilbert space of their union $ab$ is the tensor product of the two sub-Hilbert spaces of $a$ and $b$. In contrast, for the sub-Hilbert space associated with a time interval in \cref{eq:sub-H-A}, the total Hilbert space of the union of two different time intervals is generically not equal to the tensor product of the two sub-Hilbert spaces of each time interval. This invalidates the foundation for proving the strong subadditivity, and thus the TDEE $S(A)$ in the presence of time translational symmetry can have a positive second derivative $d^2S(A)/dT^2>0$, as is the case of TDEE of Anderson insulators.

\section{Conclusions and outlook}
\label{sec:Conclusions}

In this paper, we have investigated the distinct behaviors of TDEE of three classes of 1D free fermion ground states: the disorderless gapless free fermions, the band insulators, and the Anderson insulators. For band insulators, the TDEE is mainly controlled by two dimensionless parameters, $\tau\Delta$ and $\tau/\tau_0$, where $\Delta$ is the gap, $\tau_0$ is defined in \cref{eq:characteristic_time,eq:insulator-characteristic-times}, and the disorderless gapless fermions case is included by setting $\Delta=0$. In different parameter regimes, the TDEE can show volume law, logarithmic law or area law in the time interval length $T$, and we have given a semi-analytical understanding for each of these cases based on the Fisher-Hartwig conjecture for Toeplitz matrix. In particular, a band insulator can still exhibit volume-law or logarithmic law TDEE for certain $\tau$. For Anderson insulators, we numerically revealed that the scaling behavior with respect to $T$ is page-curve like (\cref{eq:anderson_insulator}), irrespective of $\tau/\tau_0$.

Focusing on the continuum time limit $\tau/\tau_0\rightarrow 0$, the distinct behaviors of TDEE $S(A)$ of the three classes of 1D free fermion ground states at large time interval length $T=\tau K$ can be summarized more concisely:

(1) For disorderless gapless fermions, $S(A)=\frac{1}{6}\ln T +\text{Const.}$ shows a \emph{logarithmic law};

(2) For band insulators with $\Delta>0$, $S(A)\simeq\text{Const.}$ shows an \emph{area law};

(3) For Anderson insulators, $S(A)=c'' T^{-\alpha}$ shows a \emph{sub-area law}, with $\alpha>0$ depending on the localization length $\xi$ and filling factor $\nu$.

Therefore, unlike the spatial entanglement entropy which shows area law for both band insulators and Anderson insulators, the TDEE shows distinct behaviors for them. This is because unlike the spatial entanglement entropy which only probes the entanglement property of the quantum state, the TDEE also implicitly probes the quantum dynamics and information spreading of excitations of Hamiltonian $H$, as can be understood from its connection with the Lieb-Robinson bound explained in \cref{sec:reduced_hilbert_gapless}.

Particularly, for disorderless gapless fermions in the continuum time limit, $S(A)=\frac{1}{6}\ln T$ has only half the coefficient of that in the Cardy-Calabrese formula of entanglement entropy $S(\ell)=\frac{1}{3}\ln \ell$ in a spatial interval of length $\ell$. Analytically, we showed that this is due to one single discontinuity in $D(\theta)$, as shown in \cref{eq:D_result_gapless_tau02}. Heuristically, the factor $2$ difference is because the TDEE is dominated by contributions from a single fermion occupation discontinuity at the Fermi energy $E_F$, while the spatial entanglement entropy is dominated by contributions from two fermion occupation discontinuities at the two Fermi momenta $\pm k_F$. This agrees with the conjecture in \cite{Liu_2024}.

Another perspective comes from the bulk-boundary duality. 
In Ref.~\cite{Nebabu_2023}, Nebabu and Qi constructed a 1+1d bulk theory dual to the 0+1d Sachdev-Ye-Kitaev (SYK) model~\cite{Sachdev1993Gapless,Kitaev2015,Maldacena2016Remarks}, and showed that the bulk dual of the boundary Majorana fermion in SYK is a chiral Majorana fermion. 
Since the SYK model describes a generalized free fermion that satisfies the Wick theorem, they are able to treat the model as free Majorana fermions and calculate the corresponding TDEE for the 0+1d SYK model, which gives $S=\frac{1}{12}\ln T$. This is the same as the entanglement entropy of a chiral Majorana fermion in a spatial interval of length $T$ in the bulk dual they constructed, which is one half of that of a non-chiral Majorana fermion. This result is similar to our findings of the TDEE here, although Ref.~\cite{Nebabu_2023} is considering a 0+1d model.

Moreover, we showed that the 2-site TDEE (2 sites in the same unit cell) for the SSH model is qualitatively distinct in topological and trivial phases. The TDEE of the topological phase is consistently larger than that of the trivial phase, which we attribute to more nonlocal Wannier functions and larger inter-cell entanglement in the topological phase.

\begin{figure}
    \centering
    \subfigure[]
    {\includegraphics[width=0.4\linewidth]{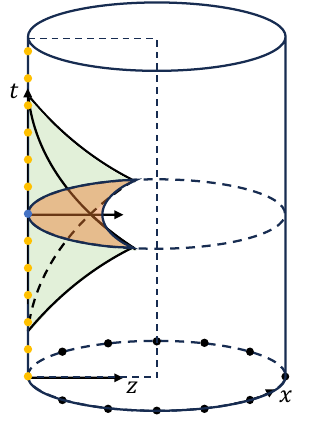}}
    \quad
    \subfigure[]
    {\includegraphics[width=0.5\linewidth]{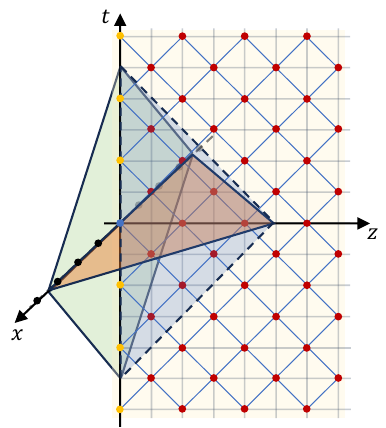}}
    \caption{Schematic illustration of a possible higher-dimensional bulk reconstruction associated with the time-direction fermion modes. 
    (a) In a holographic picture, the boundary of the cylinder corresponds to the $(1+1)$-dimensional free fermion system, and the bulk direction $z$ emerges from the holographic duality. 
    The yellow dots denote the Heisenberg-evolved fermion operators generated from the blue dot, and the green region denotes the causal diamond, which is related to the effective Lieb-Robinson-type estimate discussed above. 
    (b) A more detailed schematic bulk geometry at fixed $x=0$. 
    The red dots correspond to the bulk circuit and also label the orthonormal fermion modes generated by the time evolution of fermion operators at $x=0$, while the gray grid represents the emergent bulk circuit.}
    \label{fig:holographic dual}
\end{figure}

Based on the remarks above, an interesting open question as motivated by Ref.~\cite{Nebabu_2023} is the possibility of a holographic dual description for TDEE. Compared to Ref.~\cite{Nebabu_2023}, the key difference in our case here is that, our TDEE is calculated for a 1+1d lattice model, which should be viewed as a 1+1d boundary of a 2+1d bulk theory, as illustrated in Fig.~\ref{fig:holographic dual}. Although it is possible to construct the bulk dual for each fermion site individually, similar to the approach in Ref.~\cite{Nebabu_2023}, the crucial point here is that the slices of the dual corresponding to different sites are coupled together, forming a nontrivial and interacting bulk theory. Furthermore, for gapless free fermions, the boundary is a conformal field theory (CFT), and a natural question is to explore whether the reconstructed bulk dual geometry is asymptotically an Anti-de Sitter space.

Another intriguing direction is to explore how our setup can be realized within a circuit model. 
Several factors make this a challenging but potentially fruitful endeavor.
(i) As in Ref.~\cite{Nebabu_2023}, one could first construct the bulk dual using a tensor-network circuit and then take the continuum limit to obtain a nontrivial bulk field theory. 
(ii) A circuit model would provide a simple framework for generalizing our setup from fermionic systems to spin systems.  
However, due to the non-linear Jordan–Wigner transformation between spins and fermions, this mapping is highly nontrivial and may introduce additional constraints.
(iii) In the literature, there exist alternative definitions of temporal entanglement entropy (not for the physical quantum many-body state, as we mentioned in the introduction) different from our TDEE here, such as those obtained via Wick rotation in CFTs or using the transfer matrix formalism.  
A tensor circuit approach may serve as a unifying framework to explore the relationships between our definition and these alternatives, for example, through the study of dual-unitary circuits.
(iv) A quantum circuit realization of TDEE may be experimentally more accessible in quantum computing platforms, in which entanglement entropy may be measured more easily. 
These directions may connect our studies of TDEE to practical implementations and the broader definitions of temporal entanglement entropy of various types, bridging theory, simulation, and experiment.

\begin{figure}
    \centering
    {\includegraphics[width=0.85\linewidth]{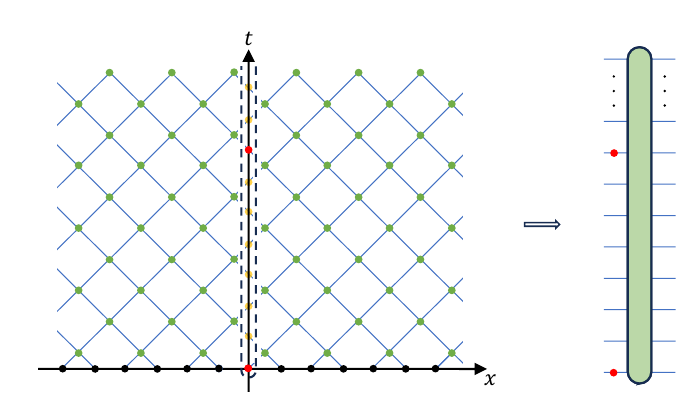}}
    \caption{Schematic illustration of a possible tensor-network connection to the Heisenberg time evolution and correlation functions along the time direction. 
    The red dots denote insertions of fermion operators $c_0$ and $c_0^\dagger$, while the green dots represent the unitary gates implementing time evolution. 
    The black dashed line indicates a cut through the circuit.
    The green block illustrates a possible effective MPS representation after contracting the circuit.}
    \label{fig:circuit_contraction}
\end{figure}

Building on the tensor-network construction, our TDEE results may also be connected to the computational complexity in the future.
Starting from a tensor-network representation of the ground state, the TDEE can be expressed in terms of correlation functions along the time direction at the fixed spatial coordinate $x=0$; equivalently, it may be viewed as a partial trace over the spatial degrees of freedom.
This reduces the problem to analyzing the correlation structure of a matrix product state (MPS) defined along time, as shown in Fig.~\ref{fig:circuit_contraction}.
A volume-law TDEE at relatively large $\tau$ may imply that the effective MPS requires a large bond dimension. Consequently, contracting the network in the spatial direction while leaving a cut along time does not, by itself, reduce the computational complexity. 
By contrast, in the continuous-time limit $\tau\rightarrow 0$, the TDEE for ground states becomes logarithm, area or sub-area laws, which is comparable to the spatial entanglement entropy. This may yield an effective time-direction MPS with a much smaller bond dimension. 

We emphasize several substantive differences between our TDEE studied in this work and the studies of influence–matrix approach in literatures~\cite{Michael2021Influence,Lerose2021Influence,Lerose2021Scaling,Giudice2022Temporal,Thoenniss2023Nonequilibrium,Lerose2023Overcoming}. 
First, we initialize in the ground state of the Hamiltonian, rather than in a product state as is customary in influence–matrix setups. 
Second, we evolve operators in the Heisenberg picture with the global unitary $U = e^{{\rm i} H \delta t}$, without a Trotter decomposition; for finite (and especially large) $\delta t$ this can yield qualitatively different behavior. 
Most importantly, our definition of TDEE is based on an \emph{orthonormal} fermion basis in the time interval (\cref{eq:orthonormal_basis}). This step of orthonormalization is absent in the influence–matrix approach, where different time points as the basis of the influence matrix are treated as an orthonormal basis to define a spurious quantum state. 
If one were to evaluate our TDEE here directly by regarding the non-orthonormal basis $\{c_i(t)\}$ as orthonormal, the result would differ and would not admit an interpretation as the entanglement entropy of a physical quantum state.

An important open problem is how to generalize the definition of TDEE to interacting systems. 
In general, entanglement entropy is only one way to characterize the properties of a state—or equivalently, of the sub-Hilbert space—in the time direction. 
The essential challenge lies in how to construct a well-defined sub-Hilbert space of a time direction interval in the presence of interactions. While the Lieb-Robinson bound provides an approximate upper bound for a time direction interval (as discussed in \cref{sec:reduced_hilbert_gapless}), it remains to be understood how to find the actual sub-Hilbert space. It is possible that in the continuum time limit $\tau\rightarrow 0$, the sub-Hilbert space of a time-direction interval will stabilize, as suggested by the numerical evidence for free fermion systems in \cref{sec:reduced_hilbert_gapless}. Recently, Ref.~\cite{kitaev2025Almost} introduced a method for deforming approximate $C^*$-algebras into exact ones, which may help formulate a controlled continuum-time limit of the sub-Hilbert space of the time direction interval and the operator algebra in it.

% \bl{I removed some of the vague ideas, to avoid referee questions, and to keep the detailed ideas to ourselves.}

%One possible direction is to use dual-unitary circuits. When ZZ coupling terms are included in the dual-unitary gates, they are equivalent to fermion-fermion interactions under the Jordan–Wigner transformation. This connection is also one of the main motivations for extending our setup to a circuit-model framework, where such interactions can be naturally implemented and studied.

%Another promising approach is to construct an approximate sub-Hilbert space. Recently, Ref.~\cite{kitaev2025Almost} introduced a method for defining approximate $C^*$-algebras, which may provide a rigorous foundation for this idea. In an interacting Hamiltonian, the time evolution causes the size of fermion operators to grow, making it impossible to represent them using a simple linear basis as in the free-fermion case. However, if we first define an approximate Hilbert subspace using these techniques, we can then investigate the properties of states within it, including their entanglement entropy.For example, we may approximate the Heisenberg evolution of many-body operators by linear combinations of fermionic creation and annihilation operators, with the coefficients determined by matching two-point correlators. Such a framework opens the door to studying TDEE in strongly interacting systems, thereby bridging free models and realistic many-body dynamics.

\begin{acknowledgments}

We thank Xiaoliang Qi, Haifeng Tang, Hong-Hao Tu, and Xueda Wen for helpful discussions. This work is supported by the National Science Foundation under award DMR-2141966.

\end{acknowledgments}

% \newpage

\appendix

\section{Analytical calculation of $D(\theta)$ and time direction entanglement entropy}
\label{sec:Background of correlation matrix and entanglement entropy}

In this section, we explicitly calculate $D(\theta)$ for specific dispersions.
Before carrying out the calculation, we first show how the function $D(\theta)$ can be used to calculate the time direction entanglement entropy, and then prove the property of $D(\theta)$ given in Eq.~\eqref{eq:Fourier_transf_D}.

Starting from Eq.~\eqref{eq:contour_integral_of_EE} in the main text, we just need the determinant of the matrix $G=(\lambda+1)I-2D$.
To determine its asymptotic behavior, we make use of the Fisher-Hartwig conjecture. 
For a Toeplitz matrix $G$ with the symbol $G(\theta)$, which has the form Eq.~\eqref{eq:FT_special_form_G}, then the asymptotic behavior of the determinant ${\rm Det}(G)$ is given by
\begin{equation}
\label{eq:asymptotic_behavior_det}
\begin{split}
    {\rm Det}(G)\sim F[f(\theta)]^{d_A}\left(\prod_{r=1}^R {d_A}^{a_r^2-b_r^2}\right)E,\\
    F[f(\theta)]=\exp{\left(\int_0^{2\pi} \frac{{\rm d}\theta}{2\pi}\ \ln{f(\theta)} \right)},
\end{split}
\end{equation}
where $E$ is a complicated function of $f(\theta)$, $a_r$, $b_r$, and $\theta_r$, but is independent of the subsystem size ${d_A}$.

Based on the asymptotic behavior of $g(\lambda)={\rm Det}(G)$, we can substitute it into Eq.~\eqref{eq:contour_integral_of_EE} to calculate the entanglement entropy. 
For different dependencies on the subsystem size ${d_A}$, the three parts in Eq.~\eqref{eq:asymptotic_behavior_det} contribute the volume-law, $\log$-law, and area-law terms of the entanglement entropy, respectively, giving $S(A)=\alpha_V d_A+\alpha_\text{log}\ln{d_A}+ \alpha_A$, where the area law coefficient $\alpha_A$ is non-universal, while the volume and log law factors are Eq.~\eqref{eq:general_EE_with_FH_conj}.
We mainly focus on the prefactors of the volume-law and $\log$-law terms.

Next, we discuss the application of the Fisher--Hartwig conjecture to the TDEE. 
As shown in Eq.~\eqref{eq:correlation_matrix_D_with_BC}, the anti-commutation matrix $B$ and the correlation matrix $C$ completely determine the TDEE through the modified correlation matrix $D$ in the orthonormal basis. 
Strictly speaking, the modified correlation matrix is not necessarily Toeplitz for a finite time interval. Here we adopt a translation-invariant convolution approximation, which is also used in our finite-$L$ calculations.
Within this approximation, we obtain $D(\theta)=C(\theta)/B^*(\theta)$ as follows.
% \begin{widetext}
\begin{equation}
\label{eq:simplify_Fourier_transf}
\begin{split}
    &\sum_l e^{-{\rm i}\theta l}(B^{*-1/2}CB^{T -1/2})_{0,l}\\
    =&\sum_l e^{-{\rm i}\theta l}\int_0^{2\pi}\frac{{\rm d}\theta_{1}}{2\pi}e^{{\rm i}l\theta_1}B^{*-1/2}(\theta_1) C(\theta_1) B^{T -1/2}(\theta_1)\\=&B^{*-1/2}(\theta) C(\theta) B^{T -1/2}(\theta).
\end{split}
\end{equation}
% \end{widetext}
Since the matrix $B$ is Hermitian, we have $B^*(\theta)=B^T(\theta)$. 
Similarly, by applying the Fourier transformation to $I=B^{*-1/2}B^{*}B^{*-1/2}$, we obtain $1=B^{*-1/2}(\theta)B^{*}(\theta)B^{*-1/2}(\theta)$, which implies $B^{*-1/2}(\theta)=(B^{*}(\theta))^{-1/2}$. 
Thus, Eq.~\eqref{eq:Fourier_transf_D} in the main text follows. 
This result greatly simplifies the calculation because the matrix $D$ is considerably more complicated than the matrices $B$ and $C$.

Actually, for the Fourier transformation of the matrices $B$ and $C$, under the Fourier transformation, we have 
\begin{eqnarray}
\label{eq:ft_C}
    C(\theta)=&&\sum_l e^{-{\rm i}\theta l}C_{0,l}=\sum_l e^{-{\rm i}\theta l} \frac{1}{L}\sum_{k}e^{{\rm i}E_k \tau l}n_F(E_k)\nonumber\\
    =&&\sum_{k} \frac{1}{L}\sum_l e^{{\rm i}(E_k\tau-\theta)l} n_F(E_k).
\end{eqnarray}
Although in many cases we can simplify Eq.~\eqref{eq:ft_C} using $\frac{1}{L}\sum_l e^{{\rm i}(E_k\tau-\theta)l}=\delta(E_k\tau-\theta)$, here this step can lead to certain subtleties.
One reason is that $\theta\in[-\pi,\pi]$ while $E_k\tau\in[-\pi\tau/\tau_0,\pi\tau/\tau_0]$. 
This indicates that $\tau=\tau_0$ is a special case, leading to a different prefactor for the entanglement entropy. 
Moreover, here we consider $l=-\frac{L-1}{2},\ldots,\frac{L-1}{2}$, which means that the number of degrees of freedom along the time direction is the same as that along the spatial direction.
Besides, in the Fisher–Hartwig conjecture, the Fourier transformation is taken with $l\in\mathbb{Z}$, but this choice leads to exact zero values for $C(\theta)$ in certain ranges of $\theta$. 
To avoid an ill-defined $D(\theta)=C(\theta)/B^*(\theta)$, we introduce a finite cutoff. 
This regularization also ensures that numerical results remain consistent with the analytical predictions.

Similarly, for the anti-commutation matrix we have
\begin{equation}
\label{eq:ft_B}
    B^*(\theta)=\sum_l e^{-{\rm i}\theta l}B^*_{0,l}=\sum_{k} \frac{1}{L}\sum_l e^{{\rm i}(E_k\tau-\theta)l}.
\end{equation}
A further subtlety arises when $\tau$ is an irrational number: in this case, the matrices $B_{0,l}$ and $C_{0,l}$ are not strictly periodic, i.e., $M_{0,l}\neq M_{0,l+L}$. 
As a result, the discrete version of the continuous variable $\theta$ is not strictly well-defined. 
Here, we may simply set $\theta=\frac{2\pi}{L}n$ with $n=-\frac{L-1}{2},\ldots,\frac{L-1}{2}$, which matches the momentum-space discretization of the free fermion model. 
This subtlety is directly related to the range of $l=-\frac{L-1}{2},\ldots,\frac{L-1}{2}$.
% because of the aperiodic nature of the functions $B_{0,l}$ and $C_{0,l}$.

% Later, we will see that for certain parameter choices, $B(\theta)$ and $C(\theta)$ are well-defined for continuous $\theta\in[-\pi,\pi]$. 
% For other parameters, however, $B(\theta)$ and $C(\theta)$ are only well-defined for discrete values of $\theta$. 
% Additionally, we always adopt a symmetric Brillouin zone, since for an asymmetric zone the result may become complex, even if the model itself possesses particle-hole symmetry.

With Eq.~\eqref{eq:ft_C} and Eq.~\eqref{eq:ft_B}, we can perform the summation over $l$ to obtain
% \begin{widetext}
\begin{equation}
\label{eq:ft_C_simplify}
\begin{split}
    C(\theta)=&\sum_{k} \frac{1}{L}\sum_l e^{{\rm i}(E_k\tau-\theta)l} n_F(E_k)\\
    % =&\sum_k \Theta(k_F-|k|)\frac{1}{L}\frac{e^{{\rm i}\frac{1-L}{2}(E_k\tau-\theta)}(1-e^{{\rm i}L(E_k\tau-\theta)})}{1-e^{{\rm i}(E_k\tau-\theta)}}\\
    =&\sum_k \Theta(k_F-|k|)\frac{\sin{\big[L(E_k\tau-\theta)/2\big]}}{L\sin{\big[(E_k\tau-\theta)/2\big]}},
\end{split}
\end{equation}
\begin{equation}
\label{eq:ft_B_simplify}
    B^*(\theta)=\sum_{k} \frac{1}{L}\sum_l e^{{\rm i}(E_k\tau-\theta)l}
    =\sum_k \frac{\sin{\big[L(E_k\tau-\theta)/2\big]}}{L\sin{\big[(E_k\tau-\theta)/2\big]}}.
\end{equation}
% \end{widetext}
It is clear that if we take the limit $L\rightarrow\infty$ directly, we recover the Dirac delta function $\delta(E_k\tau-\theta)$ in the expressions above. 
However, this would make the ratio $C(\theta)/B^*(\theta)$ ill-defined whenever $B^*(\theta)=0$, which occurs when $\tau<\tau_0$. 
Therefore, throughout the calculation we keep $L$ large but finite, and only take the limit $L\rightarrow\infty$ at the very end of the derivation.

\subsection{Free fermion with uniform hopping}
\label{sec:Free fermion with uniform hopping}

For the free fermion with uniform hopping~\cite{Liu_2024}, described by the Hamiltonian~\eqref{eq:tight_binding_model}, we consider the case $\tau=\tau_0/2=\pi/4$. 
For $E_k=\frac{8|n|}{L}-2$ with $n=-\frac{L-1}{2},\ldots,\frac{L-1}{2}$ and $L=8\mathbb{Z}+1$, the degeneracies are $1,2,\ldots,2$ for $k=0,\pm\frac{2\pi}{L},\ldots,\pm\frac{2\pi}{L}\frac{L-1}{2}$, and the energy spectrum is given by $E_k=-2,-2+\frac{8}{L},\ldots,-2+\frac{8}{L}\frac{L-1}{2}$. 
It is worth noting that for $L=4\mathbb{Z}+1$, particle-hole symmetry is absent.

% \begin{widetext}
\begin{equation}
\label{eq:ft_B_simplify_tau0}
\begin{split}
    &B^*(\theta)=\sum_{n} \frac{\sin{\big[\frac{L}{2}(E_k\tau-\theta)\big]}}{L\sin{\big[\frac{1}{2}(E_k\tau-\theta)\big]}}\\
    % =&\frac{\sin{\big[\frac{L}{2}(-\frac{\pi}{2}-\theta)\big]}}{L\sin{\big[\frac{1}{2}(-\frac{\pi}{2}-\theta)\big]}}+2\sum_{k=1}^{\frac{L-1}{2}}\frac{\sin{\big[\frac{L}{2}\big((\frac{8k}{L}-2)\frac{\pi}{4}-\theta\big)\big]}}{L\sin{\big[\frac{1}{2}\big((\frac{8k}{L}-2)\frac{\pi}{4}-\theta\big)\big]}}\\
    =&\frac{\sin{\left[\frac{L}{2}\left(-\frac{\pi}{2}-\theta\right)\right]}}{L\sin{\big[\frac{1}{2}(-\frac{\pi}{2}-\theta)\big]}}+2\sum_{k=1}^{\frac{L-1}{2}}\frac{(-1)^k\sin{\left[\frac{L}{2}\left(-\frac{\pi}{2}-\theta\right)\right]}}{L\sin{\big[\frac{1}{2}\big((\frac{8k}{L}-2)\frac{\pi}{4}-\theta\big)\big]}}.
\end{split}
\end{equation}
% \end{widetext}
Similarly, for $C(\theta)$ the only difference is that the summation runs over $k=1,\ldots,\frac{L-1}{4}$. 
Thus, we obtain
\begin{equation}
\label{eq:C/B}
    \frac{C(\theta)}{B^*(\theta)}=\frac{\frac{1}{L\sin{[\frac{1}{2}(-\frac{\pi}{2}-\theta)]}}+2\sum_{k=1}^{\frac{L-1}{4}}\frac{(-1)^k}{L\sin{[\frac{1}{2}((\frac{8k}{L}-2)\frac{\pi}{4}-\theta)]}}}{\frac{1}{L\sin{[\frac{1}{2}(-\frac{\pi}{2}-\theta)]}}+2\sum_{k=1}^{\frac{L-1}{2}}\frac{(-1)^k}{L\sin{[\frac{1}{2}((\frac{8k}{L}-2)\frac{\pi}{4}-\theta)]}}}.
\end{equation}

Next, we evaluate the summations over $k$. 
For $L\rightarrow\infty$, we have
\begin{equation}
\label{eq:simplify_summation}
\begin{split}
    &2\sum_{k=1}^{\frac{L-1}{4}}\frac{(-1)^k}{\sin{\big[\frac{1}{2}\big((\frac{8k}{L}-2)\frac{\pi}{4}-\theta\big)\big]}}\\
    =&2\sum_{k=1}^{\frac{L-1}{8}}\left(\frac{-1}{\sin{\big[\frac{(2k-1)\pi}{L}-\frac{\pi}{4}-\frac{\theta}{2}\big]}}+\frac{1}{\sin{\big[\frac{2k\pi}{L}-\frac{\pi}{4}-\frac{\theta}{2}\big]}}\right)\\
    \approx&2\sum_{k=1}^{\frac{L-1}{8}}\frac{-\pi\cos{\big[\frac{2k\pi}{L}-\frac{\pi}{4}-\frac{\theta}{2}\big]}}{L\sin^2{\big[\frac{2k\pi}{L}-\frac{\pi}{4}-\frac{\theta}{2}\big]}}
    \approx-\int_0^{\frac{\pi}{4}}{\rm d}x\,\frac{\cos{\big[x-\frac{\pi}{4}-\frac{\theta}{2}\big]}}{\sin^2{\big[x-\frac{\pi}{4}-\frac{\theta}{2}\big]}}\\
    =&\left\{\begin{aligned}
        &-\frac{1}{\sin{\frac{\theta}{2}}}+\frac{1}{\sin{\left(\frac{\theta}{2}+\frac{\pi}{4}\right)}}, &\theta\in[-\pi,-\frac{\pi}{2}]\cup[0,\pi],\\
        &\infty, &\theta\in[-\frac{\pi}{2},0].\\
    \end{aligned}\right.
\end{split}
\end{equation}
Similarly, for the summation in the denominator, we find
\begin{equation}
\label{eq:simplify_summation_den}
\begin{split}
    &2\sum_{k=1}^{\frac{L-1}{2}}\frac{(-1)^k}{\sin{\big[\frac{1}{2}\big((\frac{8k}{L}-2)\frac{\pi}{4}-\theta\big)\big]}}
    =-\int_0^{\frac{\pi}{2}}{\rm d}x\,\frac{\cos{\big[x-\frac{\pi}{4}-\frac{\theta}{2}\big]}}{\sin^2{\big[x-\frac{\pi}{4}-\frac{\theta}{2}\big]}}\\
    =&\left\{\begin{aligned}
        &\frac{1}{\cos{\left(\frac{\theta}{2}+\frac{\pi}{4}\right)}}+\frac{1}{\sin{\left(\frac{\theta}{2}+\frac{\pi}{4}\right)}}, &\theta\in[-\pi,-\frac{\pi}{2}]\cup[\frac{\pi}{2},\pi],\\
        &\infty, &\theta\in[-\frac{\pi}{2},\frac{\pi}{2}].\\
    \end{aligned}\right.
\end{split}
\end{equation}
Therefore, substituting these results into Eq.~\eqref{eq:C/B} gives Eq.~\eqref{eq:D_result_gapless_tau02} in the main text.
% The sum-to-integral approximation above is valid in the large-$L$ limit away from the poles of the summand, but is not uniform near these singularities, where the finite-$L$ expression should be retained.

Here are several remarks:
(i) Although we only show one analytical example of the time-direction entanglement entropy for $\tau<\tau_0$, we expect that the presence of a single discontinuity is a general feature for at least many cases with $\tau<\tau_0$ and for arbitrary dispersions
\footnote{In some special cases, there may be more than one discontinuity.} .
Later, we will present another analytical example that also exhibits this behavior. 
There are also many other cases where $D(\theta)$ cannot be calculated analytically, but numerical results indicate that there is still only one discontinuity. 
Even when the function $D(\theta)$, or equivalently $C(\theta)/B^*(\theta)$, is not well-defined for all $\theta\in[-\pi,\pi]$, the discrete values of $\theta$ still show only a single discontinuity.
(ii) Here, we focus on systems with an odd system size, which do not exhibit particle-hole symmetry. 
For a free fermion chain with anti-periodic boundary conditions, the discrete values of momentum can differ from those considered here.
(iii) This example is not ideal for demonstrating the single discontinuity of $D(\theta)$. 
For the analytical result in Eq.~\eqref{eq:D_result_gapless_tau02}, there is also a discontinuity at $\theta=-\pi/2$, as shown in Fig.~\ref{fig:gapless_FT_D}(a). 
We emphasize that the finite-cutoff ratio $C(\theta)/B^*(\theta)$ is used here to identify the robust discontinuity structure rather than as an exact finite-matrix correlation symbol. 
In particular, its values may lie outside $[0,1]$, and the additional discontinuity at $\theta=-\pi/2$ depends on the finite-size regularization, whereas the discontinuity at $\theta=0$ persists.
% For larger system sizes, we expect that a small discontinuity may persist near $\theta=-\pi/2$.

In the following, we consider another case that has exactly one discontinuity, free from the ambiguities discussed above. 
Here, we take $L=4\mathbb{Z}+2$, with $E_k=\frac{8}{L}\left|k-\frac{1}{2}\right|-2,l=-\frac{L}{2}+1,\ldots,\frac{L}{2}$, where each eigenenergy has a degeneracy of $2$. 
For $E_k=0$, we find $k_c=\frac{L+2}{4}\in\mathbb{Z}$, and we adopt the convention $\Theta(0)=1$. 

Plugging these expressions into $B^*(\theta)$ and $C(\theta)$, we obtain
% \begin{widetext}
\begin{equation}
\label{eq:ft_B_simplify_tau0_2n}
\begin{split}
    &B^*(\theta)=2\sum_{k=1}^{\frac{L}{2}}\frac{\sin{\left[\frac{L}{2}\left(\left(\frac{8}{L}\left(k-\frac{1}{2}\right)-2\right)\frac{\pi}{4}-\theta\right)\right]}}{L\sin{\left[\frac{1}{2}\left(\left(\frac{8}{L}\left(k-\frac{1}{2}\right)-2\right)\frac{\pi}{4}-\theta\right)\right]}}\\
    =&\frac{2}{L}f(\theta)\sum_{k=1}^{\frac{L}{2}}\frac{(-1)^k}{\sin{\left[\frac{\pi}{L}\left(k-\frac{1}{2}\right)-\frac{\pi}{4}-\frac{\theta}{2}\right]}}\\
    % =&\frac{2}{L}\sin{\left[-\frac{\pi}{2}-\frac{L}{2}\left(\frac{\pi}{2}+\theta\right)\right]}
    % \left[\sum_{l=1}^{(\frac{L}{2}-1)/2}\left(\frac{-1}{\sin{\left[\frac{\pi}{L}(2l-1-\frac{1}{2})-\frac{\pi}{4}-\frac{\theta}{2}\right]}}+\frac{1}{\sin{\left[\frac{\pi}{L}(2l-\frac{1}{2})-\frac{\pi}{4}-\frac{\theta}{2}\right]}}\right)-\frac{1}{\sin{\left[\frac{\pi}{L}(\frac{L}{2}-\frac{1}{2})-\frac{\pi}{4}-\frac{\theta}{2}\right]}}\right]\\
    % \approx&\frac{2}{L}\sin{\left[-\frac{\pi}{2}-\frac{L}{2}\left(\frac{\pi}{2}+\theta\right)\right]}\left(\sum_{l=1}^{\frac{L-2}{4}}\frac{-\pi\cos{\left[\frac{\pi(2l-1/2)}{L}-\frac{\pi}{4}-\frac{\theta}{2}\right]}}{L\sin^2{\left[\frac{\pi(2l-1/2)}{L}-\frac{\pi}{4}-\frac{\theta}{2}\right]}}-\frac{1}{\cos{\left[\frac{\pi}{4}+\frac{\theta}{2}\right]}}\right)\\
    \approx&\frac{2}{L}f(\theta)\left(-\int_0^{\frac{\pi}{4}}{\rm d}x\frac{\cos{\left[2x-\frac{\pi}{4}-\frac{\theta}{2}\right]}}{\sin^2{\left[2x-\frac{\pi}{4}-\frac{\theta}{2}\right]}}-\frac{1}{\cos{\left[\frac{\pi}{4}+\frac{\theta}{2}\right]}}\right)\\
    =&\frac{2f(\theta)}{L}\left(\frac{\sqrt{2}\cos{\frac{\theta}{2}}}{\cos{\theta}}-\frac{1}{\cos{(\frac{\pi}{4}+\frac{\theta}{2})}}\right),\theta\in[-\pi,-\frac{\pi}{2}]\cup[\frac{\pi}{2},\pi]
    % \left\{\begin{aligned}
    %     &\frac{2f(\theta)}{L}\left(\frac{\sqrt{2}\cos{\frac{\theta}{2}}}{\cos{\theta}}-\frac{1}{\cos{\left[\frac{\pi}{4}+\frac{\theta}{2}\right]}}\right), &\theta\in[-\pi,-\frac{\pi}{2}]\cup[\frac{\pi}{2},\pi],\\
    %     &\infty, &\theta\in[-\frac{\pi}{2},\frac{\pi}{2}],\\
    % \end{aligned}\right.
\end{split}
\end{equation}
where $f(\theta)=\sin{[-\frac{\pi}{2}-\frac{L}{2}(\frac{\pi}{2}+\theta)]}$ corresponds to a highly oscillating factor, and in the last line, $B^*(\theta)=\infty$ for other $\theta$.
Similarly, for $C(\theta)$ the only difference is that the summation runs over $k=1,\ldots,\frac{L+2}{4}$. 
Thus, we obtain
\begin{equation}
\label{eq:ft_C_simplify_tau0_2n}
\begin{split}
    &C(\theta)=\frac{2f(\theta)}{L}\sum_{k=1}^{\frac{L+2}{4}}\frac{(-1)^k}{\sin{\left[\frac{\pi}{L}\left(k-\frac{1}{2}\right)-\frac{\pi}{4}-\frac{\theta}{2}\right]}}\\
    % =&\frac{2}{L}\sin{\left[-\frac{\pi}{2}-\frac{L}{2}\left(\frac{\pi}{2}+\theta\right)\right]}
    % \left[\sum_{l=1}^{\frac{L-2}{8}}\left(\frac{-1}{\sin{\left[\frac{\pi}{L}(2l-1-\frac{1}{2})-\frac{\pi}{4}-\frac{\theta}{2}\right]}}+\frac{1}{\sin{\left[\frac{\pi}{L}(2l-\frac{1}{2})-\frac{\pi}{4}-\frac{\theta}{2}\right]}}\right)-\frac{1}{\sin{\left[\frac{\pi}{L}\frac{L}{4}-\frac{\pi}{4}-\frac{\theta}{2}\right]}}\right]\\
    % \approx&\frac{2}{L}\sin{\left[-\frac{\pi}{2}-\frac{L}{2}\left(\frac{\pi}{2}+\theta\right)\right]}\left(\sum_{l=1}^{\frac{L-2}{8}}\frac{-\pi\cos{\left[\frac{\pi(2l-1/2)}{L}-\frac{\pi}{4}-\frac{\theta}{2}\right]}}{L\sin^2{\left[\frac{\pi(2l-1/2)}{L}-\frac{\pi}{4}-\frac{\theta}{2}\right]}}-\frac{1}{\sin{[-\frac{\theta}{2}]}}\right)\\
    % \approx&\frac{2f(\theta)}{L}\left(-\int_0^{\frac{\pi}{8}}{\rm d}x\frac{\cos{\left[2x-\frac{\pi}{4}-\frac{\theta}{2}\right]}}{\sin^2{\left[2x-\frac{\pi}{4}-\frac{\theta}{2}\right]}}+\frac{1}{\sin{\left[\frac{\theta}{2}\right]}}\right)\\
    \approx&\frac{2f(\theta)}{L}\left(\frac{1}{2\sin{(\frac{\theta}{2}+\frac{\pi}{4})}}+\frac{1}{2\sin{\frac{\theta}{2}}}\right),\theta\in[-\pi,-\frac{\pi}{2}]\cup[0,\pi].
    % \left\{\begin{aligned}
    %     &\frac{1}{\sqrt{2}(\cos{\frac{\theta}{2}}+\sin{\frac{\theta}{2}})}+\frac{1}{2\sin{\left[\frac{\theta}{2}\right]}}, &\theta\in[-\pi,-\frac{\pi}{2}]\cup[0,\pi],\\
    %     &\infty, &\theta\in[-\frac{\pi}{2},0].\\
    % \end{aligned}\right.
\end{split}
\end{equation}
% \end{widetext}
% In the last line, $B^*(\theta)=\infty$ for other $\theta$.
In the last line, the expression in parentheses diverges for other values of $\theta$.
Here, in the second line, we have taken $L=8\mathbb{Z}+2$. 
For $L=8\mathbb{Z}+6$, the final term $1/(2\sin{\frac{\theta}{2}})$ changes its sign in the last line.

Finally, substituting these results into Eq.~\eqref{eq:C/B}, we obtain
\begin{equation}
\label{eq:C/B_result_2n}
    \frac{C(\theta)}{B^*(\theta)}=\left\{\begin{aligned}
    &\frac{\frac{1}{2\sin{(\frac{\theta}{2}+\frac{\pi}{4})}}\pm\frac{1}{2\sin{\left[\frac{\theta}{2}\right]}}}{\sqrt{2}\frac{\cos{\frac{\theta}{2}}}{\cos{\theta}}-\frac{1}{\cos{\left[\frac{\pi}{4}+\frac{\theta}{2}\right]}}}, &\theta\in[-\pi,-\frac{\pi}{2}]\cup[\frac{\pi}{2},\pi],\\
    &\Theta(-\theta), &\theta\in[-\frac{\pi}{2},\frac{\pi}{2}],\\
    % &0, &\theta\in[0,\frac{\pi}{2}],\\
    \end{aligned}\right.
\end{equation}
where the $\pm$ sign corresponds to $L=8\mathbb{Z}+2$ and $L=8\mathbb{Z}+6$, respectively. 
It is straightforward to check that Eq.~\eqref{eq:C/B_result_2n} is a continuous function at $\theta=-\frac{\pi}{2}$.

\subsection{Gapped free fermion with SSH Hamiltonian}
\label{Gapped free fermion with SSH Hamiltonian}

In the following, we calculate $D(\theta)$ for the SSH model with specific parameters. 
Using the Hamiltonian~\eqref{eq:SSH_hamiltonian_unit_cell}, we consider a linear dispersion with eigenenergies $\pm E_{k_j}=\pm\left(2-\frac{2}{L}\left(j-\frac{1}{2}\right)\right)$, where $j=1,\ldots,\frac{L}{2}$ and each energy level has a degeneracy of $2$. 
Therefore, similar to Eq.~\eqref{eq:ft_C}, we obtain
% \begin{widetext}
\begin{equation}
\label{eq:ft_C_gapped}
\begin{split}
    C(\theta)
    % =&
    % % \sum_l e^{-{\rm i}\theta l}C(l)=
    % \sum_l e^{-{\rm i}\theta l} \frac{1}{2L}\sum_{j}e^{-{\rm i}E_{k_j} \tau l}
    % % =\frac{1}{2}\sum_{k_j}\frac{1}{N}\sum_{l=-\frac{N-1}{2}}^{\frac{N-1}{2}}e^{-{\rm i}l(E_{k_j} \tau+\theta)}\\
    % =\frac{1}{2}\sum_{k_j}\frac{\sin{\left[\frac{L}{2}(E_{k_j} \tau+\theta)\right]}}{L\sin{\left[\frac{1}{2}(E_{k_j} \tau+\theta)\right]}}\\
    =\frac{1}{2}\sum_{j=1}^{\frac{L}{2}}2\frac{\sin{\left[\frac{L}{2}\left(-\left(2-\frac{2}{L}(j-\frac{1}{2})\right)\tau-\theta\right)\right]}}{L\sin{\left[\frac{1}{2}\left(-\left(2-\frac{2}{L}(j-\frac{1}{2})\right)\tau-\theta\right)\right]}}.
\end{split}
\end{equation}
If we take $\tau=\pi$, the expression simplifies to $C(\theta)\approx\frac{1}{L}\sin{(-\frac{\pi}{2}-\frac{L\theta}{2})}(-\frac{1}{2\sin{(\theta/2)}}+\frac{1}{2\cos{(\theta/2)}})$ for $\theta\in[-\pi,0]$,
% \begin{equation}
% \label{eq:ft_C_gapped_tau_pi}
% \begin{split}
%     C(\theta)=&\sin{\left[-\frac{\pi}{2}-\frac{N}{2}(2\pi+\theta)\right]}\sum_{j=1}^{\frac{N}{2}}\frac{(-1)^j}{N\sin{\left[\pi+\frac{2j-1}{2N}\pi-\frac{\theta}{2}\right]}}\\
%     % =&\frac{1}{N}\sin{\left[-\frac{\pi}{2}-\frac{N\theta}{2}\right]}\left[\sum_{l=1}^{\frac{N-2}{4}}\left(\frac{-1}{\sin{\left[\frac{4l-3}{2N}\pi-\frac{\theta}{2}-\pi\right]}}+\frac{1}{\sin{\left[\frac{4l-1}{2N}\pi-\frac{\theta}{2}-\pi\right]}}\right)-\frac{1}{\sin{\left[-\frac{\theta}{2}-\frac{\pi}{2}\right]}}\right]\\
%     % \approx&\frac{1}{N}\sin{\left[-\frac{\pi}{2}-\frac{N\theta}{2}\right]}\left[-\sum_{l=1}^{\frac{N-2}{4}}\frac{\pi\cos{\left[\frac{4l-3}{2N}\pi-\frac{\theta}{2}-\pi\right]}}{N\sin^2{\left[\frac{4l-3}{2N}\pi-\frac{\theta}{2}-\pi\right]}}-\frac{1}{\sin{\left[-\frac{\theta}{2}-\frac{\pi}{2}\right]}}\right]\\
%     \approx&\frac{1}{N}\left[-\int_{0}^{\frac{\pi}{4}}{\rm d}x\frac{\cos{\left[2x-\frac{\theta}{2}-\pi\right]}}{\sin^2{\left[2x-\frac{\theta}{2}-\pi\right]}}-\frac{1}{\sin{\left[-\frac{\theta}{2}-\frac{\pi}{2}\right]}}\right]\\
%     =&\frac{\Tilde{f}(\theta)}{N}\left(-\frac{1}{2\sin{\frac{\theta}{2}}}+\frac{1}{2\cos{\frac{\theta}{2}}}\right),\ \theta\in[-\pi,0],
%     % \left\{\begin{aligned}
%     %     &-\frac{1}{2\sin{\frac{\theta}{2}}}+\frac{1}{2\cos{\frac{\theta}{2}}}, &\theta\in[-\pi,0],\\
%     %     &\infty, &\theta\in[0,\pi],
%     % \end{aligned}\right.
% \end{split}
% \end{equation}
% \end{widetext}
where $\Tilde{f}(\theta)=\sin{(-\frac{\pi}{2}-\frac{L\theta}{2})}$ corresponds to a highly oscillating factor, and $\tilde{C}(\theta)=\infty$ for other $\theta$.
Here we have assumed $L=4\mathbb{Z}+2$. 
For $L=4\mathbb{Z}$, the second term in the final result changes sign. 

Using Eq.~\eqref{eq:B_gapped_01}, ignoring the coefficient $L^{-1}$, we have $B(\theta)=C(\theta)+C(-\theta)$, which always diverges for $\theta\in[-\pi,\pi]$. 
Thus, the ratio simplifies to $C(\theta)/B(\theta)=\Theta(\theta)$ for $\theta\in[-\pi,\pi]$, which contains two discontinuities and is therefore expected to produce a log-law entanglement entropy.

\section{Vanishing of the time-direction entanglement entropy for a large number of time points}
\label{sec:Proof of vanish of time direction entanglement entropy for large time sites}

In this section, we prove that when the number of points in the time direction interval $K>L$, where $L$ is the system size in the spatial direction, under certain special assumptions, the TDEE $S(A)$ will vanish for a system with translation symmetry $\hat{T}_x c_i \hat{T}_x^{-1}=c_{i+1}$. 

An intuitive argument is that when the number of sites in the time direction exceeds the system size, we can treat the whole system as a single subsystem, which implies there is no entanglement entropy between it and the empty complement. 
However, this argument is not strictly correct because the new basis $c_{i}(n \tau)$ for different $\tau$ is not related to the basis $c_{i}(0)$ for different $i$ by a unitary transformation.

For a system with translation symmetry $\hat{T}_x c_i \hat{T}_x^{-1}=c_{i+1}$, the wavefunction is always a plane wave, which ensures that Eq.~\eqref{eq:Toeplitz_matrix_translation_symm_C} and Eq.~\eqref{eq:Toeplitz_matrix_translation_symm_B} are valid. With $U_{jn}=\frac{1}{\sqrt{L}}e^{-{\rm i}E_j n\tau}$, we have $B^*_{mn}=(U^\dagger U)_{mn}$ and $C_{mn}=(U^\dagger \Lambda U)_{mn}$, where $\Lambda$ is a diagonal matrix with $\Lambda_{ii}=\Theta(-E_i)$. 
Denoting $n=1,\ldots,[n]$ and $j=1,\ldots,[j]$, where $[n]=K=T/\tau$ is the number of temporal sampling points and $[j]=L$ is the number of spatial single-particle modes, we decompose $U=MSN^\dagger$, where $S$ is a diagonal $[j]\times[n]$ matrix, and $M$ and $N$ are unitary matrices of sizes $[j]\times[j]$ and $[n]\times[n]$, respectively.
Then the correlation matrix is 
% \bl{What do $[n]$, $[j]$ correspond to, $K$ and $L$? If not exactly, we can say roughly what they mean.}
\begin{equation}
    D=B^{*-\frac{1}{2}}CB^{*-\frac{1}{2}}=N\frac{S^\dagger}{\sqrt{S^\dagger S}}M^\dagger \Lambda M \frac{S}{\sqrt{S^\dagger S}} N^\dagger,
\end{equation}
and its eigenvalues are the same as those of $N^\dagger D N$. 

% In the following, we assume that all sampled phases $e^{-iE_j\tau}$ are distinct, so that the matrix $U$ has maximal rank $\min([n],[j])$.
In the following, we assume that the matrix $U$ has the rank ${\rm min}([n],[j])$. When $B$ has a nontrivial null space, we consider the pseudoinverse of $B$, meaning that we only take the inverse of the linearly independent vectors in $B$.

(i) For $[j]=[n]$, $\frac{S^\dagger}{\sqrt{S^\dagger S}}$ is also a unitary matrix, which means that the eigenvalues of $D$ are the same as the eigenvalues of $\Lambda$. 
Since $\Lambda$ acts as a projection matrix with only eigenvalues $0$ and $1$, the corresponding entanglement entropy 
$S=-\sum_{i}(v_i\ln{v_i}+(1-v_i)\ln{(1-v_i)})$ 
is zero.

(ii) If $[j]>[n]$, which means the number of sites in the time direction is smaller than the system size, we have $S^\dagger=[S'^\dagger|0]$, consisting of two blocks $S'$ and $0$, where $S'$ is an $[n]\times[n]$ matrix. 
Similarly, we write $M=[M'|M'']$, where $M'$ and $M''$ are $[j]\times[n]$ and $[j]\times([j]-[n])$ matrices, respectively. 
Therefore, the eigenvalues of $D$ are the same as those of $\frac{S'^\dagger}{\sqrt{S'^\dagger S'}}M'^\dagger \Lambda M' \frac{S'}{\sqrt{S'^\dagger S'}}$. 
Since $\frac{S'}{\sqrt{S'^\dagger S'}}$ is a unitary matrix, we only need to consider the eigenvalues of $M'^\dagger \Lambda M'$. 
Here, $M'$ can be interpreted as a combination of a unitary transformation and a projection. 
% Because the unitary transformation mixes the basis vectors with eigenvalues $0$ and $1$, the eigenvalues of $M'^\dagger \Lambda M'$ will range continuously from $0$ to $1$, leading to a nonzero time-direction entanglement entropy.
In general, the unitary transformation mixes the occupied and unoccupied single-particle modes, so the eigenvalues of $M'^\dagger\Lambda M'$ can lie between 0 and 1, leading to a nonzero time-direction entanglement entropy.

(iii) If $[j]<[n]$, which means the number of sites in the time direction is larger than the system size, we have $S=[S'|0]$, consisting of two blocks $S'$ and $0$, where $S'$ is an $[j]\times[j]$ matrix. 
In this case, $\frac{S^\dagger}{\sqrt{S^\dagger S}}M^\dagger \Lambda M \frac{S}{\sqrt{S^\dagger S}}={\rm Diag}(D',0)$, where $D'$ and $0$ are $[j]\times[j]$ and $([n]-[j])\times([n]-[j])$ matrices, respectively. 
Here, $D'=\frac{S'^\dagger}{\sqrt{S'^\dagger S'}}M^\dagger \Lambda M \frac{S'}{\sqrt{S'^\dagger S'}}$. 
Since both $\frac{S'}{\sqrt{S'^\dagger S'}}$ and $M$ are unitary matrices, the eigenvalues of $D$ are only $0$ and $1$, which leads to a zero entanglement entropy.

\section{Spectral density of anti-commutation matrix}
\label{sec:Spectral density of anti-commutation matrix}

\begin{figure}
    \centering
    \subfigure[]{\includegraphics[width=0.45\linewidth]{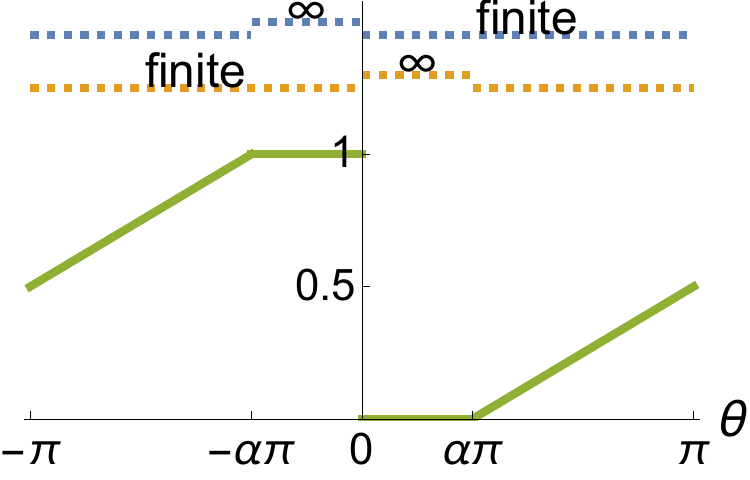}}
    \subfigure[]{\includegraphics[width=0.45\linewidth]{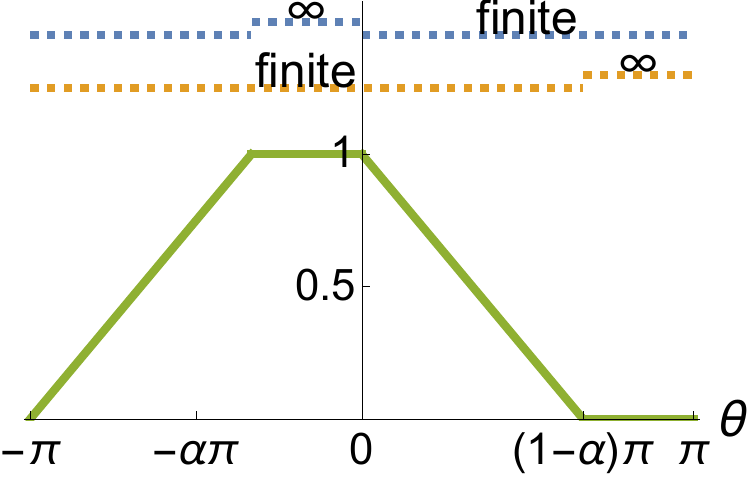}}
    \subfigure[]{\includegraphics[width=0.45\linewidth]{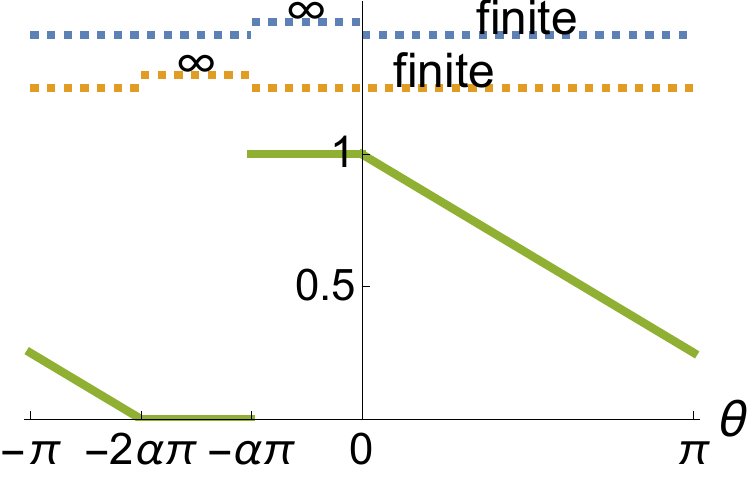}}
    \subfigure[]{\includegraphics[width=0.45\linewidth]{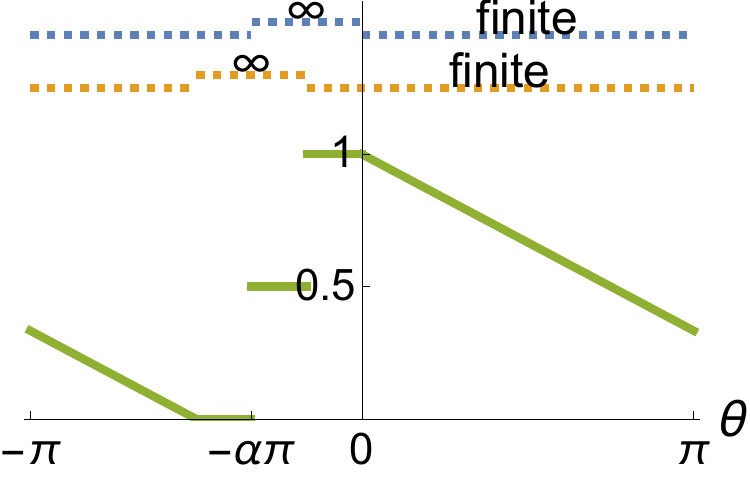}}
    \subfigure[]{\includegraphics[width=0.45\linewidth]{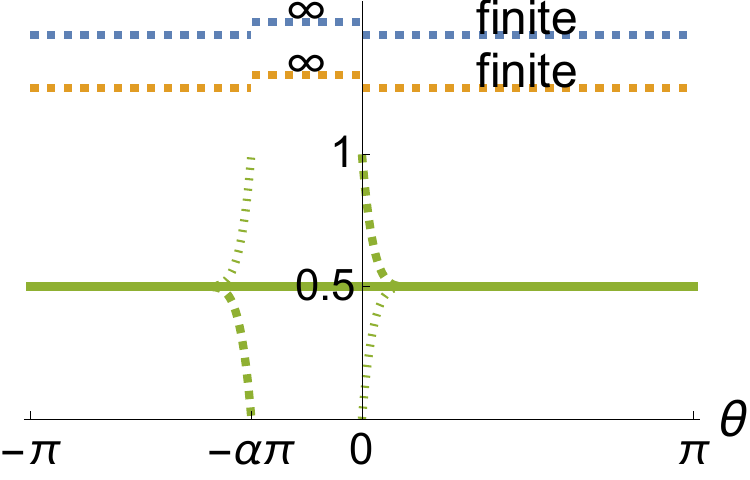}}
    \subfigure[]{\includegraphics[width=0.45\linewidth]{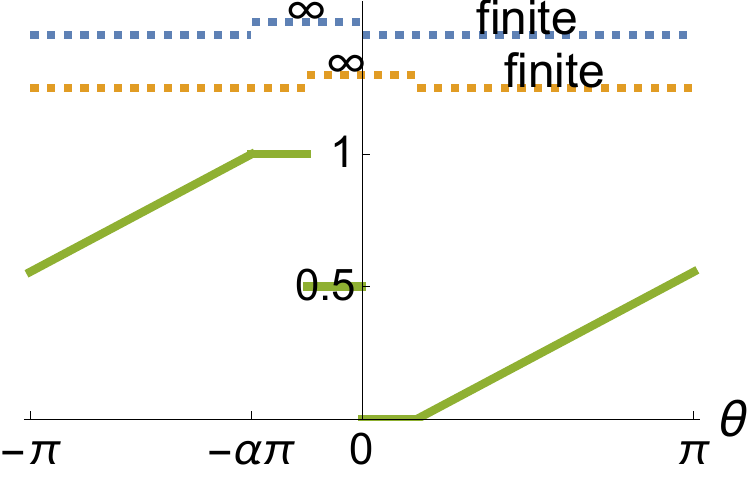}}
    \caption{The Fourier transformation of the correlation matrix $\tilde{C}_\pm(\theta)$ (yellow and blue dashed lines) and corresponding $D(\theta)$ (green line) for $\tau=\alpha\tau_0, \alpha\in(0,1)$ and (a) $\tau\Delta=0$, (b) $\tau\Delta\in(0,(2-2\alpha)\pi)$, (c) $\tau\Delta=(2-2\alpha)\pi$, (d) $\tau\Delta\in((2-2\alpha)\pi,(2-\alpha)\pi)$, (e) $\tau\Delta=(2-\alpha)\pi$, and (f) $\tau\Delta=((2-\alpha)\pi,2\pi)$ with particle-hole symmetry.}
    \label{fig:PH_symmetric_CD_1}
\end{figure}

As shown in the main text, with the anti-commutation matrix defined in Eq.~\eqref{eq:continuous_B_C}, for gapless free fermions with dispersion $E_k=-2\cos{k}$ system size $L\rightarrow\infty$ and $u=1$, we have
\begin{equation}
\label{eq:B_matrix_infinite_L}
    B_{0,n}=\int_{-\pi}^{\pi}\frac{{\rm d}k}{2\pi}e^{{\rm i}2n\tau\cos{k}}=J_0[2n\tau],
\end{equation}
where $J_0(x)$ is the zeroth-order Bessel function of the first kind.
We first take the thermodynamic limit $L\rightarrow\infty$ at fixed number of time points $K$, which allows the momentum sum to be replaced by the integral above.
We then consider the large-$K$ limit to determine the asymptotic eigenvalue distribution of the resulting Toeplitz matrix.
Viewing $B_{mn}$ as a Toeplitz matrix, we define its Fourier transform
\begin{equation}
\label{eq:B_theta_free_fermion_cos}
\begin{split}
    &B(\theta)=\sum_{n=-\infty}^{+\infty} B_{0,n} e^{-{\rm i}n\theta}=\int_0^{2\pi}\frac{{\rm d}k}{2\pi} \sum_{n=-\infty}^{+\infty}e^{{\rm i}n(2\tau\cos{k}-\theta)}\\
    =&\int_0^{2\pi}{\rm d}k\sum_{l=-\infty}^{+\infty}\delta{(-2\tau\cos{k}+\theta-2\pi l)}
    =\sum_{l=-\infty}^{+\infty} \frac{1}{|2\tau \sin{k_l}|},
\end{split}
\end{equation}
where $\cos{k_l}=\frac{\theta-2\pi l}{2\tau}\in[-1,1]$.
Therefore, $B(\theta)$ simplifies to $B(\theta)={\rm Re}[\sum_l\frac{2}{\sqrt{4\tau^2-(2\pi l-\theta)^2}}]$.
In the large-$K$ limit, the asymptotic eigenvalue distribution of the Toeplitz matrix $B$ is determined by its symbol $B(\theta)$.
The fraction of eigenvalues lying in a given range is obtained from the fraction of $\theta$ for which $B(\theta)$ lies in that range~\cite{bottcher2012introduction,bottcher2005spectral}.
Specifically, for $\tau\leq\pi/2$, only the $l=0$ branch contributes, giving
\begin{equation}
B(\theta)=
\begin{cases}
\displaystyle \frac{2}{\sqrt{4\tau^2-\theta^2}}, & |\theta|<2\tau,\\
0, & 2\tau<|\theta|\leq\pi.
\end{cases}
\end{equation}
Thus, for $\tau<\pi/2$, the support of $B(\theta)$ occupies only the arc $|\theta|<2\tau$.
By contrast, if $\tau>\frac{\pi}{2}$, the function $B(\theta)$ is defined on the whole circle $[-\pi,\pi]$, which identifies a critical point $\tau=\frac{\pi}{2}=\frac{2\pi}{E_0}=\tau_0$ with bandwidth $E_0=4u$.
This is consistent with the critical point defined in the main text.

For $\tau<\tau_0$, $B(\theta)>0$ only for $\theta\in[-2\tau,2\tau]$, which implies that the spectral density of the eigenvalues $\lambda_i$ is 
% \bl{Let's state that $\tau_0=\pi/2$, and replace $\pi/2$ below by $\tau_0$. Also, a $\Theta$ function is missing. See main text \cref{eq:density_distribution_rho}.}
\begin{equation}
\label{aeq:density_distribution_rho}
\begin{split}
    \rho(\lambda)=&\frac{1}{2\pi}\sum_{\theta,B(\theta)=\lambda}\frac{1}{|B'(\theta)|}\\
    =&\left(1-\frac{\tau}{\tau_0}\right)\delta(\lambda)+\frac{\Theta(\lambda-\tau^{-1})}{\tau_0\lambda^2\sqrt{\tau^2\lambda^2-1}}
    % \frac{4}{\pi \lambda^2\sqrt{\pi^2\lambda^2-4}},
\end{split}
\end{equation}
featuring a $\delta$-peak at $\lambda=0$ and a finite gap separating it from the nonzero part of the spectrum, whose support satisfies $\lambda\geq\tau^{-1}$~\cite{bottcher2012introduction,bottcher2005spectral}.

\section{SSH model TDEE}
\label{SSH model without particle-hole symmetry W}

In this appendix, we analyze in detail the behavior of the function $\Tilde{C}_\pm(\theta)$ of the SSH model.
We derive the function $D(\theta)$, which gives us the scaling regime diagram of the TDEE, and also the coefficients of volume-law and log-law terms.

\subsection{$W=1$}
\label{sec:SSH equal W}

\begin{figure}
    \centering
    \subfigure[]{\includegraphics[width=0.45\linewidth]{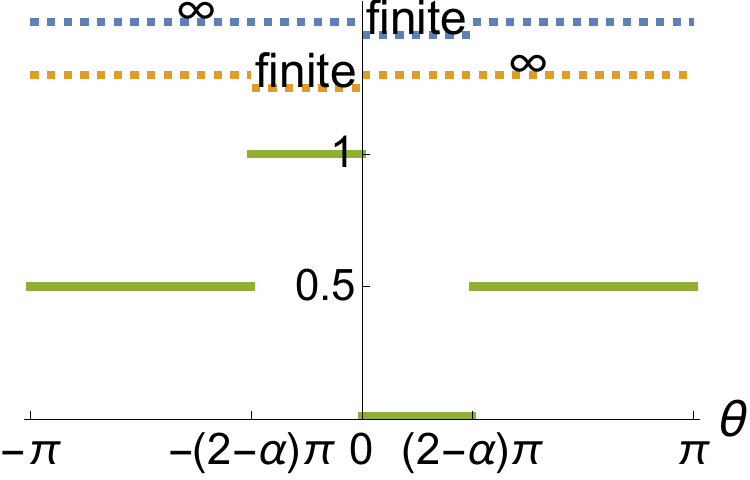}}
    \subfigure[]{\includegraphics[width=0.45\linewidth]{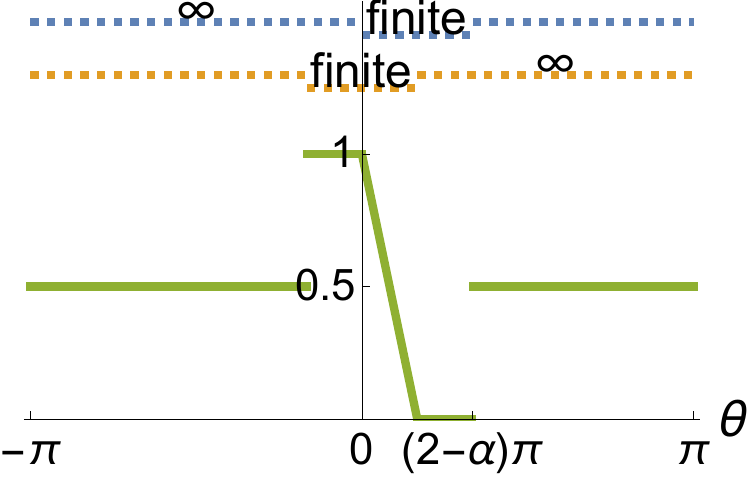}}
    \subfigure[]{\includegraphics[width=0.45\linewidth]{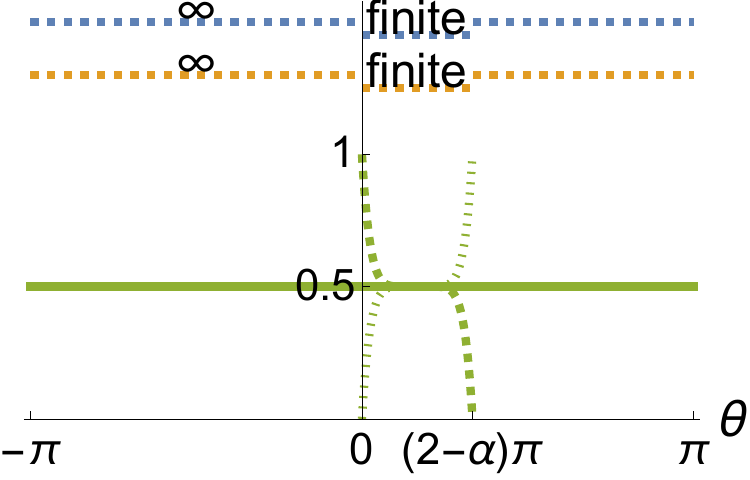}}
    \subfigure[]{\includegraphics[width=0.45\linewidth]{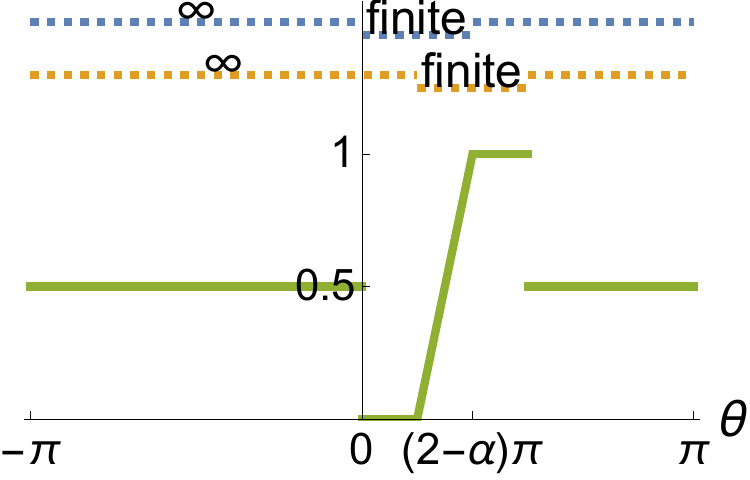}}
    \subfigure[]{\includegraphics[width=0.45\linewidth]{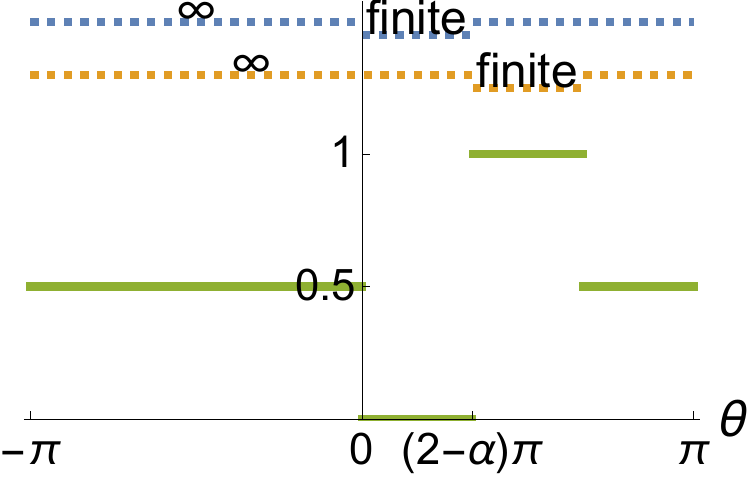}}
    \subfigure[]{\includegraphics[width=0.45\linewidth]{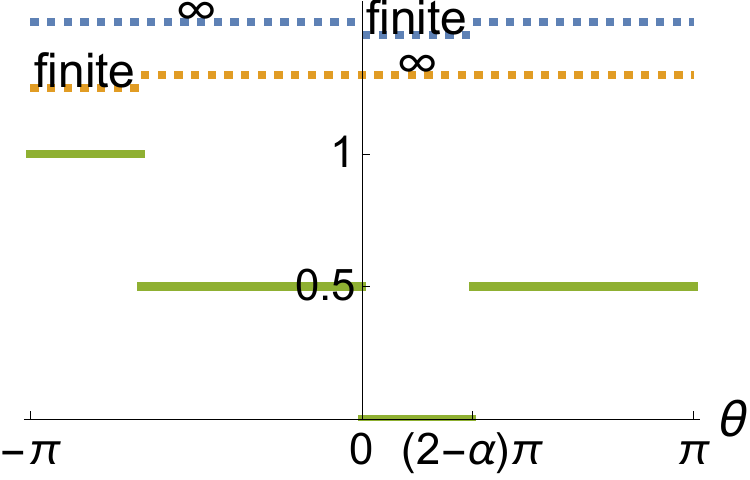}}
    \caption{The Fourier transformation of the correlation matrix $\tilde{C}_\pm(\theta)$ (yellow and blue dashed lines) and corresponding $D(\theta)$ (green line) for $\tau=\alpha\tau_0, \alpha\in(1,2)$ and (a) $\tau\Delta=0$, (b) $\tau\Delta\in(0,(2-\alpha)\pi)$, (c) $\tau\Delta=(2-\alpha)\pi$, (d) $\tau\Delta\in((2-\alpha)\pi,2(2-\alpha)\pi)$, (e) $\tau\Delta=2(2-\alpha)\pi$, and (f) $\tau\Delta=(2(2-\alpha)\pi,2\pi)$ with particle-hole symmetry.
    }
    \label{fig:PH_symmetric_CD_2}
\end{figure}

We first consider the particle-hole symmetric model with $W=1$.
Following the discussion in section~\ref{sec:Analytical understanding for gapped free fermions with the particle-hole symmetry} in the main text, we show the detailed behaviors of $\Tilde{C}_\pm(\theta)$ and $D(\theta)$ for different $\tau\Delta$ and $\tau$.

For $\tau=\alpha\tau_0^\pm<\tau_0^\pm$ with $\alpha\in(0,1)$, there are three different regimes with an increasing $\tau\Delta$.
As shown in Fig.~\ref{fig:PH_symmetric_CD_1} (a) and (c), they have a single discontinuity without $D(\theta)\equiv\frac{1}{2}$, which leads to a log-law TDEE with coefficient $\frac{1}{6}$.
Figure~\ref{fig:PH_symmetric_CD_1} (b) has no discontinuity and no $D(\theta)\equiv\frac{1}{2}$, which induces an area-law TDEE.
Figures~\ref{fig:PH_symmetric_CD_1} (d)-(f) with $D(\theta)\equiv\frac{1}{2}$ in some range of $\theta$ corresponds to a volume-law TDEE.
With different lengths of regions $\theta$ that $D(\theta)\equiv\frac{1}{2}$, the volume-law coefficients can be different, and the final results are shown in the main text Fig.~\ref{fig:PH_symmetric} (b).

For $\tau_0<\tau<2\tau_0$, as shown in Fig.~\ref{fig:PH_symmetric_CD_2}, there is always a region for $\theta$ that $D(\theta)\equiv\frac{1}{2}$, which means there is a volume-law TDEE for any $\tau\Delta$, but the coefficient varies.
The volume-law coefficient is displayed in Fig.~\ref{fig:PH_symmetric} (d) in the main text.

\subsection{$W<1$}
\label{sec:SSH small W}

We first consider $W<1$, and take $W=\frac{1}{3}$ as an example, with $\tau_0^-=\pi$ and $\tau_0^+=3\pi$.

\begin{figure}
    \centering
    \subfigure[]{\includegraphics[width=0.45\linewidth]{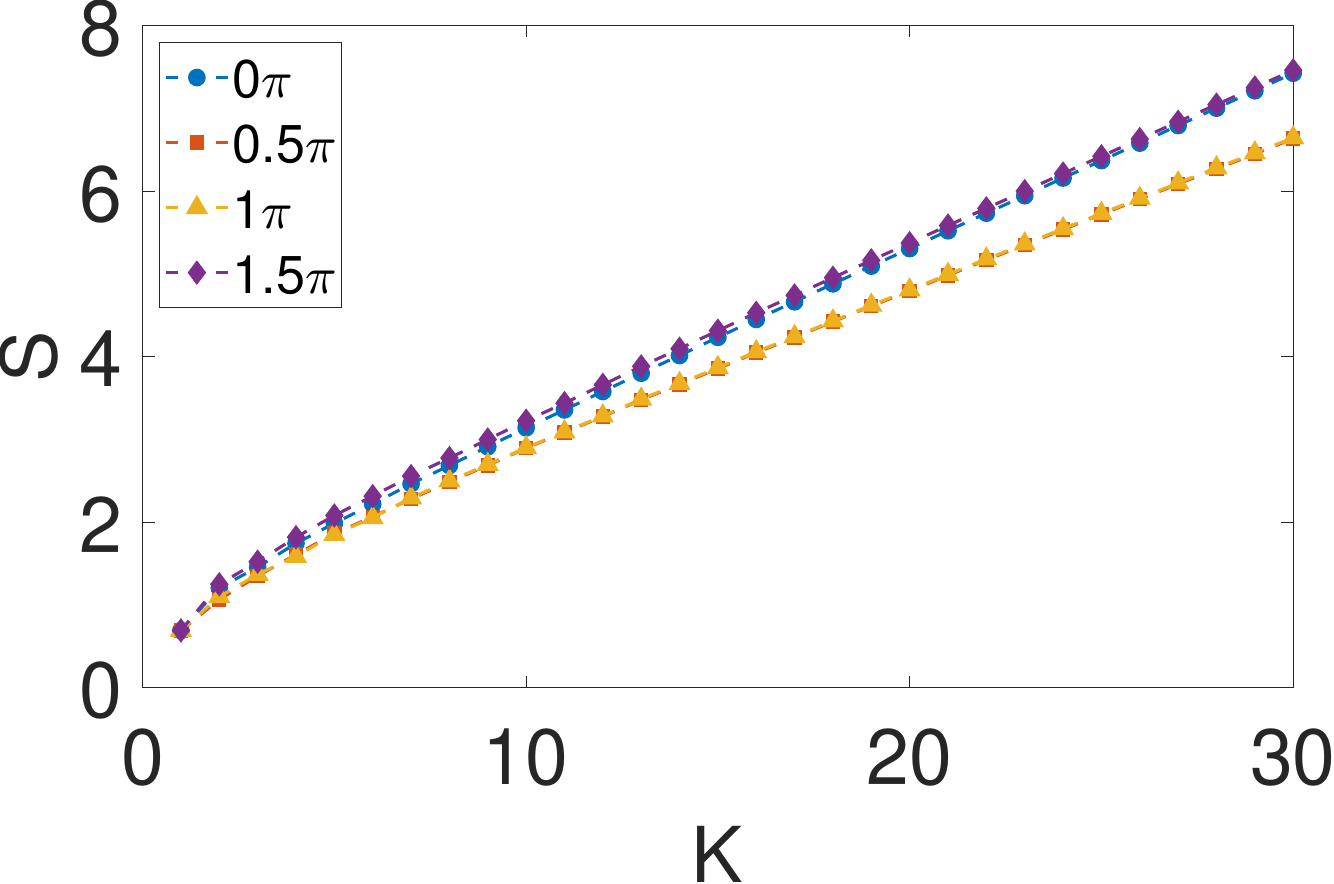}}
    \subfigure[]{\includegraphics[width=0.45\linewidth]{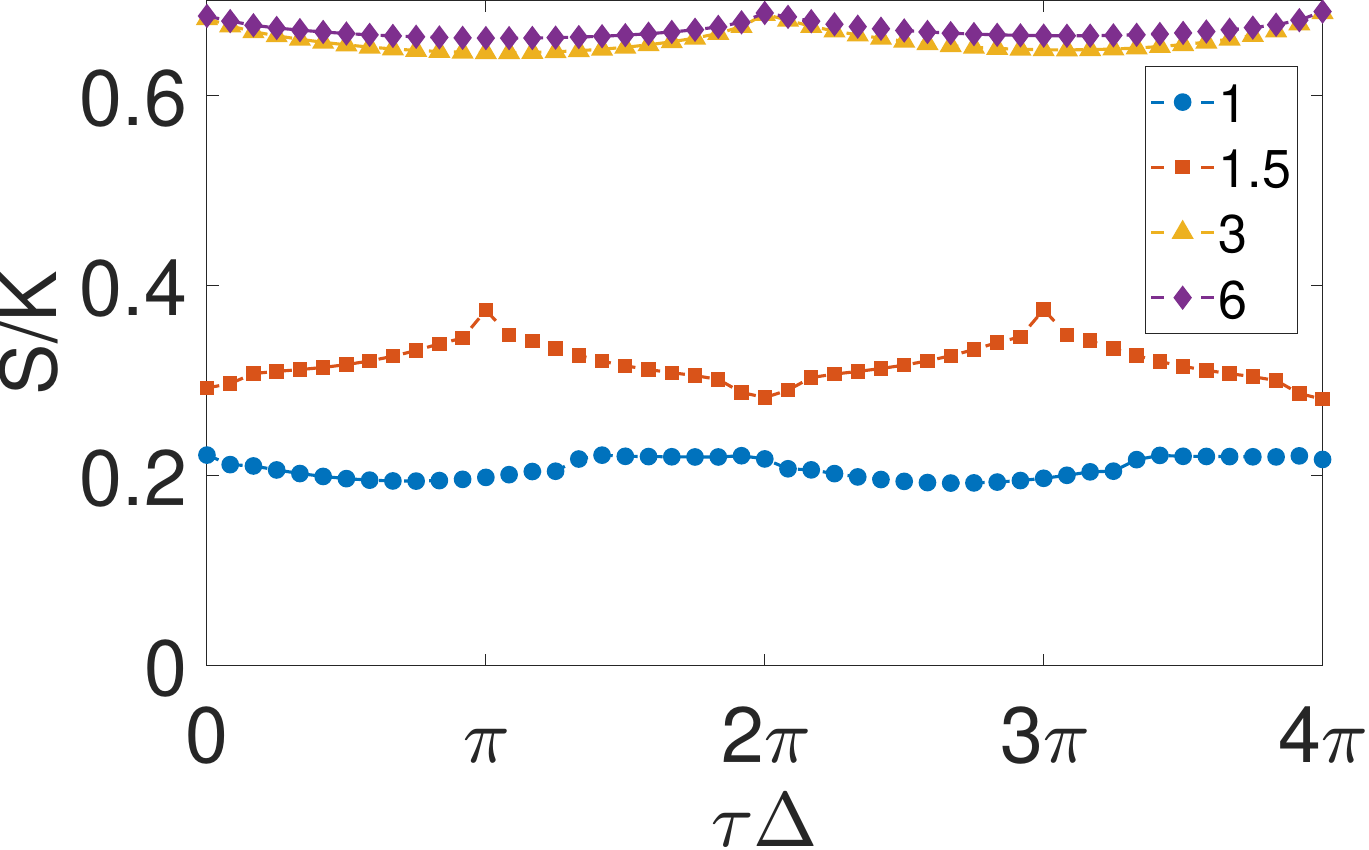}}
    \subfigure[]{\includegraphics[width=0.45\linewidth]{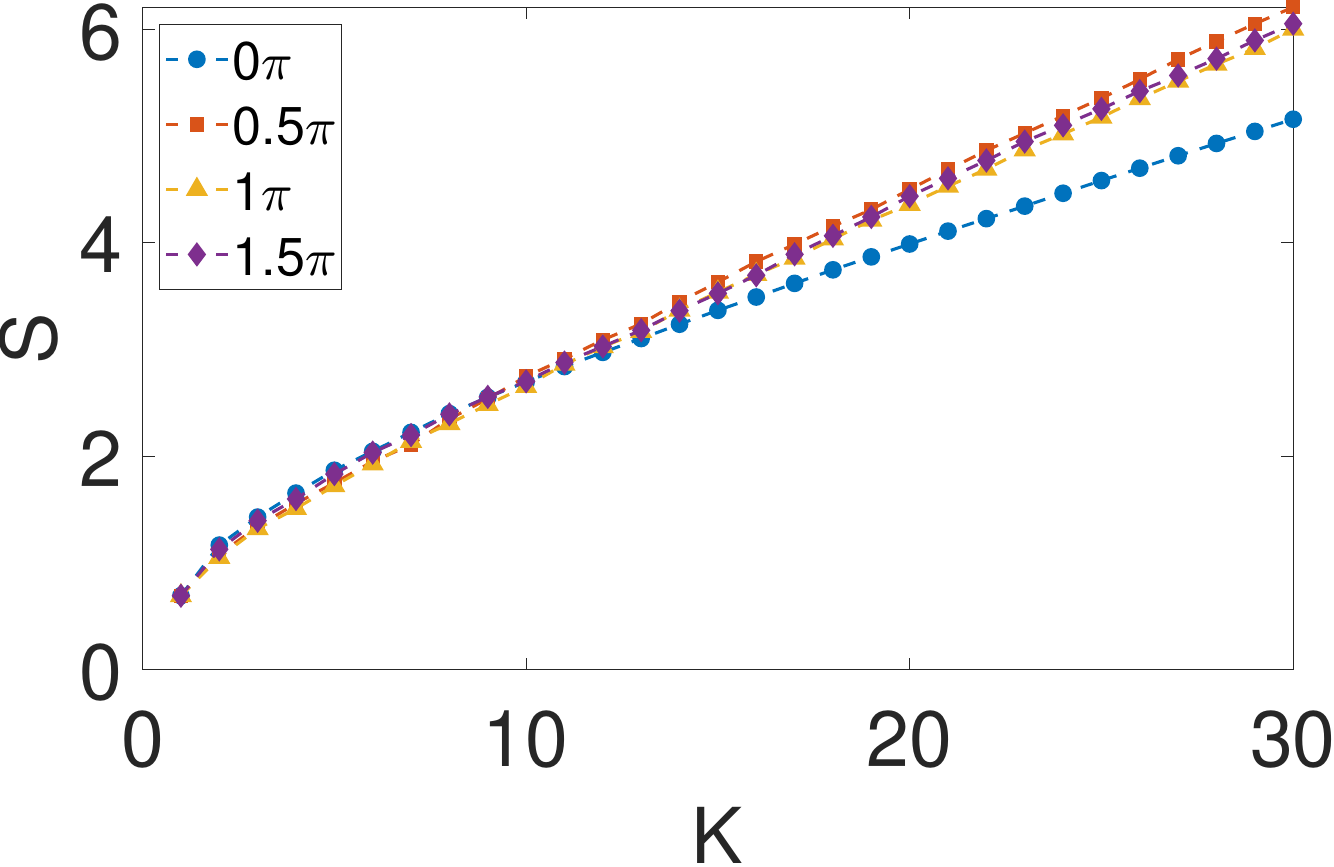}}
    \subfigure[]{\includegraphics[width=0.45\linewidth]{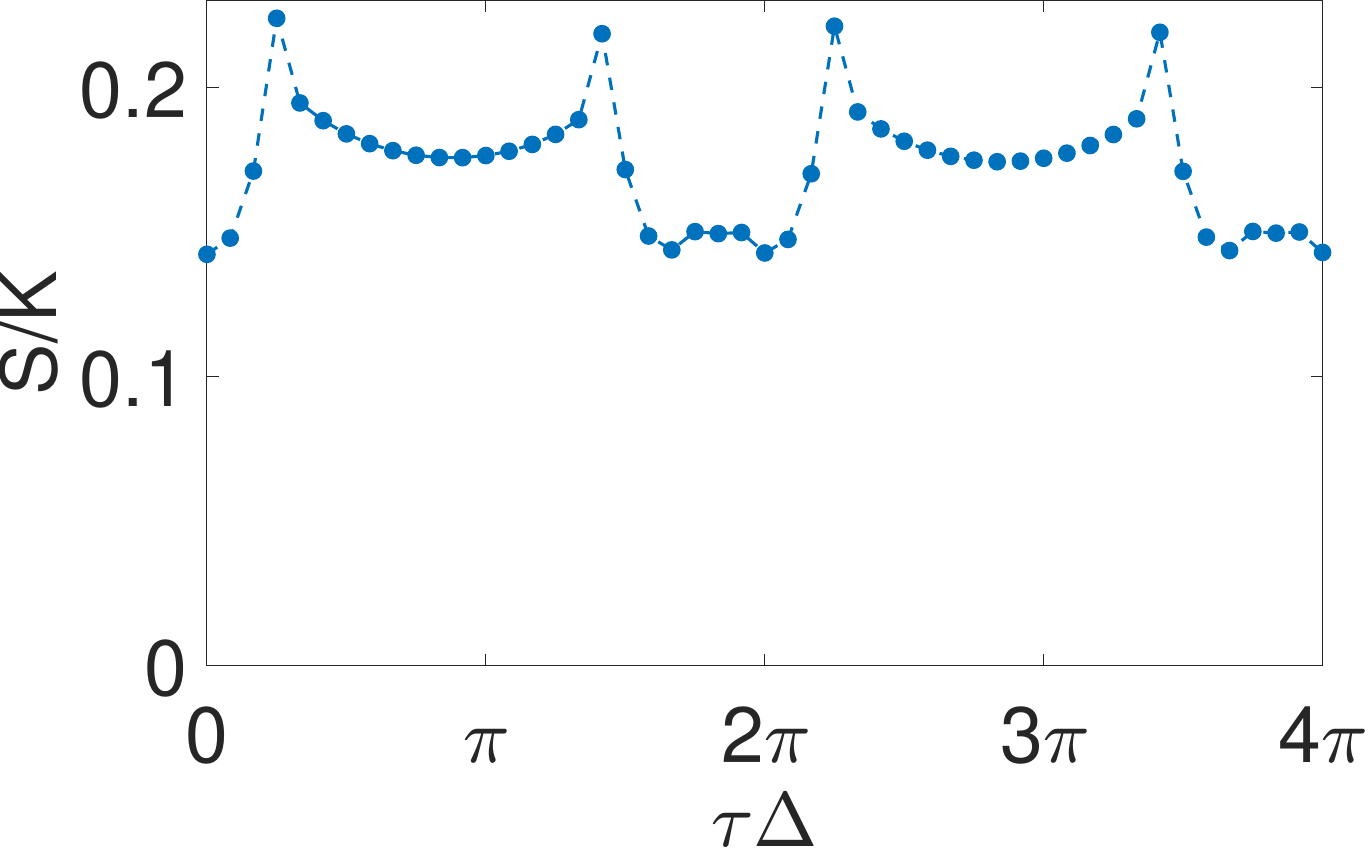}}
    \subfigure[]{\includegraphics[width=0.45\linewidth]{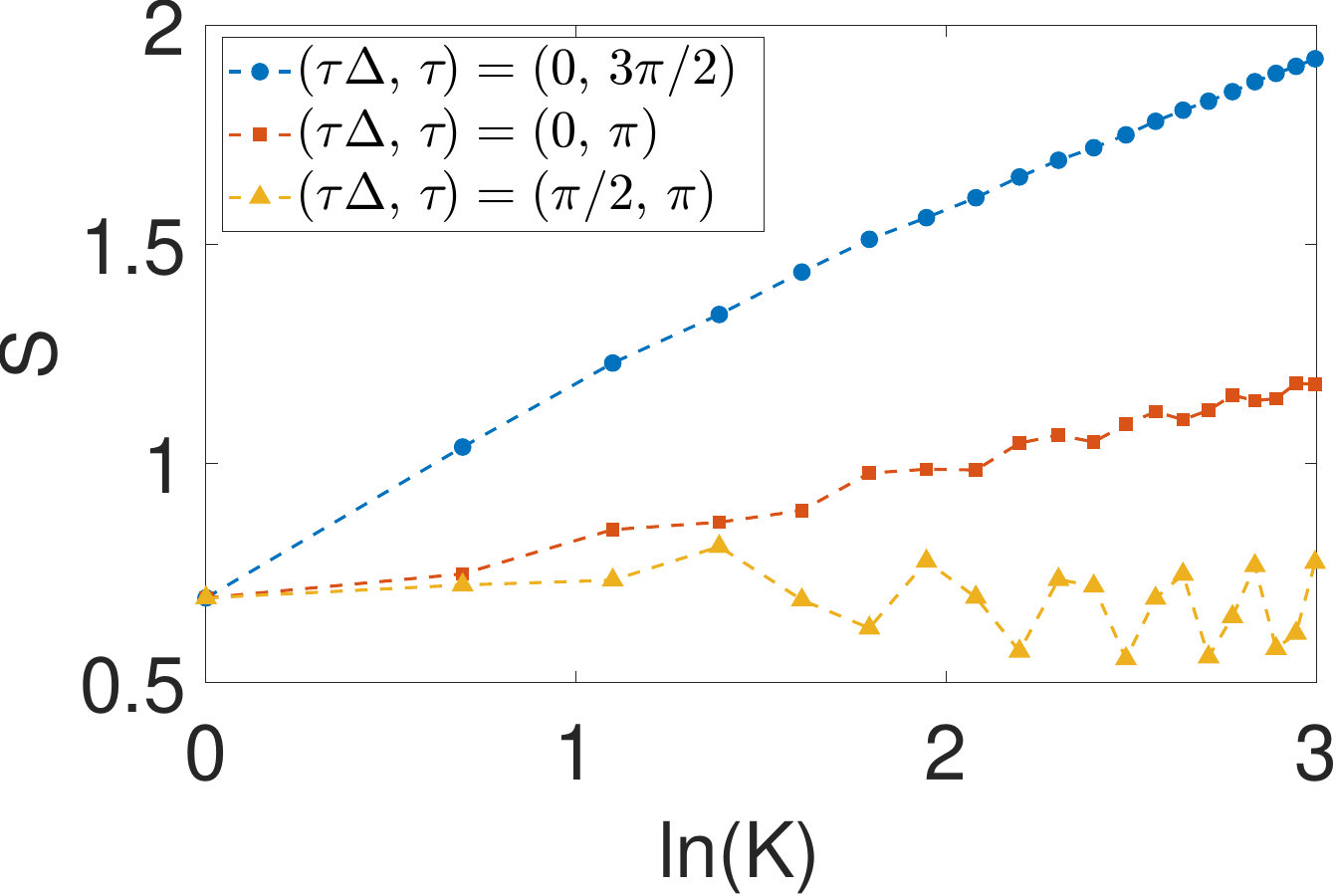}}
    \subfigure[]{\includegraphics[width=0.45\linewidth]{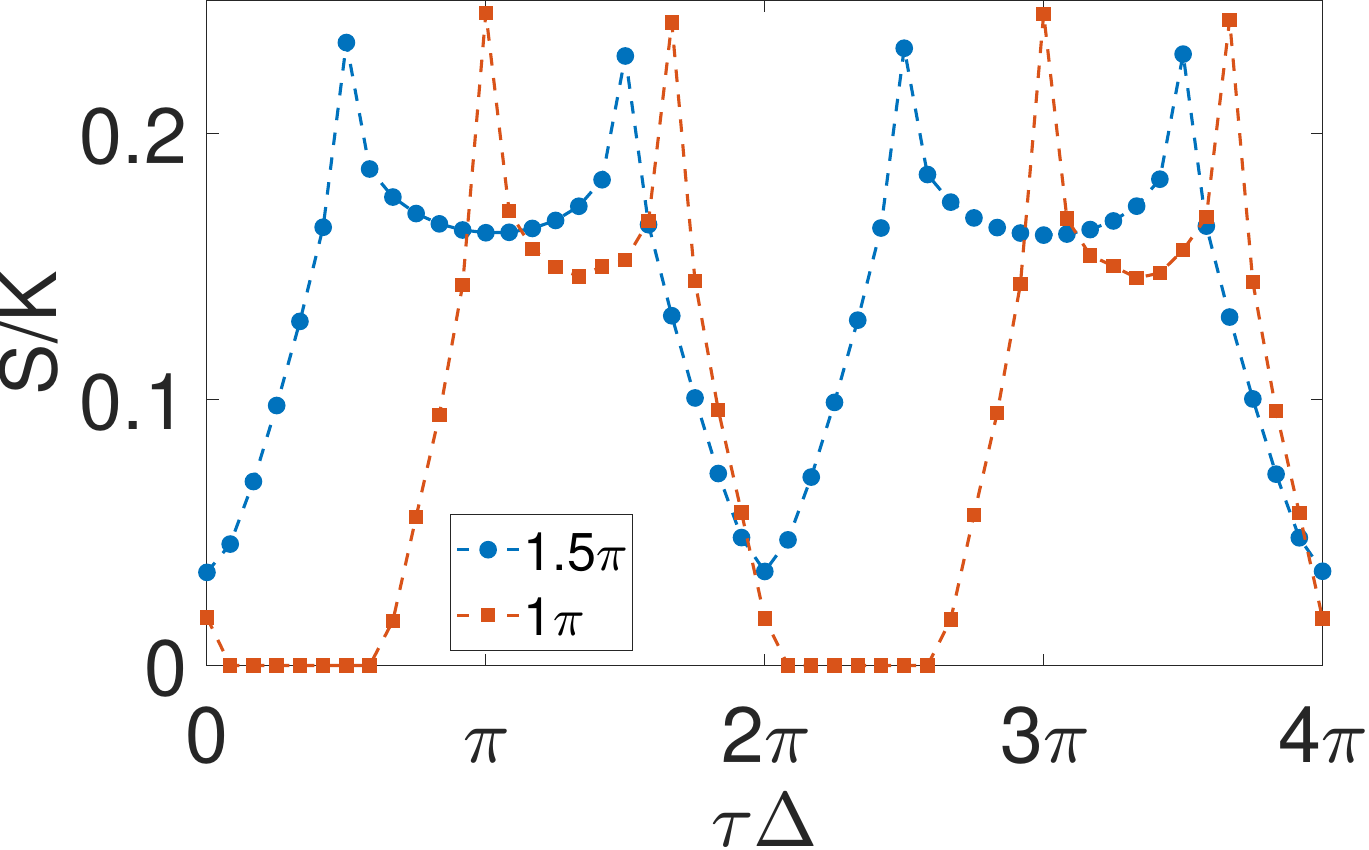}}
    \caption{$W = \frac{1}{3}$ and $L=1000$. (a) The TDEE for $\tau=2\tau_0^-$, $\tau\Delta=0, \frac{\pi}{2}, \pi, \frac{3\pi}{2}$ (legend), which shows the volume-law entanglement.
    (b) The volume-law term prefactor $S/K$ for the TDEE, with $\frac{\tau}{2\tau_0^-}=1,\frac{1}{2W},\frac{1}{W},\frac{2}{W}$ (legend), $\tau\Delta\in[0,4\pi]$.
    (c) The TDEE for $\tau=\frac{7}{4}\pi$, $\tau\Delta=0,\frac{\pi}{2},\pi,\frac{3\pi}{2}$. 
    (d) The volume-law term prefactor $S/K$ for the TDEE, with $\tau=\frac{7}{4}\pi$, $\tau\Delta\in[0,4\pi]$.
    (e) The TDEE in log-law and area-law. The log-law prefactor is $0.361, 0.186$ for $\left(\tau \Delta, \tau\right)=(0,3 \pi / 2), (0, \pi)$, respectively. 
    (f) The volume-law term prefactor $S/K$ for the TDEE, with $\tau\Delta\in[0,4\pi]$, $\tau=\frac{3}{2}\pi, \pi$ (legend).}
    % \caption{$W = \frac{1}{3}$ and $L=1000$. \textcolor{red}{$\tau_e = \tau E_-$}, where $E_-$ is the lower bandwidth. (a) The TDEE for $\frac{\tau_e}{2\tau_0^-}=1$, $\tau\Delta_e=0, \frac{\pi}{2}, \pi, \frac{3\pi}{2}$ (legend), which shows the volume-law entanglement.
    % (b) The volume-law term prefactor $S/K$ for the TDEE, with $\frac{\tau_e}{2\tau_0^-}=1,\frac{1}{2W},\frac{1}{W},\frac{2}{W}$ (legend), $\tau\Delta_e\in[0,4\pi]$.
    % (c) The TDEE for $\tau_e=\frac{7}{4}\pi$, $\tau\Delta_e=0,\frac{\pi}{2},\pi,\frac{3\pi}{2}$. 
    % (d) The volume-law term prefactor $S/K$ for the TDEE, with $\tau_{e}=\frac{7}{4}\pi$, $\tau\Delta_{e}\in[0,4\pi]$.
    % (e) The TDEE in log-law and area-law. The log-law prefactor is $0.361, 0.186$ for $\left(\tau \Delta, \tau_e\right)=(0,3 \pi / 2), (0, \pi)$, respectively. 
    % (f) The volume-law term prefactor $S/K$ for the TDEE, with $\tau\Delta_e\in[0,4\pi]$, $\tau_e=\frac{3}{2}\pi, \pi$ (legend).
    % \textcolor{red}{show log-law in figure (e) better, and show y-axes from 0 in (b), (d) and (f).
    % Also there is some symbol in (f).
    % Ploy y-axes form 0 in figure (f).
    % Check the amplitude of volume-law term, and also log-law fit.}}
    \label{fig:gapped_nonPH}
\end{figure}

\begin{figure*}
    \centering
    \subfigure[]{\includegraphics[width=0.98\linewidth]{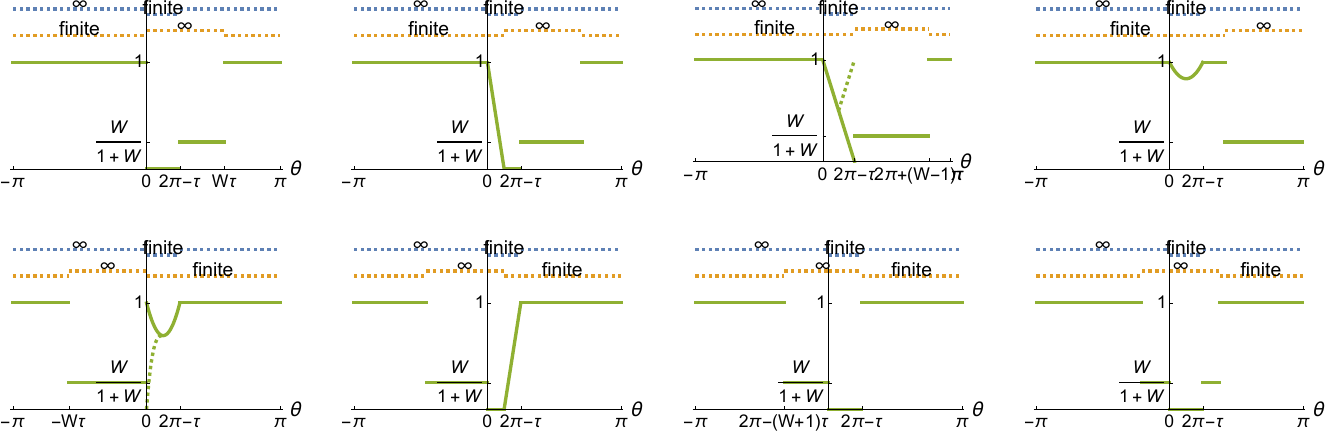}}
    \subfigure[]{\includegraphics[width=0.98\linewidth]{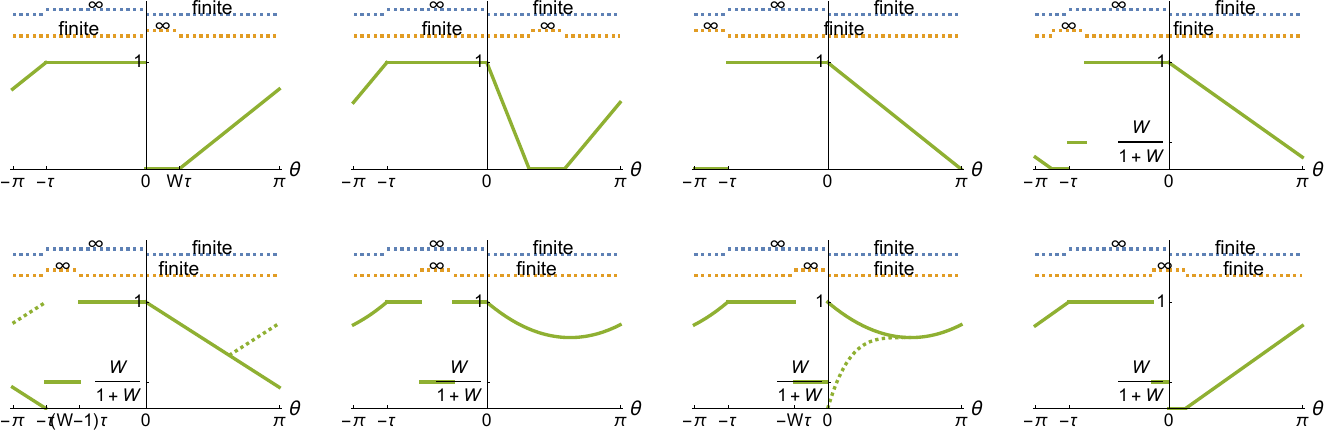}}
    \subfigure[]{\includegraphics[width=0.30\linewidth]{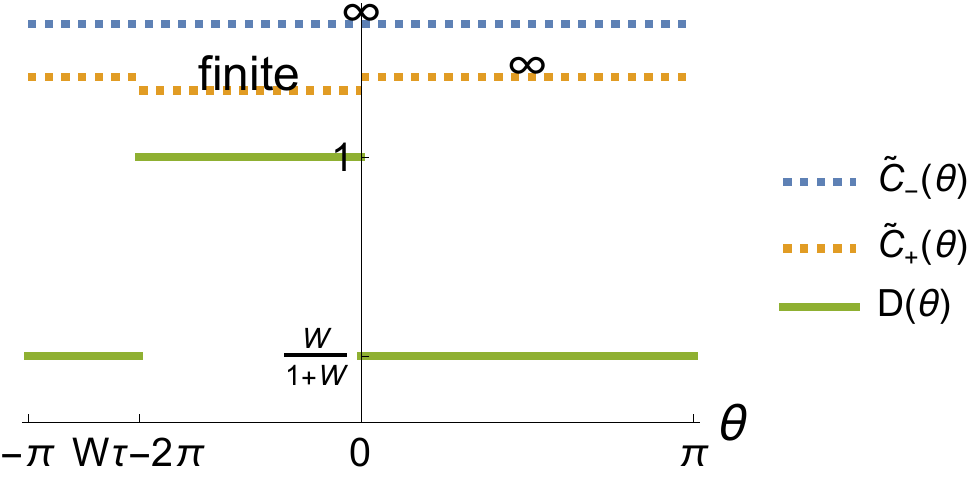}}
    \caption{The Fourier transformation of the correlation matrix $\tilde{C}_\pm(\theta)$ and corresponding $D(\theta)$ for different $\tau$ and $\tau\Delta$ without particle-hole symmetry. (a) $\tau_0<\tau<2{\rm min}(\tau_0^-,\tau_0^+)$, and $\tau\Delta=0,2\pi-\tau,2\pi-W\tau,4\pi-(W+1)\tau$ for the first, third, fifth and seventh figures. (b) $\tau<\tau_0$, and $\tau\Delta=0,2\pi-(W+1)\tau,2\pi-\tau,2\pi-W\tau$ for the first, third, fifth and seventh figures. (c) $2{\rm min}(\tau_0^-,\tau_0^+)<\tau<2{\rm max}(\tau_0^-,\tau_0^+)$ and $\tau\Delta=0$.}
    %(a) (b) $\tau=\frac{7\pi}{4}$ (c) $\tau=\frac{3\pi}{4}$
    \label{fig:nonPH_symmetric_smallW_app}
\end{figure*}

For $\tau\geq2\tau_0^-$, we plot the TDEE for $\frac{\tau}{2\tau_0^-}=1$, and $\tau\Delta=0,\frac{\pi}{2},\pi,\frac{3\pi}{2}$ in Fig.~\ref{fig:gapped_nonPH} (a).
We find that all TDEE exhibit the volume-law entanglement.
With the same $\frac{\tau}{\tau_0^+}$, the prefactors of volume-law terms are almost the same for different $\tau\Delta$.
% , and increase w.r.t. $\frac{\tau}{2\tau_0^-}$.
In Fig.~\ref{fig:gapped_nonPH} (b), we plot the prefactor of the volume-law term for $\frac{\tau}{2\tau_0^-}=1,\frac{1}{2W},\frac{1}{W},\frac{2}{W}$ and $\tau\Delta\in[0,4\pi]$, where $\frac{\tau}{2\tau_0^-}=\frac{1}{W}$ corresponds to $\tau=2\tau_0^+$.
The coefficient of the volume-law term is approximately constant for any $\tau\Delta$, increases as a function of $\frac{\tau}{2\tau_0^-}$, and saturates to $\ln{2}$ approximately for $\frac{\tau}{2\tau_0^-}\geq\frac{1}{W}$.

For $\tau_0<\tau<2\tau_0^-$ with $\tau_0=\frac{2\tau_0^- \tau_0^+}{\tau_0^-+\tau_0^+}=\frac{3\pi}{2}$, we plot the TDEE for $\tau=\frac{7\pi}{4}$ and $\tau\Delta=0,\frac{\pi}{2},\pi,\frac{3\pi}{2}$ in Fig.~\ref{fig:gapped_nonPH} (c).
We find that all TDEE exhibit the volume-law entanglement, but with different prefactors.
In Fig.~\ref{fig:gapped_nonPH} (d), we plot the prefactor of the volume-law term for $\tau\Delta\in[0,4\pi]$, which exhibits approximate $2\pi$-periodicity, and two peaks which are different from the particle-hole symmetric case before.
Moreover, approximately, the location of two peaks are $\tau\Delta\approx\frac{\pi}{4},\frac{17\pi}{12}$, and the plateau is $\tau\Delta\in[\frac{5\pi}{3},2\pi]$.

For $\tau\leq\tau_0$, we plot the TDEE for $\left(\tau \Delta, \tau\right)=(0,3 \pi / 2), (0, \pi), (\pi/2, \pi)$ in Fig.~\ref{fig:gapped_nonPH} (e).
We find that TDEE exhibits log-law entanglement for $\left(\tau \Delta, \tau\right)=(0,3 \pi / 2), (0, \pi)$ with coefficients $0.361$ and $0.186$, and area-law entanglement for $\left(\tau \Delta, \tau\right)=(\pi/2, \pi)$.
In Fig.~\ref{fig:gapped_nonPH} (f), we plot the prefactor of the volume-law term for $\tau\Delta\in[0,4\pi]$ and $\tau=\pi,\frac{3\pi}{2}$, which exhibits approximate $2\pi$-periodicity.
Although it is similar to the particle-hole symmetric case with $\tau<\tau_0$, there are two peaks which are different from the results before.
Approximately, for $\tau=\tau_0=\frac{3\pi}{2}$, two peaks are at $\tau\Delta=\frac{\pi}{2},\frac{3\pi}{2}$, and for $\tau=\pi<\tau_0$, we have two peaks at $\tau\Delta=\pi,\frac{5\pi}{3}$, with an area-law regime for $\tau\Delta\in(0,\frac{2\pi}{3})$.

% \begin{figure}
%     \centering
%     \subfigure[]{\includegraphics[width=0.9\linewidth]{nonPH_ltauc_prefactor.pdf}}
%     \subfigure[]{\includegraphics[width=0.85\linewidth]{nonPH_stauc_prefactor.pdf}}
%     \caption{Without particle-hole symmetry, the prefactors of volume-law term and phase diagrams for (a) $\tau_c<\tau<2{\rm min}(\tau_0^-,\tau_0^+)$ and (b) $\tau<\tau_c$, where the analytical results shown with dashed lines are lower bounds.}
%     %(a) (b) $\tau=\frac{7\pi}{4}$ (c) $\tau=\frac{3\pi}{4}$
%     \label{fig:nonPH_symmetric_smallW}
% \end{figure}

In the following, we present an analysis for $C_\pm(\theta)$ and $D(\theta)$, which gives the consistent prediction of TDEE with $W<1$ above. 

(i) For $\tau>2\tau_0^+$:
In this regime, $\Tilde{C}_\pm(\theta)$ diverge for all values of $\theta$ when $\Delta=0$.  
For any finite $\Delta$, which simply corresponds to a translation of $\Tilde{C}_+(\theta)$, the ratio 
$D(\theta)=(1+\varrho(E_k^+)/\varrho(E_k^-))^{-1}$ remains constant for all $\theta$.  
Here, as before, we assume a linear dispersion with a constant density of states.  
Thus, the time-direction entanglement entropy follows a volume law for any $\Delta$, with the same prefactor independent of $\Delta$.  
This analytical prediction matches the numerical results shown in Fig.~\ref{fig:gapped_nonPH} (b).

(ii) For $2\tau_0^-<\tau<2\tau_0^+$: 
In this case, $\Tilde{C}_-(\theta)$ diverges for all $\theta$ when $\Delta=0$, while $\Tilde{C}_+(\theta)$ diverges only within the range $\theta\in[0,W\tau]$ for $\Delta=0$.  
The corresponding $D(\theta)$ is illustrated in Fig.~\ref{fig:nonPH_symmetric_smallW_app} (c).  
When $\Delta$ becomes finite, $\Tilde{C}_-(\theta)$ remains unchanged, whereas $\Tilde{C}_+(\theta)$ undergoes a translation.  
As a result, $D(\theta)$ has the same overall shape, shifted along the $\theta$ axis.
Since $D(\theta)\equiv(1+\varrho(E_k^+)/\varrho(E_k^-))^{-1}$ within a region of length $W\tau$ in $\theta$, and $D(\theta)=1$ elsewhere, the TDEE continues to exhibit a volume law for any $\Delta$, but with a prefactor smaller than in the $\tau>2\tau_0^+$ case due to $W\tau<2\pi$.  
This behavior is also fully consistent with the numerical results in Fig.~\ref{fig:gapped_nonPH} (a) and (b).

(iii) For $\tau<2\tau_0^-<2\tau_0^+$:  
In this regime, there are two subregions separated by $\tau_0=\frac{2\tau_0^+\tau_0^-}{\tau_0^+ + \tau_0^-}$.
In particular, under particle-hole symmetry, we have $\tau_0=\tau_0^+=\tau_0^-$.  
We first consider the case $\tau>\tau_0$.  
When $\Delta=0$, both $\Tilde{C}_\pm(\theta)$ diverge within the ranges $\theta\in[-\tau,0]$ and $\theta\in[0,W\tau]$.  
Note that the condition $\tau=\tau_0$ corresponds to $W\tau+\tau=2\pi$.  
The behavior of $D(\theta)$ for this case is shown in the first panel of Fig.~\ref{fig:nonPH_symmetric_smallW_app} (a).  
Here, $D(\theta)\equiv (1+\varrho(E_k^+)/\varrho(E_k^-))^{-1}$ for $\theta\in[2\pi-\tau,\,W\tau]$, indicating a volume-law regime 
\footnote{Although a discontinuity occurs at $\theta=0$, the associated logarithmic term is a subleading correction compared to the dominant volume-law contribution.}.
After introducing a finite gap $\Delta$, the behavior of $D(\theta)$ changes as follows:
For $0<\tau\Delta<2\pi-\tau$, the length of the $\theta$ region where $D(\theta)\equiv (1+\varrho(E_k^+)/\varrho(E_k^-))^{-1}$ increases, leading to a larger prefactor of the volume-law term.  
For $2\pi-\tau\leq\tau\Delta\leq2\pi-W\tau$, $D(\theta)$ remains nearly invariant up to a translation, so the volume-law term stays constant.  
For $2\pi-W\tau<\tau\Delta<4\pi-(W+1)\tau$, the length of the region where $D(\theta)\equiv (1+\varrho(E_k^+)/\varrho(E_k^-))^{-1}$ decreases, causing the prefactor of the volume-law term to decrease.  
For $4\pi-(W+1)\tau<\tau\Delta<2\pi$, the length of this region remains fixed, resulting in an unchanged entanglement entropy.
The resulting scaling regime diagram and the prefactor of the volume-law term are shown in Fig.~\ref{fig:nonPH_symmetric_smallW} (a).  
These analytical predictions are in excellent agreement with the numerical results presented in Fig.~\ref{fig:gapped_nonPH} (c) and (d).

(iv) For $\tau<\tau_0$, the behavior of $D(\theta)$ with $\Delta=0$ is shown in the first panel of Fig.~\ref{fig:nonPH_symmetric_smallW_app} (b).
In this case, there is a single discontinuity at $\theta=0$, with a continuous function smoothly connecting $D(\theta)=0$ and $D(\theta)=1$.  
Therefore, the entanglement entropy follows a logarithmic law, $S=\frac{1}{6}\ln{K}$.
After introducing a finite gap $\Delta$, the behavior evolves as follows.
For $0<\tau\Delta<2\pi-(W+1)\tau$, there is no discontinuity, and two continuous functions connect $D(\theta)=0$ and $D(\theta)=1$.  
This corresponds to an area-law regime.
For $\tau\Delta=2\pi-(W+1)\tau$, a single discontinuity appears at $\theta=-\tau$, together with a continuous part connecting $D(\theta)=0$ and $D(\theta)=1$, resulting in a log-law entanglement entropy with $S=\frac{1}{6}\ln{K}$.
For $2\pi-(W+1)\tau<\tau\Delta<2\pi-\tau$, $D(\theta)\equiv (1+\varrho(E_k^+)/\varrho(E_k^-))^{-1}$ emerges over a finite range of $\theta$, and the length of this region increases linearly from $0$ to $W\tau$.  
This indicates a volume-law regime with a prefactor that grows continuously as $\tau\Delta$ increases.
For $2\pi-\tau<\tau\Delta<2\pi-W\tau$, the range of $\theta$ where $D(\theta)\equiv (1+\varrho(E_k^+)/\varrho(E_k^-))^{-1}$ remains constant, so the prefactor of the volume-law term stays fixed as $\Delta$ changes.
For $2\pi-W\tau<\tau\Delta<2\pi$, the length of this region gradually decreases to zero, causing the prefactor of the volume-law term to decrease correspondingly.
For $\tau\Delta=2\pi$, the system returns to a log-law regime, with the entanglement entropy again given by $S=\frac{1}{6}\ln{K}$.
The scaling regime diagram and the prefactor of the volume-law term are shown in Fig.~\ref{fig:nonPH_symmetric_smallW} (b).  
These analytical predictions are in excellent agreement with the numerical results presented in Fig.~\ref{fig:gapped_nonPH} (e) and (f).

\begin{figure}
    \centering
    \subfigure[]{\includegraphics[width=0.45\linewidth]{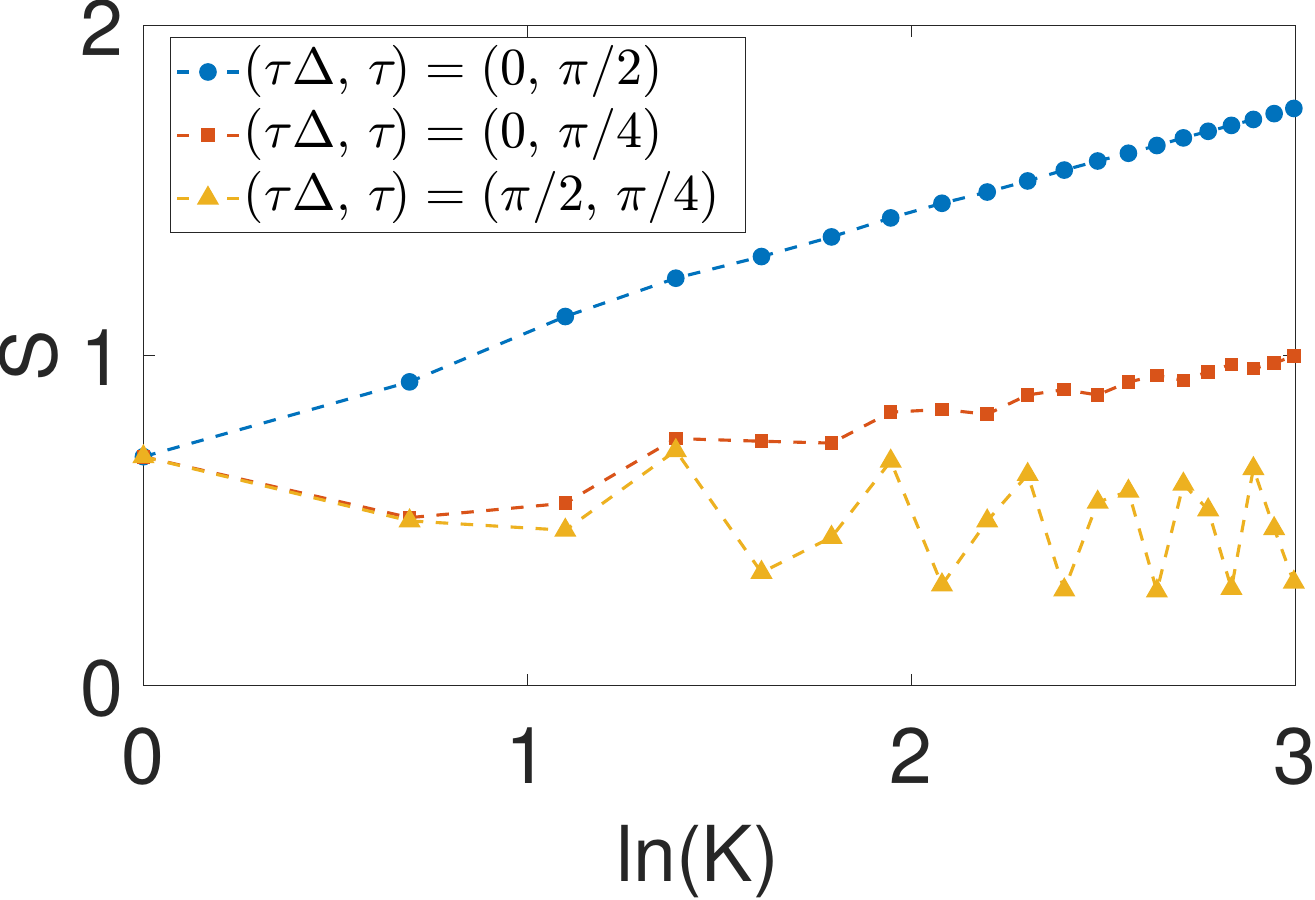}}
    \subfigure[]{\includegraphics[width=0.45\linewidth]{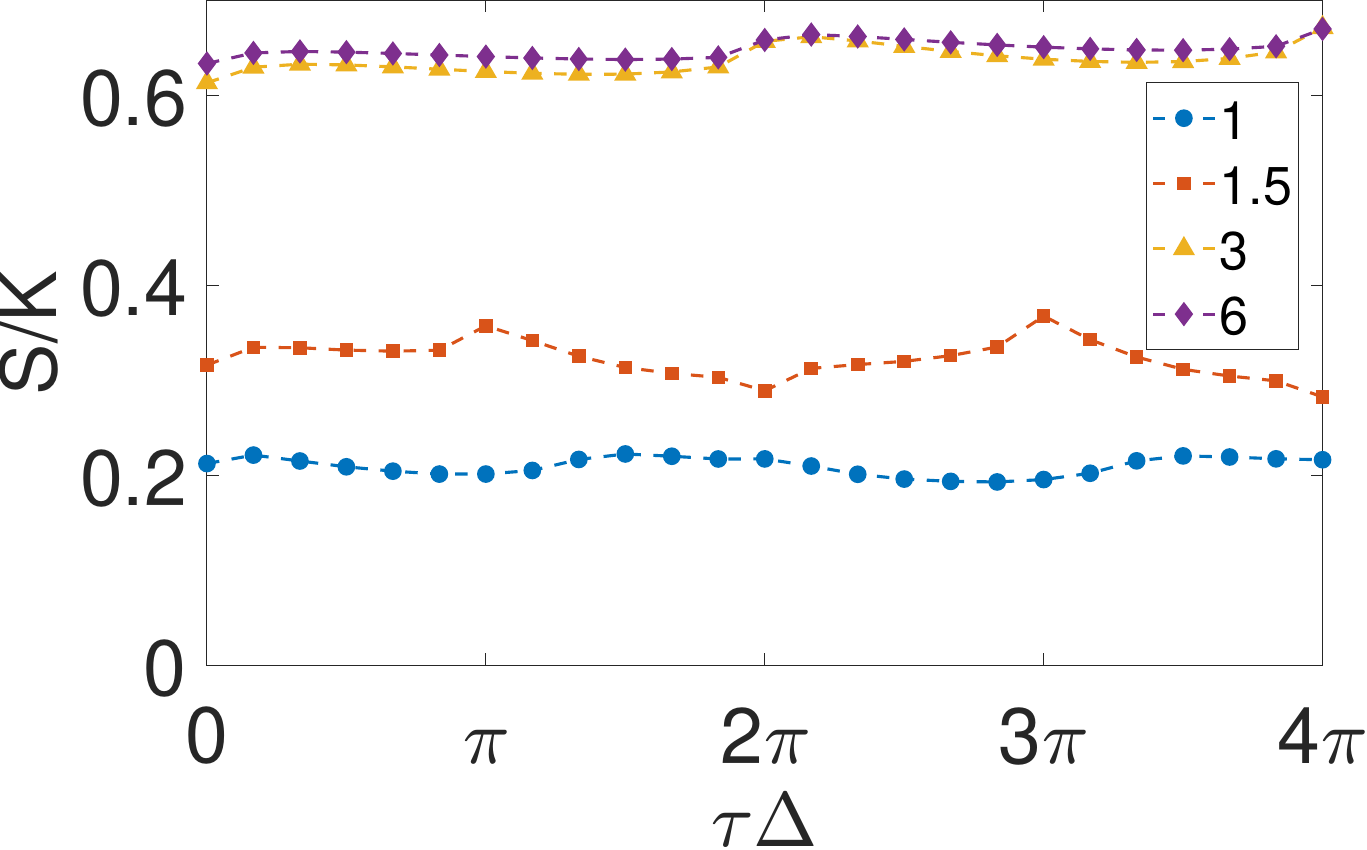}}
    \subfigure[]{\includegraphics[width=0.45\linewidth]{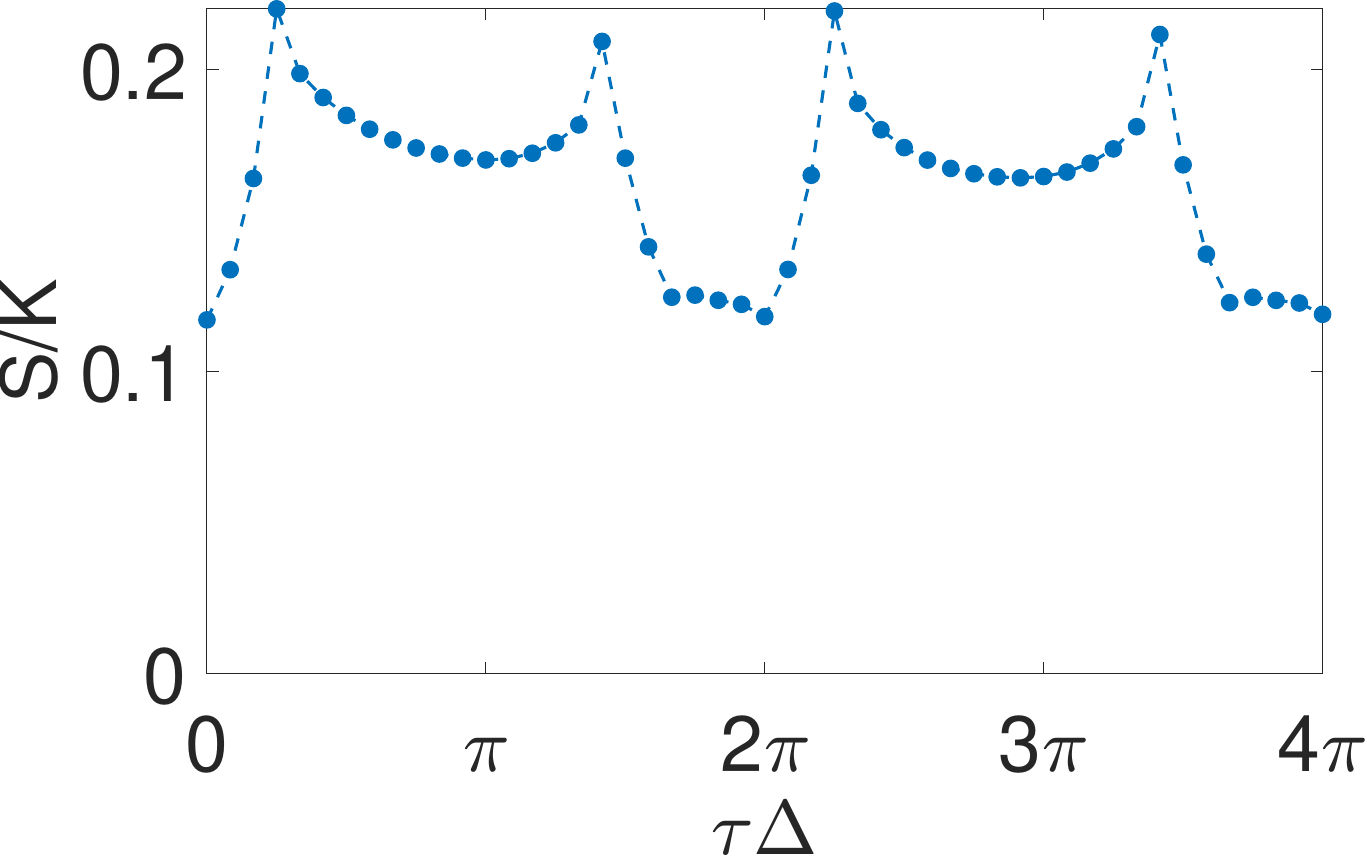}}
    \subfigure[]{\includegraphics[width=0.45\linewidth]{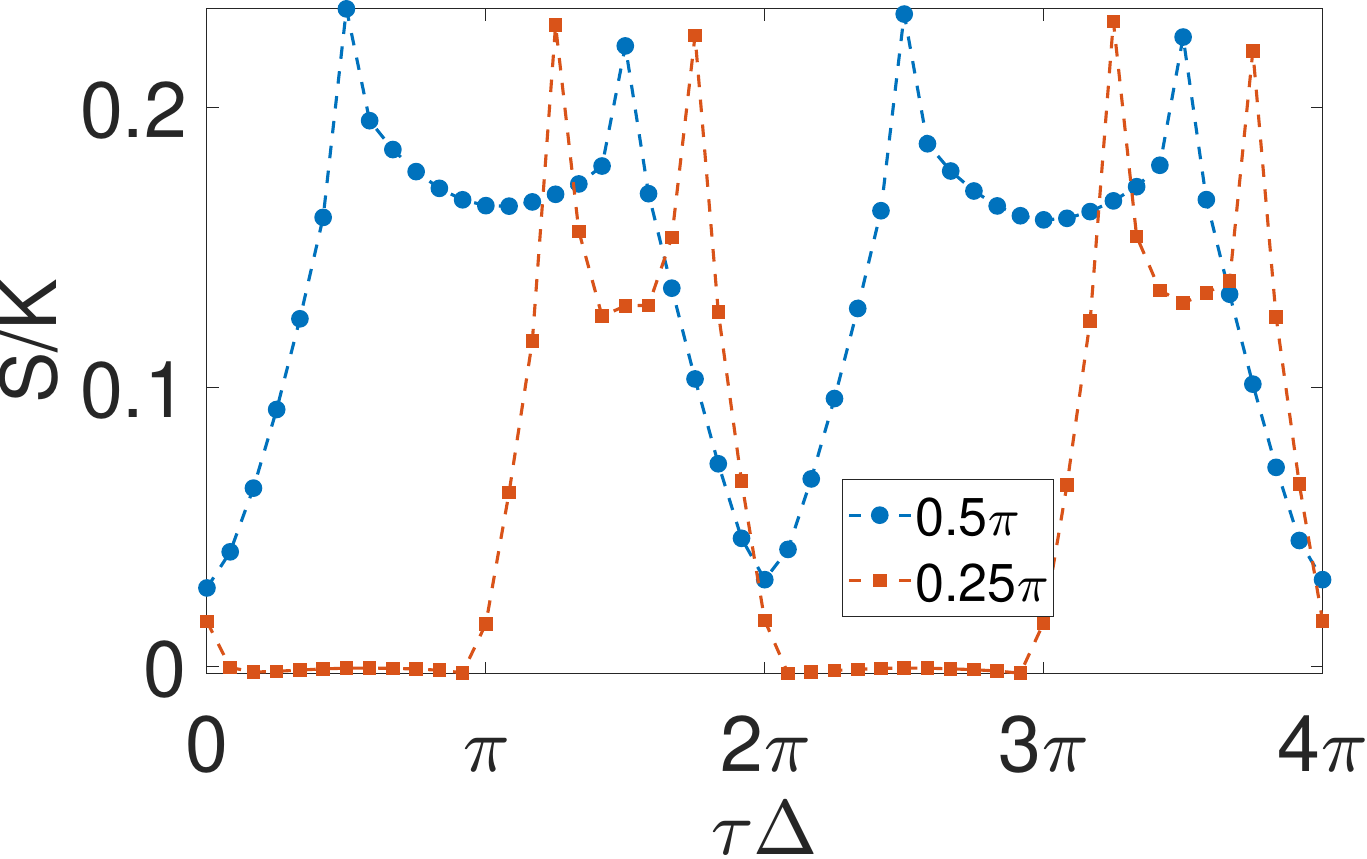}}
    \caption{$W=3$ and $L=1000$.
    (a) The TDEE in log-law and area-law. The log-law prefactor is $0.321, 0.171$ for $\left(\tau \Delta, \tau\right)=(0, \pi / 2), (0, \pi/4)$, respectively. 
    (b) The volume-law term prefactor $S/K$ for the TDEE, with $\frac{\tau}{2\tau_0^+}=1,\frac{W}{2},W,2W$ (legend), $\tau\Delta\in[0,4\pi]$.
    (c) The volume-law term prefactor $S/K$ for the TDEE, with $\tau=\frac{7}{12}\pi$, $\tau\Delta\in[0,4\pi]$.
    (d) The volume-law term prefactor $S/K$ for the TDEE, with $\tau\Delta\in[0,4\pi]$, $\tau=\frac{\pi}{2}, \frac{\pi}{4}$.}
    \label{fig:gapped_nonPH_largeW}
\end{figure}

\begin{figure*}
    \centering
    \subfigure[]{\includegraphics[width=0.98\linewidth]{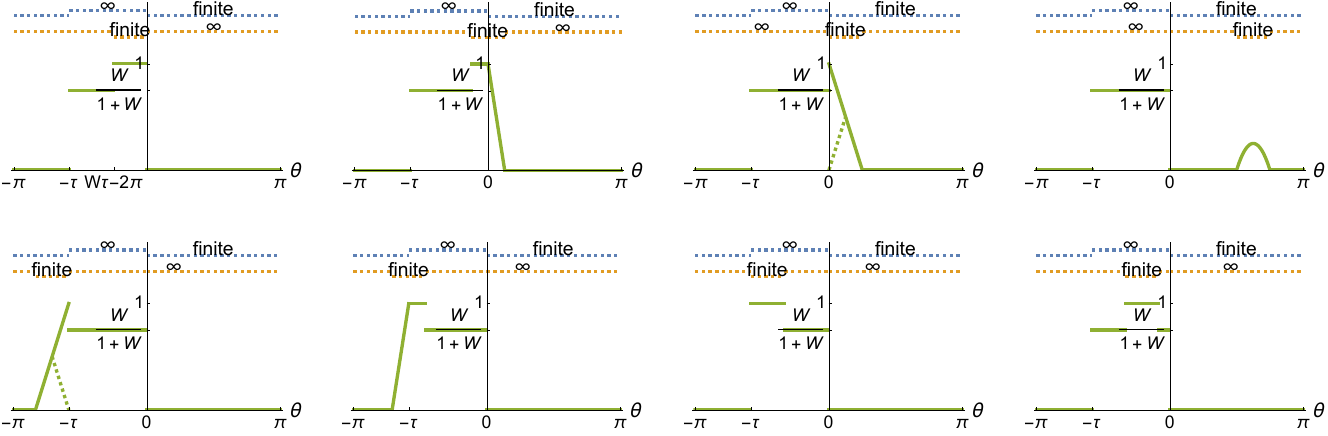}}
    \subfigure[]{\includegraphics[width=0.98\linewidth]{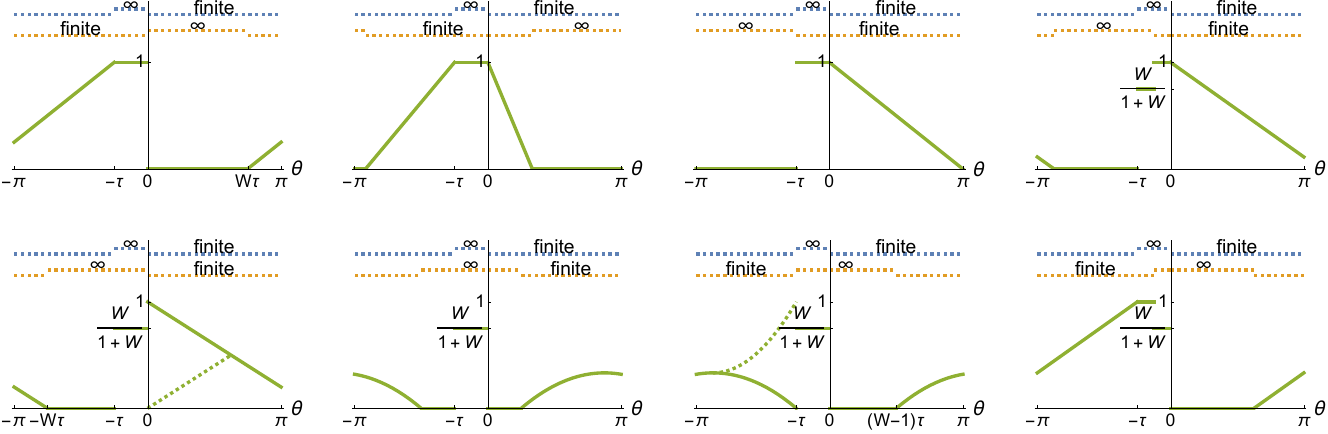}}
    \subfigure[]{\includegraphics[width=0.29\linewidth]{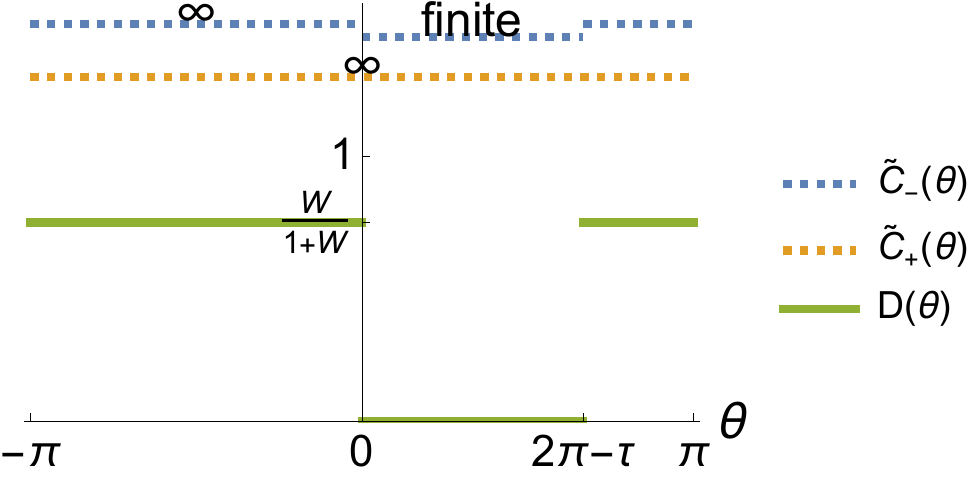}}\quad
    \subfigure[]{\includegraphics[width=0.3\linewidth]{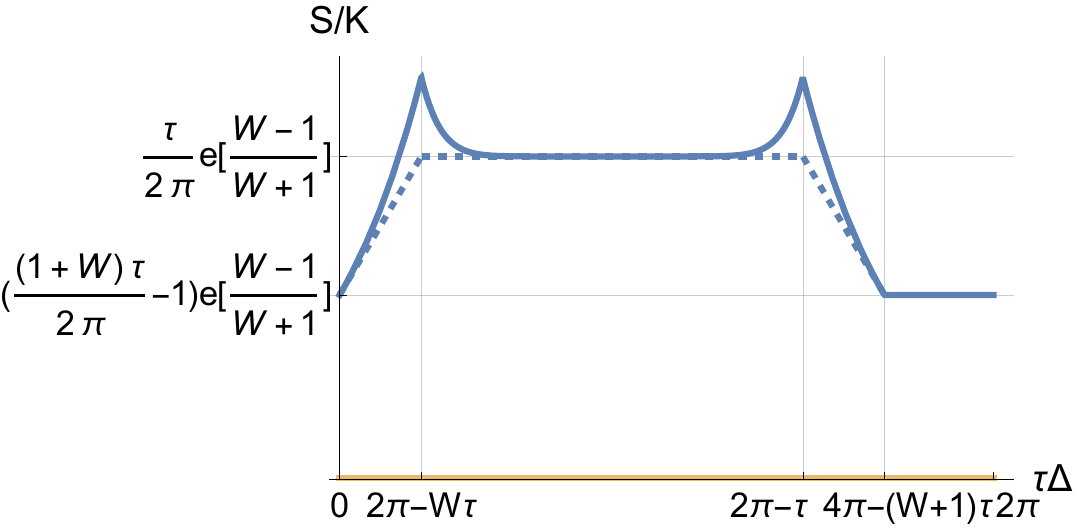}}
    \subfigure[]{\includegraphics[width=0.35\linewidth]{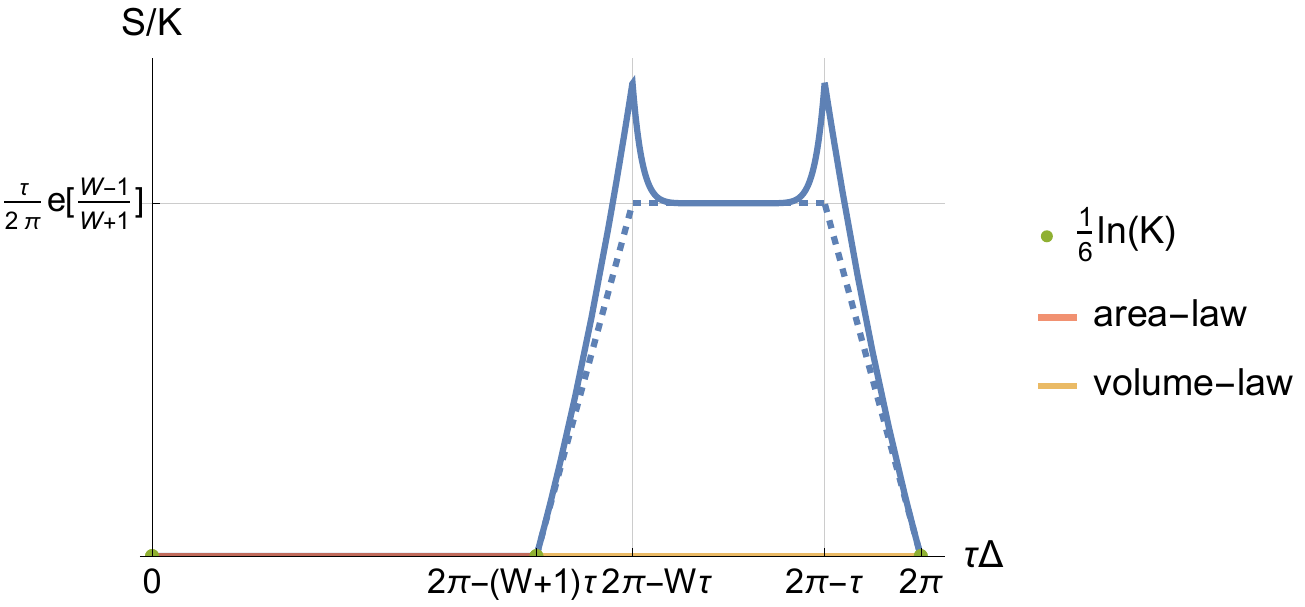}}
    \caption{The Fourier transformation of the correlation matrix $\tilde{C}_\pm(\theta)$ and corresponding $D(\theta)$ for different $\tau$ and $\tau\Delta$ without particle-hole symmetry. (a) $\tau_0<\tau<2{\rm min}(\tau_0^-,\tau_0^+)$, and $\tau\Delta=0,2\pi-W\tau,2\pi-\tau,4\pi-(W+1)\tau$ for the first, third, fifth and seventh figures. (b) $\tau<\tau_0$, and $\tau\Delta=0,2\pi-(W+1)\tau,2\pi-W\tau,2\pi-\tau$ for the first, third, fifth and seventh figures. (c) $2{\rm min}(\tau_0^-,\tau_0^+)<\tau<2{\rm max}(\tau_0^-,\tau_0^+)$ and $\tau\Delta=0$. The prefactors of volume-law term and scaling regime diagrams for (d) $\tau_0<\tau<2{\rm min}(\tau_0^-,\tau_0^+)$ and (e) $\tau<\tau_0$, where the analytical results shown with dashed lines are lower bounds.}
    %(a) (b) $\tau=\frac{7\pi}{12}$ (c) $\tau=\frac{\pi}{4}$
    \label{fig:nonPH_symmetric_largeW}
\end{figure*}

\subsection{$W>1$}
\label{sec:SSH large W}

For $W>1$, we consider $\delta_1=\frac{1}{3}, \delta_2=\frac{1}{6}$ as an example, with $W=3$.
We have $\tau_0^-=\pi$, $\tau_0^+=\frac{\pi}{3}$, and $\tau_0=\frac{\pi}{2}$.

Similar to $W<1$ above, we first consider $\tau\geq2\tau_0^+$.
% , and plot the TDEE for $\frac{\tau}{2\tau_0^+}=1$, and $\tau\Delta=0,\frac{\pi}{2},\pi,\frac{3\pi}{2}$ in Fig.~\ref{fig:gapped_nonPH_largeW} (a).
In Fig.~\ref{fig:gapped_nonPH_largeW} (b), we plot the prefactor of the volume-law term for $\frac{\tau}{2\tau_0^+}=1,\frac{W}{2},W,2W$ and $\tau\Delta\in[0,4\pi]$.
All TDEE exhibit the volume-law entanglement, and the prefactors of volume-law terms are the same for different $\tau\Delta$.
Similar to $W<1$, the coefficient increases as a function of $\frac{\tau}{2\tau_0^+}$, and saturates after $\frac{\tau}{2\tau_0^+}=W$.

For $\tau_0<\tau<2\tau_0^+$, in Fig.~\ref{fig:gapped_nonPH_largeW} (c), we plot the prefactor of the volume-law term for $\tau\Delta\in[0,4\pi]$ with $\tau=\frac{7\pi}{12}$, which is similar to $W<1$, exhibits approximate $2\pi$-periodicity, two peaks at $\tau\Delta=\frac{\pi}{4},\frac{17\pi}{12}$, and a plateau for $\tau\Delta\in[\frac{5\pi}{3},2\pi]$.

For $\tau\leq\tau_0$, we plot the prefactor of the volume-law term for $\tau\Delta\in[0,4\pi]$ with $\tau=\frac{\pi}{2},\frac{\pi}{4}$ in Fig.~\ref{fig:gapped_nonPH_largeW} (d), which exhibits approximate $2\pi$-periodicity.
Approximately, for $\tau=\tau_0=\frac{\pi}{2}$, two peaks are at $\tau\Delta=\frac{\pi}{2},\frac{3\pi}{2}$, and for $\tau=\frac{\pi}{4}<\tau_0$, we have two peaks at $\tau\Delta=\frac{5\pi}{4},\frac{7\pi}{4}$, with an area-law regime for $\tau\Delta\in(0,\pi)$.
Moreover, we also plot the TDEE with $\left(\tau \Delta, \tau\right)=(0, \pi / 2), (0, \pi/4), (\pi/2, \pi/4)$.
% It shows the log-law with coefficient $0.321, 0.171$ for the first two cases, and an area-law for the last case.
The first two cases exhibit logarithmic scaling, with fitted coefficients $0.321$ and $0.171$, respectively, while the last case shows area-law behavior.

From the numerical results above, we find that the TDEE for $W=\frac{1}{3}$ and $W=3$ is almost identical. 
Intuitively, this suggests a particle-hole transformation between $W>1$ and $W<1$, which results in the same TDEE for certain special parameters.

In the following, we present a similar analytical discussion for the TDEE with $W>1$ in the main text. 
A brief summary is given below.

(i) For $\tau>2\tau_0^-$, $D(\theta)=(1+\varrho(E_k^+)/\varrho(E_k^-))^{-1}$ is constant for any $\theta$.  
Thus, the time-direction entanglement entropy is volume-law with the same prefactor for any $\Delta$.

(ii) For $2\tau_0^+<\tau<2\tau_0^-$, as shown in Fig.~\ref{fig:nonPH_symmetric_largeW} (c), similar to the case of $W=\frac{1}{3}$, we have $D(\theta)\equiv(1+\varrho(E_k^+)/\varrho(E_k^-))^{-1}$ with the same length $\tau$ of $\theta$ for any $\Delta$.  
Therefore, the time-direction entanglement entropy is still volume-law for any $\Delta$, with a prefactor that is smaller than that in the regime $\tau>2\tau_0^-$.
It is consistent with numerical results in Fig.~\ref{fig:gapped_nonPH_largeW} (b).

(iii) For $\tau_0<\tau<2\tau_0^+<2\tau_0^-$, there is always a range of $\theta$ such that $D(\theta)\equiv (1+\varrho(E_k^+)/\varrho(E_k^-))^{-1}$, as illustrated in Fig.~\ref{fig:nonPH_symmetric_largeW} (a).  
The length of this range behaves as follows: it first increases for $\tau\Delta\in[0,2\pi-W\tau]$, remains constant for $\tau\Delta\in[2\pi-W\tau,2\pi-\tau]$, decreases for $\tau\Delta\in[2\pi-\tau,4\pi-(W+1)\tau]$, and then remains constant for $\tau\Delta\in[4\pi-(W+1)\tau,2\pi]$.  
The corresponding prefactor of the volume-law term is shown in Fig.~\ref{fig:nonPH_symmetric_largeW} (d), which is consistent with Fig.~\ref{fig:gapped_nonPH_largeW} (c).

(iv) For $\tau<\tau_0$, as shown in Fig.~\ref{fig:nonPH_symmetric_largeW} (b), there is a single discontinuity at $\theta=0$ when $\Delta=0$, which gives a logarithmic-law entanglement with $S=\frac{1}{6}\ln{K}$.  
When $0<\tau\Delta<2\pi-(W+1)\tau$, there is no discontinuity, resulting in an area-law regime.  
For $\tau\Delta=2\pi-(W+1)\tau$, there is one discontinuity at $\theta=-\tau$, yielding $S=\frac{1}{6}\ln{K}$.  
For $2\pi-(W+1)\tau<\tau\Delta<2\pi-W\tau$, $D(\theta)\equiv (1+\varrho(E_k^+)/\varrho(E_k^-))^{-1}$ for a finite range of $\theta$, and its length increases linearly from $0$ to $\tau$, leading to a volume-law regime with a growing prefactor.  
For $2\pi-W\tau<\tau\Delta<2\pi-\tau$, $D(\theta)\equiv (1+\varrho(E_k^+)/\varrho(E_k^-))^{-1}$ with a fixed length of $\theta$, meaning the prefactor of the volume-law term remains constant for different $\Delta$.  
For $2\pi-\tau<\tau\Delta<2\pi$, the corresponding length of $\theta$ decreases to zero, and the prefactor of the volume-law term decreases accordingly.  
Finally, for $\tau\Delta=2\pi$, the system returns to the logarithmic-law entanglement with $S=\frac{1}{6}\ln{K}$.  
The scaling regime diagram and the prefactor of the volume-law term are shown in Fig.~\ref{fig:nonPH_symmetric_largeW} (e).  
The analysis above is consistent with the numerical results in Fig.~\ref{fig:gapped_nonPH_largeW} (d).

Comparing the cases of $W>1$ and $W<1$, we find a duality between them, which simply exchanges $\tau$ and $W\tau$. 
For example, with $W_1W_2=1$, given $\tau_1$ and $W_1\tau_1$, we substitute them with $W_2\tau_2$ and $\tau_2$. 
Then, as $\Delta$ is tuned continuously, $\tau\Delta_i$ exhibit the same $2\pi$ periodicity and the same behavior for the prefactor of the volume-law terms.

\subsection{General solution for two-band models at half-filling}
\label{sec:General solution of TDEE for two-band models at half-filling}

Building on the analytic results for $W<1$ presented above and the discussion of the $W>1$ regime, we now summarize the general behavior of the TDEE for two-band models at half filling. 
As a reminder, the volume-law coefficient obtained analytically corresponds to a lower bound.

With the bandwidths of the conduction and valence bands $W_\pm$, the characteristic and critical times read $\tau_0^\pm$ and $\tau_0$.
Then $\tau_{\rm max}=\max{(\tau_0^+,\tau_0^-)}$ and $\tau_{\rm min}=\min{(\tau_0^+,\tau_0^-)}$.
There are four regimes in the $(\tau,\tau\Delta)$ plane, as shown in Fig.~\ref{fig:general_solution_TDEE}(a), where the color indicates the (lower bound of the) volume-law coefficient.
\begin{itemize}
    \item For $\tau>2\tau_{\rm max}$, the TDEE exhibits a volume law with coefficient $a_0=e(\frac{\tau_{\rm max}-\tau_{\rm min}}{\tau_{\rm max}+\tau_{\rm min}})$ for any $\tau$ and $\tau\Delta$.
    \item For $2\tau_{\rm min}<\tau<2\tau_{\rm max}$, the TDEE obeys a volume law with coefficient $a_1=\frac{\tau}{2\tau_{\rm max}}e(\frac{\tau_{\rm max}-\tau_{\rm min}}{\tau_{\rm max}+\tau_{\rm min}})$ for any $\tau\Delta$, as shown by the red line in Fig.~\ref{fig:general_solution_TDEE} (b).
    \item For $\tau_0<\tau<2\tau_{\rm min}$, the TDEE remains in the volume-law regime, but the coefficient varies with $\tau\Delta$ with $2\pi$ periodicity, as indicated by the green line in Fig.~\ref{fig:general_solution_TDEE} (b).
    The purple and brown dots mark $(\beta_1,a_1)$ and $(\beta_2,a_1)$, with $\beta_1=2\pi\!\left(1-\frac{\tau}{2\tau_{\rm min}}\right)$ and $\beta_2=2\pi\!\left(1-\frac{\tau}{2\tau_{\rm max}}\right)$.
    The red and yellow dots denote $(0,a_2)$ and $(\beta_1+\beta_2,a_2)$, with $a_2=\!\left(\frac{\tau}{\tau_0}-1\right)e(\frac{\tau_{\rm max}-\tau_{\rm min}}{\tau_{\rm max}+\tau_{\rm min}})$.
    \item Specifically, for $\tau=\tau_0$, the TDEE is still in the volume-law regime for any $0<\tau\Delta<2\pi$, with the coefficient shown by the yellow line in Fig.~\ref{fig:general_solution_TDEE} (b); the purple and brown dots retain the coordinates given above.
    However, at $\tau\Delta=2n\pi$ with an integer $n$, the TDEE becomes logarithmic with the coefficient $\frac{1}{3}$.
    \item For $\tau<\tau_0$, the TDEE shows a volume law for $\beta_1+\beta_2-2\pi<\tau\Delta<2\pi$, with the coefficient indicated by the blue line in Fig.~\ref{fig:general_solution_TDEE} (b); the blue dot is $(\beta_1+\beta_2-2\pi,0)$.
    The purple and brown dots have the same coordinates as above.
    For $0<\tau\Delta<\beta_1+\beta_2-2\pi$, the TDEE follows an area law.
    At the boundary between two scaling regimes, where $\tau\Delta=0$ and $\tau\Delta=\beta_1+\beta_2-2\pi$, the TDEE is logarithmic with coefficient $\frac{1}{6}$.
\end{itemize}

In the scaling regime diagram of Fig.~\ref{fig:general_solution_TDEE}(a), the volume-law regime appears in yellow and is characterized by a finite volume-law coefficient.
The area-law regime is shown in dark blue, corresponding to a vanishing volume-law coefficient.
The boundary between these regimes, indicated by the blue line, exhibits a log-law with coefficient $\frac{1}{6}$. 
At the red dots, the TDEE is also logarithmic, with coefficient $\frac{1}{3}$.

% With a fixed band gap $\Delta$, the TDEE as a function of discretization of time $\tau$ is a straight line through the origin, with the slope $\Delta^{-1}$.
With a fixed band gap $\Delta$, varying the time discretization $\tau$ traces a straight line through the origin in the $(\tau\Delta,\tau)$ parameter plane, with slope $\Delta^{-1}$.
Therefore, for the continuous time limit $\tau\rightarrow0$, the TDEE is totally determined by the region near the origin, which gives the same qualitative scaling as spatial entanglement entropy, where the TDEE is log-law with coefficient $\frac{1}{6}$ for the gapless system, and area-law for the gapped system.

\section{Reduced Hilbert space for band insulators}
\label{sec:reduced_hilbert_band_insulator}

% \bl{Put this subsection into Appendix.}

We now extend the reduced-Hilbert-space analysis of Sec.~\ref{sec:reduced_hilbert_gapless} to a gapped band insulator.
As in the gapless case, the asymptotic spectrum of the anti-commutation matrix $B$ determines the single-particle dimension $d_A=\operatorname{rank}B$ of the physical one-particle subspace generated by the time interval.

Consider a general dispersion $\epsilon(k)$ whose spectrum consists of disconnected energy bands. 
We denote the total width of the allowed energy regions by $\widetilde W=\sum_a |\mathcal E_a|$,
% \begin{equation}
% W=\sum_a |\mathcal E_a|,
% \end{equation}
where $\mathcal E_a$ denotes the energy interval occupied by the $a$-th band, and $|\mathcal E_a|$ denotes its width. For a two-band insulator with a single gap $\Delta$, this reduces to $\widetilde W=E_{\rm max}-E_{\rm min}-\Delta=E_0$.
Following the same Toeplitz analysis as in Sec.~\ref{sec:reduced_hilbert_gapless}, the symbol of the anti-commutation matrix is
\begin{equation}
\label{eq:B_theta_free_fermion_general}
B(\theta)=\int_0^{2\pi}\frac{{\rm d}k}{2\pi}
\sum_{n=-\infty}^{+\infty}
e^{-{\rm i}n(\tau\epsilon(k)+\theta)}
=\sum_{l,\alpha}
\frac{1}{|\tau\epsilon'(k_{l,\alpha})|},
\end{equation}
where $k_{l,\alpha}$ denotes all solutions of
$\tau\epsilon(k_{l,\alpha})+\theta=2\pi l$.
Although the value of $B(\theta)$ inside its support depends on the detailed dispersion, its support is determined solely by the image of the allowed energy bands under the mapping $\theta=-\tau\epsilon(k)$ modulo $2\pi$.

We first consider the regime in which the energy bands do not overlap after being mapped onto the temporal Brillouin zone. In this regime, each energy interval of width $|\mathcal E_a|$ maps to an interval of length $\tau|\mathcal E_a|$ in $\theta$. Therefore, the total support of $B(\theta)$ has length $|\operatorname{supp}B|=\tau \widetilde W$.
% \begin{equation}
% |\operatorname{supp}a|=\tau W.
% \end{equation}
The remaining fraction of the temporal Brillouin zone has $B(\theta)=0$, giving an asymptotic zero-eigenvalue fraction $1-\tau \widetilde W/(2\pi)$.
Consequently, neglecting the asymptotically vanishing eigenvalues of $B$, we estimate the effective single-particle dimension as
\begin{equation}
d_A\simeq K\frac{\tau \widetilde W}{2\pi}
=\frac{T\widetilde W}{2\pi}
=\frac{T}{\tau_0},
\qquad
\tau_0=\frac{2\pi}{\widetilde W},
\end{equation}
where $T=K\tau$ is the total duration of the time interval.

For larger $\tau$, different images of the energy bands can overlap after being folded into the temporal Brillouin zone. Such overlap can only reduce the total support of $B(\theta)$ relative to $\tau \widetilde W$. Therefore,
\begin{equation}
% d_A\lesssim K\min\left(1,\frac{\tau W}{2\pi}\right),
% \qquad
\frac{d_A}{T}\lesssim
\min\left(\frac{1}{\tau},\frac{\widetilde W}{2\pi}\right)=\min\left(\frac{1}{\tau},\frac{1}{\tau_0}\right).
\end{equation}
Thus, independent of the detailed dispersion, the number of independent physical fermionic modes generated by the time interval grows at most linearly with its duration, with an effective growth rate 
\begin{equation}
    v_{\rm LR}^{\rm eff}=1/\tau_0=\widetilde W/(2\pi).
\end{equation}
This generalizes the effective Lieb--Robinson-type bound obtained above for gapless free fermions.

Thus, the effective Lieb--Robinson-type bound constrains the growth of the reduced one-particle Hilbert-space dimension with the duration of the time interval, but does not by itself establish a spatial light cone. 
% To further relate this dimension to its spatial support, we establish the following claim.
Similar to the gapless case discussed in the main text, the reduced Hilbert space of a time interval can be further constrained by spatial symmetries. In particular, for a system with reflection symmetry about the site defining the time interval, the time-evolved fermion operators involve only mirror-symmetric combinations of spatial fermion modes. Consequently, the single-particle dimension $d_A$ is at most approximately one-half of the number of spatial orbitals within the corresponding symmetric region, up to boundary corrections. This provides an intuitive interpretation of the reduced Hilbert space in terms of the Lieb-Robinson causal diamond.

\section{Numerical calculation of $D(\theta)$ for SSH model with particle-hole symmetry}
\label{Numerical calculation of D}

\begin{figure}
    \centering
    \subfigure[]{\includegraphics[width=0.48\linewidth]{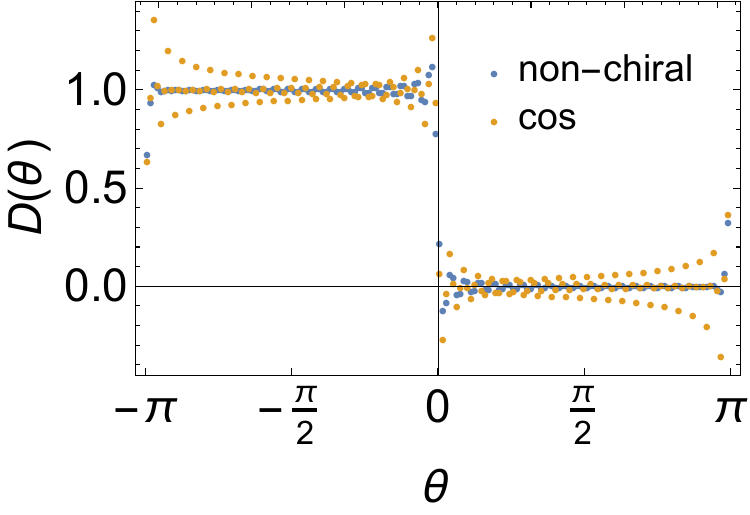}}
    \subfigure[]{\includegraphics[width=0.48\linewidth]{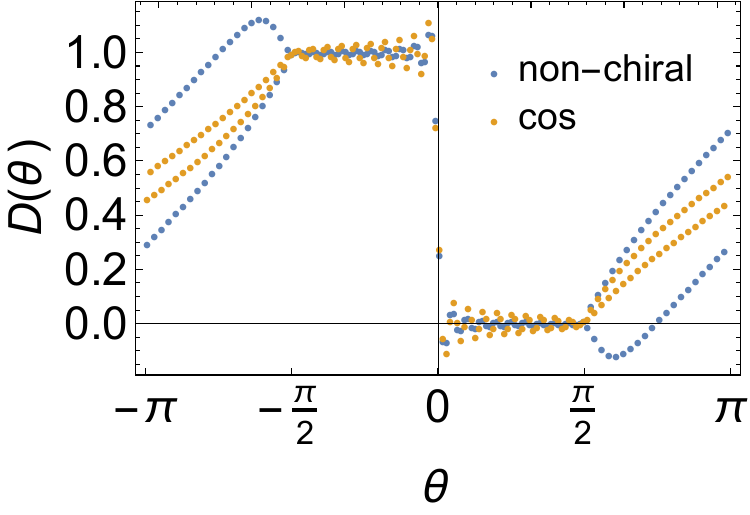}}
    \subfigure[]{\includegraphics[width=0.48\linewidth]{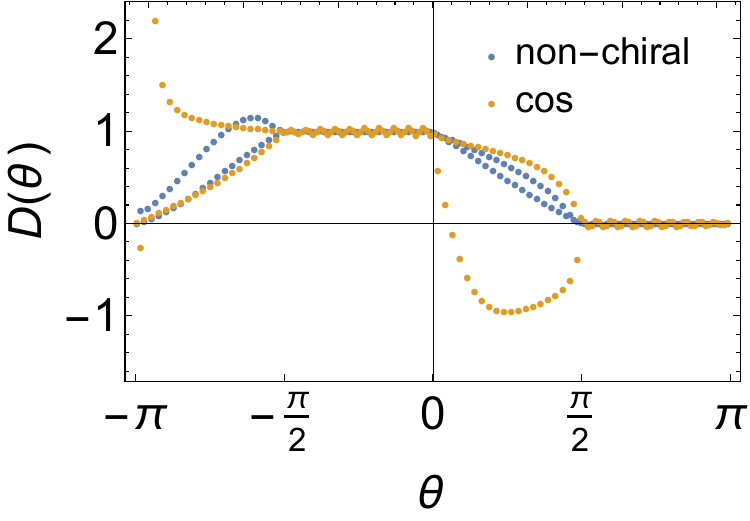}}
    \subfigure[]{\includegraphics[width=0.48\linewidth]{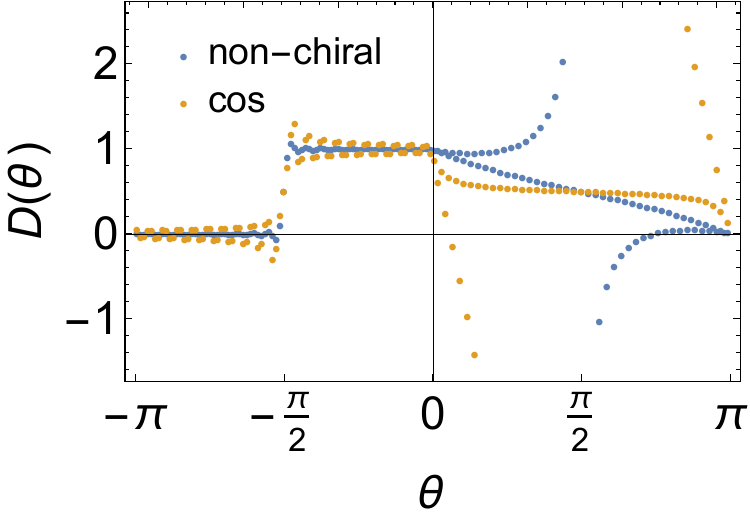}}
    \subfigure[]{\includegraphics[width=0.48\linewidth]{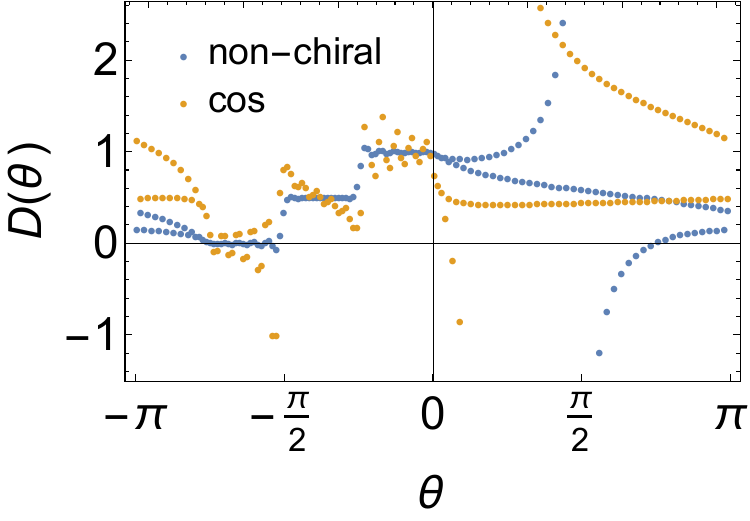}}
    \subfigure[]{\includegraphics[width=0.48\linewidth]{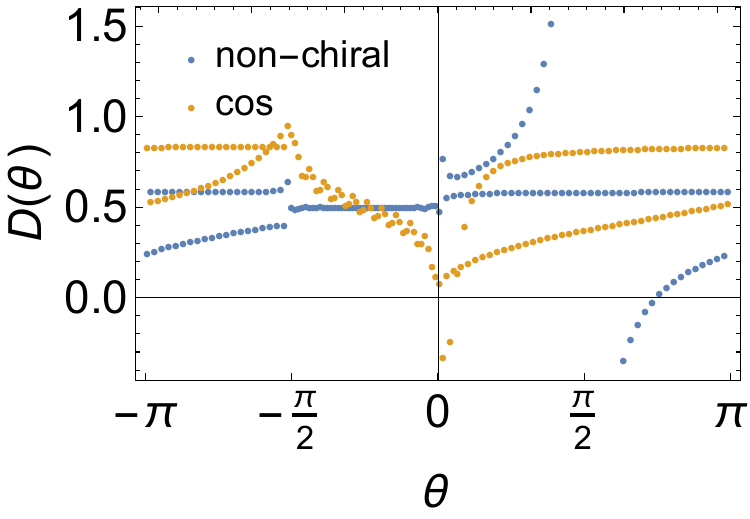}}
    \caption{Numerical results of $D(\theta)$ for the non-chiral and cosine dispersions with (a) $\tau=\tau_0,\ \tau\Delta=0$, and (b)-(f) $\tau=\frac{\tau_0}{2},\ \tau\Delta=0,\frac{\pi}{2},\pi,\frac{5\pi}{4},\frac{3\pi}{2}$.
    We take $L=83$ in (a) and $L=81$ in (b)-(f).}
    \label{fig:numerical_D}
\end{figure}

In this section, we present numerical estimates of $D(\theta)$.
In Fig.~\ref{fig:numerical_D} (a), we plot $D(\theta)$ for $\tau=\tau_0$ and $\tau\Delta=0$ with cosine and linear dispersions.
In Fig.~\ref{fig:numerical_D} (b)-(f), we plot $D(\theta)$ for the cosine and linear dispersions with $\tau=\frac{\tau_0}{2}$ and $\tau\Delta=0,\frac{\pi}{2},\pi,\frac{5\pi}{4},\frac{3\pi}{2}$.
For the linear dispersion results, they are consistent with our discussion in the main text Sec.~\ref{sec:SSH model with the particle-hole symmetry}.
For the cosine dispersion, the modification of $D(\theta)$ arises from the different density of states.

\section{Calculation of time direction entanglement entropy with different $D(\theta)$}
\label{sec:Calculation of time direction entanglement entropy with different D(theta)}

For completeness, we generalize the von Neumann entropy used in the
main text to the R\'enyi-$q$ entanglement entropy,
\begin{equation}
    S_q(A)=\frac{1}{1-q}\ln \operatorname{Tr}\rho_A^q.
\end{equation}
In terms of the eigenvalues $(1+\nu_l)/2$ of the correlation matrix,
this can be written as $S_q(A)=\sum_{l=1}^{d_A} e_q(\nu_l)$, where
\begin{equation}
    e_q(x)=\frac{1}{1-q}\ln\left[\left(\frac{1+x}{2}\right)^q+\left(\frac{1-x}{2}\right)^q\right].
\end{equation}
The von Neumann entropy used throughout the main text is recovered in
the limit $q\to1$, with $e(x)=\lim_{q\to1}e_q(x)$.

% Using the R\'enyi kernel $e_q(\lambda)$ defined above, we can use the Fisher–Hartwig conjecture to calculate the time-direction entanglement entropy~\cite{Jin_2004}. 
% The discontinuities contribute to the log-term, while the volume-law term can also be computed using this conjecture, which is also known as the Szegő–Widom theorem~\cite{Harold_1974,Ares_2022}.

Using the Rényi kernel $e_q(\lambda)$ defined above, we can calculate the time-direction entanglement entropy using the asymptotic theory of Toeplitz determinants~\cite{Jin_2004}, where the Szegő--Widom theorem determines the leading volume-law term, while the Fisher--Hartwig conjecture captures the logarithmic corrections associated with discontinuities~\cite{Harold_1974,Ares_2022}.

If $D(\theta)=(1+C_+(\theta)/C_-(\theta))^{-1}=(1+\varrho(E_k^+)/\varrho(E_k^-))^{-1}$ for some $\theta$, we expect the system to be in the volume-law regime. 
In contrast, the log-law regime appears as the leading term only when $D(\theta)=0,1$ are connected by continuous functions, with a discontinuity between $D(\theta)=1$ and $D(\theta)=0$. 
This discontinuity is analogous to the Fourier transformation of the correlation matrix along the spatial direction. 
Since the prefactor of the log-term depends only on the behavior of the discontinuity in $D(\theta)$, we can directly conclude, from known results for real-space entanglement in gapless phases, that each discontinuity between $D(\theta)=1$ and $D(\theta)=0$ contributes $\frac{1}{6}\ln{K}$
\footnote{In general, there can also be non-smooth points $\theta_0$ where $D(\theta)$ is continuous but not differentiable, which can have nontrivial contribution to the determinant of the Toeplitz matrix.
We do not consider its contribution, which is consistent with the numerical results.}.
If there is no discontinuity and $D(\theta)$ is a continuous function everywhere, we expect the system to be in the area-law regime.

Based on the assumptions above, we can compute the prefactor of the volume-law term using the Fisher–Hartwig conjecture. 
For the case of linear dispersion, our results provide a lower bound for the prefactor, as discussed in detail later.

\begin{figure}
    \centering
    \subfigure[]{\includegraphics[width=0.55\linewidth]{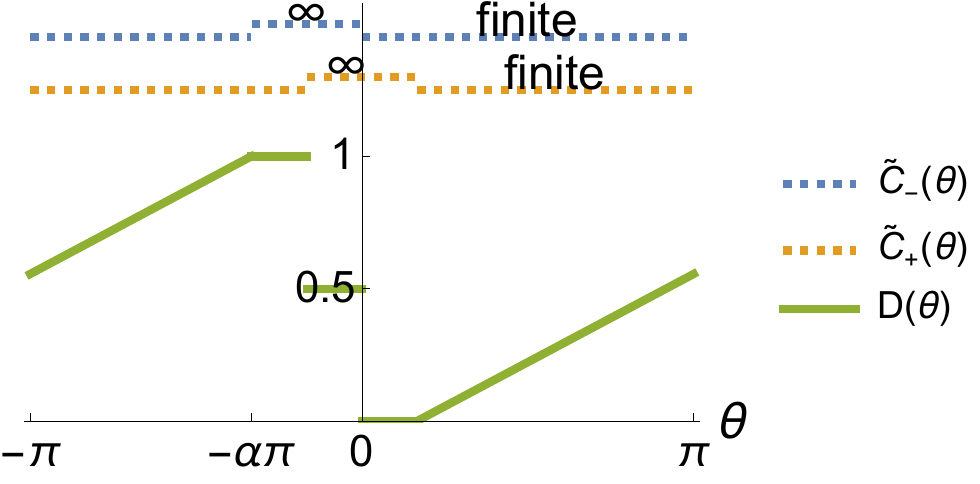}}
    \subfigure[]{\includegraphics[width=0.42\linewidth]{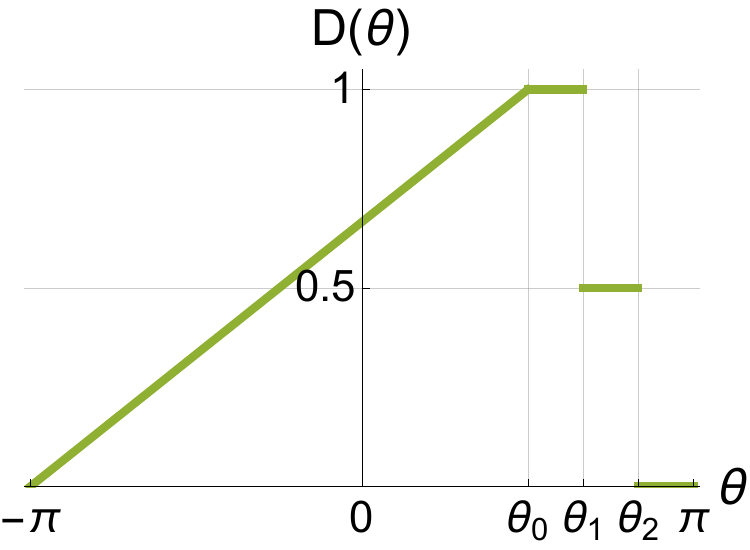}}
    \caption{An illustrative function $D(\theta)$ with a plateau of height $a$ and a continuous ramp, both of which can formally contribute to the volume-law entropy within the Toeplitz-symbol approximation.}
    \label{fig:volume_prefactor}
\end{figure}

Here, we take Fig.~\ref{fig:volume_prefactor} (a) as an example. 
Because the entanglement entropy is invariant under translation of $D(\theta)$, effectively, it is equivalent to Fig.~\ref{fig:volume_prefactor} (b).
We assume $D(\theta)=\frac{\theta+\pi}{\theta_0+\pi}$ for $\theta\in[-\pi,\theta_0]$, and $D(\theta)=1,\,a,\,0$ for $\theta\in[\theta_0,\theta_1]$, $\theta\in[\theta_1,\theta_2]$, and $\theta\in[\theta_2,\pi]$, respectively. 
Using the Fisher–Hartwig conjecture, we have
% \begin{widetext}
\begin{equation}
\label{eq:Dtheta_continuous}
\begin{split}
    &\lambda+1-2D(\theta)\\
    =&(\lambda+1-2\Tilde{D}(\theta))\sqrt{\frac{\lambda+1}{\lambda-1}}
    e^{\frac{\theta}{2\pi}\ln{\left(\frac{\lambda+1}{\lambda-1}\right)}
    -\frac{\theta_1}{2\pi}\ln{\left(\frac{\lambda+1-2a}{\lambda-1}\right)}}\\
    &\qquad\times e^{-\frac{\theta_2}{2\pi}\ln{\left(\frac{\lambda+1}{\lambda+1-2a}\right)}}\prod_{i=1,2}e^{{\rm i}b_i(\theta-\theta_i-\pi{\rm sgn}(\theta-\theta_i))}\\
    \equiv&f(\theta)\prod_{i=1,2}e^{{\rm i}b_i(\theta-\theta_i-\pi{\rm sgn}(\theta-\theta_i))},
\end{split}
\end{equation}
% \end{widetext}
where $b_1=\frac{-1}{2\pi{\rm i}}\ln{\frac{\lambda+1-2a}{\lambda-1}}$, 
$b_2=\frac{-1}{2\pi{\rm i}}\ln{\frac{\lambda+1}{\lambda+1-2a}}$, 
and $\Tilde{D}(\theta)=D(\theta)$ for $\theta\in[-\pi,\theta_0]$ while $\Tilde{D}(\theta)=1$ elsewhere. 
Here we set $a_i=0$. 

In general, $f(\theta)$ should be a smooth function of $\theta$, but in this case it is not smooth at $\theta=-\pi$ and $\theta_0$. 
Similar to the log-law regime, we assume these points are unimportant and take the corresponding $a_i=0$. 
The volume-law term is then given by
\begin{equation}
\label{eq:volume_law_prefactor}
\begin{split}
    S_q(d_A)=&\frac{1}{2\pi{\rm i}}\oint {\rm d}\lambda\, e_q(\lambda) 
    \frac{{\rm d}}{{\rm d}\lambda}\ln{\left[\exp{\left(\frac{1}{2\pi}\int_{-\pi}^\pi {\rm d}\theta \ln{f(\theta)}\right)}\right]^{d_A}}\\
    =&\frac{d_A}{2\pi{\rm i}}\oint {\rm d}\lambda\, e_q(\lambda) 
    \frac{{\rm d}}{{\rm d}\lambda}\frac{1}{2\pi}\int_{-\pi}^\pi {\rm d}\theta \ln{f(\theta)},
\end{split}
\end{equation}
where $e_q(\lambda)=\frac{1}{1-q}\ln{[(\frac{1+\lambda}{2})^q+(\frac{1-\lambda}{2})^q]}$, and the contour of the integral encloses $\lambda\in[-1,1]$.

There are four parts in $\ln{f(\theta)}$, given by
\begin{equation}
\label{eq:log_ftheta}
\begin{split}
    &\ln{f(\theta)}=\ln{(\lambda+1-2\Tilde{D}(\theta))}
    +\left(\frac{1}{2}+\frac{\theta-\theta_2}{2\pi}\right)\ln{(\lambda+1)}\\
    &\ +\left(-\frac{1}{2}-\frac{\theta-\theta_1}{2\pi}\right)\ln{(\lambda-1)}
    +\frac{\theta_2-\theta_1}{2\pi}\ln{(\lambda+1-2a)}.
\end{split}
\end{equation}
We can directly verify that the second and third terms do not contribute to the volume-law term because $e_q(\lambda=\pm1)=0$. 
This also explains why the entanglement entropy along the spatial direction in a gapless system follows a log-law. 
Therefore, we mainly focus on the first and fourth terms. 
For the fourth term, the contribution can be computed directly as $S_q^{(4)}(d_A)=\frac{\theta_2-\theta_1}{2\pi}e_q(2a-1)\cdot d_A$. 

Hence, we can calculate the corresponding volume-law prefactor for different parameters, which is shown as the dashed line in the previous section. 
As an example, for the simplest case with $\tau=\tau_0$ and particle-hole symmetry, the contribution of $D(\theta)\equiv\frac{1}{2}$ yields 
$S_1^{(4)}(d_A)=\frac{\delta\theta}{2\pi}\ln{2}\cdot d_A$, 
where $\delta\theta$ increases from $0$ at $\tau\Delta=0$ to a maximum value of $\pi$ at $\tau\Delta=\pi$, and then decreases back to $0$ at $\tau\Delta=2\pi$.

More generally, for $D(\theta)\equiv(1+\varrho(E_k^+)/\varrho(E_k^-))^{-1}\equiv a$, we have $S_q^{(4)}(d_A)=\frac{\delta\theta}{2\pi}\frac{1}{1-q}\ln{\left[a^q+(1-a)^q\right]}\cdot d_A$, which implies $S_1^{(4)}(d_A)=\frac{\delta\theta}{2\pi}(-a\ln{a}-(1-a)\ln{(1-a)})\cdot d_A$.
For $a=\frac{1}{2}$, we recover the result for a system with particle-hole symmetry. 
For systems that break particle-hole symmetry but have a linear dispersion, we instead have a general $0<a<1$. 
Moreover, for different values of $\tau\Delta$, the corresponding $\delta\theta$ will vary accordingly.

\section{Volume-law coefficient for a restricted spectral support}
\label{app:restricted_support}

In this section, we further examine the volume-law coefficient in Eq.~\eqref{eq:general_EE_with_FH_conj} of the main text, focusing on the case where the Fourier transform of the anticommutation matrix has a restricted spectral support, i.e., $B^*(\theta)$ vanishes over a finite interval.
For a regular Toeplitz symbol with full spectral support, the leading Szeg\H{o} contribution to Eq.~\eqref{eq:general_EE_with_FH_conj} can be written as
\begin{equation}
\begin{split}
    \alpha_V
    =&\oint_\Gamma\frac{{\rm d}\lambda}{2\pi i}\,e(\lambda)\frac{{\rm d}}{{\rm d}\lambda}\int_{-\pi}^{\pi}\frac{{\rm d}\theta}{2\pi}\ln\left[\lambda+1-2D(\theta)\right]\\
    =&\int_{-\pi}^{\pi}\frac{{\rm d}\theta}{2\pi}\,e\left(2D(\theta)-1\right).
\end{split}
\label{eq:volume_full_support}
\end{equation}
Here $D(\theta)=C(\theta)/B^*(\theta)$ denotes the ratio of the limiting Fourier symbols.
Jump discontinuities in $D(\theta)$ do not modify the leading Szeg\H{o} average, although they can contribute to the logarithmic corrections through the Fisher--Hartwig exponents.
This expression includes the contributions from both constant plateaus and continuously varying regions of $D(\theta)$.
However, when $B^*(\theta)$ vanishes on an interval of nonzero measure, the ordinary full-circle Szeg\H{o} asymptotics no longer apply directly.

% To understand this distinction, we consider the finite-size characteristic determinant $g_K(\lambda)$ defined in the main text.
% Using the correlation matrix $D_K=(B_K^*)^{-1/2}C_K(B_K^*)^{-1/2}$, we obtain the exact identity
% \begin{equation}
% \begin{split}
% g_K(\lambda)
% &=\det\left[(\lambda+1)I-2D_K\right]\\
% &=\frac{\det\left[(\lambda+1)B_K^*-2C_K\right]}{\det B_K^*}\\
% &=\frac{\det T_K\left[(\lambda+1)B^*(\theta)-2C(\theta)\right]}{\det T_K\left[B^*(\theta)\right]},
% \end{split}
% \label{eq:restricted_toeplitz_ratio}
% \end{equation}
% where $T_K$ denotes the Toeplitz matrix generated by the corresponding limiting Fourier symbol.
% This representation does not require the finite-size normalized correlation matrix $D_K$ itself to be Toeplitz.

When the total spectral support $E=\operatorname{supp}B^*(\theta)$ is a proper arc of the unit circle, the leading asymptotics of the determinant ratio are instead governed by the relative Szeg\H{o} theory for arc-supported Toeplitz matrices~\cite{Charlier2015Asymptotics,krasovsky2006asymptotics,Duits2019Relative}.
Under the corresponding asymptotic assumptions, the leading term takes the form
\begin{equation}
    \lim_{K\rightarrow\infty}\frac{1}{K}\ln g_K(\lambda)
    =\int_E\ln\left[\lambda+1-2D(\theta)\right]{\rm d}\mu_E(\theta),
\label{eq:restricted_relative_szego}
\end{equation}
where $K$ is the dimension of the matrix entering the characteristic determinant $g_K(\lambda)$, and ${\rm d}\mu_E(\theta)$ is the equilibrium measure associated with the total spectral support $E$.
Substituting this expression into the contour-integral representation of the entropy, we obtain
\begin{equation}
    \alpha_V=\int_E e\left(2D(\theta)-1\right){\rm d}\mu_E(\theta).
\label{eq:volume_arc_general}
\end{equation}
For full-circle support, the equilibrium measure reduces to ${\rm d}\mu_E(\theta)={\rm d}\theta/(2\pi)$, reproducing Eq.~\eqref{eq:volume_full_support}.
For a proper arc, the nonuniform equilibrium measure modifies the volume-law coefficient, even though the ratio $D(\theta)$ remains unchanged.

As an explicit example, we consider a particle-hole-symmetric band insulator at half filling with equal bandwidths $W=1$ and band energies in the intervals $[-1,0]$ and $[\Delta,\Delta+1]$.
We take $\tau=\tau_0=\pi$, with $0\le \tau\Delta\le\pi$.
In the thermodynamic limit $L\rightarrow\infty$, the folded contributions from the occupied and empty bands, up to a common normalization factor, are
\begin{equation}
\begin{split}
    C_-(\theta)&=\Theta(-\theta)\Theta(\pi+\theta),\\
    C_+(\theta)&=\Theta(\theta-\tau\Delta)\Theta(\pi-\theta)+\Theta(-\theta-\pi+\tau\Delta)\Theta(\pi+\theta).
\end{split}
\label{eq:restricted_folded_bands}
\end{equation}
Since $B^*(\theta)\propto C_-(\theta)+C_+(\theta)$ and $C(\theta)\propto C_-(\theta)$, the normalized ratio is well defined on the total spectral support $E=[-\pi,0]\cup[\tau\Delta,\pi]$ and takes the form
\begin{equation}
D(\theta)=
\begin{cases}
1, & \theta\in(-\pi+\tau\Delta,0),\\
\frac12, & \theta\in(-\pi,-\pi+\tau\Delta),\\
0, & \theta\in(\tau\Delta,\pi).
\end{cases}
\label{eq:linear_folded_D}
\end{equation}
This structure is illustrated in the second panel of Fig.~\ref{fig:PH_symmetric}(a) in the main text.
Outside $E$, both $C(\theta)$ and $B^*(\theta)$ vanish, so their ratio is undefined.
Although $E$ appears as two intervals in the representation $\theta\in[-\pi,\pi]$, it forms a single connected arc on the unit circle because $\theta=-\pi$ and $\theta=\pi$ are identified.

For a single arc $E=[-\beta,\beta]$ centered at the origin, the equilibrium measure is given by~\cite{Charlier2015Asymptotics}
\begin{equation}
    {\rm d}\mu_E(\phi)=\frac{\cos(\phi/2)}
    {2\pi\sqrt{\sin^2(\beta/2)-\sin^2(\phi/2)}}\,{\rm d}\phi.
\label{eq:single_arc_measure}
\end{equation}
For the present model, we introduce the shifted coordinate $\phi=\theta+\pi-\tau\Delta/2$ modulo $2\pi$.
Under this transformation, the total support becomes $E=[-\beta,\beta]$, with $\beta=\pi-\tau\Delta/2$, while the overlapping region where $D(\theta)=1/2$ becomes $J=[-\tau\Delta/2,\tau\Delta/2]$.
Since $e(\pm1)=0$ and $e(0)=\ln2$, only the overlapping region contributes to the leading volume-law coefficient.
Using Eq.~\eqref{eq:single_arc_measure}, we therefore find
\begin{equation}
    \alpha_V(\tau\Delta)
    =\ln2\int_{-\tau\Delta/2}^{\tau\Delta/2}{\rm d}\mu_E(\phi)
    =\frac{2\ln2}{\pi}\arcsin\left[\tan\left(\frac{\tau\Delta}{4}\right)\right].
\label{eq:linear_full_volume_coefficient}
\end{equation}
For $\pi\le \tau\Delta\le2\pi$, the result follows from the symmetry $\tau\Delta\rightarrow2\pi-\tau\Delta$, and the coefficient is periodic in $\tau\Delta$ with period $2\pi$.
We compare this analytical result with the numerical results in Fig.~\ref{fig:gapped_PH}(e) of the main text.

For comparison, retaining only the constant plateau contribution with the uniform angular measure gives
\begin{equation}
    \alpha_V^{\mathrm{plateau}}(\tau\Delta)=\frac{\tau\Delta}{2\pi}\ln2,\qquad 0\le \tau\Delta\le\pi.
\label{eq:linear_plateau_coefficient}
\end{equation}
In particular, at $\tau\Delta=\pi$, the entropy is exactly $S(A)=K\ln2$, corresponding to $\alpha_V=\ln2$.
In contrast, Eq.~\eqref{eq:linear_plateau_coefficient} gives only $\alpha_V=\ln2/2$ at this point.
This difference demonstrates the importance of the nonuniform equilibrium measure when the total spectral support is restricted to a proper arc.

The same approach can be extended to arbitrary $\tau$, $\Delta$, and unequal bandwidths by determining the corresponding folded spectral functions and their total support.
When the total support covers the full circle, Eq.~\eqref{eq:volume_full_support} applies, whereas a single proper arc is described by Eq.~\eqref{eq:volume_arc_general}.
For disconnected spectral supports, the corresponding equilibrium measure must be determined separately.
The present analysis concerns only the leading extensive term, while the logarithmic corrections require the subleading determinant asymptotics.

% \nocite{*}

\bibliography{reference.bib}% Produces the bibliography via BibTeX.

\end{document}